\documentclass[12pt]{article}
\usepackage{amsfonts}
\usepackage[dvips]{graphicx}
\usepackage[dvips,colorlinks=true]{hyperref}
\usepackage{theorem}
\usepackage{amsmath}
\usepackage{amssymb}
\usepackage{latexsym}
\usepackage{float}

\theorembodyfont{\rmfamily}

\theoremstyle{break}

\makeatletter
\renewcommand{\section}{\@startsection{section}{1}{0em}{.5\baselineskip}{1pt}{\rmfamily\bfseries\normalsize}}
\renewcommand{\subsection}{\@startsection{subsection}{2}{0em}{1\baselineskip}{1pt}{\rmfamily\itshape\normalsize}}
\newcommand{\ltmp}{\l@section}
\makeatother

\raggedright
\input{tcilatex}
\begin{document}

\date{May 23, 2017}
\title{On the multiply robust estimation of the mean of the g-functional\\
}
\author{Andrea Rotnitzky$^{1}$, James M. Robins$^{2}$ and Lucia Babino$^{3}$ 
\\
$^{1}${\small Di Tella University and CONICET, Buenos Aires, Argentina.}\\
$^{2}${\small Departments of Epidemiology and Biostatistics, Harvard T. H.
Chan School of Public Health,}\\
{\small \ Boston, MA, USA}\\
$^{3}${\small Instituto de Calculo, Facultad de Ciencias Exactas y
Naturales, }\\
{\small Universidad de Buenos Aires, Buenos Aires, Argentina.}}
\maketitle
\tableofcontents

\begin{abstract}
We study multiply robust (MR) estimators of the longitudinal g-computation
formula of Robins (1986). In the first part of this paper we review and
extend the recently proposed parametric multiply robust estimators of
Tchetgen-Tchetgen  (2009) and Molina, Rotnitzky, Sued and Robins  (2017). In
the second part of the paper we derive multiply and doubly robust estimators
that use non-parametric machine-learning (ML) estimators of nuisance
functions in lieu of parametric models. We use sample splitting to avoid the
need for Donsker conditions, thereby allowing an analyst to select the ML
algorithms of their choosing. We contrast the asymptotic behavior of our
non-parametric doubly robust and multiply robust estimators. In particular,
we derive formulas for their asymptotic bias. Examining these formulas we
conclude that although, under certain data generating laws, the rate at
which the bias of the MR estimator converges to zero can exceed that of the
DR estimator, nonetheless, under most laws, the bias of the DR and MR
estimators converge to zero at the same rate.
\end{abstract}

\vspace*{1.5in}\newpage

\thispagestyle{empty}

\newpage

\setlength{\parskip}{0ex} \setlength{\parskip}{2ex}

\newpage

\vspace*{0.5in}

\section{Introduction}

The goal of this paper is to construct multiply robust estimators of
functionals defined by the longitudinal g-computation formula (aka
g-formula) of Robins (1986) from $n$ i.i.d. observations $Z_{i},i=1,...n$.
These g-functionals are widely studied in the causal inference literature, a
leading special case being the functional corresponding to the expectation
of a counterfactual response from longitudinal data under the assumption of
no unmeasured confounding (Robins 1986, 1987, 1997). In this setting $Z_{i}$
denotes all of subject $i^{\prime }s$ observed treatment, covariate, and
outcome history over the study period. This is the third in a series of
papers on multiply robust estimation that reports on results obtained by the
authors and coworkers between 2012-2014. The first paper in the series
(Molina, Rotnitzky, Sued and Robins, 2017) is to appear in Biometrika, the
second (Babino, Rotnitzky and Robins, 2017) will hopefully appear in
Biometrics. A fourth should be available later this year.

G-functionals depend on the observed data law only through the conditional
distributions of outcome and covariate given past treatments, covariates and
outcomes. Estimation of g-functionals requires the estimation of infinite
dimensional nuisance parameters, such as a conditional mean or a conditional
density. As such, g-functionals cannot be estimated consistently under the
non-parametric model that includes all possible data laws. Therefore either
finite-dimensional parametric models or non-parametric models with
smoothness or sparsity constraints are often considered. Both parametric and
nonparametric approaches have been used by different authors to estimate the
nuisance functions.\emph{\ }

We now provide a broad overview of the paper. Consider first the parametric
case. Robins (2000, 2002) and Bang and Robins (2005) introduced a class of
iterated conditional expectation estimators for g-functionals which they
showed were doubly robust (DR), i.e. the estimators are asymptotically
linear, and thus consistent and asymptotically normal (CAN), for the
g-functional of interest provided either (i) parametric models for the
conditional laws of treatment given past treatments, outcomes and covariates
are correct for each treatment time $k,k=1,\ldots ,K,$ \ or (ii) parametric
models for certain iterated conditional expectations (ICEs)\ depending on
the conditional distributions of outcome and covariates given past
treatments, and covariates\emph{\ }are correct at each time $k$. The
estimators in Robins (2000) and and Bang and Robins (2005) were defined as
the solutions to estimating equations, while those in Robins (2002) were
plug-in estimators. Because of their similarity, we refer to all of these
estimators as B\&R estimators, although in this paper we consider only the
plug-in form. Van der Laan and Gruber (2012) and Petersen, Schwab, Gruber,
Blaser, Schomaker and van der Laan (2014) proposed DR\ estimators nearly
identical to the plug-in version of the B\&R estimator which they refer to
as Targetted Maximum Likelihood Estimators (TMLEs). See Section 4.6.2 for
additional discussion.

It has recently been shown by Molina et al. (2017) that the B\&R estimator
and thus the TMLE\ estimators confer more protection to model
misspecification than had been thought. Specifically, Molina et al. proved
that these estimators are asymptotically linear so long as the first $k\ $%
conditional treatment models are correct and the last $K-k$\ iterated
expectation models are correct for any $k\in \left\{ 1,...,K\right\} .$
Thus, the aforementioned DR\ estimators are all actually $K+1$ robust. In
section 4.6 we review this result and several $K+1$\ robust estimators.

In fact, it\ is possible to construct so-called multiply robust (MR)
estimators of the g-functional. MR estimators can be exponentially more
robust to model misspecification than the $K+1$ robust\ estimators. In
particular these estimators are asymptotically linear and thus CAN\ for the
g-functional of interest if a parametric model for either the time $k\in
\left\{ 1,...,K\right\} $ treatment probability or the time $k$ ICE is
correct, thus providing $2^{K}$ opportunities to be CAN. Tchetgen-Tchetgen
(2009) constructed an iterated augmented inverse probability weighted
(IAIPW-MR) estimator of a specific g-functional, namely the mean of an
outcome at the end of a longitudinal study with monotone missing at random
data. Molina, Rotnitzky, Sued and Robins (2017) derived a general theory for
the existence of multiply robust estimators of functionals in non or
semiparametric models whose likelihood factorizes as the product of
variation independent factors with the functional depending on just one of
these factors. Construction of an IAIPW-MR estimator of an arbitrary
g-functional follows by application of the Molina et al. general theory.
Both Tchetgen-Tchetgen (2009) and Molina et al. (2017) estimate the nuisance
high dimensional functionals parametrically. Inverse augmented IPW multiple
robust estimators that fit parametric models for the conditional treatment
probabilities and ICEs are reviewed in section 4.7.2.

Iterated augmented IPW multiply robust estimators of g-functionals are not
entirely satisfactorily because they do not respect bounds on the state
space of the g-functional of interest. To address this problem, in sections
4.7.3 and 4.7.4 we derive two classes of multiply robust iterated
conditional expectation plug-in estimators that fit parametric models for
the conditional treatment laws and the ICEs, as did Tchetgen-Tchetgen (2009)
and Molina et al. (2017).

Unfortunately, it is quite likely that all our parametric models for the $2K$
nuisance functions are misspecified. If so, parametric DR and MR estimators
will be inconsistent, motivating the need for non-parametric estimators.
Because the time-specific conditional treatment probabilities and the ICEs
are infinite dimensional conditional densities and conditional expectations
one would expect that nonparametric approaches to their estimation would be
more robust than parametric approaches discussed above. In order to discuss
the non-parametric approach we need to be more specific, as we now do\emph{. 
}

For all parametric and non-parametric doubly and multiply robust estimators
\ $\widehat{\theta }$\ of a g-functional $\theta ,$\ the difference $%
\widehat{\theta }-\theta $\ can be decomposed as the sample average of a
mean zero, finite variance, random variable $IF_{\theta }\left( P\right)
=if_{\theta }\left( Z,P\right) $\ plus a remainder $R.$\ If $R$\ is $%
o_{p}\left( n^{-1/2}\right) $\ then $\widehat{\theta }$\ is asymptotically
linear (and hence CAN) and $IF_{\theta }\left( P\right) $\ is its influence
function. Now, the remainder $R$\ can be further decomposed as the sum $%
R_{1}+R_{2}$\ of two terms; $R_{1}$\ is an empirical process term discussed
below and $R_{2}$\ is a drift term. It is well known (van der Vaart, 1998,
ch. 25) that in a nonparametric model defined solely by smoothness or
sparsity assumptions, all asymptotically linear estimators have the same
influence function $IF_{\theta }\left( P\right) .$\ It follows that $%
\widehat{\theta }$\ will be an asymptotically linear estimator of $\theta $\
if and only if both the empirical process term $R_{1}$\ and the drift $R_{2}$%
\ are $o_{p}\left( n^{-1/2}\right) .$

The exact form of the drift of an MR estimator is given in equation $\left( %
\ref{driftMR}\right) $ in Section 5.2 but is too complex to give here. For
the purpose of our introduction it suffices to point that the drift has the
following general form\emph{\ }{\large \ } 
\begin{equation}
E_{P}\left[ \sum_{k=1}^{K}\left\{ \frac{1}{h_{k}\left( Past_{k}\right) }-%
\frac{1}{\widehat{h}_{k}\left( Past_{k}\right) }\right\} \left\{ \eta
_{k}\left( Past_{k}\right) -\widehat{\eta }_{k}\left( Past_{k}\right)
\right\} \widehat{c}_{k}\left( Past_{k}\right) \right]  \label{expect}
\end{equation}%
where (i) $h_{k}\left( past_{k}\right) $ and $\eta _{k}\left(
past_{k}\right) $ are the true conditional treatment probability and ICE
function at $k$, (ii) $Past_{k}$ is the random vector denoting the data
recorded up to $k$ and $past_{k}$ denotes a possible realization of $%
Past_{k} $, (iii) $\widehat{h}_{k}\left( past_{k}\right) $ and $\widehat{%
\eta }_{k}\left( past_{k}\right) $ are estimates of $h_{k}\left(
past_{k}\right) $ and $\eta _{k}\left( past_{k}\right) ,$ (iv) $\widehat{c}%
_{k}\left( past_{k}\right) $ is an \ order 1 random variable and (v) the
functions $\widehat{\eta }_{k}\left( \cdot \right) ,$ $\widehat{h}_{k}\left(
\cdot \right) $ and $\widehat{c}_{k}\left( \cdot \right) \ $are considered
as fixed functions when taking the expectation. Note if, for every $k,$
either $\widehat{h}_{k}\left( \cdot \right) =h_{k}\left( \cdot \right) $ or $%
\widehat{\eta }_{k}\left( \cdot \right) =\eta _{k}\left( \cdot \right) ,$
then the drift is zero. This fact underlies the asymptotics of our
parametric MR estimators.

Now, non-parametric estimators of $\eta _{k}\left( past\right) $ and $%
h_{k}\left( past\right) $ cannot be $n^{1/2}$-consistent even under
smoothness or sparsity constraints. Thus, our only hope for the drift to be $%
o_{p}\left( n^{-1/2}\right) $ is that the functions $\eta _{k}\left(
past\right) $ and $h_{k}\left( past\right) $ are sufficiently smooth or
sparse in some basis so that, at each $k,$ $\widehat{h}_{k}\left(
past\right) $ converges to $h_{k}\left( past\right) $ and, $\widehat{\eta }%
_{k}\left( past\right) $ converges to $\eta _{k}\left( past\right) ,$ at
rates $n^{-\alpha _{k}}$ and $n^{-\beta _{k}}$ in such a way that the
expectation $\left( \ref{expect}\right) $ is $o_{p}\left( n^{-1/2}\right) $;
a sufficient condition is that $\alpha _{k}+\beta _{k}>1/2$ for each $k.$

The particular estimators $\widehat{h}_{k}\left( past\right) $ and $\widehat{%
\eta }_{k}\left( past\right) $ that obtain the best rates of convergence
will vary depending on the unknown smoothness or sparsity of $h_{k}\left(
past\right) $ and $\eta _{k}\left( past\right) .$ Thus one would wish to
implement various non-parametric estimators and use the data to adaptively
choose those with the fastest rates of convergence. As is well known, this
can be accomplished by using, say $J,$ machine learning algorithms to
construct candidate estimators and then using cross validation to choose the
best candidate for each of the $2K$ nuisance functions $h_{k}\left(
past\right) $ and $\eta _{k}\left( past\right) $. (Dudoit and Van der Laan,
2003). Even for $J$ \ polynomial in the sample size, this approach will
generally achieve, for each nuisance function, a rate of convergence equal
to the rate of the machine learning algorithm with the fastest convergence
rate among the $J$ algorithms. However, note that although this approach may
give the best rate of convergence of the drift to 0 given the $J$\ machine
learning algorithms, this rate could be slower than $o_{p}\left(
n^{-1/2}\right) $\ because one or more of the functions $h_{k}\left(
past\right) $\ and/or $\eta _{k}\left( past\right) $\ might not be smooth or
sparse enough.

Even when the drift is $o_{p}\left( n^{-1/2}\right) ,$ $\widehat{\theta }$
will an asymptotically linear estimator only if the empirical process term $%
R_{1}$ in the remainder $R$ is also $o_{p}\left( n^{-1/2}\right) .$ To
describe this term we need additional notation. Each of our DR and MR\
estimators $\widehat{\theta }$ are either exactly equal to, or
asypmtotically equivalent to, a sample average $\mathbb{P}_{n}\left\{
m\left( Z,\widehat{h},\widehat{\eta }\right) \right\} $ of a random variable 
$m\left( Z_{i},\widehat{h},\widehat{\eta }\right) $ that depends on subject $%
i^{\prime }s$ data and on the $K-$vectors of nuisance functions $\widehat{h}%
\equiv \left( \widehat{h}_{1},...,\widehat{h}_{K}\right) $ and $\widehat{%
\eta }\equiv \left( \widehat{\eta }_{1},...,\widehat{\eta }_{K}\right) $
obtained through the cross validation procedure described above. The
empirical process term $R$, also called the stochastic equicontinuity term,
of each of our DR and MR estimators is 
\begin{equation*}
\left\{ \mathbb{P}_{n}[m\left( Z,\widehat{h},\widehat{\eta }\right)
]-E_{P}[m\left( Z,\widehat{h},\widehat{\eta }\right) ]\right\} -\left\{ 
\mathbb{P}_{n}[m\left( Z,h,\eta \right) ]-E_{P}[m\left( Z,h,\eta \right)
]\right\}
\end{equation*}%
where the functions $\widehat{h}\ $and $\widehat{\eta }$ are again treated
as non-random when taking the expectation over $Z,$ even though they are
actually random because estimated from the same data $Z_{i},i=1,..,n.$ It is
well known that if $m\left( \cdot ,\widehat{h},\widehat{\eta }\right) $ and $%
m\left( \cdot ,h,\eta \right) $ lie in a Donsker class with probability one
and $\widehat{h}$ and $\widehat{\eta }$ are $L_{2}$-consistent for $h$\ and $%
\eta ,$ then the stochastic equicontinuity term is $o_{p}\left(
n^{-1/2}\right) $ as required.

However, for the outputs $\left( \widehat{h},\widehat{\eta }\right) $\ of an
arbitrary machine learning program, $m\left( \cdot ,\widehat{h},\widehat{%
\eta }\right) $\ cannot be assumed to lie in a Donsker class. In section 5.1
we describe how to overcome this problem by splitting the sample and using a
cross-fit estimator, a name coined in Chernozhukov, (2016). To obtain a
cross-fit estimator we first randomly split the sample into $\mathbf{U,}$
say 5, equal sized subsamples $u=1,..,\mathbf{U}$. For each split $u$ we
construct an estimator $\widehat{\theta }^{u}$of $\theta $. Then our
cross-fit MR\ estimator is 
\begin{equation*}
\widehat{\theta }_{MR}^{cf}=\mathbf{U}^{-1}\sum_{u=1}^{\mathbf{U}}\widehat{%
\theta }^{u}=\mathbf{U}^{-1}\sum_{u=1}^{\mathbf{U}}\mathbb{P}_{n/\mathbf{U}%
}^{u}\left\{ m\left( Z,\widehat{h}^{/u},\widehat{\eta }^{/u}\right) \right\}
\end{equation*}%
with $\widehat{\theta }^{u}=\mathbb{P}_{n/\mathbf{U}}^{u}\left\{ m\left( Z,%
\widehat{h}^{/u},\widehat{\eta }^{/u}\right) \right\} ,$ $\mathbb{P}_{n/%
\mathbf{U}}^{u}$ denotes the average over the $n/\mathbf{U}$ units in split $%
u$ and the $2K$ estimated functions $\widehat{h}^{/u}\equiv \left( \widehat{h%
}_{1}^{/u},...,\widehat{h}_{K}^{/u}\right) ,\widehat{\eta }^{/u}\equiv
\left( \widehat{\eta }_{1}^{/u},...,\widehat{\eta }_{K}^{/u}\right) $ are
obtained by machine learning as in the previous paragraph, but using data
only on the $n\left( \mathbf{U-1}\right) /\mathbf{U}$ units not in the $%
u^{th}$ split. When $\widehat{h}^{/u}$ and $\widehat{\eta }^{/u}$ are $L_{2}$%
-consistent for $h\ $and $\eta $ then 
\begin{equation*}
\mathbf{U}^{-1}\sum_{u=1}^{\mathbf{U}}\left[ \mathbb{P}_{n/\mathbf{U}%
}^{u}\left\{ m\left( Z,\widehat{h}^{/u},\widehat{\eta }^{/u}\right) \right\}
-\mathbb{E}\left\{ m\left( Z,\widehat{h}^{/u},\widehat{\eta }^{/u}\right)
\right\} \right] -\left[ \mathbb{P}_{n}\left\{ m\left( Z,h,\eta \right)
\right\} -E_{P}\left\{ m\left( Z,h,\eta \right) \right\} \right]
\end{equation*}%
is $o_{p}\left( n^{-1/2}\right) .$ This implies that $\left( \widehat{\theta 
}_{MR}^{cf}-\theta \right) =\mathbb{P}_{n}\left\{ IF_{\theta }\left(
P\right) \right\} +R_{2}+o_{p}\left( n^{-1/2}\right) .$ Thus, if the drift
is $o_{p}\left( n^{-1/2}\right) ,$ our machine learning cross-fit estimator $%
\widehat{\theta }_{MR}^{cf}\ $is an asymptotically linear estimator of $%
\theta ,$ where from here on we treat the terms non-parametric and machine
learning as synonyms.

Robins et al. (2008, p. 379) earlier used sample splitting to avoid the
Donsker requirement in constructing efficient asymptotically linear
estimators of functionals in non-parametric models, although their estimator
did not use cross-fitting. Subsequently, Ayygari (2010) in his 2010 Harvard
Phd. Thesis used a machine learning cross-fit estimator to obtain an
asymptotically linear estimator of the parameter $\theta $ in the
semiparametric regression model $E\left[ Y|A,X\right] =\theta A+\tau \left(
X\right) $ thereby avoiding the Donsker requirement. This work was
subsequently published as Robins et al. (2013). Zheng and van der Laan
(2010) proposed a so-called cross-validated TMLE\ that used sample-splitting
to avoid some, but perhaps not all, of the need for Donsker conditions. The
Zheng and van der Laan estimator is quite similar, but not identical, to our
doubly robust estimator of $\theta $ of Section 5, called in that section $%
\widehat{\theta }_{DR,CF,mach,bang}$. Belloni et al. (2010) proposed a
cross-fitting estimator to relax the degree of sparsity required to obtain
an asymptotically linear instrumental variable estimator. The idea that
sample-splitting and cross-fitting could be used to avoid the need for
Donsker conditions long preceded any of the above references - for example,
Van der Vaart (1998, page 391) - although the idea of explicitly combining
cross-fitting with machine learning was not emphasized.

Recall that in the parametric setting MR estimators have $2^{K}$
opportunities to be CAN for $\theta $ compared to $K+1$ opportunities for DR
estimators. In Section 5.2 we consider whether the marked advantage of MR
over DR estimators carries over to the nonparametric setting by comparing
their drifts. We find that although, under certain data generating laws, the
advantage persists, nonetheless, under most laws, the drift of the DR\ and
MR\ estimators converge to zero at the same rates and, thus, the MR
estimators advantage does not persist.

In further detail, we can approximate the drift of a nonparametric cross-fit
DR estimator $\widehat{\theta }_{DR}^{cf}$ given in $\left( \ref{driftDR}%
\right) $ of Section 5.2 by the sum of the drift of the MR estimator $%
\widehat{\theta }_{MR}^{cf}$ given above plus the quantity 
\begin{equation*}
E_{P}\left[ \sum_{1\leq j<k\leq K}\left( \frac{1}{h_{j}\left(
Past_{j}\right) }-\frac{1}{\widehat{h}_{j}\left( Past_{j}\right) }\right)
\left( \eta _{k}\left( Past_{k}\right) -\widehat{\eta }_{k}\left(
Past_{k}\right) \right) \widehat{c}_{j,k}\left( Past_{\max \left\{
k,j\right\} }\right) \right] 
\end{equation*}%
where $\widehat{c}_{j,k}\left( past_{\max \left\{ k,j\right\} }\right) $ is
a is an \ order 1 random variable.

It follows that the drift of $\widehat{\theta }_{DR}^{cf}$ has $K\left(
K-1\right) /2$ terms more than the drift than that of $\widehat{\theta }%
_{MR}^{cf}.$ However, the rates of convergence of $\widehat{\theta }%
_{MR}^{cf}$ and $\widehat{\theta }_{DR}^{cf}$ to $\theta $ are determined by
the dominating term in their drifts, i.e. the term with the slowest rate of
convergence to zero. One would generally expect that term 
\begin{equation*}
E_{P}\left[ \left( \frac{1}{h_{K}\left( Past_{K}\right) }-\frac{1}{\widehat{h%
}_{K}\left( Past_{K}\right) }\right) \left( \eta _{K}\left( Past_{K}\right) -%
\widehat{\eta }_{K}\left( Past_{K}\right) \right) \widehat{c}_{K}\left(
Past_{K}\right) \right]
\end{equation*}%
be the dominating term in both the drift of the MR and the DR estimators
because this term contains two regressions involving the entire history $%
Past_{K},$ which is a superset of the conditioning set in the regressions
involved in all other terms. Thus, one would generally expect that $\widehat{%
\theta }_{DR}^{cf}$ and $\widehat{\theta }_{MR}^{cf}$ have drifts that
converge to zero at identical rates.

However, it could happen that at the particular law $P$ that generated the
data, one of the $K\left( K-1\right) /2$\ terms appearing in the drift of $%
\widehat{\theta }_{DR}^{cf}$ $\ $but not in the drift of $\widehat{\theta }%
_{MR}^{cf}$ converges to $0$ slower than any of the terms in $\widehat{%
\theta }_{MR}^{cf}.$\ In such case, the drift of $\widehat{\theta }%
_{MR}^{cf} $ would have a faster rate of convergence to 0 than the drift of $%
\widehat{\theta }_{DR}^{cf}$. \ In particular, it could happen that $%
\widehat{\theta }_{MR}^{cf}$\ is an asymptotically linear estimator of $%
\theta \ \ $even though $\widehat{\theta }_{DR}^{cf}$\ is not. The frequency
with which the law generating the data has the drift of $\widehat{\theta }%
_{MR}^{cf}$ converging to 0 faster than the drift of $\widehat{\theta }%
_{DR}^{cf}$ may be greater for $K$\ large,\ because the ratio of the number
of terms in the drift of $\widehat{\theta }_{DR}^{cf}\,\ $compared to the
drift of $\widehat{\theta }_{MR}^{cf}$\ increases linearly with $K,$
providing an increasing number of opportunities for the drift of $\widehat{%
\theta }_{DR}^{cf}$ to dominate the drift of $\widehat{\theta }_{MR}^{cf}$.

The paper is organized as follows. In section $\left( \emph{\ref%
{representation}}\right) $ we define the g-functional, review various
representations for it and examples of its application. Section $\left( 
\emph{\ref{parametric}}\right) $\ discusses estimation of the g-functional
based on parametric models for the nuisance functions. Sections 4.1, 4.2 and
4.3 reviews non-doubly robust IPW, parametric MLE and ICE plug-in estimators
respectively. Section 4.4 and 4.5 give preliminary background on
doubly-robust estimation. Section 4.6 considers three different DR plug-in
estimators, and shows that they are, in fact, $K+1$ robust. Section 4.7
discusses 2$^{K}$ MR estimation, specifically, the theoretical background
and three particular estimators, two of which are ICE plug-in estimators.
Section $\ref{ML}$ considers non-parametric DR and MR estimation. We propose
a number of cross-fit machine learning DR and MR estimators and analyze
their asymptotic properties.

\section{Assumptions and the target of inference \label{assumptions}}

Let $Z=\left( Z_{1},\ldots ,Z_{K},L_{K+1}\right) $ where $Z_{k}=\left(
A_{k},L_{k}\right) ,k=1,\ldots ,K,$ and $A_{k}$ and $L_{k}$ are, possibly
multivariate, random vectors taking values in measurable spaces $\left( 
\mathbb{A}_{k},\mathcal{A}_{k}\right) $ and $\left( \mathbb{L}_{k},\mathcal{L%
}_{k}\right) $. Let $\mathbb{Z=\otimes }_{k=1}^{K}\left( \mathbb{A}%
_{k}\times \mathbb{L}_{k}\right) \times \mathbb{L}_{k+1}$ and $\mathcal{Z=}%
\mathbb{\otimes }_{k=1}^{K}\left( \mathcal{A}_{k}\times \mathcal{L}%
_{k}\right) \times \mathcal{L}_{K+1}$ and let $\mathcal{P}$ be the
collection of the densities of all probability measures on $\left( \mathbb{Z}%
,\mathcal{Z}\right) $ mutually absolutely continuous with respect to $\mu =%
\mathbb{\otimes }_{k=1}^{K}\left( \mu _{k}\times \mu _{k}^{\prime }\right)
\times \mu _{k+1}^{\prime },$ where for each $k,$ $\mu _{k}$ are $\mu
_{k}^{\prime }$ are measures on $\left( \mathbb{A}_{k},\mathcal{A}%
_{k}\right) $ and $\left( \mathbb{L}_{k},\mathcal{L}_{k}\right) $
respectively. For each $P\in \mathcal{P}$, we write $p=dP/d\mu ,$ and $%
p\left( z\right) =\dprod\limits_{k=0}^{K}g_{k}\left( l_{k+1}|\overline{l}%
_{k},\overline{a}_{k}\right) \dprod\limits_{k=1}^{K}h_{k}\left( a_{k}|%
\overline{l}_{k},\overline{a}_{k-1}\right) ,$ or for short $p=gh,$ where $%
g_{k}\left( l_{k+1}|\overline{l}_{k},\overline{a}_{k}\right) $ and $%
h_{k}\left( a_{k}|\overline{l}_{k},\overline{a}_{k-1}\right) $ are (versions
of) the conditional densities of $L_{k+1}$ and $A_{k}$ when $Z\sim P.$ Here
and throughout for any vector $w=\left( w_{1},\ldots ,w_{s}\right) $ and any 
$r\leq t\leq s,$ $\overline{w}_{r}^{t}\equiv \left( w_{r},\ldots
,w_{t}\right) ,\overline{w}_{r}\equiv \overline{w}_{1}^{r}$ and \underline{$%
w $}$_{r}\equiv \overline{w}_{r}^{s}.$ Furthermore, $A\perp \!\!\!\perp
B\mid C $ denotes that \thinspace $A$ and $B$ are conditionally independent
given $C, $ $\left[ K\right] $ denotes the set $\left\{ 1,\ldots ,K\right\}
,gh^{\ast }<<gh$ stands for $p^{\ast }=gh^{\ast }$ is absolutely continuous
with respect to $p=gh,$ $E_{gh}\left( \cdot \right) \,\ $stands for
expectation under $p=gh$ and, often we write $\left( \overline{L}_{K+1},%
\overline{A}_{K}\right) $ instead of $Z$ $.$

In this paper we are interested in inference about the parameter 
\begin{equation*}
\theta \left( g\right) \equiv E_{gh^{\ast }}\left\{ \psi \left( \overline{L}%
_{K+1}\right) \right\}
\end{equation*}%
based on $n$ i.i.d. copies of the random vector $Z$ with unknown
distribution $p=gh$ assumed to belong to model $\mathcal{P}$, where $\psi $
is a given real valued measurable function on $\left( \mathbb{Z},\mathcal{Z}%
\right) $ and $h_{k}^{\ast }\left( a_{k}|\overline{l}_{k},\overline{a}%
_{k-1}\right) $ is a given, i.e. known, conditional density for each $k\in %
\left[ K\right] $, such that $p^{\ast }=gh^{\ast }$ is absolutely continuous
with respect to $p=gh.$

By definition, the parameter $\theta \left( g\right) $ depends on the
unknown data generating law $p=gh$ only through $g\equiv \left( g_{0},\ldots
.,g_{K}\right) $. Explicitly, 
\begin{equation}
\theta \left( g\right) =\int \varphi \left( z\right)
\dprod\limits_{k=0}^{K}g_{k}\left( l_{k+1}|\overline{l}_{k},\overline{a}%
_{k}\right) d\mu \left( z\right)  \label{g-form}
\end{equation}%
where $\left( \overline{l}_{0},\overline{a}_{0}\right) \equiv $nill and%
\begin{equation}
\varphi \left( z\right) \equiv \left\{ \dprod\limits_{k=1}^{K}h_{k}^{\ast
}\left( a_{k}|\overline{l}_{k},\overline{a}_{k-1}\right) \right\} \psi
\left( \overline{l}_{K+1}\right) .  \label{phi}
\end{equation}%
is a known, i.e., specified, function of $z.$

The expression on the right hand side of $\left( \ref{g-form}\right) $ is
often referred to as the g-computation formula (Robins, 1986), or, the
g-formula for short. Our motivation for studying the functional $\theta
\left( g\right) $ is because special choices of $h_{k}^{\ast }$ yield $%
\theta \left( g\right) $ equal to parameters which are of interest in causal
inference and in missing data analysis. Here we give some examples.

\subsection{Examples \label{examples}}

\subsubsection{Example 1. \label{example1}}

\textit{Mean of an outcome in a longitudinal study with ignorable drop-out.
\ }Consider a longitudinal study with drop-outs. Define $L_{k}$ to be the
data vector $L_{k}^{\ast }$ that is recorded on a subject randomly selected
from a target population if the subject is still on study at the $k^{th}$
study cycle and to be equal to an arbitrary vector in $\mathbb{L}_{k},$ say $%
\varkappa _{k},\ $otherwise. Assume no subject misses the first cycle. Then $%
L_{1}=L_{1}^{\ast }.$ Let $A_{k}=1$ if the subject is on study at the $%
\left( k+1\right) ^{th}$ study cycle and $A_{k}=0$ otherwise. Thus, $%
L_{k}=A_{k-1}L_{k}^{\ast }+\left( 1-A_{k-1}\right) \varkappa _{k}$. Let $%
p=\dprod\limits_{j=0}^{K}g_{j}\dprod\limits_{j=1}^{K}h_{j}$ be the law of $%
\left( \overline{A}_{K},\overline{L}_{K+1}\right) .$ Under the missing at
random assumption that $L_{K+1}^{\ast }$ $%
\perp\!\!\!\perp A_k\mid%
$ $\left( \overline{A}_{k-1}=\overline{1},\overline{L}_{k-1}\right) $ for
each $k\in \left[ K\right] ,$ and the positivity assumption that for all $%
k\in \left[ K\right] ,$ $\Pr \left\{ h_{k}\left( 1|\overline{A}_{k-1}=1,%
\overline{L}_{k-1}\right) >0\right\} =1,$ the mean of the, potentially
missing, last cycle outcome $L_{K+1}^{\ast }$, i.e., of the outcome that
would be recorded if the study did not suffer from drop-out, equals 
\begin{equation}
E_{g_{0}}\left[ E_{g_{1}}[\ldots E_{g_{K-1}}\left\{ \left. E_{g_{K}}\left(
L_{K+1}|\overline{A}_{K}=\overline{1},\overline{L}_{K}\right) \right\vert 
\overline{A}_{K-1}=\overline{1},\overline{L}_{K-1}\right\} \ldots |\overline{%
A}_{1}=1,\overline{L}_{1}]\right] .  \label{g-mis}
\end{equation}

In this display, as well as in the expression for the positivity assumption
and throughout the rest of the paper, subscripts on $E$ are used to indicate
the sole conditional laws on which the expectation or probability depends
on. For instance, in $\left( \ref{g-mis}\right) $ the subscript $g_{K+1}$ is
a reminder that $E_{g_{K}}\left( L_{K+1}|\overline{A}_{K}=\overline{1},%
\overline{L}_{K}\right) $ depends only on $g_{K}.$

The expression in $\left( \ref{g-mis}\right) $ agrees with $\theta \left(
g\right) $ if we take $h_{k}^{\ast }\left( a_{k}|\overline{l}_{k},\overline{a%
}_{k-1}\right) =a_{k}$ and $\psi \left( \overline{l}_{K+1}\right) =l_{K+1}$
(Robins, 1986, Robins, Rotnitzky and Zhao, 1995). Note that the positivity
assumption is the same as the assumption that $gh^{\ast }<<gh.$ Note also
that because $A_{k}$ is a binary variable, $\theta \left( g\right) $
actually involves only integrals over $l_{1},\ldots ,l_{K+1}$ as, for each $%
k,$ the integral over $a_{k}$ is indeed a sum with a single non-zero
term.\medskip

\subsubsection{Example 2.}

\textit{Outcome mean under a sequence of fixed treatments.} Suppose that in
a longitudinal study $L_{k}$ denotes the vector of variables to measured at
the $k^{th}$ study cycle on a subject randomly selected from a target
population. Assume that immediately after recording $L_{k}$ the subject
decides which of the available treatments in a set $\mathcal{A}_{k}$ he will
take until the next study cycle. Let $A_{k}\in \mathcal{A}_{k}$ denote the
subject's treatment choice. Let $p=\dprod\limits_{j=0}^{K}g_{j}\dprod%
\limits_{j=1}^{K}h_{j}$ be the law of $\left( \overline{A}_{K},\overline{L}%
_{K+1}\right) .$ Also, let $L_{K+1,\overline{a}}$ be the counterfactual
outcome at the end of follow-up if, possibly contrary to fact, the subject
took treatment $\overline{A}=\overline{a^{\ast }}$ for some fixed $\overline{%
a^{\ast }}=\left( a_{1}^{\ast },\ldots ,a_{K}^{\ast }\right) .$ Contrasts of
the mean of $L_{K+1.\overline{a^{\ast }}}$ involving different $\overline{%
a^{\ast }}$ quantify treatment effects. For instance, the average treatment
effect (ATE) comparing the \textit{always on treatment} vs \textit{never on
treatment} regimes is defined as the mean of $L_{K+1.\overline{1}}$ minus
the mean of $EL_{K+1,\overline{0}}.$ Under the consistency assumption that $%
\overline{A}_{k}=\overline{a}_{k}\Rightarrow L_{k+1}=L_{k+1,\overline{a}%
_{k}} $ for all $k\in \left[ K\right] ,$ the no-unmeasured confounding
assumption that for $k\in \left[ K\right] ,$ $L_{K+1.\overline{a^{\ast }}}$ $%
\perp\!\!\!\perp A_k\mid%
$ $\left( \overline{A}_{k-1}=\overline{a^{\ast }}_{k-1},\overline{L}%
_{k-1}\right) $ and the positivity assumption that for $k\in \left[ K\right]
,$ $\Pr \left\{ h_{k}\left( a_{k}^{\ast }|\overline{a_{k-1}^{\ast }},%
\overline{L}_{k-1}\right) >0\right\} =1,$ the mean of $L_{K+1.\overline{%
a^{\ast }}}$ equals (Robins, 1986) 
\begin{equation}
E_{g_{0}}\left[ E_{g_{1}}\left[ \ldots E_{g_{K-1}}\left\{ \left.
E_{g_{K}}\left( L_{K+1}|\overline{A}_{K}=\overline{a_{K}^{\ast }},\overline{L%
}_{K}\right) \right\vert \overline{A}_{K-1}=\overline{a_{K-1}^{\ast }},%
\overline{L}_{K-1}\right\} \ldots |A_{1}=a_{1}^{\ast },L_{1}\right] \right] .
\label{g-treat}
\end{equation}

This expression agrees with $\theta \left( g\right) $ if we take $%
h_{k}^{\ast }\left( a_{k}|\overline{l}_{k},\overline{a}_{k-1}\right)
=I_{\left\{ a_{k}^{\ast }\right\} }\left( a_{k}\right) $ and $\psi \left( 
\overline{l}_{K+1}\right) =l_{K+1}$ where throughout, $I_{D}\left( x\right)
=1$ if $x\in D$ and $I_{D}\left( x\right) =0$ otherwise. Note that in
Example 1 we could arrive at the formula $\left( \ref{g-mis}\right) $ from
the formula $\left( \ref{g-treat}\right) $ if, in that example we regard $%
A_{k}$ as a sequence of time-dependent treatments indexed by $k$ and
consider estimation of the mean of $L_{K+1}$ had, contrary to fact, all
subjects followed the treatment regime specified by $a_{k}$ $=1$ for $k\in %
\left[ K\right] $; that is, the regime in which no subject had dropped-out.
Robins (1986, p. 1491; 1987a, sec. AD.5) provided additional discussion of
the usefulness of regarding missing data indicators as time-dependent
treatments.\medskip

\subsubsection{Example 3.}

\textit{Outcome mean under a non-random dynamic treatment regime. }Assume\
that the recorded data $Z$ are as in the longitudinal study of Example 2.
However, suppose that we are now interested in estimating the mean of $%
L_{K+1}$ if, contrary to fact, the entire study population followed a given 
\textit{non-random dynamic} treatment regime which stipulates that right
after study cycle $k$ and until just prior to study cycle $k+1,$ a patient
with covariate and treatment history $\left( \overline{a}_{k-1},\overline{l}%
_{k}\right) $ receives treatment $A_{k}=d_{k}\left( \overline{a}_{k-1},%
\overline{l}_{k}\right) .$ Similarly to Example 2, the average treatment
effect for comparing the two such regimes, say $d$ and $d^{\prime }$ is
defined as the mean of $L_{K+1,d}$ minus the mean of $L_{K+1,d^{\prime }}$
where for any treatment regime $d=\left\{ d_{1},\ldots ,d_{K}\right\} ,$ $%
L_{K+1,d}$ denotes the counterfactual outcome at the end of the study if,
possibly contrary to fact, the subject had followed treatment regime $d.$
Under the consistency assumption that $\overline{A}_{k}=\overline{D}_{k}$ $%
\Rightarrow L_{k+1}=L_{k+1,d},$ where for any $j\in \left[ K\right]
,D_{j}\equiv d_{j}\left( \overline{A}_{j-1},\overline{L}_{j}\right) ,$ the
no-unmeasured confounding assumption that for $k\in \left[ K\right] ,$ $%
L_{K+1,d}$ $%
\perp\!\!\!\perp A_k\mid%
$ $\left( \overline{A}_{k-1}=\overline{D}_{k-1},\overline{L}_{k-1}\right) ,$
and the positivity assumption that for $k\in \left[ K\right] ,$ $\Pr \left[
\Pr \left( A_{k}=D_{k}|\overline{A}_{k-1}=\overline{D}_{k-1},\overline{L}%
_{k-1}\right) >0\right] =1,$ the mean of $L_{K+1,d}$ is 
\begin{equation*}
E_{g_{0}}\left[ E_{g_{1}}\left[ \ldots E_{g_{K-1}}\left\{ \left.
E_{g_{K}}\left( L_{K+1}|\overline{A}_{K}=\overline{D}_{K},\overline{L}%
_{K}\right) \right\vert \overline{A}_{K-1}=\overline{D}_{K-1},\overline{L}%
_{K-1}\right\} \ldots |\overline{A}_{1}=D_{1},\overline{L}_{1}\right] \right]
.
\end{equation*}

This expression agrees with $\theta \left( g\right) $ if we take $%
h_{k}^{\ast }\left( a_{k}|\overline{l}_{k},\overline{a}_{k-1}\right)
=I_{\left\{ d_{k}\left( \overline{l}_{k},\overline{a}_{k-1}\right) \right\}
}\left( a_{k}\right) $ and $\psi \left( \overline{l}_{K+1}\right) =l_{K+1}.$
Note also that the positivity assumption is the same as the assumption that $%
gh^{\ast }<<gh.$

\medskip

\subsubsection{Example 4.}

\textit{Outcome mean under a random dynamic treatment regime. }Assume\ that
the recorded data $Z$ are as in the longitudinal study of Example 2. Suppose
that we are now interested in estimating the mean of $L_{K+1}$ if, contrary
to fact, the entire population followed a \textit{random }dynamic treatment
regime which stipulates that at study cycle $k$ a patient with covariate and
treatment history $\left( \overline{a}_{k},\overline{l}_{k+1}\right) $ is
randomized to receive treatment $A_{k+1}=a_{k}$ with probability $%
h_{k}^{\ast }\left( a_{k}|\overline{a}_{k-1},\overline{l}_{k}\right) $ where 
$a_{k}$ is in the set $\mathcal{A}_{k}$ of treatments available at time $k.$
Similarly to Example 2, the average treatment effect for comparing the two
regimes, determined by $h^{\ast }$ and $h^{\ast \ast },$ is defined as the
mean of $L_{K+1,h^{\ast }}$ minus the mean of $L_{K+1,h^{\ast \ast }}$ where
for any $h^{\ast }\equiv \left\{ h_{k}^{\ast }:k\in \left[ K\right] \right\}
,$ $L_{K+1,h^{\ast }}$ denotes the counterfactual outcome if, possibly
contrary to fact, the subject had followed the random treatment regime $%
h^{\ast }$. Under the\ consistency assumption that $\overline{A}_{h^{\ast
},k}=\overline{A}_{k}$\ \ $\Rightarrow L_{k+1}=L_{k+1,h^{\ast }}$ for all $%
k\in \left[ K\right] $ where $A_{h^{\ast },k}$ is the treatment received at
cycle $k$ when the subject follows the random regime, the no-unmeasured
confounding assumption that $L_{K+1,h^{\ast }}$ $%
\perp\!\!\!\perp A_k\mid%
$ $\left( \overline{A}_{k-1}=\overline{a}_{k},\overline{L}_{k}=\overline{l}%
_{k}\right) $ for all $\left( \overline{a}_{k},\overline{l}_{k}\right) $
such that $\dprod\limits_{j=1}^{k}h_{j}^{\ast }\left( a_{j}|\overline{a}%
_{j-1},\overline{l}_{j}\right) >0$ and the positivity assumption that $%
gh^{\ast }<<gh,$ the average treatment effect is precisely equal to $\theta
\left( g\right) $ if we take $\psi \left( \overline{l}_{K+1}\right)
=l_{K+1}. $

\section{Representations of the parameter of interest \label{representation}}

Throughout the paper we assume that $gh^{\ast }<<gh$ where $p=gh$ is the
unknown law of $\left( \overline{L}_{K+1},\overline{A}_{K}\right) .$ Robins
(1993) noted that the parameter $\theta \left( g\right) $ admits two
representations, which we now review. These representations are important as
they give rise to two distinct estimation strategies that we will review in
the next section. Define the random variables 
\begin{equation}
\pi _{j}^{\ast k}\equiv \dprod\limits_{r=j}^{k}h_{r}^{\ast }\left( A_{r}|%
\overline{L}_{r},\overline{A}_{r-1}\right)  \label{pijk}
\end{equation}%
$\pi _{j}^{k}\equiv \dprod\limits_{r=j}^{k}h_{r}\left( A_{r}|\overline{L}%
_{r},\overline{A}_{r-1}\right) ,\pi ^{k}\equiv \pi _{1}^{k}$ and $\pi ^{\ast
k}\equiv \pi _{1}^{\ast k}.$ Also, given $p=gh,$ $\overline{g}_{k}\overline{h%
}_{k}$ stands for the $\overline{p}_{k}=\dprod\limits_{j=0}^{k}g_{j}\dprod%
\limits_{j=1}^{k}h_{j}$.

\subsection{Inverse probability weighted representation of the g-functional}

The Radon-Nykodim theorem implies that 
\begin{equation}
\theta \left( g\right) =E_{gh}\left\{ \psi \left( \overline{L}_{K+1}\right)
\pi ^{\ast K}/\pi ^{K}\right\} \text{ }  \label{h}
\end{equation}

This motivates the so-called inverse probability weighted estimators of $%
\theta \left( g\right) $ discussed in section \ref{IPW-est}. Notice that
when, as in Examples 1, 2 and 3, $h_{k}^{\ast }\left( \cdot |\overline{a}%
_{k-1},\overline{l}_{k}\right) $ is the indicator of following a given
non-random treatment regime at study cycle $k,$ $\pi ^{\ast K}$ is the
indicator of having followed the regime through the entire study and $\pi
^{K}$ is the product of the conditional probabilities of following the
regime at each study cycle for a subject with recorded data $\left( 
\overline{L}_{K+1},\overline{A}_{K}\right) $. So, the right hand side of $%
\left( \ref{h}\right) $ is interpreted as the population weighted mean of $%
\psi \left( \overline{L}_{K+1}\right) $ among those subjects that follow the
regime through the entire study weighted by the inverse of their probability
of complying with the regime.

\subsection{Iterated conditional mean representation}

Robins (1986, 1997) derived another representation of $\theta \left(
g\right) $ in the form of an iterated conditional expectation. 
\begin{equation*}
\eta _{k}\left( \overline{a}_{k},\overline{l}_{k};\underline{g}_{k}\right)
\equiv \int \psi \left( \overline{L}_{K+1}\right)
\dprod\limits_{j=k+1}^{K}h_{j}^{\ast }\left( a_{j}|\overline{l}_{j},%
\overline{a}_{j-1}\right) \dprod\limits_{j=k}^{K}g_{j}\left( l_{j+1}|%
\overline{l}_{j},\overline{a}_{j}\right) \left\{
\dprod\limits_{j=k+1}^{K+1}d\mu _{j}^{\prime }\left( l_{j}\right)
\dprod\limits_{j=k+1}^{K}d\mu _{j}\left( a_{j}\right) \right\}
\end{equation*}%
for $k\in \left[ K\right] ,$ where $\dprod\limits_{j=K+1}^{K}\cdot \equiv 1$
and for any $f=\left( f_{1},\ldots ,f_{K}\right) ,\underline{f}_{k}\equiv
\left( f_{k},\ldots ,f_{K}\right) .$ Note that for any $\left( \overline{a}%
_{k},\overline{l}_{k}\right) $ such that $\dprod\limits_{j=1}^{k}h_{j}\left(
a_{j}|\overline{l}_{j},\overline{a}_{j-1}\right)
\dprod\limits_{j=0}^{k-1}g_{j}\left( l_{j+1}|\overline{l}_{j},\overline{a}%
_{j}\right) >0,$ it holds that 
\begin{equation*}
\eta _{k}\left( \overline{a}_{k},\overline{l}_{k};\underline{g}_{k}\right)
\equiv E_{\underline{g}_{k}\underline{h}_{k+1}^{\ast }}\left\{ \psi \left( 
\overline{L}_{K+1}\right) |\overline{A}_{k}=\overline{a}_{k},\overline{L}%
_{k}=\overline{l}_{k}\right\} .
\end{equation*}

Define also, $Y_{K+1}\left( \underline{g}_{K+1}\right) \equiv \psi \left( 
\overline{L}_{K+1}\right) ,$ and for $k\in \left[ K-1\right] ,$ if $\pi
^{\ast k}>0$ define 
\begin{eqnarray}
Y_{k+1}\left( \underline{g}_{k+1}\right) &\equiv &y_{k+1,\eta
_{k+1}^{g}}\left( \overline{A}_{k},\overline{L}_{k+1};\underline{g}%
_{k+1}\right)  \label{y-def} \\
&\equiv &E_{h_{k+1}^{\ast }}\left\{ \eta _{k+1}\left( \overline{A}_{k+1},%
\overline{L}_{k+1};\underline{g}_{k+1}\right) |\overline{A}_{k},\overline{L}%
_{k+1}\right\}  \notag \\
&=&\dint \eta _{k+1}\left( a_{k+1},\overline{A}_{k},\overline{L}_{k+1};%
\underline{g}_{k+1}\right) h_{k+1}^{\ast }\left( a_{k+1}|\overline{A}_{k},%
\overline{L}_{k+1}\right) d\mu _{k+1}\left( a_{k+1}\right)  \notag
\end{eqnarray}

It immediately follows that for any $k\in \left[ K\right] ,$ if $\pi ^{\ast
k}>0$ \ then 
\begin{equation*}
\eta _{k}\left( \overline{A}_{k},\overline{L}_{k};\underline{g}_{k}\right)
=E_{g_{k}}\left\{ Y_{k+1}\left( \underline{g}_{k+1}\right) |\overline{A}_{k},%
\overline{L}_{k}\right\} ,
\end{equation*}%
and 
\begin{equation}
\theta \left( g\right) =E_{g_{0}}\left\{ Y_{1}\left( \underline{g}%
_{1}\right) \right\} .  \label{iden1}
\end{equation}

\subsubsection{Interpretation of $\protect\eta _{k}$\ and $Y_{k}\left( 
\protect\underline{g}_{k}\right) $ in Example 1.}

Under the conditional independence and positivity assumptions made in this
example, $\eta _{k}\left( \overline{a}_{k},\overline{l}_{k};\underline{g}%
_{k}\right) $ evaluated at $\overline{a}_{k}=\overline{1}$ coincides with $%
E\left( L_{K+1}^{\ast }|\overline{A}_{k}=\overline{1},\overline{L}_{k}=%
\overline{l}_{k}\right) ,$ i.e., the mean of the intended outcome $%
L_{K+1}^{\ast }$ among subjects that are still on study at study cycle $k+1,$
i.e., with $\overline{A}_{k}=\overline{1},$ and that have recorded past $%
\overline{L}_{k}=\overline{l}_{k}$ up to time $t_{k}$ (Robins, 1986, 1997).
Note that by the assumed conditional independence $L_{K+1}^{\ast }$ $%
\perp\!\!\!\perp A_k\mid%
$ $\left( \overline{A}_{k-1}=\overline{1},\overline{L}_{k-1}\right) $, this
conditional mean is the same as $E\left( L_{K+1}^{\ast }|\overline{A}_{k-1}=%
\overline{1},\overline{L}_{k}=\overline{l}_{k}\right) .$ So, we can
interpret $\eta _{k}\left( \overline{a}_{k}=\overline{1},\overline{l}_{k};%
\underline{g}_{k}\right) $ as the best predictor of $L_{K+1}^{\ast }\,\ $for
subjects that are still on study at study cycle $k$ given the observed data $%
\overline{L}_{k}=\overline{l}_{k}.$ On the other hand, $\eta _{k}\left( 
\overline{a}_{k},\overline{l}_{k};\underline{g}_{k}\right) $ has no
meaningful interpretation when $a_{j}=0$ for some $j<k.$ Nevertheless, we
need not worry about this interpretation because the values taken by the
function $\eta _{k}\left( \overline{a}_{k},\overline{l}_{k};\underline{g}%
_{k}\right) $ when $\overline{a}_{k}\not=\overline{1}$ are irrelevant. This
is because $\theta \left( g\right) $ does not depend on them. To interpret $%
Y_{k}\left( \underline{g}_{k}\right) $ notice that this is only defined for
units with $\pi ^{\ast k}>0,$ i.e., for units with $\overline{A}_{k-1}=%
\overline{1}.$ For these units $Y_{k}\left( \underline{g}_{k}\right) $
equals $\eta _{k}\left( \overline{a}_{k}=\overline{1},\overline{L}_{k};%
\underline{g}_{k}\right) $ because in this example $h_{k}^{\ast }\left(
a_{k}|\overline{A}_{k-1}=\overline{1},\overline{L}_{k}\right)
=a_{k}.\medskip $

\subsubsection{Interpretation of $\protect\eta _{k}$\ and $Y_{k}\left( 
\protect\underline{g}_{k}\right) $ in Example 2.}

Consider, for some fixed $\overline{a^{\ast }}=\left( a_{1}^{\ast },\ldots
,a_{K}^{\ast }\right) ,$ the mean of $L_{K+1.\overline{a^{\ast }}}$. This
equals $\theta \left( g\right) $ under the consistency, no-unmeasured
confounding and positivity assumptions when we take $h_{k}^{\ast }\left(
a_{k}|\overline{A}_{k-1}=\overline{1},\overline{L}_{k}\right) =I_{\left\{
a_{k}^{\ast }\right\} }\left( a_{k}\right) .$ The interpretation of $\eta
_{k}\left( \overline{a}_{k},\overline{l}_{k};\underline{g}_{k}\right) $ for $%
\overline{a}_{k}=\overline{a}_{k}^{\ast }$ is identical to the one just
given for Example 1, replacing $\overline{1}$ with $\overline{a}_{k}^{\ast }$
and $L_{K+1}^{\ast }$with $L_{K+1,\overline{a}^{\ast }}.$ For $\overline{a}%
_{k}\not=\overline{a}_{k}^{\ast },$ the interpretation of $\eta _{k}\left( 
\overline{a}_{k},\overline{l}_{k};\underline{g}_{k}\right) $ is irrelevant
because, just as in Example 1, $\theta \left( g\right) \,\ $does not depend
on the values taken by $\eta _{k}\left( \overline{a}_{k},\overline{l}_{k};%
\underline{g}_{k}\right) \,\ $when $\overline{a}_{k}\not=\overline{a}%
_{k}^{\ast }$. Also, in analogy to Example 1, $Y_{k}\left( \underline{g}%
_{k}\right) $ is defined only for units with $\overline{A}_{k-1}=\overline{a}%
_{k-1}^{\ast }.$ For these units, $Y_{k}\left( \underline{g}_{k}\right) $
equals to $\eta _{k}\left( \overline{a}_{k}=\overline{a}_{k}^{\ast },%
\overline{L}_{k};\underline{g}_{k}\right) .$

\subsubsection{Interpretation of $\protect\eta _{k}$\ and $Y_{k}\left( 
\protect\underline{g}_{k}\right) $ in Example 3.}

Under the consistency, conditional independence and positivity assumptions
made in this example, the function $\eta _{k}\left( \overline{a}_{k},%
\overline{l}_{k};\underline{g}_{k}\right) $ evaluated at $\overline{a}_{k}=%
\overline{d}_{k}\left( \overline{a}_{k-1},\overline{l}_{k}\right) $ where $%
\overline{d}_{k}\left( \overline{a}_{k-1},\overline{l}_{k}\right) =\left[
d_{1}\left( l_{1}\right) ,d_{2}\left( a_{1},\overline{l}_{2}\right) \ldots
,d_{k}\left( \overline{a}_{k-1},\overline{l}_{k}\right) \right] ,$ coincides
with $E\left\{ L_{K+1,d}|\overline{A}_{k}=\overline{d}_{k}\left( \overline{A}%
_{k-1},\overline{l}_{k-1}\right) ,\overline{L}_{k}=\overline{l}_{k}\right\}
, $ i.e. the mean of the counterfactual outcome $L_{K+1,d}$ among subjects
that remain compliers with the treatment regime $d$ at the $k+1^{th}$ cycle,
and that have recorded past $\overline{L}_{k}=\overline{l}_{k}$ (Robins,
1986, 1997). As in Examples 1 and 2, the interpretation of $\eta _{k}\left( 
\overline{A}_{k},\overline{L}_{k};\underline{g}_{k}\right) $ when $\overline{%
A}_{k}\not=\overline{d}_{k}\left( \overline{A}_{k-1},\overline{L}_{k}\right) 
$ $\ $is irrelevant since $\theta \left( g\right) $ does not depend on it.
Also, in analogy to Example 1, $Y_{k}\left( \underline{g}_{k}\right) $ is
only defined for units with $\overline{A}_{k-1}=\overline{d}_{k-1}\left( 
\overline{A}_{k-2},\overline{L}_{k-1}\right) .$ For these units, $%
Y_{k}\left( \underline{g}_{k}\right) $ equals to $\eta _{k}\left( \overline{A%
}_{k}=\overline{d}_{k}\left( \overline{A}_{k-1},\overline{l}_{k}\right) ,%
\overline{L}_{k};\underline{g}_{k}\right) .$

\subsubsection{Interpretation of $\protect\eta _{k}$\ and $Y_{k}\left( 
\protect\underline{g}_{k}\right) $ in Example 4.}

Under the consistency, conditional independence and positivity assumptions
made in this example, $\eta _{k}\left( \overline{a}_{k},\overline{l}_{k};%
\underline{g}_{k}\right) $ coincides with the $E\left( L_{K+1,h^{\ast }}|%
\overline{A}_{k}=\overline{a}_{k},\overline{L}_{k}=\overline{l}_{k}\right) ,$
i.e., the mean of the counterfactual outcome $L_{K+1,h^{\ast }}$ among
subjects that received, in the real world, treatment $\overline{A}_{k}\,=%
\overline{a}_{k}\ $up to cycle $k$ and have recorded past outcomes $%
\overline{L}_{k}=\overline{l}_{k}.$ Unlike the preceding examples, if $%
h^{\ast }$ assigns positive probability to all possible treatment values $%
a_{k},$ then $\theta \left( g\right) $ depends on the values $\eta
_{k}\left( \overline{a}_{k},\overline{l}_{k};\underline{g}_{k}\right) $ for
all $\left( \overline{a}_{k},\overline{l}_{k}\right) $. This is because,
unlike the preceding examples, here $\pi ^{\ast k}>0$ w.p.1. It follows from
definition $\left( \ref{pijk}\right) \,$that $\pi ^{\ast k}\equiv \pi
_{1}^{\ast k}$ is the product of the conditional probabilities given past $%
L^{\prime }s$ and treatments, the subject receives the treatments $%
A_{j},j=1,\ldots ,k,$ that he/she actually received when he/she follows
regime $h^{\ast }.$ Also, $Y_{k}\left( \underline{g}_{k}\right) \equiv
E_{h_{k+1}^{\ast }}\left\{ \eta _{k+1}\left( \overline{A}_{k+1},\overline{L}%
_{k+1};\underline{g}_{k+1}\right) |\overline{A}_{k},\overline{L}%
_{k+1}\right\} $ is equal to $E\left( L_{K+1,h^{\ast }}|\overline{A}_{k-1}=%
\overline{a}_{k-1},\overline{L}_{k}=\overline{l}_{k}\right) ,\,\ $i.e. the
mean of the counterfactual outcome $L_{K+1,h^{\ast }}$ among subjects that
received, in the real world, treatment $\overline{A}_{k-1}\,=\overline{a}%
_{k-1}\ $up to cycle $k-1$ and have recorded past outcomes $\overline{L}_{k}=%
\overline{l}_{k}$ up to an including cycle $k.$

\section{Estimation based on parametric models for the nuisance functions 
\label{parametric}}

\subsection{Inverse probability weighting estimation\label{IPW-est}}

Uniform consistent estimation of $\theta \left( g\right) $\ under the large
non-parametric model $P$\ cannot be carried out due to the curse of
dimensionality. Both in theory and in practice one is forced to consider a
dimension reducing plan. One such plan is motivated from display $\left( \ref%
{h}\right) .$ Specifically, suppose that for each $k\in \left[ K\right] $ we
postulate a smooth parametric class for $h_{k},$ say,%
\begin{equation}
\mathcal{C}_{k}=\left\{ h_{k,\alpha _{k}}\in \mathcal{V}_{k}:\alpha _{k}\in
\Xi _{k}\right\}  \label{parh}
\end{equation}%
where $\Xi _{k}$ is a subset of a Euclidean space and $\mathcal{V}_{k}$ is
the set of all conditional densities $p_{A_{k}|\overline{L}_{k},\overline{A}%
_{k-1}}$ for probability measures in $\mathcal{P}$. We can then compute the
estimator 
\begin{equation*}
\widehat{\theta }_{IPW}\equiv \mathbb{P}_{n}\left\{ \psi \left( \overline{L}%
_{K+1}\right) \pi ^{\ast K}/\widehat{\pi }^{K}\right\}
\end{equation*}%
where throughout $\widehat{\pi }_{j}^{k}\equiv \dprod\limits_{r=1}^{k}%
\overline{h}_{r,\widehat{\alpha }_{ML}},\widehat{\pi }^{k}\equiv \widehat{%
\pi }_{1}^{k},$ $\widehat{\alpha }_{ML}=\left( \widehat{\alpha }%
_{1,ML},\ldots ,\widehat{\alpha }_{K,ML}\right) ,$ $\widehat{\alpha }%
_{k,ML}=\arg \max_{\alpha _{k}\in \Xi _{k}}\mathbb{P}_{n}\left\{ \log
h_{k,\alpha _{k}}\right\} $ and $\mathbb{P}_{n}\left( \cdot \right) $ is the
empirical mean operator, i.e. $\mathbb{P}_{n}\left( V\right)
=n^{-1}\sum\limits_{i=1}^{n}V_{i}.$ Under regularity conditions, $\widehat{%
\theta }_{IPW}$ is consistent and asymptotically normal, throughout CAN,
i.e. $\sqrt{n}\left\{ \widehat{\theta }_{IPW}-\theta \left( g\right)
\right\} $ converges to a mean zero normal random variable provided $p$ is
in the submodel $\cap _{k=1}^{K}\mathcal{H}_{k}$ of $\mathcal{P}$ where 
\begin{equation}
\mathcal{H}_{k}\equiv \left\{ p\in \mathcal{P}\text{ }:h_{k}\in \mathcal{C}%
_{k}\right\}  \label{Hk}
\end{equation}

\subsection{Fully parametric maximum likelihood estimation}

Suppose that we postulate parametric models for each $g_{k},$ say $\left\{
g_{k,\xi _{k}}\in \mathcal{U}_{k}:\xi _{k}\in \digamma _{k}\right\} $ where $%
\digamma _{k}$ is some Euclidean space, $k=1,\ldots ,K$ and compute the
maximum likelihood estimator $\theta \left( \widehat{g}_{ML}\right) $ of $%
\theta \left( g\right) $ where $\widehat{g}_{ML}\equiv \left( g_{0,n},g_{%
\widehat{\xi }_{ML}}\right) ,g_{\xi }\equiv \left( g_{1,\xi _{1}},\ldots
,g_{K,\xi _{K}}\right) ,$ $\widehat{\xi }_{ML}=\left( \widehat{\xi }%
_{1,ML},\ldots ,\widehat{\xi }_{K,ML}\right) ,$ $\widehat{\xi }_{k,ML}=\arg
\max_{\xi _{k}\in \digamma _{k}}\mathbb{P}_{n}\left\{ \log g_{k,\alpha
_{k}}\right\} $ and $g_{0,n}$ is the empirical law of $L_{1}.$ The plug-in
estimator $\theta \left( \widehat{g}_{ML}\right) $ is CAN for $\theta \left(
g\right) $ if the postulated parametric models are correct.

\subsection{Iterated regression estimation \label{ITR}}

One can construct estimators of $\theta \left( g\right) $ that are CAN under
semiparametric, rather than parametric models for $g.$ The representation $%
\left( \ref{iden1}\right) $ of $\theta \left( g\right) $ and the recursion
to arrive at $Y_{1}\left( \underline{g}_{1}\right) $ motivates a dimension
reducing plan in which estimation of $\theta \left( g\right) $ is conducted
assuming that, for each each $k\in \left[ K\right] $, the map 
\begin{equation}
\left( \overline{a}_{k},\overline{l}_{k}\right) \in \text{Posit}%
_{k}\rightarrow \eta _{k}\left( \overline{a}_{k},\overline{l}_{k};\underline{%
g}_{k}\right) ,  \label{map}
\end{equation}%
with domain the set 
\begin{equation*}
\text{Posit}_{k}\equiv \left\{ \left( \overline{a}_{k},\overline{l}%
_{k}\right) :h_{j}^{\ast }\left( a_{j}|\overline{a}_{j-1},\overline{l}%
_{j}\right) >0,j=1,\ldots ,k-1\right\} ,
\end{equation*}%
of possible histories $\left( \overline{a}_{k-1},\overline{l}_{k}\right) $
under $h^{\ast },$ belongs to the parametric class 
\begin{equation}
\mathcal{R}_{k}=\left\{ \eta _{k,\tau _{k}}\in \mathcal{D}_{k}:\eta _{k,\tau
_{k}}\left( \overline{a}_{k},\overline{l}_{k}\right) =\Psi \left\{ \tau
_{k}^{T}s_{k}\left( \overline{a}_{k},\overline{l}_{k}\right) \right\} :\tau
_{k}\in \Upsilon _{k},\right\} ,  \label{parg}
\end{equation}%
where $\mathcal{D}_{k}$ is the set of all real valued functions with domain
in Posit$_{k}$, $\Psi $ is a canonical link in a generalized linear model, $%
s_{k}$ is a known function and $\tau _{k}$ an unknown parameter, with $%
\Upsilon _{k}$ a subset of a Euclidean space. Define 
\begin{equation}
\mathcal{G}_{k}\equiv \left\{ p\in \mathcal{P}\text{ }:\eta _{k}\left( \cdot
,\cdot ;\underline{g}_{k}\right) \in \mathcal{R}_{k}\right\} ,k\in \left[ K%
\right] ,  \label{Gk}
\end{equation}%
and the estimator 
\begin{equation*}
\widehat{\theta }_{\mathcal{G}}\equiv \mathbb{P}_{n}\left( \widehat{Y}%
_{1}\right) ,
\end{equation*}
where $\widehat{Y}_{1}$ is the output of the following recursive algorithm.

\begin{description}
\item[Algorithm 1.] Set $\widehat{Y}_{K+1}\equiv \psi \left( \overline{L}%
_{K+1}\right) $ and recursively, for $k=K,K-1,\ldots ,1,$

\begin{description}
\item[a) ] Estimate $\tau _{k}$ indexing the regression model 
\begin{equation*}
\eta _{k,\tau _{k}}\left( \overline{A}_{k},\overline{L}_{k}\right) \equiv
\Psi \left\{ \tau _{k}^{T}s_{k}\left( \overline{A}_{k},\overline{L}%
_{k}\right) \right\} ,
\end{equation*}%
for $E\left( \widehat{Y}_{k+1}|\overline{A}_{k},\overline{L}_{k}\right) $
restricted to units verifying $\pi ^{\ast k}>0$ with $\widehat{\tau }_{k,%
\mathcal{G}}$ solving%
\begin{equation}
\mathbb{P}_{n}\left[ \pi ^{\ast k}s_{k}\left( \overline{A}_{k},\overline{L}%
_{k}\right) \left\{ \widehat{Y}_{k+1}-\Psi \left\{ \tau _{k}^{T}s_{k}\left( 
\overline{A}_{k},\overline{L}_{k}\right) \right\} \right\} \right] =0.
\label{non-rob1}
\end{equation}

\item[b) ] For units with $\pi ^{\ast k-1}>0,$ compute 
\begin{equation*}
\widehat{Y}_{k}\equiv y_{k,\widehat{\tau }_{k,\mathcal{G}}}\left( \overline{A%
}_{k-1},\overline{L}_{k}\right) \equiv \int h_{k}^{\ast }\left( a_{k}|%
\overline{A}_{k-1},\overline{L}_{k}\right) \eta _{k,\widehat{\tau }_{k,%
\mathcal{G}}}\left( a_{k},\overline{A}_{k-1},\overline{L}_{k}\right) d\mu
_{k}\left( a_{k}\right) .
\end{equation*}
\end{description}
\end{description}

Note that if, as in Examples 1-3, for each $k\in \left[ K\right] ,$ $%
h_{k}^{\ast }$ is an indicator function, then $\pi ^{\ast k}$ is also an
indicator function. In such case, the factor $\pi ^{\ast k}$ ensures that
estimation of $\tau _{k}$ is based only on subjects with $\pi ^{\ast k}=1.$
In Examples 1-3, subjects with $\pi ^{\ast k}=1$ are those that remain
compliers at cycle $k.$ The estimator $\widehat{\tau }_{k,\mathcal{G}}$
coincides with the estimator obtained from fitting, by iteratively
reweighted least squares (IRLS),\ the regression model $\Psi \left\{ \tau
_{k}^{T}s_{k}\left( \overline{A}_{k},\overline{L}_{k}\right) \right\} $
restricted to those subjects$.$ Note also that when, as in Examples 1-3, $%
h_{k}^{\ast }$ is an indicator function, the integral in step (b) of the
algorithm is equal to the function $\eta _{k,\widehat{\tau }_{k,\mathcal{G}%
}}\left( a_{k},\overline{A}_{k-1},\overline{L}_{k}\right) $ evaluated at the
value of $a_{k}$ for which $h_{k}^{\ast }\left( a_{k}|\overline{A}_{k-1},%
\overline{L}_{k}\right) =1.$ Thus, for instance, in Examples 1 and 3, $y_{k,%
\widehat{\tau }_{k,\mathcal{G}}}\left( \overline{A}_{k-1},\overline{L}%
_{k}\right) $ is equal to $\eta _{k,\widehat{\tau }_{k,\mathcal{G}}}\left( 1,%
\overline{A}_{k-1},\overline{L}_{k}\right) $ and $\eta _{k,\widehat{\tau }%
_{k,\mathcal{G}}}\left( d_{k}\left( \overline{A}_{k-1},\overline{L}%
_{k}\right) ,\overline{A}_{k-1},\overline{L}_{k}\right) $ respectively.

Whether or not $\pi ^{\ast k}$\ is binary, we note that the equation $\left( %
\ref{non-rob1}\right) $\ will have a unique solution when $\psi \left( 
\overline{L}_{K+1}\right) $\ falls in the range of $\Psi \left( \cdot
\right) .$ For $k=K,$ the equation solved by the estimator $\widehat{\tau }%
_{K,\mathcal{G}}$ \ agrees with the score equation from the fit of a
generalized linear model with canonical link except that each individual
contribution is weighted $\pi ^{\ast k}$ and should therefore be the
maximizer of the weighted log-likelihood for the associated exponential
family model with outcome $\psi \left( \overline{L}_{K+1}\right) $. For $%
k<K, $\ the estimating equation $\left( \ref{non-rob1}\right) $\ is again a
weighted score equation, under the same generalized linear model with the
same canonical link function, but for the pseudo-outcome\ $\widehat{Y}%
_{k+1}. $ This pseudo-outcome\ falls in the range of $\Psi \left( \cdot
\right) $\ because, by construction, it agrees with the conditional mean of $%
\Psi \left\{ \widehat{\tau }_{k+1,\mathcal{G}}^{T}s_{k+1}\left( \overline{A}%
_{k+1},\overline{L}_{k+1}\right) \right\} $\ given $\left( \overline{A}_{k},%
\overline{L}_{k+1}\right) $\ under $h_{k}^{\ast }.$ Thus, for $k<K,$ the
equation $\left( \ref{non-rob1}\right) $ has a unique solution.

We note that when one specifies parametric models $\mathcal{R}_{k}$ for $%
\eta _{k}^{g}$ there is the possibility that the resulting models $\mathcal{G%
}_{k}$ are incompatible. We do not discuss this issue in this paper. Molina
et. al. (2017) give a careful discussion of the topic and Babino et. al.
(2017) propose a modeling strategy which avoids model incompatibility.

\bigskip

To analyze the asymptotic behavior of $\widehat{\theta }_{\mathcal{G}}\,\ $%
and of several of the forthcoming estimators, we define for any $\eta
_{k}\left( \overline{A}_{k},\overline{L}_{k}\right) ,k\in \left[ K\right] ,$ 
\begin{eqnarray*}
y_{k,\eta _{k}}\left( \overline{A}_{k-1},\overline{L}_{k}\right) &\equiv
&E_{h_{k}^{\ast }}\left\{ \eta _{k}\left( \overline{A}_{k},\overline{L}%
_{k}\right) |\overline{A}_{k-1},\overline{L}_{k}\right\} \\
&=&\int \eta _{k}\left( a_{k},\overline{A}_{k-1},\overline{L}_{k}\right)
h_{k}^{\ast }\left( a_{k}|\overline{A}_{k-1},\overline{L}_{k}\right) d\mu
_{k}\left( a_{k}\right) ,
\end{eqnarray*}%
and 
\begin{eqnarray*}
\Delta _{k}\left( \eta _{k},\eta _{k+1};g_{k}\right) &\equiv &\pi ^{\ast k} 
\left[ \eta _{k}\left( \overline{A}_{k},\overline{L}_{k}\right) -E_{g_{k}}%
\left[ \left. E_{h_{k+1}^{\ast }}\left\{ \eta _{k+1}\left( \overline{A}%
_{k+1},\overline{L}_{k+1}\right) |\overline{A}_{k},\overline{L}%
_{k+1}\right\} \right\vert \overline{A}_{k},\overline{L}_{k}\right] \right]
\\
&=&\pi ^{\ast k}\left[ \eta _{k}\left( \overline{A}_{k},\overline{L}%
_{k}\right) -E_{g_{k}}\left\{ \left. y_{k+1,\eta _{k+1}}\left( \overline{A}%
_{k},\overline{L}_{k+1}\right) \right\vert \overline{A}_{k},\overline{L}%
_{k}\right\} \right]
\end{eqnarray*}%
with $y_{K+1,\eta _{K+1}}\left( \overline{A}_{K},\overline{L}_{K+1}\right)
\equiv \psi \left( \overline{L}_{K+1}\right) .$ Note that 
\begin{equation}
\Delta _{k}\left( \eta _{k},\eta _{k+1};g_{k}\right) =0\text{ if }\eta
_{j}=\eta _{j}^{g}\text{ for }j=k,k+1.  \label{delta-true}
\end{equation}%
where here, and sometimes in what follows, we write, for short, $\eta
_{j}^{g}\left( \cdot ,\cdot \right) \ $instead of $\eta _{j}\left( \cdot
,\cdot ;\underline{g}_{j}\right) .$

We further define for any $\eta =\left( \eta _{1},\ldots ,\eta _{K}\right) $
and any $p=gh,$%
\begin{equation*}
d^{g}\left( \eta \right) \equiv \sum_{k=1}^{K}E_{\overline{g}_{k-1},%
\overline{h}_{k}}\left\{ \frac{1}{\pi ^{k}}\Delta _{k}\left( \eta _{k},\eta
_{k+1};g_{k}\right) \right\}
\end{equation*}%
where, recall $\pi ^{k}\equiv \dprod\limits_{j=1}^{k}h_{j}\left( A_{j}|%
\overline{A}_{j-1},\overline{L}_{j}\right) .$ Note that $d^{g}\left( \eta
\right) $ does not depend on $h$ because each expectation $E_{\overline{g}%
_{k-1},\overline{h}_{k}}\left\{ \frac{1}{\pi ^{k}}\Delta _{k}\left( \eta
_{k},\eta _{k+1};g_{k}\right) \right\} $ is not a function of $\overline{h}%
_{k}.$

In the Appendix we show the following result.

\textbf{Lemma 1:} For any $\eta _{k}\left( \overline{A}_{k},\overline{L}%
_{k}\right) ,k\in \left[ K\right] ,$ it holds that 
\begin{equation*}
E_{g_{1}}\left\{ y_{1,\eta _{1}}\left( L_{1}\right) \right\} -\theta \left(
g\right) =d^{g}\left( \eta \right)
\end{equation*}

To facilitate the analysis of the limiting distribution of $\widehat{\theta }%
_{\mathcal{G}}$ we make the following notational conventions and
definitions. For any function $t\left( Z;h,\eta \right) $ which depends on
some or all the components of $h=\left( h_{1},\ldots ,h_{K}\right) $ and $%
\eta =\left( \eta _{1},\ldots ,\eta _{K}\right) ,$ and any data dependent
functions $\widehat{h}$ and $\widehat{\eta },$ $\widehat{E}_{gh}\left\{
t\left( Z;\widehat{h},\widehat{\eta }\right) \right\} $ stands for
expectation under $p=gh$ regarding $\widehat{h}$ and $\widehat{\eta }$ as
non-random functions, that is%
\begin{equation*}
\widehat{E}_{gh}\left\{ t\left( Z;\widehat{h},\widehat{\eta }\right)
\right\} \equiv \int t\left( z;\widehat{h},\widehat{\eta }\right) p\left(
z\right) d\mu \left( z\right)
\end{equation*}%
With this definition%
\begin{equation*}
d^{g}\left( \widehat{\eta }\right) =\sum_{k=1}^{K}\widehat{E}_{\overline{g}%
_{k-1},\overline{h}_{k}}\left\{ \frac{1}{\pi ^{k}}\Delta _{k}\left( \widehat{%
\eta }_{k},\widehat{\eta }_{k+1};g_{k}\right) \right\}
\end{equation*}

We are now ready to study the limiting behavior of $\widehat{\theta }_{%
\mathcal{G}}.$ Letting $\widehat{\eta }_{\mathcal{G}}\equiv \left( \widehat{%
\eta }_{1,\mathcal{G}},\ldots ,\widehat{\eta }_{K,\mathcal{G}}\right) $
where $\widehat{\eta }_{1,\mathcal{G}}\equiv \eta _{k,\widehat{\tau }_{k,%
\mathcal{G}}},$ Lemma 1 immediately implies the following representation for 
$\widehat{\theta }_{\mathcal{G}}.$

\begin{equation}
\widehat{\theta }_{\mathcal{G}}-\theta \left( g\right) =\mathbb{P}%
_{n}\left\{ y_{1,\widehat{\eta }_{1,\mathcal{G}}}\left( L_{1}\right)
\right\} -\widehat{E}_{g_{1}}\left\{ y_{1,\widehat{\eta }_{1,\mathcal{G}%
}}\left( L_{1}\right) \right\} +d^{g}\left( \widehat{\eta }_{\mathcal{G}%
}\right) .  \label{decomp11}
\end{equation}

To analyze the limiting distribution of $\widehat{\theta }_{\mathcal{G}}$ we
first note that the vector $\widehat{\tau }_{\mathcal{G}}\equiv \left( 
\widehat{\tau }_{1,\mathcal{G}},\ldots ,\widehat{\tau }_{K,\mathcal{G}%
}\right) $ solves a joint system of estimating equations, so under
regularity conditions, it has a probability limit under any $p\in \mathcal{P}
$ which we denote with $\tau _{\lim ,\mathcal{G}}\left( p\right) \equiv
\left( \tau _{1,\lim ,\mathcal{G}}\left( p\right) ,\ldots ,\tau _{K,\lim ,%
\mathcal{G}}\left( p\right) \right) .$ Furthermore, $\left\{ \widehat{\tau }%
_{\mathcal{G}}-\tau _{\lim ,\mathcal{G}}\left( p\right) \right\} $ is
asymptotically linear. In addition, under regularity conditions, the map $%
\tau \rightarrow d^{g}\left( \eta _{\tau }\right) $ is differentiable. Then,
writing $\eta _{k,\lim \mathcal{G}}\left( p\right) \equiv \eta _{k,\tau
_{\lim \mathcal{G}}\left( p\right) },k\in \left[ K\right] ,$ we conclude
that 
\begin{equation*}
d^{g}\left( \widehat{\eta }_{\mathcal{G}}\right) -d^{g}\left[ \eta _{\lim ,%
\mathcal{G}}\left( p\right) \right] \text{ is asymptotically linear. }
\end{equation*}

Furthermore, under regularity conditions, $y_{1,\widehat{\eta }_{1,\mathcal{G%
}}}$ and $y_{1,\eta _{1,\lim \mathcal{G}}}$ fall in a Donsker class, so 
\begin{equation*}
\mathbb{P}_{n}\left\{ y_{1,\widehat{\eta }_{1,\mathcal{G}}}\left(
L_{1}\right) \right\} -\widehat{E}_{g_{1}}\left\{ y_{1,\widehat{\eta }_{1,%
\mathcal{G}}}\left( L_{1}\right) \right\} =\mathbb{P}_{n}\left\{ y_{1,\eta
_{1,\lim \mathcal{G}}}\left( L_{1}\right) \right\} -\widehat{E}%
_{g_{1}}\left\{ y_{1,\eta _{1,\lim \mathcal{G}}}\left( L_{1}\right) \right\}
+o_{p}\left( n^{-1/2}\right) \text{ }
\end{equation*}%
is asymptotically linear. The representation $\left( \ref{decomp11}\right) $
then implies that 
\begin{equation*}
\widehat{\theta }_{\mathcal{G}}-\theta \left( g\right) -d^{g}\left[ \eta
_{\lim ,\mathcal{G}}\left( p\right) \right] \text{ is asymptotically linear.}
\end{equation*}

To establish that $\widehat{\theta }_{\mathcal{G}}$ is CAN under model $\cap
_{k=1}^{K}\mathcal{G}_{k}$ it then suffices to show that 
\begin{equation}
d^{g}\left[ \eta _{\lim ,\mathcal{G}}\left( p\right) \right] =0\text{ if }%
p\in \cap _{k=1}^{K}\mathcal{G}_{k}  \label{dlim}
\end{equation}%
This fact is a consequence of the following result.

\textbf{Proposition 1: }Under regularity conditions, 
\begin{equation}
\eta _{k,\lim ,\mathcal{G}}\left( p\right) =\eta _{k}^{g}\text{ if }p\in
\cap _{j=k}^{K}\mathcal{G}_{j}  \label{result1}
\end{equation}

Proposition 1 and $\left( \ref{delta-true}\right) $ now imply that for all $%
k\in \left[ K\right] ,$ 
\begin{equation*}
\Delta _{k}\left( \eta _{k,\lim ,\mathcal{G}}\left( p\right) ,\eta
_{k+1,\lim ,\mathcal{G}}\left( p\right) ;g_{k}\right) =0\text{ if }p\in \cap
_{j=1}^{K}\mathcal{G}_{j}
\end{equation*}%
and therefore that $\left( \ref{dlim}\right) $ holds.

\begin{description}
\item \textbf{Proof of Proposition 1: } By reverse induction in $k$. Suppose
first that $k=K.$\ Assume $p=gh\in $ $\mathcal{G}_{K}$. Then, \thinspace $%
E_{g_{K}}\left\{ \psi \left( \overline{L}_{K+1}\right) |\overline{A}_{K},%
\overline{L}_{K}\right\} =\eta _{K,\tau _{K}\left( g_{K}\right) }\left( 
\overline{A}_{K},\overline{L}_{K}\right) $ for some $\tau _{K}\left(
g_{K}\right) $ and therefore the equation $\left( \ref{non-rob1}\right) $ is
an unbiased estimating equation for $\tau _{K}\left( g_{K}\right) $ since $%
\widehat{Y}_{K+1}=\psi \left( \overline{L}_{K+1}\right) .$ Consequently,
under standard regularity conditions for $M$- estimators, the probability
limit $\tau _{K,\lim ,\mathcal{G}}$ of $\widehat{\tau }_{K,\mathcal{G}}$ is
equal to $\tau _{K}\left( g_{K}\right) $ which, in turn, implies that $%
\left( \mathbf{\ref{result1}}\right) $ holds for $k=K.\medskip $ \newline
Suppose next that $\left( \mathbf{\ref{result1}}\right) $ holds for $%
k=K,\ldots ,j+1.$ Noticing that, by construction, $\widehat{Y}_{j+1}=y_{j+1,%
\widehat{\eta }_{j+1,\mathcal{G}}}\left( \overline{A}_{k},\overline{L}%
_{k+1}\right) ,$ we conclude that $\widehat{\tau }_{j,\mathcal{G}}$ solves%
\begin{equation*}
0=\mathbb{P}_{n}\left[ \pi ^{\ast j}s_{j}\left( \overline{A}_{j},\overline{L}%
_{j}\right) \left\{ y_{j+1,\eta _{j+1,\lim ,\mathcal{G}}}\left( \overline{A}%
_{k},\overline{L}_{k+1}\right) -\Psi \left\{ \tau _{j}^{T}s_{j}\left( 
\overline{A}_{j},\overline{L}_{j}\right) \right\} \right\} \right]
+o_{p}\left( 1\right)
\end{equation*}%
Suppose $p=gh\in $ $\cap _{k=j}^{K}\mathcal{G}_{k}.$ Then, by the inductive
hypothesis $y_{j+1,\eta _{j+1,\lim ,\mathcal{G}}}\left( \overline{A}_{j},%
\overline{L}_{j+1}\right) =Y_{j+1}\left( \underline{g}_{j+1}\right) $. Thus, 
$E_{g_{j}}\left\{ \left. y_{j+1,\eta _{j+1,\lim ,\mathcal{G}}}\left( 
\overline{A}_{j},\overline{L}_{j+1}\right) \right\vert \overline{A}_{j},%
\overline{L}_{j}\right\} =\eta _{j}^{g}\left( \overline{A}_{j},\overline{L}%
_{j}\right) .$ Furthermore, since $p\in \mathcal{G}_{j}$ then $\eta
_{j}^{g}=\eta _{j,\tau _{j}\left( g_{j}\right) }$ for some $\tau _{j}\left(
g_{j}\right) $ and therefore the population equation%
\begin{equation*}
E_{\overline{g}_{j},\overline{h}_{j}}\left[ \pi ^{\ast j}s_{j}\left( 
\overline{A}_{j},\overline{L}_{j}\right) \left\{ y_{k+1,\eta _{k+1,\lim ,%
\mathcal{G}}}\left( \overline{A}_{k},\overline{L}_{k+1}\right) -\Psi \left\{
\tau _{j}^{T}s_{j}\left( \overline{A}_{j},\overline{L}_{j}\right) \right\}
\right\} \right] =0
\end{equation*}%
\ref{WITR}is solved at $\tau _{j}=\tau _{j}\left( g_{j}\right) .$ Then,
under regularity conditions for the consistency of $M-$ estimators, the
probability limit $\tau _{j,\lim ,\mathcal{G}}$ of $\widehat{\tau }_{j,%
\mathcal{G}}$ is equal to $\tau _{j}\left( g_{j}\right) ,$ which shows $%
\left( \mathbf{\ref{result1}}\right) $ holds for $k=j.$
\end{description}

\subsection{Weighted iterated regression\label{WITR}}

Suppose that in Algorithm 1 we replace step (a) with a procedure that
estimates, $\tau _{k}$ by \textit{weighted} IRLS, i.e. with $\widehat{\tau }%
_{k,\omega }$ solving 
\begin{equation}
\mathbb{P}_{n}\left[ \pi ^{\ast k}\omega _{k}\left( \overline{A}_{k},%
\overline{L}_{k}\right) s_{k}\left( \overline{A}_{k},\overline{L}_{k}\right)
\left\{ \widehat{Y}_{k+1,\omega }-\Psi \left\{ \tau _{k}^{T}s_{k}\left( 
\overline{A}_{k},\overline{L}_{k}\right) \right\} \right\} \right] =0
\label{itwr-eq}
\end{equation}%
for some user specified scalar function $\omega _{k}\left( \overline{A}_{k},%
\overline{L}_{k}\right) ,$ and where for each $k,$ $\widehat{Y}_{k,\omega }$
is defined as $\widehat{Y}_{k}$ in step (b) of Algorithm 1 but with $%
\widehat{\tau }_{k,\omega }$ instead of $\widehat{\tau }_{k,\mathcal{G}}.$
The resulting estimator $\widehat{\theta }_{\omega }\equiv P_{n}\left[ y_{1,%
\widehat{\tau }_{1,\omega }}\left( \overline{L}_{1}\right) \right] $ is also
CAN for $\theta \left( g\right) $ under regularity conditions if $p\in \cap
_{j=1}^{K}\mathcal{G}_{j}$. In fact, the same holds even if $\omega
_{k}\left( \overline{A}_{k},\overline{L}_{k}\right) =\omega _{k,\widehat{%
\alpha }_{ML}}\left( \overline{A}_{k},\overline{L}_{k}\right) $ depends on
the maximum likelihood estimator $\widehat{\alpha }_{ML}$ of $\alpha $
defined in section \ref{IPW-est}. Specifically, to analyze the limiting
distribution of $\widehat{\theta }_{\omega }$ where we allow the possibility
that $\omega _{k}=\omega _{k,\widehat{\alpha }_{ML}},$ note that regardless
of the validity of any of the models $\mathcal{H}_{k}$ or $\mathcal{G}_{k},$ 
$\left( \widehat{\tau }_{\omega },\widehat{\alpha }_{ML}\right) $ is
ultimately an $M$-estimator and as such, under regularity conditions, it has
a probability limit $\left( \tau _{\lim ,\omega }\left( p\right) ,\alpha
_{\lim }\left( h\right) \right) .$ Furthermore, $\left\{ \widehat{\tau }%
_{\omega }-\tau _{\lim ,\omega }\left( p\right) \right\} $ is asymptotically
linear. Then, with $\eta _{k,\lim ,\omega }\left( p\right) \equiv \eta
_{k,\tau _{\lim ,\omega }\left( p\right) },k\in \left[ K\right] ,$ we reason
as in the preceding section and conclude that 
\begin{equation*}
\widehat{\theta }_{\omega }-\theta \left( g\right) -d^{g}\left[ \eta _{\lim
,\omega }\left( p\right) \right] \text{ is asymptotically linear}
\end{equation*}

An argument essentially identical to that given for the proof of Proposition
1 shows that, under regularity conditions 
\begin{equation}
\eta _{k,\lim ,\omega }\left( p\right) =\eta _{k}^{g}\text{ if }p\in \cap
_{j=k}^{K}\mathcal{G}_{j}  \label{result2}
\end{equation}%
and consequently, that $d^{g}\left[ \eta _{\lim ,\omega }\left( p\right) %
\right] =0$ and thus, that $\widehat{\theta }_{\omega }$ is CAN for $\theta
\left( g\right) ,$ if $p\in \cap _{k=1}^{K}\mathcal{G}_{k}.$

We will argue in sections $\left( \ref{w-reg-dr}\right) $ and $\left( \ref%
{w-reg-mr}\right) $ that a particular choice of weights $\omega _{k},$
namely, $\omega _{k}=1/\widehat{\pi }^{k}$ where 
\begin{equation*}
\widehat{\pi }^{k}\equiv \widehat{\pi }^{k}\left( \overline{A}_{k},\overline{%
L}_{k}\right) \equiv \dprod\limits_{j=1}^{k}h_{j,\widehat{\alpha }%
_{j,ML}}\left( A_{j}|\overline{A}_{j-1},\overline{L}_{j}\right) ,
\end{equation*}%
yields estimators of $\theta $ that are CAN under a model larger than $\cap
_{j=k}^{K}\mathcal{G}_{j}.$

For ease of reference, we denote the estimators $\widehat{\tau }_{k,\omega }$
and $\widehat{\theta }_{\omega }$ using $\omega _{k}=1/\widehat{\pi }^{k}$
as $\widehat{\tau }_{k,reg}$ and $\widehat{\theta }_{reg},$ and the pseudo
outcome $\widehat{Y}_{k+1,\omega }$ in equation $\left( \ref{itwr-eq}\right) 
$ as $\widehat{Y}_{k+1,reg}$

\subsection{Doubly robust estimation by iterated regression\label%
{sec:DR-estimation}}

When $A_{k}$ is binary and $h_{k}^{\ast }\left( a_{k}|\overline{l}_{k},%
\overline{a}_{k-1}\right) =a_{k}$ $\ $as in Example 1, Bang and Robins
(2005) (throughout B\&R) proposed another iterated regression algorithm
which, they argued, remarkably returns a so-called doubly robust estimator
of $\theta \left( g\right) $ in the union model $\left( \cap _{k=1}^{K}%
\mathcal{H}_{k}\right) \cup $ $\left( \cap _{k=1}^{K}\mathcal{G}_{k}\right)
. $ This is an estimator that is CAN when $p$ is in $\left( \cap _{k=1}^{K}%
\mathcal{H}_{k}\right) \cup $ $\left( \cap _{k=1}^{K}\mathcal{G}_{k}\right) $%
, equivalently the estimator is CAN when either the models for all the $%
h_{k} $ are correct, or the models for all the $\eta _{k}$ are correct, but
not necessarily both. Here we generalize the construction of the B\&R
estimator to arbitrary conditional densities $h_{k}^{\ast }\left( a_{k}|%
\overline{l}_{k},\overline{a}_{k-1}\right) .$\thinspace The construction
starts with the computation of the maximum likelihood estimator $\widehat{%
\alpha }_{ML}$ as above. Next, one considers the extended parametric class 
\begin{equation*}
\mathcal{R}_{k}^{ext}=\left\{ \eta _{k,\upsilon _{k}}\in \mathcal{D}%
_{k}:\eta _{k,\upsilon _{k}}\left( \overline{a}_{k},\overline{l}_{k}\right)
=\Psi \left\{ \tau _{k}^{T}s_{k}\left( \overline{a}_{k},\overline{l}%
_{k}\right) +\lambda _{k}\widehat{\pi }^{k}\left( \overline{a}_{k},\overline{%
l}_{k}\right) ^{-1}\right\} ,\upsilon _{k}\equiv \left( \tau _{k},\lambda
_{k}\right) \in \Upsilon _{k}\times \mathbb{R}\right\}
\end{equation*}%
and subsequently applies Algorithm 1 to the extended model $\mathcal{R}%
_{k}^{ext}$. Specifically,

\begin{description}
\item[Algorithm 2 \textit{(Robins, 2002, Bang and Robins, 2005)}] Set $%
\widetilde{Y}_{K+1}\equiv \psi \left( \overline{L}_{K+1}\right) $ and
recursively, for $k=K,K-1,\ldots ,1,$

\begin{description}
\item[a)] Estimate $\upsilon _{k}\equiv \left( \tau _{k},\lambda _{k}\right) 
$ indexing the regression model 
\begin{equation*}
\eta _{k,\upsilon _{k}}\left( \overline{A}_{k},\overline{L}_{k}\right)
\equiv \Psi \left\{ \tau _{k}^{T}s_{k}\left( \overline{A}_{k},\overline{L}%
_{k}\right) +\lambda _{k}\left( 1/\widehat{\pi }^{k}\right) \right\}
\end{equation*}%
for $E\left( \widetilde{Y}_{k+1}|\overline{A}_{k},\overline{L}_{k},\right) $
restricted to units verifying $\pi ^{\ast k}>0$ with $\widetilde{\upsilon }%
_{k}\equiv \left( \widetilde{\tau }_{k},\widetilde{\lambda }_{k}\right) $
solving%
\begin{equation}
\mathbb{P}_{n}\left[ \pi ^{\ast k}\left[ 
\begin{array}{c}
s_{k}\left( \overline{A}_{k},\overline{L}_{k}\right) \\ 
1/\widehat{\pi }^{k}%
\end{array}%
\right] \left[ \widetilde{Y}_{k+1}-\Psi \left\{ \tau _{k}^{T}s_{k}\left( 
\overline{A}_{k},\overline{L}_{k}\right) +\lambda _{k}\left( 1/\widehat{\pi }%
^{k}\right) \right\} \right] \right] =0  \label{bang}
\end{equation}

\item[b) ] For units with $\pi ^{\ast k-1}>0,$ compute 
\begin{equation*}
\widetilde{Y}_{k}\equiv y_{k,\widetilde{\upsilon }_{k}}\left( \overline{A}%
_{k-1},\overline{L}_{k}\right) =\int h_{k}^{\ast }\left( a_{k}|\overline{A}%
_{k-1},\overline{L}_{k}\right) \eta _{k,\widetilde{\upsilon }_{k}}\left(
a_{k},\overline{A}_{k-1},\overline{L}_{k}\right) d\mu _{k}\left(
a_{k}\right) .
\end{equation*}
\end{description}
\end{description}

Finally, estimate $\theta \left( g\right) $ with $\widehat{\theta }_{Bang}=%
\mathbb{P}_{n}\left( \widetilde{Y}_{1}\right) .$

To analyze the limit distribution of $\widehat{\theta }_{Bang}$ and argue
that it is CAN under model $\left( \cap _{k=1}^{K}\mathcal{H}_{k}\right)
\cup $ $\left( \cap _{k=1}^{K}\mathcal{G}_{k}\right) $ we define $\widetilde{%
\eta }_{k}\equiv \eta _{k,\widetilde{\upsilon }_{k}}$ for $k\in \left[ K%
\right] $ and $\widetilde{\eta }\equiv \left( \widetilde{\eta }_{1},\ldots ,%
\widetilde{\eta }_{K}\right) $. Invoking Lemma 1 we obtain

\begin{equation}
\widehat{\theta }_{Bang}-\theta \left( g\right) \equiv \mathbb{P}_{n}\left\{
y_{1,\widetilde{\eta }_{1}}\left( L_{1}\right) \right\} -\widehat{E}%
_{g_{1}}\left\{ y_{1,\widetilde{\eta }_{1}}\left( L_{1}\right) \right\}
+d^{g}\left( \widetilde{\eta }\right)  \label{bang-exp}
\end{equation}

As in our analysis of the distribution of $\widehat{\theta }_{\mathcal{G}},$
to analyze the limiting distribution of $\widehat{\theta }_{Bang}$ we start
by noting that the vectors $\widetilde{\upsilon }\equiv \left( \widetilde{%
\upsilon }_{1},\ldots ,\widetilde{\upsilon }_{K}\right) $ and $\widehat{%
\alpha }_{ML}$ ultimately solve a joint system of estimating equations, so
under regularity conditions, $\widetilde{\upsilon }$ has a probability limit 
$v_{\lim }\left( p\right) \equiv \left( v_{1,\lim }\left( p\right) ,\ldots
,v_{K,\lim }\left( p\right) \right) $ under any $p\in \mathcal{P}.$
Furthermore, $\left\{ \widetilde{\upsilon }-v_{\lim }\left( p\right)
\right\} $ is asymptotically linear. Then, under regularity conditions that
imply the differentiability of the path $v\rightarrow d^{g}\left( \eta
_{\upsilon }\right) $ where $\eta _{\upsilon }\equiv \left( \eta
_{1,\upsilon },\ldots ,\eta _{K,\upsilon }\right) $, letting $\eta _{k,\lim
,Bang}\left( p\right) \equiv \eta _{k,v_{k,\lim }\left( p\right) },$ we have
that 
\begin{equation*}
d^{g}\left( \widetilde{\eta }\right) -d^{g}\left[ \eta _{\lim ,Bang}\left(
p\right) \right] \text{ is asymptotically linear. }
\end{equation*}

Furthermore, under regularity conditions, if $y_{1,\widetilde{\eta }_{1}}$
and $y_{1,\eta _{\lim ,Bang}\left( p\right) }$ fall in a Donsker class, then 
\begin{equation*}
\mathbb{P}_{n}\left\{ y_{1,\widetilde{\eta }_{1}}\left( L_{1}\right)
\right\} -\widehat{E}_{g_{1}}\left\{ y_{1,\widetilde{\eta }_{1}}\left(
L_{1}\right) \right\} =\mathbb{P}_{n}\left\{ y_{1,\eta _{\lim ,Bang}\left(
p\right) }\left( L_{1}\right) \right\} -E_{g_{1}}\left\{ y_{1,\eta _{\lim
,Bang}\left( p\right) }\left( L_{1}\right) \right\} +o_{p}\left(
n^{-1/2}\right) \text{ }
\end{equation*}%
is asymptotically linear. So, from expansion $\left( \ref{bang-exp}\right) ,$
we conclude that 
\begin{equation*}
\widehat{\theta }_{Bang}-\theta \left( g\right) -d^{g}\left[ \eta _{\lim
,Bang}\left( p\right) \right] \text{ is asymptotically linear.}
\end{equation*}

Now, because the limit values $v_{\lim }\left( p\right) $ and $\alpha _{\lim
}\left( h\right) \equiv \left( \alpha _{1,\lim }\left( h_{1}\right) ,\ldots
,\alpha _{K,\lim }\left( h_{K}\right) \right) $ of $\widetilde{\upsilon }$
and $\widehat{\alpha }_{ML}$ satisfy the population version of $\left( \ref%
{bang}\right) $ (i.e. with $\mathbb{P}_{n}$ replaced by $E_{gh}$ and all the
estimators replaced by their probability limits), then in particular, the
second row of equation $\left( \ref{bang}\right) $ implies that 
\begin{equation}
E_{\overline{g}_{k-1},\overline{h}_{k}}\left\{ \frac{1}{\pi _{\lim
}^{k}\left( \overline{h}_{k}\right) }\Delta _{k}\left( \eta _{k,\lim
,Bang}\left( p\right) ,\eta _{k+1,\lim ,Bang}\left( p\right) ;g_{k}\right)
\right\} =0,  \label{key-1}
\end{equation}%
where $\pi _{\lim }^{k}\left( \overline{h}_{k}\right) \equiv
\dprod\limits_{j=1}^{k}h_{j,\alpha _{j,\lim }\left( h_{j}\right) }\left(
A_{j}|\overline{A}_{j-1},\overline{L}_{j}\right) .$ Then, with $h_{\lim
}\left( h\right) \equiv \left( h_{1,\alpha _{1,\lim }\left( h_{1}\right)
},\ldots ,h_{K,\alpha _{K,\lim }\left( h_{K}\right) }\right) $, 
\begin{equation}
d^{g}\left[ \eta _{\lim ,Bang}\left( p\right) \right] =c^{p}\left[ h_{\lim
}\left( h\right) ,\eta _{\lim ,Bang}\left( p\right) \right] ,
\label{good-bang}
\end{equation}%
where for any $h^{\dag }=\left( h_{1}^{\dag },\ldots ,h_{K}^{\dag }\right) $
and $\eta ^{\dag }=\left( \eta _{1}^{\dag },\ldots ,\eta _{K}^{\dag }\right)
,$%
\begin{equation*}
c^{p}\left( h^{\dag },\eta ^{\dag }\right) \equiv \sum_{k=1}^{K}E_{\overline{%
g}_{k-1},\overline{h}_{k}}\left[ \left\{ \frac{1}{\pi ^{k}}-\frac{1}{\pi
^{\dag k}}\right\} \Delta _{k}\left( \eta _{k}^{\dag },\eta _{k+1}^{\dag
};g_{k}\right) \right]
\end{equation*}%
with $\pi ^{\dag k}\equiv \dprod\limits_{j=1}^{k}h_{j}^{\dag }\left( A_{j}|%
\overline{A}_{j-1},\overline{L}_{j}\right) .$ Note that, unlike $d^{g}\left(
\eta ^{\dag }\right) ,c^{p}\left( h^{\dag },\eta ^{\dag }\right) $ depends
on $p=gh$ not only through $g$ but also through $h.$

From $\left( \ref{good-bang}\right) $ we conclude that $\widehat{\theta }%
_{Bang}$ is CAN under $\left( \cap _{k=1}^{K}\mathcal{H}_{k}\right) \cup
\left( \cap _{k=1}^{K}\mathcal{G}_{k}\right) $ provided $c^{p}\left[ h_{\lim
}\left( h\right) ,\eta _{\lim ,Bang}\left( p\right) \right] =0$ for $p\in
\cap _{k=1}^{K}\mathcal{H}_{k}$ and for $p\in \cap _{k=1}^{K}\mathcal{G}%
_{k}. $

That $c^{p}\left[ h_{\lim }\left( h\right) ,\eta _{\lim ,Bang}\left(
p\right) \right] =0$ for $p\in \cap _{k=1}^{K}\mathcal{H}_{k}$ follows
immediately after recognizing that, under regularity conditions, the MLE $%
\widehat{\alpha }_{j,ML}$ is consistent so when $p\in \cap _{k=1}^{K}%
\mathcal{H}_{k},$ $h_{j,\alpha _{j,\lim }\left( h_{j}\right) }=h_{j}$ for
all $j.$

On the other hand, 
\begin{equation}
\eta _{k,\lim \text{,}Bang}\left( p\right) =\eta _{k}^{g}\text{ if }p\in
\cap _{j=k}^{K}\mathcal{G}_{j}  \label{lim-bang}
\end{equation}%
This result follows essentially along the same lines of the proof of $\left( %
\ref{result1}\right) $, upon noticing that when $p\in \mathcal{G}_{k}$ then $%
p$ also belongs to $\mathcal{G}_{k}^{\text{ext}}$ where\ $\mathcal{G}_{k}^{%
\text{ext}}$ is defined like $\mathcal{G}_{k}$ but with $\mathcal{R}%
_{k}^{ext}$ instead of $\mathcal{R}_{k}.$

We therefore conclude from $\left( \ref{delta-true}\right) $ and $\left( \ref%
{lim-bang}\right) $ that if $p\in \cap _{j=1}^{K}\mathcal{G}_{j},$ $\Delta
_{k}\left( \eta _{k,\lim }\left( p\right) ,\eta _{k+1,\lim }\left( p\right)
;g_{k}\right) =0$ for all $k\in \left[ K\right] ,$ and therefore $d^{g}\left[
\eta _{\lim ,Bang}\left( p\right) \right] =0.$

\subsection{K+1 - multiply robust estimation \label{sec:more-DR-estimation}}

\subsubsection{The Bang and Robins estimator is $K+1$ - multiply robust}

Surprisingly, a closer examination at the analysis of the asymptotic
properties of the Bang and Robins estimator in the preceding subsection
reveals, as we will argue next, that 
\begin{equation}
c^{p}\left[ h_{\lim }\left( h\right) ,\eta _{\lim ,Bang}\left( p\right) %
\right] =0\text{ if }p=gh\in \cup _{j=1}^{K+1}\left[ \left( \cap _{k=1}^{j-1}%
\mathcal{H}_{k}\right) \cap \left( \cap _{k=j}^{K}\mathcal{G}_{k}\right) %
\right]  \label{c-bang}
\end{equation}%
where $\cap _{k=1}^{0}\mathcal{H}_{k}\mathcal{\equiv }\cap _{k=K+1}^{K}%
\mathcal{G}_{k}\mathcal{\equiv P}$. The assertion in $\left( \ref{c-bang}%
\right) $ implies that under regularity conditions, $\widehat{\theta }%
_{Bang} $ is CAN not just under model $\left( \cap _{k=1}^{K}\mathcal{H}%
_{k}\right) $ $\cup \left( \cap _{k=1}^{K}\mathcal{G}_{k}\right) $ but also
under the lager model $\cup _{j=1}^{K+1}\left[ \left( \cap _{k=1}^{j-1}%
\mathcal{H}_{k}\right) \cap \left( \cap _{k=j}^{K}\mathcal{G}_{k}\right) %
\right] .$ This fact, that went unnoticed in B\&R, is a special case of a
general result on doubly robust estimation in factorized likelihood models
discussed in Molina et. al. (2017). Thus $\widehat{\theta }_{Bang}\,\ $%
confers even more robustness to model misspecification than that claimed in
B\&R, for it is CAN for $\theta \left( g\right) $ not only when one of the
following occurs, (i) the models for all the $h_{k}$ are correct, or (ii)
the models for all the $\eta _{k}$ are correct, but also when (iii) for some 
$j\in \left[ K-1\right] \,\ $the models for $h_{k},1\leq k\leq j$ and the
models for $\eta _{k},j+1\leq k\leq K$ are all correct. We designate an
estimator that is CAN whenever (i), (ii) or (iii) holds, a $\left(
K+1\right) -$ multiply robust estimator.

To show $\left( \ref{c-bang}\right) ,$ suppose that for some $j\in \left[ K%
\right] $, $p\in \left( \cap _{k=1}^{j-1}\mathcal{H}_{k}\right) \cap \left(
\cap _{k=j}^{K}\mathcal{G}_{k}\right) .$ Because $p\in \cap _{k=1}^{j-1}%
\mathcal{H}_{k},$ $\pi _{\lim }^{k}\left( \overline{h}_{k}\right) =\pi ^{k}$
for $k=1,\ldots ,j-1,$ so the first $j-1$ terms in the sum involved in $c^{p}%
\left[ h_{\lim }\left( h\right) ,\eta _{\lim ,Bang}\left( p\right) \right] $
are 0. On the other hand, when $p\in \cap _{k=j}^{K}\mathcal{G}_{k},$ it
follows from $\left( \ref{lim-bang}\right) $ and $\left( \ref{delta-true}%
\right) $ that $\Delta _{k}\left( \eta _{k,\lim }\left( p\right) ,\eta
_{k+1,\lim }\left( p\right) ;g_{k}\right) =0$ for $k=j,\ldots ,K$ so the
last $K-j+1$ terms of the summation involved in $c^{p}\left[ h_{\lim }\left(
h\right) ,\eta _{\lim ,Bang}\left( p\right) \right] $ also vanish.

\subsubsection{The greedy iterated fit $K+1$ - multiply robust estimators 
\label{other-mr}}

The B\&R estimator is not the only $K+1$ multiply robust estimator in model $%
\cup _{j=1}^{K+1}\left[ \left( \cap _{k=1}^{j-1}\mathcal{H}_{k}\right) \cap
\left( \cap _{k=j}^{K}\mathcal{G}_{k}\right) \right] .$ In fact, an
examination of the steps followed in the analysis of the preceding
subsection reveals that any estimator, say $\widehat{\widehat{\theta }},$ of 
$\theta $ that admits the expansion 
\begin{equation*}
\widehat{\widehat{\theta }}-\theta \left( g\right) =\mathbb{P}_{n}\left\{
y_{1,\widehat{\widehat{\eta }}_{1}}\left( L_{1}\right) \right\} -\widehat{E}%
_{g_{1}}\left\{ y_{1,\widehat{\widehat{\eta }}_{1}}\left( L_{1}\right)
\right\} +d^{g}\left( \widehat{\widehat{\eta }}\right)
\end{equation*}%
and verifies

\begin{description}
\item[1)] $\mathbb{P}_{n}\left\{ y_{1,\widehat{\widehat{\eta }}_{1}}\left(
L_{1}\right) \right\} -\widehat{E}_{g_{1}}\left\{ y_{1,\widehat{\widehat{%
\eta }}_{1}}\left( L_{1}\right) \right\} $ is asymptotically linear and,

\item[2)] $d^{g}\left( \widehat{\widehat{\eta }}\right) -c^{p}\left[ h_{\lim
}\left( h\right) ,\eta _{\lim }\left( p\right) \right] $ is asymptotically
linear for some $\left( h_{\lim }\left( h\right) ,\eta _{\lim }\left(
p\right) \right) \,\ $satisfying

\begin{description}
\item[i)] $\eta _{k,\lim }\left( p\right) =\eta _{k}^{g}$ when $p\in \cap
_{j=k}^{K}\mathcal{G}_{j},$ and

\item[ii)] $h_{k,\lim }\left( h\right) =h_{k}$ when $p\in \mathcal{H}_{k}$
\end{description}
\end{description}

will be $K+1$ multiply robust in model $\cup _{j=1}^{K+1}\left[ \left( \cap
_{k=1}^{j-1}\mathcal{H}_{k}\right) \cap \left( \cap _{k=j}^{K}\mathcal{G}%
_{k}\right) \right] $.

We now describe two estimators which satisfy these conditions. The first is
the output of a slight modification of Algorithm 2, whereby the parameters $%
\tau _{k}$ and $\lambda _{k}$ of the extended model $\mathcal{R}_{k}^{ext}$
are estimated greedily: first $\tau _{k}$ is estimated under the original
model $\mathcal{R}_{k}$ and next $\lambda _{k}$\ is estimated under $%
\mathcal{R}_{k}^{ext}$ but assuming $\tau _{k}$ is fixed and known and equal
to its estimated value. In the book Targeted Learning (2011), van der Laan
and Rose, emphasize the utility of such a greedy version of the B\&R plug-in
estimator, as a greedy fit makes it easy to replace parametric estimators of 
$\eta _{k,\tau _{k}}\left( \overline{A}_{k},\overline{L}_{k}\right) $\ by
more data adaptive machine learning estimators.

\begin{description}
\item[Algorithm 3. (\textit{Greedy iterated regression fit).}] Set $\overset{%
\smile }{Y}_{K+1}\equiv \psi \left( \overline{L}_{K+1}\right) $ and for $%
k=K,K-1,\ldots ,1,$

\begin{description}
\item[a.1) ] Estimate $\tau _{k}$ indexing the regression model 
\begin{equation*}
\eta _{k,\tau _{k}}\left( \overline{A}_{k},\overline{L}_{k}\right) \equiv
\Psi \left\{ \tau _{k}^{T}s_{k}\left( \overline{A}_{k},\overline{L}%
_{k}\right) \right\}
\end{equation*}%
for $E\left( \overset{\smile }{Y}_{k+1}|\overline{A}_{k},\overline{L}%
_{k}\right) $ restricted to units verifying $\pi ^{\ast k}>0$ with $\overset{%
\smile }{\tau }_{k}$ solving%
\begin{equation*}
\mathbb{P}_{n}\left[ \pi ^{\ast k}s_{k}\left( \overline{A}_{k},\overline{L}%
_{k}\right) \left\{ \overset{\smile }{Y}_{k+1}-\Psi \left\{ \tau
_{k}^{T}s_{k}\left( \overline{A}_{k},\overline{L}_{k}\right) \right\}
\right\} \right] =0
\end{equation*}

\item[a.2) ] Based on units with $\pi ^{\ast k}>0,$ estimate $\lambda _{k}$
indexing the regression model for $E\left( \overset{\smile }{Y}_{k+1}|%
\overline{A}_{k},\overline{L}_{k}\right) :$ $\eta _{k,\overset{\smile }{\tau 
}_{k},\lambda _{K}}\left( \overline{A}_{k},\overline{L}_{k}\right) \equiv
\Psi \left\{ \overset{\smile }{\tau }_{k}^{T}s_{k}\left( \overline{A}_{k},%
\overline{L}_{k}\right) +\lambda _{k}\left( 1/\widehat{\pi }^{k}\right)
\right\} $ which has offset $\overset{\smile }{\tau }_{k}^{T}s_{k}\left( 
\overline{A}_{k},\overline{L}_{k}\right) $ with $\overset{\smile }{\lambda }%
_{k}$ solving%
\begin{equation*}
\mathbb{P}_{n}\left[ \pi ^{\ast k}\left( 1/\widehat{\pi }^{k}\right) \left[ 
\overset{\smile }{Y}_{k+1}-\Psi \left\{ \overset{\smile }{\tau }%
_{k}^{T}s_{k}\left( \overline{A}_{k},\overline{L}_{k}\right) +\lambda
_{k}\left( 1/\widehat{\pi }^{k}\right) \right\} \right] \right] =0
\end{equation*}

\item[b) ] For units with $\pi ^{\ast k-1}>0,$ compute 
\begin{eqnarray*}
\overset{\smile }{Y}_{k} &\equiv &y_{k,\overset{\smile }{\tau }_{k},\overset{%
\smile }{\lambda }_{k}}\left( \overline{A}_{k-1},\overline{L}_{k}\right) \\
&\equiv &\int h_{k}^{\ast }\left( a_{k}|\overline{A}_{k-1},\overline{L}%
_{k}\right) \Psi \left\{ \overset{\smile }{\tau }_{k}^{T}s_{k}\left( a\,_{k},%
\overline{A}_{k-1},\overline{L}_{k}\right) +\overset{\smile }{\lambda }_{k}%
\widehat{\pi }^{k}\left( a\,_{k},\overline{A}_{k-1},\overline{L}_{k}\right)
^{-1}\right\} d\mu _{k}\left( a_{k}\right) .
\end{eqnarray*}
\end{description}
\end{description}

Finally, $\widehat{\theta }_{greed}=\mathbb{P}_{n}\left( \overset{\smile }{Y}%
_{1}\right) .$

To analyze the limiting behavior of \bigskip $\widehat{\theta }_{greed},$ we
define, for $k\in \left[ K\right] ,$ $\overset{\smile }{\eta }_{k}\left( 
\overline{A}_{k},\overline{L}_{k}\right) \equiv \overset{\smile }{\eta }_{k,%
\overset{\smile }{\tau },\overset{\smile }{\lambda }}\left( \overline{A}_{k},%
\overline{L}_{k}\right) \equiv \Psi \left\{ \overset{\smile }{\tau }%
_{k}^{T}s_{k}\left( \overline{A}_{k},\overline{L}_{k}\right) +\overset{%
\smile }{\lambda }_{k}\widehat{\pi }^{k}\left( \overline{A}_{k},\overline{L}%
_{k}\right) ^{-1}\right\} .$ Then, by Lemma 1, 
\begin{equation}
\widehat{\theta }_{greed}-\theta \left( g\right) \equiv \mathbb{P}%
_{n}\left\{ y_{1,\overset{\smile }{\eta }_{1}}\left( L_{1}\right) \right\} -%
\widehat{E}_{g_{1}}\left\{ y_{1,\overset{\smile }{\eta }_{1}}\left(
L_{1}\right) \right\} +d^{g}\left( \overset{\smile }{\eta }\right)
\label{exp-greed}
\end{equation}

As in our analyses of the distributions of $\widehat{\theta }_{\mathcal{G}}$
and $\widehat{\theta }_{Bang},$ to analyze the limiting distribution of $%
\widehat{\theta }_{greed}$ we start by noting that the vectors $\overset{%
\smile }{\tau }\equiv \left( \overset{\smile }{\tau }_{1},\ldots ,\overset{%
\smile }{\tau }_{K}\right) $ $,\overset{\smile }{\lambda }\equiv \left( 
\overset{\smile }{\lambda }_{1},\ldots ,\overset{\smile }{\lambda }%
_{K}\right) $ and $\widehat{\alpha }_{ML}$ solve a joint system of
estimating equations, so under regularity conditions, $\left( \overset{%
\smile }{\tau },\overset{\smile }{\lambda }\right) $ has a probability limit
under any $p\in \mathcal{P}$ which we denote with $\left( \tau ,\lambda
\right) _{\lim }\left( p\right) .$ Letting $\eta _{k,\lim ,greed}\left(
p\right) \equiv \eta _{k,\left( \tau ,\lambda \right) _{\lim }\left(
p\right) }$ we conclude that if the map $\left( \tau ,\lambda \right)
\rightarrow d^{g}\left( \eta _{\left( \tau ,\lambda \right) }\right) $ is
differentiable, then 
\begin{equation*}
d^{g}\left( \overset{\smile }{\eta }\right) -d^{g}\left[ \eta _{k,\lim
,greed}\left( p\right) \right] \text{ is asymptotically linear.}
\end{equation*}

Furthermore, if $y_{1,\overset{\smile }{\eta }_{1}}$ and $y_{1,\eta _{1,\lim
,greed}\left( p\right) }\,\ $fall in a Donsker class, then 
\begin{equation*}
\mathbb{P}_{n}\left\{ y_{1,\overset{\smile }{\eta }_{1}}\left( L_{1}\right)
\right\} -\widehat{E}_{g_{1}}\left\{ y_{1,\overset{\smile }{\eta }%
_{1}}\left( L_{1}\right) \right\} =\mathbb{P}_{n}\left\{ y_{1,\eta _{1,\lim
,greed}\left( p\right) }\left( L_{1}\right) \right\} -E_{g_{1}}\left\{
y_{1,\eta _{1,\lim ,greed}\left( p\right) }\left( L_{1}\right) \right\}
+o_{p}\left( n^{-1/2}\right) \text{ }
\end{equation*}%
is asymptotically linear. So, from expansion $\left( \ref{exp-greed}\right)
, $ we conclude that 
\begin{equation*}
\widehat{\theta }_{greed}-\theta \left( g\right) -d^{g}\left[ \eta _{k,\lim
,greed}\left( p\right) \right] \text{ is asymptotically linear}
\end{equation*}%
Just as for Algorithm 2, the inclusion of the covariate $1/\widehat{\pi }%
^{k} $ in the extended model fitted in step (a.2) of Algorithm 3 implies
that 
\begin{equation}
d^{g}\left[ \eta _{k,\lim ,greed}\left( p\right) \right] =c^{p}\left[
h_{\lim }\left( h\right) ,\eta _{k,\lim ,greed}\left( p\right) \right]
\label{greed}
\end{equation}%
So the $K+1$ multiply robustness of $\widehat{\theta }_{greed}$ in model $%
\cup _{j=1}^{K+1}\left[ \left( \cap _{k=1}^{j-1}\mathcal{H}_{k}\right) \cap
\left( \cap _{k=j}^{K}\mathcal{G}_{k}\right) \right] $ follows because 
\begin{equation*}
\eta _{k,\lim ,greed}\left( p\right) =\eta _{k}^{g}\text{ if }p\in \cap
_{j=k}^{K}\mathcal{G}_{j}
\end{equation*}

This result, whose proof we omit, follows essentially along the lines of the
proof of \ref{lim-bang}.

\subsubsection{The inverse probability weighted regression $K+1$ - MR
estimators \label{w-reg-dr}}

Here we will argue that the weighted-iterated regression estimators $%
\widehat{\theta }_{reg}$ defined like the estimator $\widehat{\theta }%
_{\omega }$ of section \ref{WITR} using weights $\omega _{k}=1/\widehat{\pi }%
_{k},$ is also $K+1$- multiply robust in model $\cup _{j=1}^{K+1}\left[
\left( \cap _{k=1}^{j-1}\mathcal{H}_{k}\right) \cap \left( \cap _{k=j}^{K}%
\mathcal{G}_{k}\right) \right] ,$ provided one of the components of $%
s_{k}\left( \overline{A}_{k},\overline{L}_{k}\right) $ is the constant 1.

Note that, unlike in Algorithms 2 or 3, to compute $\widehat{\theta }_{reg}$
we do not include the covariate $1/\widehat{\pi }^{k}$ in an extended
regression model. However, by using weights $\omega _{k}=1/\widehat{\pi }%
^{k} $ in equation $\left( \ref{itwr-eq}\right) $ and requiring that the
vector $s_{k}\left( \overline{A}_{k},\overline{L}_{k}\right) $ includes the
component 1, we ensure that 
\begin{equation}
d^{g}\left[ \eta _{\lim ,reg}\left( p\right) \right] =c^{p}\left[ h_{\lim
}\left( h\right) ,\eta _{\lim ,reg}\left( p\right) \right]  \label{good-reg}
\end{equation}%
where $\eta _{\lim ,reg}\left( p\right) =\eta _{k,\tau _{k,\lim ,reg}\left(
p\right) }$ with $\tau _{k,\lim ,reg}\left( p\right) $ the probability limit
of $\widehat{\tau }_{k,reg}.$ Furthermore, the same argument as in the proof
of $\left( \ref{result1}\right) $ shows that $\eta _{k,\lim ,reg}\left(
p\right) =\eta _{k}^{g}$ when $p\in \cap _{j=k}^{K}\mathcal{G}_{j}$. Thus,
the requirement (2.i) in the conditions listed at the beginning of section %
\ref{other-mr}. Since requirement (2.ii) holds as well, we conclude that
under regularity conditions, $\widehat{\theta }_{reg}$ $\ $is CAN in model $%
\cup _{j=1}^{K+1}\left[ \left( \cap _{k=1}^{j-1}\mathcal{H}_{k}\right) \cap
\left( \cap _{k=j}^{K}\mathcal{G}_{k}\right) \right] .$

\subsection{$\,2^{K}$ - multiply robust estimation\label%
{sec:2K-MR-estimation}}

\subsubsection{Theoretical results background}

Remarkably, it is possible to construct estimators of $\theta \left(
g\right) $ which are CAN under the even larger model $\cap _{k=1}^{K}\left( 
\mathcal{H}_{k}\cup \mathcal{G}_{k}\right) $ than $\cup _{j=1}^{K+1}\left[
\left( \cap _{k=1}^{j-1}\mathcal{H}_{k}\right) \cap \left( \cap _{k=j}^{K}%
\mathcal{G}_{k}\right) \right] .$ Such estimators, which we designate as $%
2^{K}$-multiply robust, confer even more protection against model
misspecification than $\widehat{\theta }_{bang},\widehat{\theta }_{greed},$ $%
\widehat{\theta }_{reg}.$ Their construction is motivated by the following
theoretical result in Molina et. al. (2017).

Given $h_{k}\left( A_{k}|\overline{A}_{k-1},\overline{L}_{k}\right) $ and $%
\eta _{k}\left( \overline{A}_{k},\overline{L}_{k}\right) ,k\in \left[ K%
\right] ,$ define for each $j\in \left[ K\right] $ the random variable 
\begin{eqnarray}
Q_{j}\left( \overline{h}_{j}^{K},\overline{\eta }_{j}^{K}\right) &\equiv &%
\frac{\pi _{j}^{\ast K}}{\pi _{j}^{K}}\psi \left( \overline{L}_{K+1}\right)
-\sum_{k=j}^{K}\left\{ \frac{\pi _{j}^{\ast k}}{\pi _{j}^{k}}\eta _{k}\left( 
\overline{A}_{k},\overline{L}_{k}\right) -\frac{\pi _{j}^{\ast \left(
k-1\right) }}{\pi _{j}^{\left( k-1\right) }}y_{k,\eta _{k}}\left( \overline{A%
}_{k-1},\overline{L}_{k}\right) \right\}  \label{qj} \\
&=&y_{j,\eta _{j}}\left( \overline{A}_{j-1},\overline{L}_{j}\right)
+\sum_{k=j}^{K}\frac{\pi _{j}^{\ast k}}{\pi _{j}^{k}}\left\{ y_{k+1,\eta
_{k+1}}\left( \overline{A}_{k},\overline{L}_{k+1}\right) -\eta _{k}\left( 
\overline{A}_{k},\overline{L}_{k}\right) \right\}  \notag \\
&=&\frac{h_{j}^{\ast }\left( A_{j}|\overline{A}_{j-1},\overline{L}%
_{j}\right) }{h_{j}\left( A_{j}|\overline{A}_{j-1},\overline{L}_{j}\right) }%
\left\{ Q_{j+1}\left( \overline{h}_{j+1}^{K},\overline{\eta }%
_{j+1}^{K}\right) -\eta _{j}\left( \overline{A}_{j},\overline{L}_{j}\right)
\right\} +y_{j,\eta _{j}}\left( \overline{A}_{j-1},\overline{L}_{j}\right) .
\notag
\end{eqnarray}%
where $y_{K+1,\eta _{K}}\left( \overline{A}_{K},\overline{L}_{K+1}\right)
\equiv \psi \left( \overline{L}_{K+1}\right) .$

Lemma 6 of Molina et al. (2017) implies that if for each $k\in \left[ K%
\right] $ either $h_{k}^{\dag }=h_{k}$ or $\eta _{k}^{\dag }=\eta
_{k}^{g},\,\ $then 
\begin{equation}
\theta \left( g\right) =E_{gh}\left[ Q_{1}\left( \overline{h}_{1}^{\dag K},%
\overline{\eta }_{1}^{\dag K}\right) \right]  \label{a1}
\end{equation}%
and, if for each $k\in \left\{ j+1,\ldots ,K\right\} $ either $h_{k}^{\dag
}=h_{k}$ or $\eta _{k}^{\dag }=\eta _{k}^{g}$ then 
\begin{equation}
\eta _{j}^{g}\left( \overline{A}_{j},\overline{L}_{j}\right) =E_{gh}\left\{
\left. Q_{j+1}\left( \overline{h}_{j+1}^{\dag K},\overline{\eta }%
_{j+1}^{\dag K}\right) \right\vert \overline{A}_{j},\overline{L}_{j}\right\}
\label{a2}
\end{equation}

Now, define for arbitrary $\eta _{k}^{\dag }\equiv \eta _{k}^{\dag }\left( 
\overline{A}_{k},\overline{L}_{k}\right) ,h_{k}^{\dag }\equiv h_{k}^{\dag
}\left( A_{k}|\overline{A}_{k-1},\overline{L}_{k}\right) $ and $p=gh,$

\begin{equation*}
a^{p}\left( h^{\dag },\eta ^{\dag }\right) \equiv
\sum_{k=1}^{K}E_{gh}\left\{ \frac{\pi ^{\ast \left( k-1\right) }}{\pi ^{\dag
\left( k-1\right) }}\left( \frac{h_{k}^{\ast }}{h_{k}}-\frac{h_{k}^{\ast }}{%
h_{k}^{\dag }}\right) \left( \eta _{k}^{\dag }-\eta _{k}^{g}\right) \right\}
\end{equation*}%
and for any unit satisfying $\pi ^{\ast j}>0$ define 
\begin{equation*}
a_{j}^{p}\left( \overline{h}_{j+1}^{\dag K},\eta _{j+1}^{\dag ^{K}};%
\overline{A}_{j},\overline{L}_{j}\right) \equiv \sum_{k=j+1}^{K}E_{%
\underline{g}_{j},\underline{h}_{j+1}}\left\{ \left. \frac{\pi _{j+1}^{\ast
\left( k-1\right) }}{\pi _{j+1}^{\dag \left( k-1\right) }}\left( \frac{%
h_{k}^{\ast }}{h_{k}}-\frac{h_{k}^{\ast }}{h_{k}^{\dag }}\right) \left( \eta
_{k}^{\dag }-\eta _{k}^{g}\right) \right\vert \overline{A}_{j},\overline{L}%
_{j}\right\}
\end{equation*}%
where $\pi ^{\dag \left( k-1\right) }\equiv
\dprod\limits_{j=1}^{k-1}h_{j}^{\dag }\left( A_{j}|\overline{A}_{j-1},%
\overline{L}_{j}\right) $ and $h_{k}\equiv h_{k}\left( A_{k}|\overline{A}%
_{k-1},\overline{L}_{k}\right) .$ In the Appendix we show the following
Lemma.

\textbf{Lemma 2:} Define $Q_{K}\left( \overline{h}_{K+1}^{\dag K},\overline{%
\eta }_{K+1}^{\dag K}\right) \equiv \psi \left( \overline{L}_{K+1}\right) $
and $\sum_{k=K+1}^{K}\left( \cdot \right) =0.$ The following holds:

i) for arbitrary $\eta _{k}^{\dag },h_{k}^{\dag }$ and $p=gh,k\in \left[ K%
\right] ,$ \ 
\begin{equation}
E_{gh}\left\{ Q_{1}\left( \overline{h}_{1}^{\dag K},\overline{\eta }%
_{1}^{\dag K}\right) \right\} -\theta \left( g\right) =a^{p}\left( h^{\dag
},\eta ^{\dag }\right)  \label{bias1}
\end{equation}

ii) for any $j\in \left[ K\right] $ and arbitrary $h_{k}^{\dag }$ and $\eta
_{k}^{\dag },$ $k\in \left\{ j+1,\ldots ,K\right\} ,$ if $\pi ^{\ast j}>0$
then 
\begin{equation}
E_{\underline{g}_{j},\underline{h}_{j+1}}\left\{ \left. Q_{j+1}\left( 
\overline{h}_{j+1}^{\dag K},\overline{\eta }_{j+1}^{\dag K}\right)
\right\vert \overline{A}_{j},\overline{L}_{j}\right\} -\eta _{j}^{g}\left( 
\overline{A}_{j},\overline{L}_{j}\right) =a_{j}^{p}\left( \overline{h}%
_{j+1}^{\dag K},\eta _{j+1}^{\dag ^{K}};\overline{A}_{j},\overline{L}%
_{j}\right)  \label{bias00}
\end{equation}

By Lemma 2, the right hand side in display $\left( \ref{bias1}\right) $ is
equal to the bias of $\mathbb{P}_{n}\left[ Q_{1}\left( \overline{h}%
_{1}^{\dag K},\overline{\eta }_{1}^{\dag K}\right) \right] $ as an estimator
of $\theta \left( g\right) $ for fixed functions $h_{k}^{\dag }$ and $\eta
_{k}^{\dag },$ $k\in \left[ K\right] .$ We see that it is comprised by a sum
of $K$ terms. Each term is equal to 0 if either $h_{k}^{\dag }$ $=h_{k}$ or $%
\eta _{k}^{\dag }=\eta _{k}^{g}.$

\subsubsection{Iterated regressions of multiply robust outcomes}

The theoretical results of the preceding subsection suggest that the
estimator $\widehat{\theta }_{MR}\equiv \mathbb{P}_{n}\left( \widehat{Q}%
_{1}\right) $ where $\widehat{Q}_{1}$ is the random variable returned by the
following algorithm is, under regularity conditions, $2^{K}$-multiply robust
CAN for $\theta \left( g\right) \,\ $in model $\cap _{k=1}^{K}\left( 
\mathcal{H}_{k}\cup \mathcal{G}_{k}\right) .$

\begin{description}
\item[Algorithm 4. \textit{(Iterated regression of multiply robust outcomes)}%
] Set $\widehat{Q}_{K+1}\equiv \psi \left( \overline{L}_{K+1}\right) $ and
for $k=K,K-1,\ldots ,1,$

\begin{description}
\item[a) ] Estimate $\tau _{k}$ indexing the regression model 
\begin{equation*}
\eta _{k,\tau _{k}}\left( \overline{A}_{k},\overline{L}_{k}\right) \equiv
\Psi \left\{ \tau _{k}^{T}s_{k}\left( \overline{A}_{k},\overline{L}%
_{k}\right) \right\}
\end{equation*}%
for $E\left( \widehat{Q}_{k+1}|\overline{A}_{k},\overline{L}_{k}\right) $
restricted to units verifying $\pi ^{\ast k}>0$ with $\widehat{\tau }_{k,MR}$
solving%
\begin{equation}
\mathbb{P}_{n}\left[ \pi ^{\ast k}s_{k}\left( \overline{A}_{k},\overline{L}%
_{k}\right) \left\{ \widehat{Q}_{k+1}-\Psi \left\{ \tau _{k}^{T}s_{k}\left( 
\overline{A}_{k},\overline{L}_{k}\right) \right\} \right\} \right] =0
\label{b1}
\end{equation}

\item[b) ] For units with $\pi ^{\ast k-1}>0,$ compute 
\begin{equation*}
\text{ }\widehat{Y}_{k,MR}\equiv \widehat{y}_{k,MR}\left( \overline{A}_{k-1},%
\overline{L}_{k}\right) \equiv \dint h_{k}^{\ast }\left( a_{k}|\overline{A}%
_{k-1},\overline{L}_{k}\right) \eta _{k,\widehat{\tau }_{k,MR}^{T}}\left(
a_{k},\overline{A}_{k-1},\overline{L}_{k}\right) d\mu _{k}\left( a_{k}\right)
\end{equation*}%
and%
\begin{equation*}
\widehat{Q}_{k}\equiv \frac{h_{k}^{\ast }}{\widehat{h}_{k}}\left[ \widehat{Q}%
_{k+1}-\eta _{k,\widehat{\tau }_{k,MR}^{T}}\left( \overline{A}_{k},\overline{%
L}_{k}\right) \right] +\widehat{Y}_{k,MR}.
\end{equation*}
\end{description}
\end{description}

Tchetgen-Tchetgen (2009) proposed Algorithm 4 for estimation of the mean of
an outcome at the end of longitudinal study with monotone missing at random
data, i.e. for the target parameter $\theta \left( g\right) $ of example 1.
The estimator $\widehat{\theta }_{MR}$ from Algorithm 4 for the mean of $%
\theta \left( g\right) $ an arbitrary g-functional follows by applying the
general theory for constructing multiply robust estimating functions
discussed in Molina et al. (2017).

To analyze the limiting distribution of $\widehat{\theta }_{MR}\,\ $and that
of several estimators that we shall introduce later, define for any $\eta
_{j}^{\dag }$ and $h_{j}^{\dag },j\in \left[ K\right] ,$ and all $k\in \left[
K\right] ,$

\medskip 
\begin{equation}
\Gamma _{k}\left( \overline{h}_{k+1}^{\dag K},\overline{\eta }_{k}^{\dag
K};g_{k}\right) \equiv \pi ^{\ast k}\left[ \eta _{k}^{\dag }\left( \overline{%
A}_{k},\overline{L}_{k}\right) -E_{g_{k}}\left\{ \left. Q_{k+1}\left( 
\overline{h}_{k+1}^{\dag K},\overline{\eta }_{k+1}^{\dag K}\right)
\right\vert \overline{A}_{k},\overline{L}_{k}\right\} \right]
\end{equation}%
where, recall, $Q_{K}\left( \overline{h}_{K+1}^{\dag K},\overline{\eta }%
_{K+1}^{\dag K}\right) \equiv \psi \left( \overline{L}_{K+1}\right) .$

From Lemma 2 we have that if $\pi ^{\ast k}>0$ then 
\begin{equation}
\Gamma _{k}\left( \overline{h}_{k+1}^{\dag K},\overline{\eta }_{k}^{\dag
K};g_{k}\right) =0\text{ if }\eta _{k}^{\dag }=\eta _{k}^{g}\text{ and for }%
j>k,\text{either }\eta _{j}^{\dag }=\eta _{j}^{g}\text{ or }h_{k}^{\dag
}=h_{k}^{g}  \label{gamma}
\end{equation}%
We further define 
\begin{equation*}
b^{p}\left( h^{\dag },\eta ^{\dag }\right) \equiv \sum_{k=1}^{K}E_{gh}\left[ 
\frac{1}{\pi _{1}^{\left( k-1\right) }}\left( \frac{1}{h_{k}}-\frac{1}{%
h_{k}^{\dag }}\right) \Gamma _{k}\left( \overline{h}_{k+1}^{\dag K},%
\overline{\eta }_{k}^{\dag K};g_{k}\right) \right] .
\end{equation*}

In the Appendix we show the following useful decompositions:

\textbf{Lemma 3:} For any $\eta _{j}^{\dag },h_{j}^{\dag },j\in \left[ K%
\right] ,$ it holds that $a^{p}\left( h^{\dag },\eta ^{\dag }\right)
=b^{p}\left( h^{\dag },\eta ^{\dag }\right) =c^{p}\left( h^{\dag },\eta
^{\dag }\right) $

The identity $a^{p}\left( h^{\dag },\eta ^{\dag }\right) =c^{p}\left(
h^{\dag },\eta ^{\dag }\right) $ will be helpful in our analysis, in section %
\ref{ML}, of machine learning doubly and multiply robust estimators. Aside
from this, it is interesting to note that we could have arrived at the
identities $\left( \ref{good-bang}\right) ,$ $\left( \ref{greed}\right) $
and $\left( \ref{good-reg}\right) $ by noticing that indeed because of the
special way in which the iterated regression functions $\eta _{k}^{g}$ are
estimated, it just happens that the doubly robust estimators defined earlier 
$\widehat{\theta }_{Bang},\widehat{\theta }_{greed}$ and $\widehat{\theta }%
_{reg}$ satisfy $\widehat{\theta }_{Bang}=\mathbb{P}_{n}\left[ Q_{1}\left( 
\widehat{\overline{h}}_{1}^{K},\widehat{\overline{\eta }}_{Bang,1}^{K}%
\right) \right] ,\widehat{\theta }_{greed}=\mathbb{P}_{n}\left[ Q_{1}\left( 
\widehat{\overline{h}}_{1}^{K},\widehat{\overline{\eta }}_{greed,1}^{K}%
\right) \right] $ and $\widehat{\theta }_{reg}=\mathbb{P}_{n}\left[
Q_{1}\left( \widehat{\overline{h}}_{1}^{K},\widehat{\overline{\eta }}%
_{reg,1}^{K}\right) \right] .$

The identity $a^{p}\left( h^{\dag },\eta ^{\dag }\right) =b^{p}\left(
h^{\dag },\eta ^{\dag }\right) $ and Lemma 2 immediately imply the following
representation for $\widehat{\theta }_{MR}.$

\begin{equation}
\widehat{\theta }_{MR}-\theta \left( g\right) =\mathbb{P}_{n}\left\{
Q_{1}\left( \widehat{\overline{h}}_{1}^{K},\widehat{\overline{\eta }}%
_{MR,1}^{K}\right) \right\} -\widehat{E}_{g_{1}}\left\{ Q_{1}\left( \widehat{%
\overline{h}}_{1}^{K},\widehat{\overline{\eta }}_{MR,1}^{K}\right) \right\}
+b^{p}\left( \widehat{h},\widehat{\eta }_{MR}\right)  \label{decom1}
\end{equation}%
where $\widehat{\eta }_{MR}\equiv \left( \widehat{\eta }_{1,MR},\ldots ,%
\widehat{\eta }_{K,MR}\right) $ with $\widehat{\eta }_{k,MR}\equiv \eta _{k,%
\widehat{\tau }_{k,MR}}.$

Just as we reasoned earlier, to analyze the limiting distribution of $%
\widehat{\theta }_{MR}$ we first note that the vectors $\widehat{\tau }%
_{MR}\equiv \left( \widehat{\tau }_{1,MR},\ldots ,\widehat{\tau }%
_{K,MR}\right) $ and $\widehat{\alpha }_{ML}$ ultimately solve a joint
system of estimating equations, so under regularity conditions, they have a
probability limit under any $p\in \mathcal{P}$ which we denote with $\tau
_{\lim ,MR}\left( p\right) $ and $\alpha _{\lim }\left( \overline{h}\right)
. $ Furthermore, $\left\{ \widehat{\tau }_{MR}-\tau _{\lim ,MR}\left(
p\right) \right\} $ is asymptotically linear. Then, letting $\eta _{k,\lim
,MR}\left( p\right) \equiv \eta _{k,\tau _{\lim ,MR}\left( p\right) },$ if
the map $\left( \alpha ,\tau \right) \rightarrow b^{p}\left( h_{\alpha
},\eta _{\tau }\right) $\ is differentiable we have that 
\begin{equation*}
b^{p}\left( \widehat{h},\widehat{\eta }_{MR}\right) -b^{p}\left[ h_{\lim
},\eta _{\lim ,MR}\left( p\right) \right] \text{ is asymptotically linear. }
\end{equation*}

Furthermore, if $\widehat{\overline{h}}_{1}^{K},\widehat{\overline{\eta }}%
_{MR,1}^{K},h_{\lim }\left( h\right) $ and $\eta _{\lim ,MR}\left( p\right) $
fall in a Donsker class, then 
\begin{eqnarray*}
&&\left. \mathbb{P}_{n}\left\{ Q_{1}\left( \widehat{\overline{h}}_{1}^{K},%
\widehat{\overline{\eta }}_{MR,1}^{K}\right) \right\} -E_{g_{1}}\left\{
Q_{1}\left( \widehat{\overline{h}}_{1}^{K},\widehat{\overline{\eta }}%
_{MR,1}^{K}\right) \right\} =\right. \\
&=&\mathbb{P}_{n}\left\{ Q_{1}\left( \overline{h}_{\lim ,1}^{K}\left(
h\right) ,\widehat{\overline{\eta }}_{\lim .MR,1}^{K}\left( p\right) \right)
\right\} -E_{g_{1}}\left\{ Q_{1}\left( \overline{h}_{\lim ,1}^{K}\left(
h\right) ,\widehat{\overline{\eta }}_{\lim .MR,1}^{K}\left( p\right) \right)
\right\} +o_{p}\left( n^{-1/2}\right) \text{ }
\end{eqnarray*}%
is asymptotically linear.

The representation $\left( \ref{decom1}\right) $ then implies that 
\begin{equation*}
\widehat{\theta }_{MR}-\theta \left( g\right) -b^{p}\left[ h_{\lim },\eta
_{\lim ,MR}\left( p\right) \right] \text{ is asymptotically linear}
\end{equation*}%
Below we show that, under regularity conditions, 
\begin{equation}
b^{p}\left[ h_{\lim },\eta _{\lim ,MR}\left( p\right) \right] =0\text{ if }%
p\in \cap _{k=1}^{K}\left( \mathcal{H}_{k}\cup \mathcal{G}_{k}\right)
\label{imp}
\end{equation}%
which then establishes that, under regularity conditions, $\widehat{\theta }%
_{MR}$ is CAN under model $\cap _{k=1}^{K}\left( \mathcal{H}_{k}\cup 
\mathcal{G}_{k}\right) .$

The assertion $\left( \ref{imp}\right) $ is essentially a consequence of the
following proposition.

\textbf{Proposition 2:} 
\begin{equation}
\eta _{k,\lim ,MR}\left( p\right) =\eta _{k}^{g}\text{ if }p\in \mathcal{G}%
_{k}\cap \left[ \cap _{j=k+1}^{K}\left( \mathcal{H}_{k}\cup \mathcal{G}%
_{k}\right) \right]  \label{etak}
\end{equation}%
which we now show by induction.

\begin{description}
\item \textbf{Proof of Proposition 2.} By reverse induction in $k.$ For $%
k=K, $ step (a) of Algorithm 4 is the same as step (a) of Algorithm 1, so $%
\eta _{K,\lim ,MR}\left( p\right) =\eta _{K,\lim ,\mathcal{G}}\left(
p\right) $ and consequently, under regularity conditions, $\left( \ref{etak}%
\right) $ holds for $k=K$ as was already established in the proof of
Proposition 1. \newline
Next, suppose that $\left( \ref{etak}\right) $ holds for $k=K,\ldots ,j+1.$
Noticing that, by construction, $\widehat{Q}_{j+1}=Q_{j+1}\left( \widehat{%
\overline{h}}_{,j+1}^{K},\widehat{\overline{\eta }}_{MR,j+1}^{K}\right) ,$
we conclude that $\widehat{\tau }_{j,MR}$ solves%
\begin{equation*}
0=\mathbb{P}_{n}\left[ \pi ^{\ast j}s_{j}\left( \overline{A}_{j},\overline{L}%
_{j}\right) \left\{ Q_{j+1}\left( \overline{h}_{\lim ,j+1}^{K},\overline{%
\eta }_{\lim ,j+1}^{K}\right) -\Psi \left\{ \tau _{j}^{T}s_{j}\left( 
\overline{A}_{j},\overline{L}_{j}\right) \right\} \right\} \right]
+o_{p}\left( 1\right)
\end{equation*}%
Suppose $p\in \mathcal{G}_{j}\cap \left[ \cap _{k=j+1}^{K}\left( \mathcal{H}%
_{k}\cup \mathcal{G}_{k}\right) \right] .$ Then, for each $k=j+1,\ldots ,K,$
either $p\in \mathcal{H}_{k}$ or $p\in \mathcal{G}_{k}.$ If $p\in \mathcal{G}%
_{k}$ then since $p$ also belongs to $\cap _{r=k+1}^{K}\left( \mathcal{H}%
_{r}\cup \mathcal{G}_{r}\right) $ we have by inductive hypothesis that $\eta
_{k,\lim ,MR}\left( p\right) =\eta _{k}^{g}.$ If $p\in \mathcal{H}_{k},$
then $h_{k,\lim }=h_{k}.$ Thus, for every $k=j+1,\ldots ,K,h_{k,\lim }=h_{k}$
or $\eta _{k,\lim ,MR}\left( p\right) =\eta _{k}^{g}.$ Consequently, by part
(ii) of Lemma 2, $E_{g_{j}}\left\{ \left. Q_{j+1}\left( \overline{h}_{\lim
,j+1}^{K},\overline{\eta }_{\lim ,j+1}^{K}\right) \right\vert \overline{A}%
_{j},\overline{L}_{j}\right\} =\eta _{j}^{g}\left( \overline{A}_{j},%
\overline{L}_{j}\right) .$ Furthermore, since $p\in \mathcal{G}_{j},\eta
_{j}^{g}=\eta _{j,\tau _{j}\left( g_{j}\right) }$ for some $\tau _{j}\left(
g_{j}\right) $ and therefore the equation%
\begin{equation*}
E_{\overline{g}_{j},\overline{h}_{j}}\left[ \pi ^{\ast j}s_{j}\left( 
\overline{A}_{j},\overline{L}_{j}\right) \left\{ Q_{j+1}\left( \overline{h}%
_{\lim ,j+1}^{K},\overline{\eta }_{\lim ,j+1}^{K}\right) -\Psi \left\{ \tau
_{j}^{T}s_{j}\left( \overline{A}_{j},\overline{L}_{j}\right) \right\}
\right\} \right] =0
\end{equation*}%
is solved at $\tau _{j}=\tau _{j}\left( g_{j}\right) .$ Then, under
regularity conditions for the consistency of $M-$ estimators, the
probability limit $\tau _{j,\lim ,MR}$ of $\widehat{\tau }_{j,MR}$ is equal
to $\tau _{j}\left( g_{j}\right) $ which shows $\left( \mathbf{\ref{etak}}%
\right) $ holds for $k=j.$
\end{description}

Having shown $\left( \mathbf{\ref{etak}}\right) $ we now show that $\left( %
\ref{imp}\right) $ holds by proving that for $p\in \cap _{r=k}^{K}\left( 
\mathcal{H}_{r}\cup \mathcal{G}_{r}\right) $ it holds that%
\begin{equation}
E_{\overline{g}_{k-1},\overline{h}_{k}}\left[ \frac{1}{\pi ^{\left(
j-1\right) }}\left( \frac{1}{h_{k}}-\frac{1}{h_{k,\lim }}\right) \Gamma
_{k}\left( \overline{h}_{\lim ,k+1}^{K},\overline{\eta }_{\lim
,MR,k}^{K}\left( p\right) ;g_{k}\right) \right] =0  \label{ii}
\end{equation}

Suppose then that $p\in \cap _{r=k}^{K}\left( \mathcal{H}_{r}\cup \mathcal{G}%
_{r}\right) .$ If $p\in \mathcal{H}_{k}\,\ $then $h_{k,\lim }=h_{k}$ and
thus $\left( \ref{ii}\right) $ holds. If $p\not\in \mathcal{H}_{k}$ then $%
p\in \mathcal{G}_{k}\cap \left[ \cap _{r=k+1}^{K}\left( \mathcal{H}_{r}\cup 
\mathcal{G}_{r}\right) \right] $. Then, by $\left( \ref{etak}\right) ,$ $%
\eta _{k,\lim ,MR}\left( p\right) =\eta _{k}^{g}.$ In addition, for $%
r=k+1,\ldots ,K,$ either $p\in \mathcal{H}_{r}\,\ $in which case $h_{r,\lim
}=h_{r}$ or $p\in \mathcal{G}_{r}\cap \left[ \cap _{s=r+1}^{K}\left( 
\mathcal{H}_{s}\cup \mathcal{G}_{s}\right) \right] $ in which case, again by 
$\left( \ref{etak}\right) ,\eta _{r,\lim ,MR}\left( p\right) =\eta _{r}^{g}.$
Thus, we conclude that when $p\in \mathcal{G}_{k}\cap \left[ \cap
_{r=k+1}^{K}\left( \mathcal{H}_{r}\cup \mathcal{G}_{r}\right) \right] ,$ $%
\eta _{k,\lim ,MR}\left( p\right) =\eta _{k}^{g}$ and for $r=k+1,\ldots ,K,$
either $\eta _{r,\lim ,MR}\left( p\right) =\eta _{r}^{g}$ or $h_{r,\lim
}=h_{r}.$ Thus, by $\left( \ref{gamma}\right) ,$ $\Gamma _{k}\left( 
\overline{h}_{\lim ,k+1}^{K},\overline{\eta }_{\lim ,MR,k}^{K}\left(
p\right) ;g_{k}\right) =0,$ which then implies that $\left( \ref{ii}\right) $
holds.

\begin{description}
\item[Remark 1. \textit{(Another K+1 - multiply robust estimator)}] By
estimating in Algorithm 4 each $\tau _{k}$ regressing $\widehat{Q}_{k+1}$ on 
$s_{k}\left( \overline{A}_{k},\overline{L}_{k}\right) $ we ensure that our
estimator $\widehat{\eta }_{k,MR}$ converges to $\eta _{k,\lim ,MR}\left(
p\right) $ satisfying $\left( \ref{etak}\right) $. Suppose instead that we
estimate $\tau _{k}$ with the estimator $\widehat{\tau }_{k,\mathcal{G}}$ of
Algorithm 1, but we estimate $\theta \left( g\right) $ with 
\begin{equation}
\widehat{\theta }_{DR}=\mathbb{P}_{n}\left\{ Q_{1}\left( \widehat{\overline{h%
}}_{1}^{K},\widehat{\overline{\eta }}_{1,\mathcal{G}}^{K}\right) \right\}
\label{theta_dr}
\end{equation}%
where, recall from section \ref{ITR}, $\widehat{\eta }_{\mathcal{G}}\equiv
\left( \widehat{\eta }_{1,\mathcal{G}},\ldots ,\widehat{\eta }_{K,\mathcal{G}%
}\right) $ and $\widehat{\eta }_{k,\mathcal{G}}\equiv \eta _{k,\widehat{\tau 
}_{k,\mathcal{G}}}.$ Then, with $\eta _{\lim ,\mathcal{G}}\left( p\right)
\equiv \eta _{k,\tau _{k,\lim ,\mathcal{G}}}$ defined as in section \ref{ITR}%
, we have that under regularity conditions, 
\begin{equation*}
\widehat{\theta }_{DR}-\theta \left( g\right) -b^{p}\left[ h_{\lim },\eta
_{\lim ,\mathcal{G}}\left( p\right) \right] \text{ is asymptotically linear.}
\end{equation*}%
However, unlike $b^{p}\left[ h_{\lim },\eta _{\lim ,MR}\left( p\right) %
\right] ,$ $b^{p}\left[ h_{\lim },\eta _{\lim ,\mathcal{G}}\left( p\right) %
\right] $ is not equal to 0 for all $p\in \cap _{k=1}^{K}\left( \mathcal{H}%
_{k}\cup \mathcal{G}_{k}\right) $ because by estimating $\tau _{k}$ with $%
\widehat{\tau }_{k,\mathcal{G}}$ we only ensure that $\eta _{k,\lim ,%
\mathcal{G}}\left( p\right) =\eta _{k}^{g}$ if $p\in \cap _{j=k}^{K}\mathcal{%
G}_{k}$, as established in $\left( \ref{result1}\right) ,$ but not
necessarily for $p$ in the bigger model $\mathcal{G}_{k}\cap \left[ \cap
_{j=k+1}^{K}\left( \mathcal{H}_{k}\cup \mathcal{G}_{k}\right) \right] $.
Yet, $\left( \ref{result1}\right) $ does imply that $b^{p}\left[ h_{\lim
},\eta _{\lim ,\mathcal{G}}\left( p\right) \right] =0$ for $p\in \cup
_{j=1}^{K+1}\left[ \left( \cap _{k=1}^{j-1}\mathcal{H}_{k}\right) \cap
\left( \cap _{k=j}^{K}\mathcal{G}_{k}\right) \right] .\,\ $This implies
that, under regularity conditions, $\widehat{\theta }_{DR}$ is another $K+1$
multiply robust estimator.
\end{description}

Algorithm 4 may not always be feasible. Specifically, if the link function $%
\Psi $ takes values in a bounded space, it may happen that the equation $%
\left( \ref{b1}\right) $ does not have a solution as the values of $\widehat{%
Q}_{k+1,MR}$ can be arbitrarily large. Such will be the case whenever $%
\widehat{\pi }_{k+1}^{K}$ is very close to 0 for some sample units. In fact,
even if we had succeeded in computing $\widehat{\tau }_{k,MR}$ for all $k,$
we may still face the possibility that $\widehat{\theta }_{MR}$ falls
outside the parameter space for $\theta \left( g\right) .$ For instance, if
the parameter space for $\theta \left( g\right) $ is the interval $\left(
-\sigma ,\sigma \right) $ for some $\sigma >0$ (a situation which occurs
when $\left\vert \varphi \left( z\right) \right\vert <\sigma $ for all $z$
where $\varphi \left( z\right) $ is defined in \ref{phi}$),$ $\widehat{%
\theta }_{MR}$ may fall outside the interval $\left( -\sigma ,\sigma \right) 
$ if for some units in the sample $\widehat{\pi }^{K}$ is very close to 0.
In the next subsection we discuss a number of ways in which this problem can
be overcome.

\subsubsection{Inverse probability weighted iterated regression. \label%
{w-reg-mr}}

There exist a number of ways to overcome the issues with unbounded outcomes
in Algorithm 4. In fact, remarkably, whenever for each $j\in \left[ K\right]
,$ $s_{j}\left( \overline{a}_{j},\overline{l}_{j}\right) $ can be decomposed
as 
\begin{equation}
s_{j}\left( \overline{a}_{j},\overline{l}_{j}\right) =\left[ 
\begin{array}{c}
b_{j}\left( \overline{a}_{j},\overline{l}_{j}\right) \\ 
s_{j-1}\left( \overline{a}_{j-1},\overline{l}_{j-1}\right)%
\end{array}%
\right]  \label{large}
\end{equation}%
for some known, possibly vector valued, function $b_{j},$ where $s_{0}\left( 
\overline{a}_{0},\overline{l}_{0}\right) \equiv 1,$ it just happens that the
weighted iterated regression estimator $\widehat{\theta }_{reg}$ of section %
\ref{WITR} using weights $\omega _{k}\left( \overline{A}_{k},\overline{L}%
_{k}\right) =1/\widehat{\pi }^{k}$ is $2^{K}$-multiply robust, i.e. it is
CAN\ for $\theta \left( g\right) $ in model $\cap _{k=1}^{K}\left( \mathcal{H%
}_{k}\cup \mathcal{G}_{k}\right) .$ Note that if $\left( \ref{large}\right)
\,\ $does not hold it it is always possible to enlarge the parametric class $%
\mathcal{R}_{j}$ by adding to the covariate vector $s_{j}\left( \overline{A}%
_{j},\overline{L}_{j}\right) $ the components of $s_{j-1}\left( \overline{A}%
_{j-1},\overline{L}_{j-1}\right) $ that are not in $s_{j}\left( \overline{A}%
_{j},\overline{L}_{j}\right) $ so as to ensure that $\left( \ref{large}%
\right) $ holds. Unlike\ the outcomes in step (a) of Algorithm 4, by
construction, the outcomes $\widehat{Y}_{k+1,reg}\equiv \widehat{Y}%
_{k+1,\omega },k\in \left[ K-1\right] ,$ being regressed to obtain the
estimator $\widehat{\tau }_{k,reg}$ are guaranteed to fall in the range of
the link function $\Psi \left( \cdot \right) .$Thus, so long as the sample
space of $\psi \left( \overline{L}_{K+1}\right) $ falls in the range of $%
\Psi \left( \cdot \right) ,$ the equation $\left( \ref{itwr-eq}\right) $
with $\omega _{k}\left( \overline{A}_{k},\overline{L}_{k}\right) =1/\widehat{%
\pi }^{k}$ and with $\widehat{Y}_{k+1,reg}$ replacing $\widehat{Y}%
_{k+1,\omega },$ is guaranteed to have a solution for all $k\in \left[ K%
\right] .$ Furthermore, if the range of $\Psi \left( \cdot \right) $ and
sample space of $\psi \left( \overline{L}_{K+1}\right) $ agree, then the
estimator $\widehat{\theta }_{reg}$ is guaranteed to fall in the sample
space of $\theta \left( g\right) $.

To see why $\widehat{\theta }_{reg}$ is $2^{K}$-multiply robust when $%
s_{j}\left( \overline{A}_{j},\overline{L}_{j}\right) $ is a sub-vector of $%
s_{k}\left( \overline{A}_{k},\overline{L}_{k}\right) $ for any $k>j,$ notice
that in such case $\widehat{\tau }_{k,reg}^{T}$ satisfies 
\begin{equation*}
\mathbb{P}_{n}\left[ \frac{\pi ^{\ast j}}{\widehat{\pi }^{j}}s_{j}\left( 
\overline{A}_{j},\overline{L}_{j}\right) \sum_{k=j+1}^{K}\left\{ \left( 
\widehat{Y}_{k+1,reg}-\widehat{\eta }_{k,reg}\right) \frac{\pi _{j+1}^{\ast
k}}{\widehat{\pi }_{j+1}^{k}}\right\} \right] =0
\end{equation*}%
where $\widehat{\eta }_{k,reg}\equiv \Psi \left\{ \widehat{\tau }%
_{k,reg}^{T}s_{k}\left( \overline{A}_{k},\overline{L}_{k}\right) \right\} .$
Thus, for any $j\in \left[ K\right] ,\widehat{\tau }_{j,reg}$ solves 
\begin{eqnarray*}
0 &=&\mathbb{P}_{n}\left\{ \frac{\pi ^{\ast j}}{\widehat{\pi }^{j}}%
s_{j}\left( \overline{A}_{j},\overline{L}_{j}\right) \left[ \widehat{Y}%
_{j+1,reg}+\sum_{k=j+1}^{K}\left\{ \left( \widehat{Y}_{k+1,reg}-\widehat{%
\eta }_{k,reg}\right) \frac{\pi _{j+1}^{\ast k}}{\widehat{\pi }_{j+1}^{k}}%
\right\} -\Psi \left( \tau _{j}^{T}s_{j}\left( \overline{A}_{j},\overline{L}%
_{j}\right) \right) \right] \right\} \\
&=&\mathbb{P}_{n}\left[ \frac{\pi ^{\ast j}}{\widehat{\pi }^{j}}s_{j}\left( 
\overline{A}_{j},\overline{L}_{j}\right) \left\{ \widehat{Q}_{j+1,reg}-\Psi
\left( \tau _{j}^{T}s_{j}\left( \overline{A}_{j},\overline{L}_{j}\right)
\right) \right\} \right]
\end{eqnarray*}%
where for any $j\in \left[ K-1\right] ,$ $\widehat{Q}_{j+1,reg}\equiv
Q_{j+1}\left( \overline{\widehat{h}}_{j+1}^{K},\overline{\widehat{\eta }}%
_{reg,j+1}^{K}\right) .$ Also, 
\begin{eqnarray*}
\widehat{\theta }_{reg} &=&\mathbb{P}_{n}\left[ \widehat{Y}%
_{1,reg}+\sum_{k=1}^{K}\frac{\pi ^{\ast k}}{\widehat{\pi }^{k}}\left( 
\widehat{Y}_{k+1,reg}-\widehat{\eta }_{k,reg}\right) \right] \\
&=&\mathbb{P}_{n}\left( \widehat{Q}_{1,reg}\right)
\end{eqnarray*}%
Thus, $\widehat{\theta }_{reg}$ is indeed the result of fitting Algorithm 4
except that in equation $\left( \ref{b1}\right) $ $\pi ^{\ast k}s_{k}\left( 
\overline{A}_{k},\overline{L}_{k}\right) /\widehat{\pi }^{k}$ instead of $%
s_{k}\left( \overline{A}_{k},\overline{L}_{k}\right) $ multiplies the
difference $\left\{ \widehat{Q}_{k+1}-\Psi \left( \tau _{k}^{T}s_{k}\left( 
\overline{A}_{k},\overline{L}_{k}\right) \right) \right\} .$ The proof that $%
\widehat{\theta }_{reg}$ is CAN for $\theta \left( g\right) $ under $\cap
_{k=1}^{K}\left( \mathcal{H}_{k}\cup \mathcal{G}_{k}\right) $ is essentially
the same as the proof that $\widehat{\theta }_{MR}$ is CAN under the same
model. Note, however, that the variances of the limiting mean zero normal
distributions of $\widehat{\theta }_{MR}$ and $\widehat{\theta }_{reg}$ are
not the same because $n^{1/2}\left\{ b^{p}\left( \widehat{h},\widehat{\eta }%
_{MR}\right) -b^{p}\left[ h_{\lim },\eta _{\lim ,MR}\left( p\right) \right]
\right\} $ and $n^{1/2}\left\{ b^{p}\left( \widehat{h},\widehat{\eta }%
_{reg}\right) -b^{p}\left[ h_{\lim },\eta _{\lim ,reg}\left( p\right) \right]
\right\} $ converge to different mean zero normal distributions.

\subsubsection{Greedy-fit multiply robust iterated regression.}

The estimator $\widehat{\theta }_{reg}$ of section \ref{w-reg-mr} requires
that one fits models $\mathcal{R}_{k}$ given in $\left( \ref{parg}\right) $
whose dimension grow with $k.$ When $K$ is large, step (a) of the algorithm
would then require the fit by (weighted) IRLS of a large model and thus may
result in numerical instability problems. The following extension of the
greedy fit Algorithm 3, results in a $2^{K}$-multiply robust estimator of $%
\theta \left( g\right) $ which does not require that the models $\mathcal{R}%
_{k}$ be of growing dimension. Furthermore, the estimation procedure never
requires the fit of a model whose parameter has dimension larger than $\max
\left\{ \dim \left( \tau _{k}\right) :k\in \left[ K\right] \right\} ,$ where 
$\tau _{k}$ is the parameter indexing model $\mathcal{R}_{k}.$ In what
follows $s_{0}\left( \overline{A}_{0},\overline{L}_{0}\right) \equiv 1.$

\begin{description}
\item[Algorithm 5. \textit{(Multiply robust estimation by greedy fit
iterated regression)}] \bigskip For $j\in \left[ K\right] $ set $\widehat{Y}%
_{K+1}^{\left( j\right) }=\psi \left( \overline{L}_{K+1}\right) ,$ define $%
s_{0}\left( \overline{A}_{0},\overline{L}_{0}\right) \equiv 1,$ and for any $%
k=K,K-1,\ldots ,1,$ repeat

\begin{description}
\item[a) ] Estimate $\tau _{k}$ indexing the regression model 
\begin{equation*}
\eta _{k,\tau _{k}}\left( \overline{A}_{k},\overline{L}_{k}\right) \equiv
\Psi \left\{ \tau _{k}^{T}s_{k}\left( \overline{A}_{k},\overline{L}%
_{k}\right) \right\}
\end{equation*}%
for $E\left( \widehat{Y}_{k+1}^{\left( k\right) }|\overline{A}_{k},\overline{%
L}_{k}\right) $ restricted to units verifying $\pi ^{\ast k}>0$ with $%
\widehat{\tau }_{k,greed}$ solving%
\begin{equation}
\mathbb{P}_{n}\left[ \pi ^{\ast k}s_{k}\left( \overline{A}_{k},\overline{L}%
_{k}\right) \left\{ \widehat{Y}_{k+1}^{\left( k\right) }-\Psi \left\{ \tau
_{k}^{T}s_{k}\left( \overline{A}_{k},\overline{L}_{k}\right) \right\}
\right\} \right] =0
\end{equation}

\item[b) ] For $j=k-1,k-2\ldots .,0,\,\ $repeat \{

\begin{description}
\item[i)] Estimate the parameter $\lambda _{k}^{\left( j\right) }$ indexing
the regression model\ \ 
\begin{equation*}
\left. \eta _{k,\lambda _{k}^{\left( j\right) }}^{\left( j\right) }\left( 
\overline{A}_{k},\overline{L}_{k}\right) \equiv \Psi \left\{ \widehat{\tau }%
_{k,greed}^{T}s_{k}\left( \overline{A}_{k},\overline{L}_{k}\right)
+\sum_{u=j+1}^{k-1}\widehat{\lambda }_{k}^{\left( u\right) }\frac{%
s_{u}\left( \overline{A}_{u},\overline{L}_{u}\right) }{\widehat{\pi }%
_{u+1}^{k}}+\lambda _{k}^{\left( j\right) }\frac{s_{j}\left( \overline{A}%
_{j},\overline{L}_{j}\right) }{\widehat{\pi }_{j+1}^{k}}\right\} \right.
\end{equation*}%
for $E\left( \widehat{Y}_{k+1}^{\left( j\right) }|\overline{A}_{k},\overline{%
L}_{k}\right) $ restricted to units verifying $\pi ^{\ast k}>0\,\ $with $%
\widehat{\lambda }_{k}^{\left( j\right) }$ solving 
\begin{equation}
\mathbb{P}_{n}\left\{ \pi ^{\ast k}\frac{s_{j}\left( \overline{A}_{j},%
\overline{L}_{j}\right) }{\widehat{\pi }_{j+1}^{k}}\left( \widehat{Y}%
_{k+1}^{\left( j\right) }-\eta _{k,\lambda _{k}^{\left( j\right) }}^{\left(
j\right) }\left( \overline{A}_{k},\overline{L}_{k}\right) \right) \right\} =0
\label{lambda}
\end{equation}%
Let $\widehat{\eta }_{k}^{\left( j\right) }\left( \overline{A}_{k},\overline{%
L}_{k}\right) \equiv \eta _{k,\widehat{\lambda }_{k}^{\left( j\right)
}}^{\left( j\right) }\left( \overline{A}_{k},\overline{L}_{k}\right) .$

\item[ii)] For units with $\pi ^{\ast k-1}>0,$ compute$\ $%
\begin{eqnarray*}
\widehat{Y}_{k}^{\left( j\right) } &\equiv &y_{k,\widehat{\eta }_{k}^{\left(
j\right) }}\left( \overline{A}_{k-1},\overline{L}_{k}\right) \\
&\equiv &\dint h_{k}^{\ast }\left( a_{k}|\overline{A}_{k-1},\overline{L}%
_{k}\right) \eta _{k,\widehat{\lambda }_{k}^{\left( j\right) }}^{\left(
j\right) }\left( a_{k},\overline{A}_{k-1},\overline{L}_{k}\right) d\mu
_{k}\left( a_{k}\right)
\end{eqnarray*}
\end{description}
\end{description}
\end{description}

\ \ \ \ \ \ \ \ \ \ \ \ \ \ \ \ \ \ \ \ \}

Finally, $\widehat{\theta }_{MR,greed}=\mathbb{P}_{n}\left( \widehat{Y}%
_{1}^{\left( 0\right) }\right) .$

By construction, each $\widehat{\tau }_{j,greed}$ is the estimated
coefficient in a regression on $\left( \overline{A}_{j},\overline{L}%
_{j}\right) $ of the outcome $\widehat{Y}_{j+1}^{\left( j\right) }$ with
weights $\pi ^{\ast j}$. On the other hand, step (b) of the algorithm (the
fit of the extended model) ensures precisely that $\widehat{\tau }_{j,greed}$
is also the estimated coefficient in a regression on $\left( \overline{A}%
_{j},\overline{L}_{j}\right) $ of the pseudo outcome $Q_{j+1}\left( 
\overline{\widehat{h}}_{j+1}^{K},\overline{\widehat{\eta }}_{j+1}^{\left(
j\right) ,K}\right) $ with weights $\pi ^{\ast j}.$ Specifically, by step
(a) $\widehat{\tau }_{j,greed}$ satisfies 
\begin{equation}
0=\mathbb{P}_{n}\left[ \pi ^{\ast j}s_{j}\left( \overline{A}_{j},\overline{L}%
_{j}\right) \left\{ \widehat{Y}_{j+1}^{\left( j\right) }-\eta _{j,\widehat{%
\tau }_{j,greed}}\right\} \right]  \label{aa}
\end{equation}%
Because, by step (b.i), for all $j<k\leq K,$%
\begin{equation*}
\mathbb{P}_{n}\left[ \pi ^{\ast j}s_{j}\left( \overline{A}_{j},\overline{L}%
_{j}\right) \left\{ \left( \widehat{Y}_{k+1}^{\left( j\right) }-\widehat{%
\eta }_{k}^{\left( j\right) }\right) \left( \pi _{j+1}^{\ast k}/\widehat{\pi 
}_{j+1}^{k}\right) \right\} \right] =0
\end{equation*}%
where,\ recall, $\widehat{\eta }_{k}^{\left( j\right) }\equiv \eta _{k,%
\widehat{\lambda }_{k}^{\left( j\right) }}^{\left( j\right) },$ then $\left( %
\ref{aa}\right) $ implies that 
\begin{equation*}
0=\mathbb{P}_{n}\left[ \pi ^{\ast j}s_{j}\left( \overline{A}_{j},\overline{L}%
_{j}\right) \left\{ \widehat{Y}_{j+1}^{\left( j\right) }-\eta _{j,\widehat{%
\tau }_{j,greed}}\right\} \right] +\mathbb{P}_{n}\left[ \sum_{k=j+1}^{K}\pi
^{\ast j}s_{j}\left( \overline{A}_{j},\overline{L}_{j}\right) \left\{ \left( 
\widehat{Y}_{k+1}^{\left( j\right) }-\widehat{\eta }_{k}^{\left( j\right)
}\right) \left( \pi _{j+1}^{\ast k}/\widehat{\pi }_{j+1}^{k}\right) \right\} %
\right]
\end{equation*}%
This last equality, in turn, is equal to 
\begin{equation}
0=\mathbb{P}_{n}\left[ \pi ^{\ast j}s_{j}\left( \overline{A}_{j},\overline{L}%
_{j}\right) \left\{ Q_{j+1}\left( \overline{\widehat{h}}_{j+1}^{K},\overline{%
\widehat{\eta }}_{j+1}^{\left( j\right) ,K}\right) -\eta _{j,\widehat{\tau }%
_{j,greed}}\left( \overline{A}_{j},\overline{L}_{j}\right) \right\} \right]
\label{bb}
\end{equation}

Furthermore, 
\begin{equation}
\widehat{\theta }_{MR,greed}=\mathbb{P}_{n}\left( \widehat{Y}_{1}^{\left(
0\right) }\right) =\mathbb{P}_{n}\left[ Q_{j+1}\left( \overline{\widehat{h}}%
_{1}^{K},\overline{\widehat{\eta }}_{1}^{\left( 0\right) ,K}\right) \right]
\label{bbgreed}
\end{equation}%
An analysis similar to that conducted for $\widehat{\theta }_{MR}$ now shows
that $\widehat{\theta }_{MR,greed}$ is $2^{K}$-multiply robust CAN for $%
\theta \left( g\right) $ in model $\cap _{k=1}^{K}\left( \mathcal{G}_{k}\cup 
\mathcal{H}_{k}\right) .$

\subsubsection{The multiply robust estimators in the missing data example 1}

We now illustrate the implementation of the estimators $\widehat{\theta }%
_{reg},\widehat{\theta }_{MR}$ and $\widehat{\theta }_{MR,greed}$ for $K=2$
in Example 1, assuming $\psi \left( \overline{L}_{K+1}\right) =L_{3}$ and $%
L_{3}$ is a binary outcome. In model $\mathcal{R}_{k},k=1,2,$ we need only
consider a parametric class for $\eta _{k}\left( \overline{l}_{k};\underline{%
g}_{k}\right) \equiv \eta _{k}\left( \overline{a}_{k}=1,\overline{l}_{k};%
\underline{g}_{k}\right) ,k=1,2,$ because in the algorithms that compute of $%
\widehat{\theta }_{reg},\widehat{\theta }_{MR}$ and $\widehat{\theta }%
_{MR,greed}$, the regression in step (a) is restricted to subjects with $\pi
^{\ast k}=1,$ i.e. with $\overline{A}_{k}=\overline{1}.$ As indicated in
section \ref{example1}, under the assumptions of Example 1, $\eta _{k}\left( 
\overline{l}_{k};\underline{g}_{k}\right) $ coincides with $E\left(
L_{3}^{\ast }|\overline{A}_{k}=\overline{1},\overline{L}_{k}=\overline{l}%
_{k}\right) ,k=1,2,$ where, recall, $L_{3}^{\ast }$ is the intended,
possibly unobserved outcome. If $L_{3}^{\ast }$ is binary, a natural
parametric model $\eta _{k,\tau _{k}}\left( \overline{l}_{k}\right) $ for $%
\eta _{k}\left( \overline{l}_{k};\underline{g}_{k}\right) $ is then a
logistic regression model 
\begin{equation*}
\eta _{k,\tau _{k}}\left( \overline{l}_{k}\right) =\text{expit}\left\{ \tau
_{k}^{T}s_{k}\left( \overline{l}_{k}\right) \right\}
\end{equation*}%
for a vector $s_{k}\left( \overline{l}_{k}\right) $ of given functions of $%
\overline{l}_{k}$ which includes one entry with the constant 1.

The calculation of $\widehat{\theta }_{reg},\widehat{\theta }_{MR}$ and $%
\widehat{\theta }_{MR,greed}$ requires that we first fit by maximum
likelihood parametric models for each $h_{k}.$ Since $A_{k}$ is binary and
by assumption, $A_{k}=0\Rightarrow A_{j}=0$ for $j>k,$ then a natural
parametric model for $h_{k}\left( a_{k}|\overline{a}_{k-1},\overline{l}%
_{k}\right) $ is 
\begin{equation*}
h_{k,\alpha _{k}}\left( a_{k}|\overline{a}_{k-1},\overline{l}_{k}\right)
=a_{k-1}exp\left\{ a_{k}\alpha _{k}^{T}r_{k}\left( \overline{l}_{k}\right)
\right\} /\left\{ 1+exp\left\{ \alpha _{k}^{T}r_{k}\left( \overline{l}%
_{k}\right) \right\} \right\}
\end{equation*}%
where $r_{k}\left( \overline{l}_{k}\right) $ is a vector of specified
functions of $\overline{l}_{k}$. That is, we assume that the probability of
response at cycle $k+1$ among those still in study at cycle $k$ follows a
logistic regression with covariate vector $r_{k}\left( \overline{l}%
_{k}\right) .$ Because by definition, $A_{0}=1$ the estimator $h_{1,\widehat{%
\alpha }_{ML,1}}\left( 1|\overline{A}_{0},\overline{L}_{1}\right) $ is the
fitted value from the logistic regression of the binary outcome $A_{1}$ on
the covariate vector $r_{1}\left( \overline{L}_{1}\right) $ among the entire
study participants. On the other hand, $h_{2,\widehat{\alpha }_{ML,2}}\left(
1|\overline{A}_{1},\overline{L}_{2}\right) $ is equal to 0 for subjects for
whom $L_{2}^{\ast }$ is missing, i.e. for which $A_{1}=0,$ and it is equal
to the fitted value from the logistic regression of outcome $A_{2}$ on
covariates $r_{2}\left( \overline{L}_{2}\right) $ among subjects for which $%
L_{2}^{\ast }$ is observed. In what follows we describe the three
algorithms. To simplify notation, we use the shortcuts $\widehat{h}%
_{1}\equiv h_{1,\widehat{\alpha }_{ML,1}}\left( 1|\overline{L}_{1}\right) $
and $\widehat{h}_{2}\equiv h_{2,\widehat{\alpha }_{ML,2}}\left( 1|A_{1}=1,%
\overline{L}_{2}\right) .$

In what follows we explain in detail the algorithm to compute $\widehat{%
\theta }_{MR,greed}$. To avoid repetition, the algorithms for computing $%
\widehat{\theta }_{reg},\widehat{\theta }_{MR}\,\ $are given with less detail

\begin{description}
\item[Calculation of $\widehat{\protect\theta }_{greed}.$ \textit{(Greedy
fit multiply robust estimation)}] 

\item[Steps for $k=2$] 

\item[(a)] This step requires that we use subjects with $\pi ^{\ast 2}>0.$
These are precisely the subjects with $A_{2}=1,$ i.e. the subjects that did
not dropped from the study. This step of the algorithm requires that we fit
model 
\begin{equation*}
\eta _{2,\tau _{2}}\left( \overline{A}_{2},\overline{L}_{2}\right) \equiv 
\text{expit}\left\{ \tau _{2}^{T}s_{2}\left( \overline{A}_{2},\overline{L}%
_{2}\right) \right\}
\end{equation*}%
just using subjects with $A_{2}=1.$ Because subjects with $A_{2}=1$ must
necessarily have $A_{1}=1,$ then the relevant model that we need to estimate
is 
\begin{equation*}
\eta _{2,\tau _{2}}\left( \overline{A}_{2}=1,\overline{L}_{2}\right) \equiv 
\text{expit}\left\{ \tau _{2}^{T}s_{2}\left( \overline{A}_{2}=1,\overline{L}%
_{2}\right) \right\}
\end{equation*}%
If, as we indicated at the start of this section, in a slight abuse of
notation we write $s_{2}\left( \overline{L}_{2}\right) \equiv s_{2}\left( 
\overline{A}_{2}=1,\overline{L}_{2}\right) ,$ then this step of the
algorithm boils down to computing the logistic regression estimator $%
\widehat{\tau }_{2,greed}$ for the outcome $\widehat{Y}_{3}^{\left( 2\right)
}\equiv L_{3}\,$on the covariate vector $s_{2}\left( \overline{L}_{2}\right) 
$ just using subjects $A_{2}=1.$ That is, $\widehat{\tau }_{2,greed}$
satisfies%
\begin{equation}
\mathbb{P}_{n}\left( A_{2}s_{2}\left( \overline{L}_{2}\right) \left[ 
\widehat{Y}_{3}^{\left( 2\right) }-\text{expit}\left\{ \widehat{\tau }%
_{2,greed}^{T}s_{2}\left( \overline{L}_{2}\right) \right\} \right] \right) =0
\label{g1}
\end{equation}

\item[Step (b) for $k=2,j=1$] 

\begin{description}
\item[(i)] In this step we are required to use again only subjects with $\pi
^{\ast 2}>0,$ so we continue to restrict the calculations to subjects with $%
A_{2}=1$. Using these subjects we are required to fit the logistic
regression model $\eta _{2,\lambda _{2}^{\left( 1\right) }}^{\left( 1\right)
}\left( \overline{A}_{2},\overline{L}_{2}\right) $ for $E\left( \widehat{Y}%
_{3}^{\left( 1\right) }|\overline{A}_{2},\overline{L}_{2}\right) $ where $%
\widehat{Y}_{3}^{\left( 1\right) }\equiv L_{3},$ 
\begin{eqnarray*}
\eta _{2,\lambda _{2}^{\left( 1\right) }}^{\left( 1\right) }\left( \overline{%
A}_{2},\overline{L}_{2}\right) &\equiv &\text{expit}\left\{ \widehat{\tau }%
_{2,greed}^{T}s_{2}\left( \overline{A}_{2},\overline{L}_{2}\right) +\lambda
_{2}^{\left( 1\right) }\frac{s_{1}\left( \overline{A}_{1},\overline{L}%
_{1}\right) }{\widehat{\pi }_{2}^{2}}\right\} \\
&\equiv &\text{expit}\left\{ \widehat{\tau }_{2,greed}^{T}s_{2}\left( 
\overline{A}_{2},\overline{L}_{2}\right) +\lambda _{2}^{\left( 1\right) }%
\frac{s_{1}\left( \overline{A}_{1},\overline{L}_{1}\right) }{\widehat{h}_{2}}%
\right\} ,
\end{eqnarray*}%
$\widehat{\tau }_{2,greed}^{T}s_{2}\left( \overline{A}_{2},\overline{L}%
_{2}\right) $ is an offset and $\lambda _{2}^{\left( 1\right) }$ is the
unknown parameter. Once again, because we are only using subjects with $%
\overline{A}_{2}=\overline{1},$ then we only care to fit the model for $%
E\left( \widehat{Y}_{3}^{\left( 1\right) }|\overline{A}_{2}=1,\overline{L}%
_{2}\right) :$%
\begin{equation*}
\eta _{2,\lambda _{2}^{\left( 1\right) }}^{\left( 1\right) }\left( \overline{%
A}_{2}=1,\overline{L}_{2}\right) \equiv \text{expit}\left\{ \widehat{\tau }%
_{2,greed}^{T}s_{2}\left( \overline{L}_{2}\right) +\lambda _{2}^{\left(
1\right) }\frac{s_{1}\left( \overline{L}_{1}\right) }{\widehat{h}_{2}}%
\right\}
\end{equation*}%
where $s_{k}\left( \overline{L}_{k}\right) \equiv s_{k}\left( \overline{A}%
_{k}=1,\overline{L}_{k}\right) ,k=1,2.\,\ $Thus, the estimator $\widehat{%
\lambda }_{2}^{\left( 1\right) }$ satisfies%
\begin{equation}
\mathbb{P}_{n}\left( A_{2}\frac{s_{1}\left( L_{1}\right) }{\widehat{h}_{2}}%
\left[ \widehat{Y}_{3}^{\left( 1\right) }-\text{expit}\left\{ \widehat{\tau }%
_{2,greed}^{T}s_{2}\left( \overline{L}_{2}\right) +\widehat{\lambda }%
_{2}^{\left( 1\right) T}\frac{s_{1}\left( L_{1}\right) }{\widehat{h}_{2}}%
\right\} \right] \right) =0  \label{g2}
\end{equation}

\item[(ii)] This step is calculated using only subjects with $\pi ^{\ast
1}>0\,,$ i.e. with $A_{1}=1.$ For these subjects we must compute%
\begin{eqnarray*}
\widehat{Y}_{2}^{\left( 1\right) } &\equiv &y_{2,\eta _{2,\widehat{\lambda }%
_{2}^{\left( 1\right) }}^{\left( 1\right) }}\left( A_{1},\overline{L}%
_{2}\right) \\
&\equiv &\int \eta _{2,\widehat{\lambda }_{2}^{\left( 1\right) }}^{\left(
1\right) }\left( A_{1},a_{2},\overline{L}_{2}\right) h_{2}^{\ast }\left(
a_{2}|A_{1},\overline{L}_{2}\right) d\mu \left( a_{2}\right) \\
&=&\sum_{a_{2}=0}^{1}\eta _{2,\widehat{\lambda }_{2}^{\left( 1\right)
}}^{\left( 1\right) }\left( A_{1},a_{2},\overline{L}_{2}\right) h_{2}^{\ast
}\left( a_{2}|A_{1},\overline{L}_{2}\right)
\end{eqnarray*}%
Now, because we are only doing the calculation for subjects with $A_{1}=1,$
and because $h_{2}^{\ast }\left( a_{2}|A_{1}=1,\overline{L}_{2}\right)
=a_{2},$ $\ $the last display simplifies to 
\begin{eqnarray*}
\widehat{Y}_{2}^{\left( 1\right) } &=&\eta _{2,\widehat{\lambda }%
_{2}^{\left( 1\right) }}^{\left( 1\right) }\left( A_{1}=1,A_{2}=1,\overline{L%
}_{2}\right) \\
&=&\text{expit}\left\{ \widehat{\tau }_{2,greed}^{T}s_{2}\left( \overline{L}%
_{2}\right) +\widehat{\lambda }_{2}^{\left( 1\right) }s_{1}\left(
L_{1}\right) /\widehat{h}_{2}\right\}
\end{eqnarray*}
$\ $
\end{description}

\item[Step (b) for $k=2,j=0$] 

\begin{description}
\item[(i)] In this step we are required to use again only subjects with $\pi
^{\ast 2}>0,$ so we continue to restrict the calculations to subjects with $%
A_{2}=1$. Using these subjects we are now required to fit the logistic
regression model $\eta _{2,\lambda _{2}^{\left( 0\right) }}^{\left( 0\right)
}\left( \overline{A}_{2},\overline{L}_{2}\right) $ for $E\left( \widehat{Y}%
_{3}^{\left( 0\right) }|\overline{A}_{2},\overline{L}_{2}\right) $ where $%
\widehat{Y}_{3}^{\left( 0\right) }\equiv L_{3},$ 
\begin{eqnarray*}
\eta _{2,\lambda _{2}^{\left( 0\right) }}^{\left( 0\right) }\left( \overline{%
A}_{2},\overline{L}_{2}\right) &\equiv &\text{expit}\left\{ \widehat{\tau }%
_{2,greed}^{T}s_{2}\left( \overline{A}_{2},\overline{L}_{2}\right) +\widehat{%
\lambda }_{2}^{\left( 1\right) }\frac{s_{1}\left( \overline{A}_{1},\overline{%
L}_{1}\right) }{\widehat{\pi }_{2}^{2}}+\lambda _{2}^{\left( 0\right) }\frac{%
1}{\widehat{\pi }_{1}^{2}}\right\} \\
&\equiv &\text{expit}\left\{ \widehat{\tau }_{2,greed}^{T}s_{2}\left( 
\overline{A}_{2},\overline{L}_{2}\right) +\widehat{\lambda }_{2}^{\left(
1\right) }\frac{s_{1}\left( \overline{A}_{1},\overline{L}_{1}\right) }{%
\widehat{h}_{2}}+\lambda _{2}^{\left( 0\right) }\frac{1}{\widehat{h}_{1}%
\widehat{h}_{2}}\right\} ,
\end{eqnarray*}%
$\widehat{\tau }_{2,greed}^{T}s_{2}\left( \overline{A}_{2},\overline{L}%
_{2}\right) +\widehat{\lambda }_{2}^{\left( 1\right) }\frac{s_{1}\left( 
\overline{A}_{2},\overline{L}_{2}\right) }{\widehat{h}_{2}}$ is an offset
and $\lambda _{2}^{\left( 0\right) }$ is the unknown parameter. Once again,
because we are only using subjects with $\overline{A}_{2}=\overline{1},$
then we only care to fit the model for $E\left( \widehat{Y}_{3}^{\left(
0\right) }|\overline{A}_{2}=1,\overline{L}_{2}\right) :$%
\begin{equation*}
\eta _{2,\lambda _{2}^{\left( 0\right) }}^{\left( 0\right) }\left( \overline{%
A}_{2}=1,\overline{L}_{2}\right) \equiv \text{expit}\left\{ \widehat{\tau }%
_{2,greed}^{T}s_{2}\left( \overline{L}_{2}\right) +\widehat{\lambda }%
_{2}^{\left( 1\right) }\frac{s_{1}\left( \overline{L}_{1}\right) }{\widehat{h%
}_{2}}+\lambda _{2}^{\left( 0\right) }\frac{1}{\widehat{h}_{1}\widehat{h}_{2}%
}\right\}
\end{equation*}%
where $s_{k}\left( \overline{L}_{k}\right) \equiv s_{k}\left( \overline{A}%
_{k}=1,\overline{L}_{k}\right) ,k=1,2.\,\ $Thus, the estimator $\widehat{%
\lambda }_{2}^{\left( 0\right) }$ satisfies%
\begin{equation}
\mathbb{P}_{n}\left( A_{2}\frac{1}{\widehat{h}_{1}\widehat{h}_{2}}\left[ 
\widehat{Y}_{3}^{\left( 0\right) }-\text{expit}\left\{ \widehat{\tau }%
_{2,greed}^{T}s_{2}\left( \overline{L}_{2}\right) +\widehat{\lambda }%
_{2}^{\left( 1\right) T}\frac{s_{1}\left( L_{1}\right) }{\widehat{h}_{2}}+%
\widehat{\lambda }_{2}^{\left( 0\right) }\frac{1}{\widehat{h}_{1}\widehat{h}%
_{2}}\right\} \right] \right) =0  \label{g3}
\end{equation}

\item[(ii)] This step is calculated using only subjects with $\pi ^{\ast
1}>0\,,$ i.e. with $A_{1}=1.$ For these subjects we must compute%
\begin{eqnarray*}
\widehat{Y}_{2}^{\left( 0\right) } &\equiv &y_{2,\eta _{2,\widehat{\lambda }%
_{2}^{\left( 0\right) }}^{\left( 0\right) }}\left( A_{1},\overline{L}%
_{2}\right) \\
&\equiv &\int \eta _{2,\widehat{\lambda }_{2}^{\left( 0\right) }}^{\left(
0\right) }\left( A_{1},a_{2},\overline{L}_{2}\right) h_{2}^{\ast }\left(
a_{2}|A_{1},\overline{L}_{2}\right) d\mu \left( a_{2}\right) \\
&=&\sum_{a_{2}=0}^{1}\eta _{2,\widehat{\lambda }_{2}^{\left( 0\right)
}}^{\left( 0\right) }\left( A_{1},a_{2},\overline{L}_{2}\right) h_{2}^{\ast
}\left( a_{2}|A_{1},\overline{L}_{2}\right)
\end{eqnarray*}%
Now, because we are only doing the calculation for subjects with $A_{1}=1,$
and because $h_{2}^{\ast }\left( a_{2}|A_{1}=1,\overline{L}_{2}\right)
=a_{2},$ $\ $the last display simplifies to 
\begin{eqnarray}
\widehat{Y}_{2}^{\left( 0\right) } &=&\eta _{2,\widehat{\lambda }%
_{2}^{\left( 0\right) }}^{\left( 0\right) }\left( A_{1}=1,A_{2}=1,\overline{L%
}_{2}\right)  \label{st} \\
&=&\text{expit}\left\{ \widehat{\tau }_{2,greed}^{T}s_{2}\left( \overline{L}%
_{2}\right) +\widehat{\lambda }_{2}^{\left( 1\right) }s_{1}\left(
L_{1}\right) /\widehat{h}_{2}+\widehat{\lambda }_{2}^{\left( 0\right)
}1/\left( \widehat{h}_{1}\widehat{h}_{2}\right) \right\}  \notag
\end{eqnarray}
\end{description}

\item[Steps for $k=1$] 

\item[(a)] This step requires that we use subjects with $\pi ^{\ast 1}>0.$
These are precisely the subjects with $A_{1}=1.$ This step of the algorithm
requires that we fit the model for $E\left( \widehat{Y}_{2}^{\left( 1\right)
}|\overline{A}_{1},\overline{L}_{1}\right) :$ 
\begin{equation*}
\eta _{1,\tau _{1}}\left( \overline{A}_{1},\overline{L}_{1}\right) \equiv 
\text{expit}\left\{ \tau _{1}^{T}s_{1}\left( \overline{A}_{1},\overline{L}%
_{1}\right) \right\}
\end{equation*}%
just using subjects with $A_{1}=1.$ Then the model we care to estimate is
actually%
\begin{equation*}
\eta _{1,\tau _{1}}\left( \overline{A}_{1}=1,\overline{L}_{1}\right) \equiv 
\text{expit}\left\{ \tau _{1}^{T}s_{1}\left( \overline{A}_{1}=1,\overline{L}%
_{1}\right) \right\}
\end{equation*}%
Writing $s_{1}\left( \overline{L}_{1}\right) \equiv s_{1}\left( \overline{A}%
_{1}=1,\overline{L}_{1}\right) ,$ then this step of the algorithm boils down
to computing the logistic regression estimator $\widehat{\tau }_{1,greed}$
for the outcome $\widehat{Y}_{2}^{\left( 1\right) }\equiv L_{3}\,$on the
covariate vector $s_{1}\left( \overline{L}_{1}\right) $ just using subjects $%
A_{1}=1.$ Then, the estimator $\widehat{\tau }_{1,greed}$ satisfies%
\begin{equation}
\mathbb{P}_{n}\left[ A_{1}s_{1}\left( \overline{L}_{1}\right) \left\{ 
\widehat{Y}_{2}^{\left( 1\right) }-\text{expit}\left\{ \widehat{\tau }%
_{1,greed}^{T}s_{1}\left( \overline{L}_{1}\right) \right\} \right\} \right]
=0  \label{g4}
\end{equation}

\item[Step (b) for $k=1,$ $j=0$] 

\begin{description}
\item[(i)] In this step we are required to use again only subjects with $\pi
^{\ast 1}>0,$ so we continue to restrict the calculations to subjects with $%
A_{1}=1$. Using these subjects we are required to fit the logistic
regression model $\eta _{1,\lambda _{1}^{\left( 0\right) }}^{\left( 0\right)
}\left( \overline{A}_{1},\overline{L}_{1}\right) $ for $E\left( \widehat{Y}%
_{2}^{\left( 0\right) }|\overline{A}_{1},\overline{L}_{1}\right) $ where $%
\widehat{Y}_{2}^{\left( 0\right) }$ was calculated in $\left( \ref{st}%
\right) $ and 
\begin{eqnarray*}
\eta _{1,\lambda _{1}^{\left( 0\right) }}^{\left( 0\right) }\left( \overline{%
A}_{1},\overline{L}_{1}\right) &\equiv &\text{expit}\left\{ \widehat{\tau }%
_{1,greed}^{T}s_{1}\left( \overline{A}_{1},\overline{L}_{1}\right) +\lambda
_{1}^{\left( 0\right) }\frac{1}{\widehat{\pi }_{1}^{1}}\right\} \\
&\equiv &\text{expit}\left\{ \widehat{\tau }_{1,greed}^{T}s_{1}\left( 
\overline{A}_{1},\overline{L}_{1}\right) +\lambda _{1}^{\left( 0\right) }%
\frac{1}{\widehat{h}_{1}}\right\}
\end{eqnarray*}%
$\widehat{\tau }_{1,greed}^{T}s_{1}\left( \overline{A}_{1},\overline{L}%
_{1}\right) $ is an offset and $\lambda _{1}^{\left( 0\right) }$ is the
unknown parameter. Once again, because we are only using subjects with $%
\overline{A}_{1}=\overline{1},$ then we only care to fit the model for $%
E\left( \widehat{Y}_{2}^{\left( 0\right) }|\overline{A}_{1}=1,\overline{L}%
_{1}\right) :$%
\begin{equation*}
\eta _{1,\lambda _{1}^{\left( 0\right) }}^{\left( 0\right) }\left( \overline{%
A}_{1}=1,\overline{L}_{1}\right) \equiv \text{expit}\left\{ \widehat{\tau }%
_{1,greed}^{T}s_{1}\left( \overline{L}_{1}\right) +\lambda _{1}^{\left(
0\right) }\frac{1}{\widehat{h}_{1}}\right\}
\end{equation*}%
where $s_{1}\left( \overline{L}_{1}\right) \equiv s_{1}\left( \overline{A}%
_{1}=1,\overline{L}_{1}\right) .\,\ $Thus, the estimator $\widehat{\lambda }%
_{1}^{\left( 0\right) }$ satisfies%
\begin{equation}
\mathbb{P}_{n}\left( A_{1}\frac{1}{\widehat{h}_{1}}\left[ \widehat{Y}%
_{2}^{\left( 0\right) }-\text{expit}\left\{ \widehat{\tau }%
_{1,greed}^{T}s_{1}\left( \overline{L}_{1}\right) +\widehat{\lambda }%
_{1}^{\left( 0\right) }\frac{1}{\widehat{h}_{1}}\right\} \right] \right) =0
\label{g5}
\end{equation}

\item[(ii)] This step is calculated using only subjects with $\pi ^{\ast
0}>0\,,$ i.e. all subjects in the sample because by definition, $\pi ^{\ast
0}=1$. For all subjects we must then compute%
\begin{eqnarray*}
\widehat{Y}_{1}^{\left( 0\right) } &\equiv &y_{1,\eta _{1,\widehat{\lambda }%
_{1}^{\left( 0\right) }}^{\left( 0\right) }}\left( \overline{L}_{1}\right) \\
&\equiv &\int \eta _{1,\widehat{\lambda }_{1}^{\left( 0\right) }}^{\left(
0\right) }\left( a_{1},\overline{L}_{1}\right) h_{1}^{\ast }\left( a_{1}|%
\overline{L}_{1}\right) d\mu \left( a_{1}\right) \\
&=&\sum_{a_{2}=0}^{1}\eta _{1,\widehat{\lambda }_{1}^{\left( 0\right)
}}^{\left( 0\right) }\left( a_{1},\overline{L}_{1}\right) h_{1}^{\ast
}\left( a_{1}|\overline{L}_{1}\right)
\end{eqnarray*}%
Now, because $h_{1}^{\ast }\left( a_{1}|\overline{L}_{1}\right) =a_{1},$ $\ $%
the last display simplifies to 
\begin{eqnarray*}
\widehat{Y}_{1}^{\left( 0\right) } &=&\eta _{1,\widehat{\lambda }%
_{1}^{\left( 0\right) }}^{\left( 0\right) }\left( A_{1}=1,\overline{L}%
_{1}\right) \\
&=&\text{expit}\left\{ \widehat{\tau }_{1,greed}^{T}s_{1}\left( \overline{L}%
_{1}\right) +\widehat{\lambda }_{1}^{\left( 0\right) }1/\widehat{h}%
_{1}\right\}
\end{eqnarray*}
\end{description}
\end{description}

Finally, the estimator $\widehat{\theta }_{MR,greed}$ is $\mathbb{P}%
_{n}\left( \widehat{Y}_{1}^{\left( 0\right) }\right) .$ That is, $\widehat{%
\theta }_{MR,greed}$ satisfies 
\begin{equation}
\mathbb{P}_{n}\left( \widehat{Y}_{1}^{\left( 0\right) }-\widehat{\theta }%
_{MR,greed}\right) =0  \label{g6}
\end{equation}%
Note that the outcomes $\widehat{Y}_{j}^{\left( k\right) }$ being regressed
at each iteration of the algorithm are bounded between 0 and 1. But, unlike
for the computation of $\widehat{\theta }_{reg}$ given below, to compute $%
\widehat{\theta }_{MR,greed}$ we do require that $s_{1}\left( \overline{L}%
_{1}\right) $ be a subvector of $s_{2}\left( \overline{L}_{2}\right) $ nor
that $1$ be a component of $s_{1}\left( \overline{L}_{1}\right) $ and $%
s_{2}\left( \overline{L}_{2}\right) .$

We now derive equations $\left( \ref{bb}\right) $ and $\left( \ref{bbgreed}%
\right) $ for this example. To arrive at $\left( \ref{bbgreed}\right) $ we
sum the equations $\left( \ref{g3}\right) ,\left( \ref{g5}\right) $ and $%
\left( \ref{g6}\right) ,$ i.e. the equations in which the outcome has a
superscript $\left( 0\right) ,$ and obtain 
\begin{eqnarray*}
0 &=&\mathbb{P}_{n}\left( \left[ \widehat{Y}_{1}^{\left( 0\right) }-\widehat{%
\theta }_{MR,greed}\right] +\frac{A_{1}}{\widehat{h}_{1}}\left[ \widehat{Y}%
_{2}^{\left( 0\right) }-\text{expit}\left\{ \widehat{\tau }%
_{1,greed}^{T}s_{1}\left( \overline{L}_{1}\right) +\widehat{\lambda }%
_{1}^{\left( 0\right) }\frac{1}{\widehat{h}_{1}}\right\} \right] \right. \\
&&\left. +\frac{A_{2}}{\widehat{h}_{1}\widehat{h}_{2}}\left[ \widehat{Y}%
_{3}^{\left( 0\right) }-\text{expit}\left\{ \widehat{\tau }%
_{2,greed}^{T}s_{2}\left( \overline{L}_{2}\right) +\widehat{\lambda }%
_{2}^{\left( 1\right) T}\frac{s_{1}\left( L_{1}\right) }{\widehat{h}_{2}}+%
\widehat{\lambda }_{2}^{\left( 0\right) }\frac{1}{\widehat{h}_{1}\widehat{h}%
_{2}}\right\} \right] \right) \\
&=&\mathbb{P}_{n}\left( \left[ y_{1,\eta _{1,\widehat{\lambda }_{1}^{\left(
0\right) }}^{\left( 0\right) }}\left( L_{1}\right) -\widehat{\theta }%
_{MR,greed}\right] +\frac{A_{1}}{\widehat{h}_{1}}\left[ y_{2,\eta _{2,%
\widehat{\lambda }_{2}^{\left( 0\right) }}^{\left( 0\right) }}\left( A_{1}=1,%
\overline{L}_{2}\right) -\eta _{1,\widehat{\lambda }_{1}^{\left( 0\right)
}}^{\left( 0\right) }\left( A_{1}=1,\overline{L}_{1}\right) \right] \right.
\\
&&\left. +\frac{A_{1}A_{2}}{\widehat{h}_{1}\widehat{h}_{2}}\left[ L_{3}-\eta
_{2,\widehat{\lambda }_{2}^{\left( 0\right) }}^{\left( 0\right) }\left( 
\overline{A}_{2}=\overline{1},\overline{L}_{2}\right) \right] \right) \\
&=&\mathbb{P}_{n}\left[ \frac{A_{1}A_{2}}{\widehat{h}_{1}\widehat{h}_{2}}%
L_{3}-\left\{ \frac{A_{1}}{\widehat{h}_{1}}\eta _{1,\widehat{\lambda }%
_{1}^{\left( 0\right) }}^{\left( 0\right) }\left( \overline{A}_{1},\overline{%
L}_{1}\right) -y_{1,\eta _{1,\widehat{\lambda }_{1}^{\left( 0\right)
}}^{\left( 0\right) }}\left( L_{1}\right) \right\} \right. \\
&&\left. -\left\{ \frac{A_{1}A_{2}}{\widehat{h}_{1}\widehat{h}_{2}}\eta _{2,%
\widehat{\lambda }_{2}^{\left( 0\right) }}^{\left( 0\right) }\left( 
\overline{A}_{2},\overline{L}_{2}\right) -\frac{A_{1}}{\widehat{h}_{1}}%
y_{2,\eta _{2,\widehat{\lambda }_{2}^{\left( 0\right) }}^{\left( 0\right)
}}\left( A_{1},\overline{L}_{2}\right) \right\} \right] -\widehat{\theta }%
_{MR,greed} \\
&=&\mathbb{P}_{n}\left[ \frac{\pi ^{\ast 2}}{\widehat{\pi }^{2}}%
L_{3}-\sum_{k=1}^{2}\left\{ \frac{\pi ^{\ast k}}{\widehat{\pi }^{k}}\eta _{k,%
\widehat{\lambda }_{k}^{\left( 0\right) }}^{\left( 0\right) }\left( 
\overline{A}_{k},\overline{L}_{k}\right) -\frac{\pi ^{\ast \left( k-1\right)
}}{\widehat{\pi }^{\left( k-1\right) }}y_{k,\eta _{k,\widehat{\lambda }%
_{k}^{\left( 0\right) }}^{\left( 0\right) }}\left( \overline{A}_{k-1},%
\overline{L}_{k}\right) \right\} \right] -\widehat{\theta }_{MR,greed} \\
&=&\mathbb{P}_{n}\left[ Q_{1}\left( \overline{\widehat{h}},\overline{%
\widehat{\eta }}^{\left( 0\right) }\right) \right] -\widehat{\theta }%
_{MR,greed}
\end{eqnarray*}%
where $\overline{\widehat{\eta }}^{\left( 0\right) }\equiv \left( \eta _{1,%
\widehat{\lambda }_{1}^{\left( 0\right) }}^{\left( 0\right) },\eta _{2,%
\widehat{\lambda }_{2}^{\left( 0\right) }}^{\left( 0\right) }\right) .$

Likewise, to arrive at equation $\left( \ref{bb}\right) $ we sum equations $%
\left( \ref{g2}\right) $ and $\left( \ref{g4}\right) $%
\begin{eqnarray*}
0 &=&\mathbb{P}_{n}\left( A_{1}s_{1}\left( \overline{L}_{1}\right) \left[ 
\widehat{Y}_{2}^{\left( 1\right) }-\text{expit}\left\{ \widehat{\tau }%
_{1,greed}^{T}s_{1}\left( \overline{L}_{1}\right) \right\} \right] +\right.
\\
&&\left. +A_{2}\frac{s_{1}\left( \overline{L}_{1}\right) }{\widehat{h}_{2}}%
\left[ \widehat{Y}_{3}^{\left( 1\right) }-\text{expit}\left\{ \widehat{\tau }%
_{2,greed}^{T}s_{2}\left( \overline{L}_{2}\right) +\widehat{\lambda }%
_{2}^{\left( 1\right) T}\frac{s_{1}\left( L_{1}\right) }{\widehat{h}_{2}}%
\right\} \right] \right) \\
&=&\mathbb{P}_{n}\left( A_{1}s_{1}\left( \overline{L}_{1}\right) \left[
y_{2,\eta _{2,\widehat{\lambda }_{2}^{\left( 1\right) }}^{\left( 1\right)
}}\left( A_{1}=1,\overline{L}_{2}\right) -\text{expit}\left\{ \widehat{\tau }%
_{1,greed}^{T}s_{1}\left( \overline{L}_{1}\right) \right\} \right] +\right.
\\
&&\left. +A_{2}\frac{s_{1}\left( \overline{L}_{1}\right) }{\widehat{h}_{2}}%
\left[ L_{3}-\eta _{2,\widehat{\lambda }_{2}^{\left( 1\right) }}^{\left(
1\right) }\left( \overline{A}_{2}=\overline{1},\overline{L}_{2}\right) %
\right] \right) \\
&=&\mathbb{P}_{n}\left( A_{1}s_{1}\left( \overline{A}_{1}=1,\overline{L}%
_{1}\right) \left[ \frac{A_{2}}{\widehat{h}_{2}}L_{3}-\left\{ \frac{A_{2}}{%
\widehat{h}_{2}}\eta _{2,\widehat{\lambda }_{2}^{\left( 1\right) }}^{\left(
1\right) }\left( \overline{A}_{2},\overline{L}_{2}\right) -y_{2,\eta _{2,%
\widehat{\lambda }_{2}^{\left( 1\right) }}^{\left( 1\right) }}\left( A_{1},%
\overline{L}_{2}\right) \right\} \right] -\right. \\
&&\left. -\eta _{1,\widehat{\tau }_{1,greed}}\left( \overline{A}_{1},%
\overline{L}_{j}\right) \right) \\
&=&\mathbb{P}_{n}\left( \pi ^{\ast 1}s_{1}\left( \overline{A}_{1},\overline{L%
}_{1}\right) \left[ \frac{\pi _{2}^{\ast 2}}{\widehat{\pi }_{2}^{2}}%
L_{3}-\left\{ \frac{\pi _{2}^{\ast 2}}{\widehat{\pi }_{2}^{2}}\eta _{2,%
\widehat{\lambda }_{2}^{\left( 1\right) }}^{\left( 1\right) }\left( 
\overline{A}_{2},\overline{L}_{2}\right) -\frac{\pi _{2}^{\ast 1}}{\widehat{%
\pi }_{2}^{1}}y_{2,\eta _{2,\widehat{\lambda }_{2}^{\left( 1\right)
}}^{\left( 1\right) }}\left( A_{1},\overline{L}_{2}\right) \right\} \right]
\right. \\
&&\left. -\eta _{1,\widehat{\tau }_{1,greed}}\left( \overline{A}_{1},%
\overline{L}_{j}\right) \right) \\
&=&\mathbb{P}_{n}\left[ \pi ^{\ast 1}s_{1}\left( \overline{A}_{1},\overline{L%
}_{1}\right) \left\{ Q_{2}\left( \overline{\widehat{h}}_{2}^{2},\overline{%
\widehat{\eta }}_{2}^{\left( 1\right) ,2}\right) -\eta _{1,\widehat{\tau }%
_{1,greed}}\left( \overline{A}_{1},\overline{L}_{j}\right) \right\} \right]
\end{eqnarray*}%
where $\overline{\widehat{h}}_{2}^{2}\equiv \widehat{h}_{2}$ and $\overline{%
\widehat{\eta }}_{2}^{\left( 1\right) ,2}\equiv \eta _{2,\widehat{\lambda }%
_{2}^{\left( 1\right) }}^{\left( 1\right) }.$

\begin{description}
\item[Calculation of $\widehat{\protect\theta }_{reg}.$ (\textit{Weighted
iterated regression).}] 

\item[Steps for $k=2$] 

\item[(a)] Using subjects with $A_{2}=1,$ compute the weighted logistic
regression estimator $\widehat{\tau }_{2,reg}$ for the outcome $L_{3}\,$on
the covariate vector $s_{2}\left( \overline{L}_{2}\right) $ with weight $1/%
\widehat{\pi }^{2}=1/\left( \widehat{h}_{1}\widehat{h}_{2}\right) .$ That
is, $\widehat{\tau }_{2,reg}$ solves%
\begin{equation}
\mathbb{P}_{n}\left\{ A_{2}\frac{s_{2}\left( \overline{L}_{2}\right) }{%
\widehat{h}_{1}\widehat{h}_{2}}\left[ L_{3}-\text{expit}\left\{ \tau
_{2}^{T}s_{2}\left( \overline{L}_{2}\right) \right\} \right] \right\} =0
\label{reg1}
\end{equation}

\item[(b)] For each subject with $A_{1}=1\,\ $compute $\widehat{Y}%
_{2,reg}\equiv $ expit$\left\{ \widehat{\tau }_{2,reg}^{T}s_{2}\left( 
\overline{L}_{2}\right) \right\} .$

\item[Steps for $k=1$] 

\item[(a)] Using subjects with $A_{1}=1,$ compute the weighted logistic
regression estimator $\widehat{\tau }_{1,reg}$ for the outcome $\widehat{Y}%
_{2,reg}\,$\ on the covariate vector $s_{1}\left( L_{1}\right) $ with weight 
$1/\widehat{\pi }^{1}=1/\widehat{h}_{1}$ $.$ That is, $\widehat{\tau }%
_{1,reg}$ solves%
\begin{equation}
\mathbb{P}_{n}\left[ A_{1}\frac{s_{1}\left( L_{1}\right) }{\widehat{h}_{1}}%
\left\{ \widehat{Y}_{2,reg}-\text{expit}\left\{ \tau _{1}^{T}s_{1}\left(
L_{1}\right) \right\} \right\} \right] =0  \label{reg2}
\end{equation}

\item[(b)] For all study subjects compute $\widehat{Y}_{1,reg}\equiv $ expit$%
\left\{ \widehat{\tau }_{1,reg}^{T}s_{1}\left( \overline{L}_{1}\right)
\right\} .$
\end{description}

The estimator $\widehat{\theta }_{reg}$ is $\mathbb{P}_{n}\left( \widehat{Y}%
_{1,reg}\right) .$ As indicated earlier, the estimator $\widehat{\theta }%
_{reg}$ is multiply robust so long as $s_{1}\left( L_{1}\right) $ is a
subvector of $s_{2}\left( \overline{L}_{2}\right) $ and 1 is an entry of
both $s_{1}\left( \overline{L}_{1}\right) $ and $s_{2}\left( \overline{L}%
_{2}\right) $. Note that the outcomes $L_{3}$ and $\widehat{Y}_{2,reg}$ in
step (a) of each iteration are bounded between 0 and 1 and consequently, the
equations $\left( \ref{reg1}\right) $ and $\left( \ref{reg2}\right) $ always
have a solution. In addition, because $\widehat{Y}_{1,reg}$ is also bounded
between 0 and 1, the estimator $\widehat{\theta }_{reg}$ is guaranteed to
fall between 0 and 1.

We now turn to the application of Algorithm 4. As indicated earlier, the
algorithm applied to the present example returns precisely the estimator of $%
\theta \left( g\right) $ derived by by Tchetgen-Tchetgen (2009).

\begin{description}
\item[Calculation of $\widehat{\protect\theta }_{MR}.$ \textit{(Iterated
regression of multiply robust outcomes)} ] 

\item[Steps for $k=2$] 

\item[(a)] Using subjects with $A_{2}=1,$ compute the logistic regression
estimator $\widehat{\tau }_{2,MR}$ for the outcome $L_{3}\,$on the covariate
vector $s_{2}\left( \overline{L}_{2}\right) .$ That is, $\widehat{\tau }%
_{2,MR}$ solves%
\begin{equation}
\mathbb{P}_{n}\left[ A_{2}s_{2}\left( \overline{L}_{2}\right) \left\{ L_{3}-%
\text{expit}\left\{ \tau _{2}^{T}s_{2}\left( \overline{L}_{2}\right)
\right\} \right\} \right] =0
\end{equation}

\item[(b)] For each subject with $A_{1}=1\,\ $compute 
\begin{equation*}
\text{ }\widehat{Q}_{2}\equiv \frac{A_{2}}{\widehat{h}_{2}}\left\{ L_{3}-%
\text{expit}\left( \widehat{\tau }_{2,MR}^{T}s_{2}\left( \overline{L}%
_{2}\right) \right) \right\} +\text{expit}\left( \widehat{\tau }%
_{2,MR}^{T}s_{2}\left( \overline{L}_{2}\right) \right)
\end{equation*}

\item[Steps for $k=1$] 

\item[(a)] Using subjects with $A_{1}=1,$ compute the logistic regression
estimator $\widehat{\tau }_{1,MR}$ for the outcome $\widehat{Q}_{2}$ $\,$on
the covariate vector $s_{1}\left( \overline{L}_{1}\right) .$ That is, $%
\widehat{\tau }_{1,MR}$ solves%
\begin{equation}
\mathbb{P}_{n}\left[ A_{1}s_{1}\left( \overline{L}_{1}\right) \left\{ 
\widehat{Q}_{2}-\text{expit}\left\{ \tau _{1}^{T}s_{1}\left( L_{1}\right)
\right\} \right\} \right] =0  \label{q-eq}
\end{equation}

\item[(b)] For all subjects$\,\ $compute 
\begin{equation*}
\text{ }\widehat{Q}_{1}\equiv \frac{A_{1}}{\widehat{h}_{1}}\left\{ \widehat{Q%
}_{2}-\text{expit}\left( \widehat{\tau }_{1,MR}^{T}s_{1}\left( \overline{L}%
_{1}\right) \right) \right\} +\text{expit}\left( \widehat{\tau }%
_{1,MR}^{T}s_{1}\left( \overline{L}_{1}\right) \right)
\end{equation*}%
$.$
\end{description}

Finally $\widehat{\theta }_{MR}=\mathbb{P}_{n}\left( \text{ }\widehat{Q}%
_{1}\right) .$ Note that the outcome $\widehat{Q}_{2}\,\ $in step (a) of the
second iteration (i.e. corresponding to $k=1),$ unlike the outcome $\widehat{%
Y}_{2,reg}$ of the preceding, is not guaranteed to be between 0 and 1 since $%
1/\widehat{h}_{2}$ can be arbitrarily large. Consequently, it is possible
that equation $\left( \ref{q-eq}\right) $ would not have a solution. Even if
a solution $\widehat{\tau }_{1,MR}$ is found, there is no guarantee that the
estimator $\widehat{\theta }_{MR}$ would fall between $0$ and 1 since $%
\widehat{Q}_{1}$ can be arbitrarily large.

\section{Machine learning $K+1\,\ $and $2^{K}$ multiply robust estimators 
\label{ML}}

So far we have considered estimation of the functions $\eta _{k}^{g}$ and $%
h_{k}$ under parametric working models. We will now consider extending some
of the estimators in the preceding sections to allow for more flexible
estimation of $\eta _{k}$ and $h_{k}$ by machine learning algorithms.

Our machine learning estimators will use sample splitting because the true
functions $\eta _{k}^{g}$ and $h_{k}$, and/or the machine learning
estimators of any of them may fail to fall in a Donsker class. We thus
randomly split the sample into $\mathbf{U}$ equal sized subsamples indexed $%
u=1,..,\mathbf{U}$, where $\mathbf{U}$ is a small fixed number, say $5.$ We
refer to the set of sample units in the $u^{th}$ sample split as the $u^{th}$
validation sample and the set comprised by the remaining sample units as the 
$u^{th}$ training sample. We let $\mathbb{P}_{n}^{v,u}$ be the empirical
distribution of the subjects in the $u^{th}$ validation sample and $\mathbb{P%
}_{n}^{t,u}$ be the empirical distribution of the subjects in the $u^{th}$
training sample. We consider machine learning generalizations of earlier
doubly robust and multiple robust estimators of 
\begin{eqnarray*}
\theta \left( g\right) &\equiv &E_{gh^{\ast }}\left\{ \psi \left( \overline{L%
}_{K+1}\right) \right\} \\
&=&\int \psi \left( \overline{l}_{K+1}\right)
\dprod\limits_{k=1}^{K}h_{k}^{\ast }\left( a_{k}|\overline{l}_{k},\overline{a%
}_{k-1}\right) \dprod\limits_{k=0}^{K}g_{k}\left( l_{k+1}|\overline{l}_{k},%
\overline{a}_{k}\right) d\mu \left( z\right)
\end{eqnarray*}%
defined as 
\begin{equation*}
\widehat{\theta }_{DR,CF,mach}=\mathbf{U}^{-1}\sum_{u=1}^{\mathbf{U}}\mathbb{%
P}_{n}^{v,u}\left\{ Q_{1}\left( \widehat{\overline{h}}_{mach}^{\left(
t,u\right) },\widehat{\overline{\eta }}_{mach}^{\left( t,u\right) }\right)
\right\} ,
\end{equation*}%
\begin{equation*}
\widehat{\theta }_{DR,CF,mach,bang}\equiv \mathbf{U}^{-1}\sum_{u=1}^{\mathbf{%
U}}\mathbb{P}_{n}^{v,u}\left( y_{1,\widehat{\eta }_{1,mach,bang}^{t,u}}^{u}%
\left( L_{1}\right) \right) ,
\end{equation*}%
\begin{equation*}
\widehat{\theta }_{DR,CF,mach,reg}\equiv \mathbf{U}^{-1}\sum_{u=1}^{\mathbf{U%
}}\mathbb{P}_{n}^{v,u}\left( y_{1,\widehat{\eta }_{1,mach,reg}^{t,u}}^{u}%
\left( L_{1}\right) \right) ,
\end{equation*}%
\begin{equation*}
\widehat{\theta }_{MR,CF,mach}=\mathbf{U}^{-1}\sum_{u=1}^{\mathbf{U}}\mathbb{%
P}_{n}^{v,u}\left\{ Q_{1}\left( \widetilde{\overline{h}}_{mach}^{\left(
t,u\right) },\widetilde{\overline{\eta }}_{mach}^{\left( t,u\right) }\right)
\right\} ,
\end{equation*}%
\begin{equation*}
\widehat{\theta }_{MR,CF,mach,bang}\equiv \mathbf{U}^{-1}\sum_{u=1}^{\mathbf{%
U}}\mathbb{P}_{n}^{v,u}\left( y_{1,\widetilde{\eta }%
_{1,mach,bang}^{t,u}}^{u}\left( L_{1}\right) \right)
\end{equation*}%
\begin{equation*}
\widehat{\theta }_{MR,CF,mach,reg}\equiv \mathbf{U}^{-1}\sum_{u=1}^{\mathbf{U%
}}\mathbb{P}_{n}^{v,u}\left( y_{1,\widetilde{\eta }_{1,mach,reg}^{t,u}}^{u}%
\left( L_{1}\right) \right)
\end{equation*}%
where, recall, for any $\overline{h}\equiv \left( h_{1},...,h_{K}\right) $
and $\overline{\eta }\equiv \left( \eta _{1},\ldots ,\eta _{K}\right) ,$ 
\begin{equation*}
Q_{1}\left( \overline{h},\overline{\eta }\right) \equiv \frac{\pi ^{\ast K}}{%
\pi ^{K}}\psi \left( \overline{L}_{K+1}\right) -\sum_{k=1}^{K}\left\{ \frac{%
\pi ^{\ast k}}{\pi ^{k}}\eta _{k}\left( \overline{A}_{k},\overline{L}%
_{k}\right) -\frac{\pi ^{\ast \left( k-1\right) }}{\pi ^{\left( k-1\right) }}%
y_{k,\eta _{k}}\left( \overline{A}_{k-1},\overline{L}_{k}\right) \right\} ,
\end{equation*}%
with%
\begin{equation*}
y_{k,\eta _{k}}\left( \overline{A}_{k-1},\overline{L}_{k}\right) \equiv \int
h_{k}^{\ast }\left( a_{k}|\overline{A}_{k-1},\overline{L}_{k}\right) \eta
_{k}\left( a_{k},\overline{A}_{k-1},\overline{L}_{k}\right) d\mu _{k}\left(
a_{k}\right) ,
\end{equation*}%
\begin{equation*}
\pi ^{\ast k}\equiv \prod\limits_{j=1}^{k}h_{j}^{\ast }\text{ and }\pi
^{k}\equiv \prod\limits_{j=1}^{k}h_{j},
\end{equation*}%
and where $\widehat{\overline{h}}_{mach}\equiv \left( \widehat{h}%
_{1,mach}^{\left( t,u\right) },\ldots ,\widehat{h}_{K,mach}^{\left(
t,u\right) }\right) $ is the output of part (a), $y_{1,\widehat{\eta }%
_{1,mach}^{t,u}}^{u}$ and $\widehat{\overline{\eta }}_{mach}^{\left(
t,u\right) }\equiv \left( \widehat{\eta }_{1,mach}^{\left( t,u\right)
},\ldots ,\widehat{\eta }_{K,mach}^{\left( t,u\right) }\right) $ are outputs
of part (b), $\widetilde{\overline{\eta }}_{mach}^{\left( t,u\right) }\equiv
\left( \widetilde{\eta }_{1,mach}^{\left( t,u\right) },\ldots ,\widetilde{%
\eta }_{K,mach}^{\left( t,u\right) }\right) $ is the output of part (c), $%
y_{1,\widetilde{\eta }_{1,mach,bang}^{t,u}}^{u}$ and $y_{1,\widetilde{\eta }%
_{1,mach,reg}^{t,u}}^{u}$ are the outputs of step (d), and $y_{1,\widehat{%
\eta }_{1,mach,bang}^{t,u}}^{u}$ and $y_{1,\widehat{\eta }%
_{1,mach,reg}^{t,u}}^{u}$ are the outputs of step (e) of the following
algorithm:

\begin{description}
\item[Algorithm 6. \textit{(Cross--Fitting Machine Learning Multiple Robust
estimation)}] For $u=1,\ldots ,\mathbf{U}$ , in the $u^{th}$ training sample
run steps a)-c) and in the validation sample run steps d) and e)

\item[a)] For $k=K,K-1,\ldots ,1,$ a preferred machine learning algorithm to
estimate $h_{k}.$ Let $\widehat{h}_{k}^{\left( t,u\right) }$ be the output
of the algorithm and let $\widehat{\pi }^{\left( t,u\right) ,k}\equiv 
\widehat{h}_{1}^{\left( t,u\right) }\times \ldots \times \widehat{h}%
_{k}^{\left( t,u\right) }.$

\item[b)] Set $\widehat{Y}_{K+1,mach}^{u}\equiv \psi \left( \overline{L}%
_{K+1}\right) $ and for $k=K,K-1,\ldots ,1,$ repeat,

\begin{description}
\item[b.1)] Compute $\widehat{\eta }_{k,mach}^{t,u}\left( \cdot ,\cdot
\right) ,$ the output of a preferred machine learning algorithm for
estimating $E\left( \left. \widehat{Y}_{k+1,mach}^{u}\right\vert \overline{A}%
_{k},\overline{L}_{k}\right) $.

\item[b.2) ] For units with $\pi ^{\ast k-1}>0,$ compute 
\begin{equation*}
\widehat{Y}_{k,mach}^{u}\equiv y_{k,\widehat{\eta }_{k,mach}^{t,u}}^{u}%
\left( \overline{A}_{k-1},\overline{L}_{k}\right) \equiv \int h_{k}^{\ast
}\left( a_{k}|\overline{A}_{k-1},\overline{L}_{k}\right) \widehat{\eta }%
_{k,mach}^{t,u}\left( a_{k},\overline{A}_{k-1},\overline{L}_{k}\right) d\mu
_{k}\left( a_{k}\right) .
\end{equation*}
\end{description}

\item[c)] Set $\widetilde{Q}_{K+1,mach}^{u}\equiv \psi \left( \overline{L}%
_{K+1}\right) $ and for $k=K,K-1,\ldots ,1,$ repeat

\begin{description}
\item[c.1)] Compute $\widetilde{\eta }_{k,mach}^{t,u}\left( \cdot ,\cdot
\right) ,$ the output of a preferred machine learning algorithm for
estimating $E\left( \left. \widetilde{Q}_{k+1,mach}^{u}\right\vert \overline{%
A}_{k},\overline{L}_{k}\right) $.

\item[c.2) ] For units with $\pi ^{\ast k-1}>0,$ compute 
\begin{equation*}
\widetilde{Y}_{k,mach}^{u}\equiv y_{k,\widetilde{\eta }_{k,mach}^{t,u}}^{u}%
\left( \overline{A}_{k-1},\overline{L}_{k}\right) \equiv \int h_{k}^{\ast
}\left( a_{k}|\overline{A}_{k-1},\overline{L}_{k}\right) \widetilde{\eta }%
_{k,mach}^{t,u}\left( a_{k},\overline{A}_{k-1},\overline{L}_{k}\right) d\mu
_{k}\left( a_{k}\right) .
\end{equation*}%
and%
\begin{equation*}
\widetilde{Q}_{k,mach}^{u}\equiv \frac{h_{k}^{\ast }}{\widehat{h}_{k}^{t,u}}%
\left[ \widetilde{Q}_{k+1,mach}^{u}-\widetilde{\eta }_{k,mach}^{t,u}\left( 
\overline{A}_{k},\overline{L}_{k}\right) \right] +\widetilde{Y}_{k,mach}^{u}.
\end{equation*}
\end{description}

\item[d. ] For units in the validation sample, set $\widetilde{Y}%
_{K+1,mach}^{u}\equiv \psi \left( \overline{L}_{K+1}\right) $ and for $%
k=K,K-1,\ldots ,1,$ repeat,

\begin{description}
\item[d.1 ] Based on validation units with $\pi ^{\ast k}>0,$ estimate $%
\lambda _{k}$ and $\beta _{k}$ indexing the regression models $\Psi \left\{
\Psi ^{-1}\left[ \widetilde{\eta }_{k,mach}^{t,u}\right] +\lambda _{k}\left(
1/\widehat{\pi }_{mach}^{\left( t,u\right) ,k}\right) \right\} $ and $\Psi
\left\{ \Psi ^{-1}\left[ \widetilde{\eta }_{k,mach}^{t,u}\right] +\beta
_{k}\right\} $ for $E\left( \widetilde{Y}_{k+1,mach,bang}^{u}|\overline{A}%
_{k},\overline{L}_{k}\right) $ and $E\left( \widetilde{Y}_{k+1,mach,reg}^{u}|%
\overline{A}_{k},\overline{L}_{k}\right) ,$ which have offset $\Psi ^{-1}%
\left[ \widetilde{\eta }_{k,mach}^{t,u}\right] ,$ with $\widetilde{\lambda }%
_{k,mach}$ and $\widetilde{\beta }_{k,mach}$ solving%
\begin{equation*}
\mathbb{P}_{n}^{v,u}\left[ \pi ^{\ast k}\left( 1/\widehat{\pi }%
_{mach}^{\left( t,u\right) ,k}\right) \left[ \widetilde{Y}%
_{k+1,mach,bang}^{u}-\Psi \left\{ \Psi ^{-1}\left[ \widetilde{\eta }%
_{k,mach}^{t,u}\right] +\lambda _{k}\left( 1/\widehat{\pi }_{mach}^{\left(
t,u\right) ,k}\right) \right\} \right] \right] =0
\end{equation*}%
\begin{equation*}
\mathbb{P}_{n}^{v,u}\left[ \pi ^{\ast k}\left( 1/\widehat{\pi }%
_{mach}^{\left( t,u\right) ,k}\right) \left[ \widetilde{Y}%
_{k+1,mach,reg}^{u}-\Psi \left\{ \Psi ^{-1}\left[ \widetilde{\eta }%
_{k,mach}^{t,u}\right] +\beta _{k}\right\} \right] \right] =0
\end{equation*}

and set $\widetilde{\eta }_{k,mach,bang}^{u}\left( \overline{A}_{k},%
\overline{L}_{k}\right) =\Psi \left\{ \Psi ^{-1}\left[ \widetilde{\eta }%
_{k,mach}^{t,u}\right] +\widetilde{\lambda }_{k,mach}\left( 1/\widehat{\pi }%
_{mach}^{\left( t,u\right) ,k}\right) \right\} $ and $\widetilde{\eta }%
_{k,mach,reg}^{u}\left( \overline{A}_{k},\overline{L}_{k}\right) =\Psi
\left\{ \Psi ^{-1}\left[ \widetilde{\eta }_{k,mach}^{t,u}\right] +\widetilde{%
\beta }_{k,mach}\right\} $

\item[d.2] For validation sample units with $\pi ^{\ast k-1}>0,$ compute 
\begin{eqnarray*}
\widetilde{Y}_{k,mach,bang}^{u} &\equiv &y_{k,\widetilde{\eta }%
_{k,mach,bang}^{u}}^{u}\left( \overline{A}_{k-1},\overline{L}_{k}\right) \\
&\equiv &\int h_{k}^{\ast }\left( a_{k}|\overline{A}_{k-1},\overline{L}%
_{k}\right) \widetilde{\eta }_{k,mach,bang}^{u}\left( a_{k},\overline{A}%
_{k-1},\overline{L}_{k}\right) d\mu _{k}\left( a_{k}\right)
\end{eqnarray*}%
and 
\begin{eqnarray*}
\widetilde{Y}_{k,mach,reg}^{u} &\equiv &y_{k,\widetilde{\eta }%
_{k,mach,reg}^{u}}^{u}\left( \overline{A}_{k-1},\overline{L}_{k}\right) \\
&\equiv &\int h_{k}^{\ast }\left( a_{k}|\overline{A}_{k-1},\overline{L}%
_{k}\right) \widetilde{\eta }_{k,mach,reg}^{u}\left( a_{k},\overline{A}%
_{k-1},\overline{L}_{k}\right) d\mu _{k}\left( a_{k}\right)
\end{eqnarray*}
\end{description}

\item[e. ] For units in the validation sample set $\widehat{Y}%
_{K+1,mach}^{u}\equiv \psi \left( \overline{L}_{K+1}\right) $ and for $%
k=K,K-1,\ldots ,1,$ repeat steps d.1) and d.2) but with $\widehat{\eta }%
_{k,mach}^{t,u}$ replacing $\widetilde{\eta }_{k,mach}^{t,u}$ and renaming $%
\widetilde{\eta }_{k,mach,bang}^{u}$ as $\widehat{\eta }_{k,mach,bang}^{u},%
\widetilde{\eta }_{k,mach,reg}^{u}$ as $\widehat{\eta }_{k,mach,reg}^{u},$ $%
\widetilde{Y}_{k,mach,bang}^{u}$ as $\widehat{Y}_{k,mach,bang}^{u},$ and $%
\widetilde{Y}_{k,mach,reg}^{u}$ as $\widehat{Y}_{k,mach,reg}^{u}.$
\end{description}

\subsection{Asymptotic theory of cross-fitting estimators: preliminary
background \label{prelim-backg}}

To study the asymptotic properties of the above estimators we will now
formulate a general and unified notation. Given a random sample $\mathcal{S}%
=\left\{ Z_{1},\ldots ,Z_{n}\right\} $ comprised of $n$ i.i.d. copies of a
random vector $Z$ from an unknown law $P$ with density $p$ with respect to
an underlying measure$,$ and given two subsamples $\mathcal{S}_{t}$ and $%
\mathcal{S}_{v}$ of $\mathcal{S},$ let $\mathbb{P}_{n}^{j}$ denote the
empirical distribution of sample $\mathcal{S}_{j},j=t,v.$ Let $m\left(
z,r^{\dag }\right) $ be a given function of $z$ and $r^{\dag }$ where $%
r^{\dag }\equiv r^{\dag }\left( \cdot \right) :z\rightarrow r^{\dag }\left(
z\right) $ is some map on the sample space of $Z\,,\ $and let $\mu \left(
P\right) \equiv E_{P}\left[ m(O,r\left( P\right) )\right] $ where $r\left(
P\right) \left( \cdot \right) :z\mapsto r\left( P\right) \left( z\right) $
is a map that depends on $P$. Define $M\left( r\right) \equiv m(Z,r).$
Consider an estimator $\widehat{\mu }\equiv \mathbb{P}_{n}^{v}[M\left( 
\widehat{r}^{t}\right) ]\equiv \mathbb{P}_{n}^{v}[m\left( Z,\widehat{r}%
^{t}\right) ]$ of $\mu \left( P\right) $ that depends on $\widehat{r}%
^{t}\equiv $ $\widehat{r}\left( \mathbb{P}_{n}^{t}\right) \left( \cdot
\right) ,$ an estimator of $r\left( P\right) \left( \cdot \right) $ based on
data from $\mathcal{S}_{t}.$ Consider the decomposition%
\begin{equation*}
\mathbb{P}_{n}^{v}[M\left( \widehat{r}^{t}\right) ]-\mu \left( P\right) =%
\mathbb{P}_{n}^{v}[M\left( \widehat{r}^{t}\right) ]-E^{v}[M\left( \widehat{r}%
^{t}\right) ]+E^{v}[M\left( \widehat{r}^{t}\right) ]-\mu \left( P\right)
\end{equation*}%
where throughout $E^{v}[\cdot ]$ stands for the population expectation
operator that regards the data from $\mathcal{S}_{t}$ as fixed, i.e.
non-random, e.g. $E^{v}[M\left( \widehat{r}^{t}\right) ]$ stands for $\left.
E_{P}\left[ M\left( r\right) \right] \right\vert _{r=\widehat{r}^{t}}\equiv
\int m\left( z,\widehat{r}^{t}\right) dP\left( z\right) .$ Note that $%
E^{v}[M\left( \widehat{r}^{t}\right) ]$ is random as it depends on the data
from the random sub-sample $\mathcal{S}_{t}$.

We refer to $E^{v}[M\left( \widehat{r}^{t}\right) ]-\mu \left( P\right) $ as
the drift. We refer to $\mathbb{P}_{n}^{v}[M\left( \widehat{r}\right)
]-E^{v}[M\left( \widehat{r}^{t}\right) ]$ as the centered term.

Consider now estimation of $\mu \left( P\right) $ by sample splitting.
Specifically, randomly partition the sample into $\mathbf{U}$ equal sized
subsamples indexed by $u=1,..,\mathbf{U,}$\ where $\mathbf{U}$ is a small
fixed number. For a given $u,$ let $\mathcal{S}_{v,u}$ denote the set of
sample units in the $u^{th}$ partition. Call $\mathcal{S}_{v,u}$ the $u^{th}$
validation sample. Let $\mathcal{S}_{t,u}\,\ \,$denote the set comprised by
the remaining sample units, call it the $u^{th}$ training sample. As before,
we let $\mathbb{P}_{n}^{j,u}$ be the empirical distribution of the data in $%
\mathcal{S}_{j,u}\,,j=v,t.\ $

The split-specific estimator of $\mu \left( P\right) $ is given by 
\begin{equation*}
\widehat{\mu }\left( \widehat{r}^{t}\right) =\mathbb{P}_{n}^{v}[M\left( 
\widehat{r}^{t}\right) ]
\end{equation*}%
where $\widehat{r}^{t}=\widehat{r}\left( \mathbb{P}_{n}^{t}\right) \left(
\cdot \right) $ is some estimator of $r\left( P\right) \left( \cdot \right) $
based on data in the training sample and where, by convention, we eliminate
the superscript $u$ when we refer to a single split. One of our goals in
this section is to study the asymptotic properties of the cross-fitting (CF)
estimator $\widehat{\mu }^{cf}$ obtained as the average of estimators $%
\widehat{\mu }\left( \widehat{r}^{t}\right) $ over all $\mathbf{U}$
validation samples, that is, 
\begin{equation*}
\widehat{\mu }^{cf}=\mathbf{U}^{-1}\sum_{u=1}^{\mathbf{U}}\mathbb{P}%
_{n}^{v,u}[M\left( \widehat{r}^{t,u}\right) ]
\end{equation*}%
Consider now a single split. Henceforth, we suppose that there exists a
bounded function $r^{\ast }\left( P\right) \left( \cdot \right) ,$ not
necessarily equal to $r\left( P\right) \left( \cdot \right) ,$ such that for 
$\widehat{r}=\widehat{r}\left( \mathbb{P}_{n}\right) \left( \cdot \right) $
it holds that 
\begin{equation*}
\int \left[ m\left( z,\widehat{r}^{t}\right) -m\left( z,r^{\ast }\right) %
\right] ^{2}dP\left( z\right) \rightarrow _{P}0\text{ as }n\rightarrow \infty
\end{equation*}

Then, with $n_{v}$ denoting the cardinality of $\mathcal{S}_{v},$ 
\begin{equation*}
\sqrt{n_{v}}\left\{ \mathbb{P}_{n}^{v}[M\left( \widehat{r}^{t}\right)
]-E^{v}[M\left( \widehat{r}^{t}\right) ]\right\} =\sqrt{n_{v}}\left\{ 
\mathbb{P}_{n}^{v}[M\left( r^{\ast }\right) ]-E^{v}[M\left( r^{\ast }\right)
]\right\} +o_{p}\left( 1\right)
\end{equation*}%
as $n\rightarrow \infty $ as is well known (see van der Vaart, 1998).

Hence as $n\rightarrow \infty ,$ we have $\ $%
\begin{eqnarray*}
\sqrt{n_{v}}\left\{ \mathbb{P}_{n}^{v}[M\left( \widehat{r}^{t}\right) ]-\mu
\left( P\right) ]\right\} &=&\sqrt{n_{v}}\left\{ \mathbb{P}_{n}^{v}[M\left(
r^{\ast }\right) ]-E^{v}[M\left( r^{\ast }\right) ]\right\} \\
&&+\sqrt{n_{v}}\left\{ P^{v}[M\left( \widehat{r}^{t}\right) ]-\mu \left(
P\right) \right\} +o_{p}\left( 1\right) ,\ 
\end{eqnarray*}

Thus if $E^{v}[M\left( \widehat{r}^{t}\right) ]-\mu \left( P\right)
=o_{p}\left( 1/\sqrt{n_{v}}\right) $, we can conclude that $\widehat{\mu }%
\left( \widehat{r}^{t}\right) =\mathbb{P}_{n}^{v}[M\left( \widehat{r}%
^{t}\right) ]$ is an asymptotically linear estimator of $\mu \left( P\right) 
$, and thus, since $\mathbb{P}_{n}[M\left( r^{\ast }\right) ]=\mathbf{U}%
^{-1}\sum_{u=1}^{\mathbf{U}}\mathbb{P}_{n}^{v,u}[M\left( r^{\ast }\right) ],$
we conclude that $\widehat{\mu }^{cf}$ is an asymptotically linear estimator
of $\mu \left( P\right) $ with influence function $M\left( r^{\ast }\right)
. $ That is, as $n$ goes to $\infty $

\begin{equation*}
\sqrt{n}\left\{ \widehat{\mu }^{cf}-\mu \left( P\right) \right\} =\sqrt{n}%
\mathbb{P}_{n}[M\left( r^{\ast }\right) ]+o_{p}\left( 1\right) .
\end{equation*}

\subsection{Analysis of the drifts of the machine learning DR and MR
estimators \label{general-drift}}

We will now apply the generic formulation of the preceding section to
compare the distribution of machine learning doubly robust and multiply
robust estimators of $\theta \left( g\right) .$ To do so, we begin by
comparing the asymptotic properties of $\widehat{\theta }_{DR,CF,mach}$ and $%
\widehat{\theta }_{MR,CF,mach}.$

First note that $\theta \left( g\right) =P\left[ Q_{1}\left( \overline{h},%
\overline{\eta }^{g}\right) \right] ,$ so that in our general formulation of
asymptotic theory we identify $r\left( P\right) $ with $\left( \overline{h},%
\overline{\eta }^{g}\right) $ and for any $r^{\dag }=\left( \overline{h}%
^{\dag },\overline{\eta }^{\dag }\right) $ we define 
\begin{eqnarray*}
m\left( Z;r^{\dag }\right) &\equiv &M\left( r^{\dag }\right) \\
&\equiv &Q_{1}\left( \overline{h}^{\dag },\overline{\eta }^{\dag }\right) \\
&\equiv &\frac{\pi ^{\ast K}}{\pi ^{\dag K}}\psi \left( \overline{L}%
_{K+1}\right) -\sum_{k=1}^{K}\left\{ \frac{\pi ^{\ast k}}{\pi ^{\dag k}}\eta
_{k}^{\dag }\left( \overline{A}_{k},\overline{L}_{k}\right) -\frac{\pi
^{\ast \left( k-1\right) }}{\pi ^{\dag \left( k-1\right) }}y_{k,\eta
_{k}^{\dag }}\left( \overline{A}_{k-1},\overline{L}_{k}\right) \right\}
\end{eqnarray*}

The estimator $\widehat{\theta }_{DR,CF,mach}$ is the average of $\mathbb{P}%
_{n}^{v,u}\left\{ Q_{1}\left( \widehat{\overline{h}}^{\left( t,u\right) },%
\widehat{\overline{\eta }}_{mach}^{\left( t,u\right) }\right) \right\} $
over $u=1,\ldots ,\mathbf{U}$, so $\mathbb{P}_{n}^{v}[M\left( \widehat{r}%
^{t}\right) ]$ is just a split specific 
\begin{equation*}
\widehat{\theta }_{DR,mach}\equiv \mathbb{P}_{n}^{v}\left\{ Q_{1}\left( 
\widehat{\overline{h}}^{t},\widehat{\overline{\eta }}_{mach}^{t}\right)
\right\}
\end{equation*}%
where, recall that by convention we eliminate the superscript $u$ when
referring to a generic split.

Likewise, when studying the limit law of $\widehat{\theta }_{MR,CF,mach},$ $%
\mathbb{P}_{n}^{v}[M\left( \widehat{r}^{t}\right) ]$ is\ equal to 
\begin{equation*}
\widehat{\theta }_{MR,mach}\equiv \mathbb{P}_{n}^{v}\left\{ Q_{1}\left( 
\widehat{\overline{h}}^{t},\widetilde{\overline{\eta }}_{mach}^{t}\right)
\right\}
\end{equation*}

We are interested in investigating the rates of convergence to 0 of the
drifts of $\widehat{\theta }_{DR,CF,mach}$ and $\widehat{\theta }%
_{MR,CF,mach}.$ In view of the discussion of the preceding section, it
suffices to study the rates of the drifts of the single split estimators $%
\widehat{\theta }_{DR,mach}$ and $\widehat{\theta }_{MR,mach}.$ Notice that
these drifts are $E^{v}\left[ Q_{1}\left( \widehat{\overline{h}}^{t},%
\widehat{\overline{\eta }}_{mach}^{t}\right) \right] -\theta \left( g\right) 
$ and $E^{v}\left[ Q_{1}\left( \widehat{\overline{h}}^{t},\widetilde{%
\overline{\eta }}_{mach}^{t}\right) \right] -\theta \left( g\right) $ which,
by Lemma 3, can be expressed as $a^{p}\left( h^{\dag },\eta ^{\dag }\right)
=b^{p}\left( h^{\dag },\eta ^{\dag }\right) =c^{p}\left( h^{\dag },\eta
^{\dag }\right) $ evaluated at $\left( \widehat{\overline{h}}^{t},\widehat{%
\overline{\eta }}_{mach}^{t}\right) $ or $\left( \widehat{\overline{h}}^{t},%
\widetilde{\overline{\eta }}_{mach}^{t}\right) .$ We will exploit these
formulae appropriately to make manifest the difference in the orders of the
drifts of $\widehat{\theta }_{DR,mach}$ and $\widehat{\theta }_{MR,mach}$.

Using $c^{p}\left( h^{\dag },\eta ^{\dag }\right) $ applied to $\left(
h^{\dag },\eta ^{\dag }\right) =\left( \widehat{\overline{h}}^{t},\widehat{%
\overline{\eta }}_{mach}^{t}\right) $ we obtain the following expression for
the drift of $\widehat{\theta }_{DR,mach}$ 
\begin{eqnarray}
&&\left. E^{v}\left[ Q_{1}\left( \widehat{\overline{h}}^{t},\widehat{%
\overline{\eta }}_{mach}^{t}\right) \right] -\theta \left( g\right) =\right.
\label{driftDR} \\
&=&\sum_{k=1}^{K}E_{\overline{g}_{k-1},\overline{h}_{k}}\left[ \left\{ \frac{%
\pi ^{\ast k}}{\pi ^{k}}-\frac{\pi ^{\ast k}}{\widehat{\pi }^{k}}\right\} %
\left[ \widehat{\eta }_{k,mach}^{t}-E_{g_{k}}\left\{ \left. y_{k+1,\widehat{%
\eta }_{k+1,mach}^{t}}\left( \overline{A}_{k},\overline{L}_{k+1}\right)
\right\vert \overline{A}_{k},\overline{L}_{k}\right\} \right] \right]  \notag
\end{eqnarray}%
where $y_{K+1,\widehat{\eta }_{K+1,mach}^{t}}\left( \overline{A}_{K},%
\overline{L}_{K+1}\right) \equiv \psi \left( \overline{L}_{K+1}\right) .$
Using the identity for any $k\in \left[ K\right] $ and any $\left( \widehat{h%
}_{1},...,\widehat{h}_{K}\right) ,$%
\begin{equation*}
\frac{\pi ^{\ast k}}{\pi ^{k}}-\frac{\pi ^{\ast k}}{\widehat{\pi }^{k}}%
=\sum_{j=1}^{k}\frac{\pi ^{\ast j-1}}{\pi ^{j-1}}\left\{ \frac{h_{j}^{\ast }%
}{h_{j}}-\frac{h_{j}^{\ast }}{\widehat{h}_{j}}\right\} \frac{\pi
_{j+1}^{\ast K}}{\widehat{\pi }_{j+1}^{K}}
\end{equation*}%
we arrive at 
\begin{eqnarray*}
&&\left. E^{v}\left[ Q_{1}\left( \widehat{\overline{h}}^{t},\widehat{%
\overline{\eta }}_{mach}^{t}\right) \right] -\theta \left( g\right) =\right.
\\
&=&\sum_{k=1}^{K}E^{v}\left[ \frac{\pi ^{\ast K}}{\pi ^{k-1}\widehat{\pi }%
_{k+1}^{K}}\left\{ \frac{h_{k}^{\ast }}{h_{k}}-\frac{h_{k}^{\ast }}{\widehat{%
h}_{k}}\right\} \left[ \widehat{\eta }_{k,mach}^{t}-E_{g_{k}}\left\{ \left.
y_{k+1,\widehat{\eta }_{k+1,mach}^{t}}\left( \overline{A}_{k},\overline{L}%
_{k+1}\right) \right\vert \overline{A}_{k},\overline{L}_{k}\right\} \right] %
\right] \\
&&+\sum_{1\leq j<k\leq K}E^{v}\left[ \frac{\pi ^{\ast K}}{\pi ^{j-1}\widehat{%
\pi }_{j+1}^{K}}\left\{ \frac{h_{j}^{\ast }}{h_{j}}-\frac{h_{j}^{\ast }}{%
\widehat{h}_{j}}\right\} \left[ \widehat{\eta }_{k,mach}^{t}-E_{g_{k}}\left%
\{ \left. y_{k+1,\widehat{\eta }_{k+1,mach}^{t}}\left( \overline{A}_{k},%
\overline{L}_{k+1}\right) \right\vert \overline{A}_{k},\overline{L}%
_{k}\right\} \right] \right] .
\end{eqnarray*}

Likewise, using the formula $b^{p}\left( h^{\dag },\eta ^{\dag }\right) $
applied to $\left( \overline{h}^{\dag },\overline{\eta }^{\dag }\right)
=\left( \widehat{\overline{h}}^{t},\widetilde{\overline{\eta }}%
_{mach}^{t}\right) $ we obtain the following expression for the drift of $%
\widehat{\theta }_{MR,mach}$ 
\begin{eqnarray}
&&\left. E^{v}\left[ Q_{1}\left( \widehat{\overline{h}}^{t},\widetilde{%
\overline{\eta }}_{mach}^{t}\right) \right] -\theta \left( g\right) =\right.
\label{driftMR} \\
&=&\sum_{k=1}^{K}E^{v}\left[ \frac{\pi ^{\ast \left( k-1\right) }}{\pi
_{1}^{\left( k-1\right) }}\left( \frac{h_{k}^{\ast }}{h_{k}}-\frac{%
h_{k}^{\ast }}{\widehat{h}_{k}}\right) \left[ \widetilde{\eta }%
_{k,mach}^{t}-E_{g_{k}}\left\{ \left. \widetilde{Q}_{k+1,mach}\right\vert 
\overline{A}_{k},\overline{L}_{k}\right\} \right] \right]  \notag
\end{eqnarray}

with $\widetilde{Q}_{K+1,mach}\equiv \psi \left( \overline{L}_{K+1}\right) .$

Note that $\widehat{\eta }_{k,mach}^{t}-E_{g_{k}}\left\{ \left. y_{k+1,%
\widehat{\eta }_{k+1,mach}^{t}}\left( \overline{A}_{k},\overline{L}%
_{k+1}\right) \right\vert \overline{A}_{k},\overline{L}_{k}\right\} $ is
equal to the residual $\widehat{E}_{mach}^{t}\left\{ \left. y_{k+1,\widehat{%
\eta }_{k+1,mach}^{t}}\left( \overline{A}_{k},\overline{L}_{k+1}\right)
\right\vert \overline{A}_{k},\overline{L}_{k}\right\} -E_{g_{k}}\left\{
\left. y_{k+1,\widehat{\eta }_{k+1,mach}^{t}}\left( \overline{A}_{k},%
\overline{L}_{k+1}\right) \right\vert \overline{A}_{k},\overline{L}%
_{k}\right\} $ and $\widetilde{\eta }_{k,mach}^{t}-E_{g_{k}}\left\{ \left. 
\widetilde{Q}_{k+1,mach}\right\vert \overline{A}_{k},\overline{L}%
_{k}\right\} $ is equal to the residual $\widehat{E}_{mach}^{t}\left\{
\left. \widetilde{Q}_{k+1,mach}\right\vert \overline{A}_{k},\overline{L}%
_{k}\right\} -E_{g_{k}}\left\{ \left. \widetilde{Q}_{k+1,mach}\right\vert 
\overline{A}_{k},\overline{L}_{k}\right\} $ where for any $W=w\left( 
\overline{A}_{k},\overline{L}_{k+1}\right) ,$ $\widehat{E}_{mach}\left\{ W|%
\overline{A}_{k},\overline{L}_{k}\right\} $ is the machine learning
estimator of the true expectation $E_{g_{k}}\left( W|\overline{A}_{k},%
\overline{L}_{k}\right) .$ Note also that had we used the expression $%
b^{p}\left( h^{\dag },\eta ^{\dag }\right) $ to represent the drift of $%
\widehat{\theta }_{DR,mach}$, this would have resulted in an expression
involving the differences $\widehat{\overline{\eta }}_{k,mach}^{t}-E_{g_{k}}%
\left\{ \left. \widehat{Q}_{k+1,mach}\right\vert \overline{A}_{k},\overline{L%
}_{k}\right\} =\widehat{E}_{mach}^{t}\left\{ \left. y_{k+1,\widehat{\eta }%
_{k+1,mach}^{t}}\left( \overline{A}_{k},\overline{L}_{k+1}\right)
\right\vert \overline{A}_{k},\overline{L}_{k}\right\} -E_{g_{k}}\left\{
\left. \widehat{Q}_{k+1,mach}\right\vert \overline{A}_{k},\overline{L}%
_{k}\right\} $ with $\widehat{Q}_{k+1}$ defined iteratively for $%
k=K-1,K-2,...,0,$ as $\widehat{Q}_{k+1,mach}\equiv \frac{h_{k+1}^{\ast }}{%
\widehat{h}_{k+1}^{t,u}}\left[ \widetilde{Q}_{k+2,mach}^{u}-\widehat{\eta }%
_{k+1,mach}^{t,u}\left( \overline{A}_{k+1},\overline{L}_{k+1}\right) \right]
+\widehat{Y}_{k+1,mach}^{u}.$ Because these differences are not the
residuals from applying the machine learning algorithm to the outcome $%
y_{k+1,\widehat{\eta }_{k+1,mach}^{t}}\left( \overline{A}_{k},\overline{L}%
_{k+1}\right) ,$ using the expression $b^{p}\left( h^{\dag },\eta ^{\dag
}\right) $ to represent the drift of $\widehat{\theta }_{DR,mach}$ would
have made the structure of the drift less transparent. Likewise, a similar
situation would arise if we use the expression $c^{p}\left( h^{\dag },\eta
^{\dag }\right) $ to represent the drift of $\widehat{\theta }_{MR,mach}.$

\ Although the drift of $\widehat{\theta }_{DR,mach}$ has many more terms
than the drift of $\widehat{\theta }_{MR,mach}$, at this level of generality
it does not seem possible to quantitatively compare the size of the drifts
of the two estimators when the precise machine learning algorithm being used
is not further specified. In the special case that the machine learning
algorithm is a linear operator, direct and easily interpretable comparisons
become possible. These are discussed in the next subsection. In particular,
we will argue that if $\eta _{k}^{g}$ and the true $h_{k}$ lie in specific
smoothness classes, then we can quantify the rates of convergence of the
drifts to zero. Our analysis relies on a specific representation for $%
a^{p}\left( \text{ }h,\eta \text{ }\right) ,$ given in the next subsection,
when $\eta =\left( \eta _{1},...,\eta _{K}\right) $ takes two special forms
which mimic the forms that the estimators $\widehat{\eta }_{k+1,mach}$ and $%
\widetilde{\eta }_{k+1,mach}$ take when the machine learning algorithms used
are linear operators.

\subsubsection{Analysis when the ML algorithms used are linear operators.}

We will now argue that if $\eta _{k}^{g}$ and the true $h_{k}$ lie in
specific smoothness classes, then we can quantify the rates of convergence
of the drifts to zero. Our analysis relies on a specific representation for $%
a^{p}\left( \text{ }h,\eta \text{ }\right) ,$ given in the next Theorem,
when $\eta =\left( \eta _{1},...,\eta _{K}\right) $ takes two special forms
which mimic the forms that the estimators $\widehat{\eta }_{k+1,mach}$ and $%
\widetilde{\eta }_{k+1,mach}$ take when the machine learning algorithms used
are linear operators. To state the Theorem, we must first define a number of
objects, which we now do.

Given $h^{\dagger }=\left( h_{1}^{\dagger },\ldots ,h_{K}^{\dagger }\right)
, $ define for $0\leq j<u\leq K,$%
\begin{equation*}
\nabla _{j,u}\equiv \frac{\pi _{j+1}^{\ast u-1}}{\pi _{j+1}^{\dagger ,u-1}}%
\left( \frac{h_{u}^{\ast }}{h_{u}}-\frac{h_{u}^{\ast }}{h_{u}^{\dagger }}%
\right)
\end{equation*}

Given linear operators $\Pi ^{j}\left[ \cdot \right] :L_{2}\left(
Q_{j}\right) \rightarrow L_{2}\left( P_{j}\right) ,j\in \left[ K\right] ,$
where $Q_{j}$ and $P_{j}$ are the laws of $\left( \overline{A}_{j},\overline{%
L}_{j+1}\right) $ and $\left( \overline{A}_{j},\overline{L}_{j}\right) $
respectively, we define the following operators

\begin{enumerate}
\item for $j=1,...,K-1,$ 
\begin{equation*}
\Pi _{DR}^{j}\left[ \cdot \right] =\Pi ^{j}\left\{ E_{p}\left( \left. \frac{%
h_{j+1}^{\ast }}{h_{j+1}}\text{ }\cdot \text{ }\right\vert \overline{A}_{j},%
\overline{L}_{j+1}\right) \right\}
\end{equation*}

\item for $1\leq j<k\leq $ $K-1,$ 
\begin{equation*}
\Pi _{DR,j,k}\left[ \cdot \right] =\Pi _{DR}^{j}\circ \ldots \circ \Pi
_{DR}^{k-1}\left[ \cdot \right]
\end{equation*}%
where $\circ $ denotes the composition operation. Note that $\Pi _{DR,j,j+1}%
\left[ \cdot \right] =\Pi _{DR}^{j}\left[ \cdot \right] $.

\item for $1\leq j<u\leq $ $K,$ 
\begin{equation*}
\Pi _{MR,j,u}\left[ \cdot \right] \equiv \Pi ^{j}\left[ E_{p}\left( \left.
\nabla _{j,u}\cdot \right\vert \overline{A}_{j},\overline{L}_{j+1}\right) %
\right]
\end{equation*}%
Note that 
\begin{equation*}
\Pi _{MR,j,j+1}\left[ \cdot \right] =\Pi ^{j}\left[ E_{p}\left\{ \left.
\left( \frac{h_{j+1}^{\ast }}{h_{j+1}}-\frac{h_{j+1}^{\ast }}{h_{j+1}^{\dag }%
}\right) \cdot \right\vert \overline{A}_{j},\overline{L}_{j+1}\right\} %
\right]
\end{equation*}

\item For $1\leq r_{1}<r_{2}<\ldots <r_{u}\leq $ $K,$ 
\begin{equation*}
\Pi _{MR,r_{1},r_{2},...,r_{u}}\left[ \cdot \right] \equiv \Pi
_{MR,r_{1},r_{2}}\circ \ldots \circ \Pi _{MR,r_{u-2},r_{u-1}}\circ \Pi
_{MR,r_{u-1},r_{u}}\left[ \cdot \right]
\end{equation*}
\end{enumerate}

Next, define the following random variables:

\begin{enumerate}
\item[a.] for $j=1,...,K,$ define 
\begin{equation*}
\eta _{j,DR}\equiv \eta _{j,DR}\left( \overline{A}_{j},\overline{L}%
_{j}\right) \equiv \Pi ^{j}\left[ y_{j+1,\eta _{j+1}^{g}}\left( \overline{A}%
_{j},\overline{L}_{j+1}\right) \right]
\end{equation*}

\item[b.] for $j=K,K-1,...,1,$ recursively define%
\begin{equation*}
\widehat{\eta }_{j,DR}\equiv \widehat{\eta }_{j,DR}\left( \overline{A}_{j},%
\overline{L}_{j}\right) \equiv \Pi ^{j}\left[ y_{j+1,\widehat{\eta }%
_{j+1,DR}}\left( \overline{A}_{j},\overline{L}_{j+1}\right) \right]
\end{equation*}

\item[c.] Given $h^{\dagger }=\left( h_{1}^{\dagger },\ldots ,h_{K}^{\dagger
}\right) ,$ for $j=K,K-1,...,1,$ recursively define 
\begin{equation*}
\widetilde{\eta }_{j,MR}\equiv \widetilde{\eta }_{j,MR}\left( \overline{A}%
_{j},\overline{L}_{j}\right) \equiv \Pi ^{j}\left[ Q_{j+1}\left( \overline{h}%
_{j+1}^{\dagger K},\overline{\widetilde{\eta }}_{j+1,MR}^{K}\right) \right]
\end{equation*}%
where $Q\left( \overline{h}_{K+1}^{\dagger K},\overline{\widetilde{\eta }}%
_{j,MR}^{K}\right) \equiv \psi \left( \overline{L}_{K+1}\right) .$

\item[d.] Given $h^{\dagger }=\left( h_{1}^{\dagger },\ldots ,h_{K}^{\dagger
}\right) ,$ for $j=K,K-1,...,1,$ recursively define 
\begin{eqnarray*}
\eta _{j,MR} &\equiv &\eta _{j,MR}\left( \overline{A}_{j},\overline{L}%
_{j}\right) \\
&\equiv &\eta _{j,DR}+\Pi ^{j}\left[ Q_{j+1}\left( \overline{h}%
_{j+1}^{\dagger K},\overline{\widetilde{\eta }}_{j+1,MR}^{K}\right)
-E_{p}\left\{ \left. Q_{j+1}\left( \overline{h}_{j+1}^{\dagger K},\overline{%
\widetilde{\eta }}_{j+1,MR}^{K}\right) \right\vert \overline{A}_{j},%
\overline{L}_{j+1}\right\} \right] .
\end{eqnarray*}
\end{enumerate}

\bigskip

The following Theorem gives special representations for $a^{p}\left( h,\eta
\right) $ when $\eta =$ $\widehat{\eta }_{DR}$ and $\eta =$ $\widetilde{\eta 
}_{DR}.$

\textbf{Theorem 1. }Let $\widehat{\eta }_{DR}\equiv \left( \widehat{\eta }%
_{1,DR},...,\widehat{\eta }_{K,DR}\right) $ and $\widetilde{\eta }%
_{DR}\equiv \left( \widetilde{\eta }_{1,DR},...,\widetilde{\eta }%
_{K,DR}\right) $ where $\widehat{\eta }_{j,DR}$ and $\widetilde{\eta }%
_{j,DR} $, $j\in \left[ K\right] $ are the random variables defined in (b)
and (c) above and $h^{\dagger }\equiv \left( h_{1}^{\dagger },\ldots
,h_{K}^{\dagger }\right) ,$ with $h_{k}^{\dagger }$ an arbitrary density for
the law of $A_{k}$ given $\left( \overline{A}_{k-1},\overline{L}_{k}\right) $%
, $k\in \left[ K\right] .$ The following identities hold.

\begin{enumerate}
\item for $k\in \left[ K\right] $ 
\begin{equation}
\widehat{\eta }_{k,DR}-\eta _{k}^{g}=\eta _{k,DR}-\eta
_{k}^{g}+\dsum\limits_{j=k+1}^{K}\Pi _{DR,k,j}\left[ \eta _{j,DR}-\eta
_{j}^{g}\right]  \label{dr-theo1}
\end{equation}

\item \ 
\begin{eqnarray*}
a^{p}\left( h^{\dagger },\widehat{\eta }_{DR}\right) &\equiv
&\dsum\limits_{k=1}^{K}E_{p}\left\{ \frac{\pi ^{\ast k-1}}{\widehat{\pi }%
^{k-1}}\left( \frac{h_{k}^{\ast }}{h_{k}}-\frac{h_{k}^{\ast }}{%
h_{k}^{\dagger }}\right) \left( \widehat{\eta }_{k,DR}-\eta _{k}^{g}\right)
\right\} \\
&=&\dsum\limits_{k=1}^{K}E_{p}\left\{ \frac{\pi ^{\ast k-1}}{\pi ^{\dagger
k-1}}\left( \frac{h_{k}^{\ast }}{h_{k}}-\frac{h_{k}^{\ast }}{h_{k}^{\dagger }%
}\right) \left( \eta _{k,DR}-\eta _{k}^{g}\right) \right\} \\
&&+\dsum\limits_{1\leq k<j\leq K}E_{p}\left\{ \frac{\pi ^{\ast k-1}}{\pi
^{\dagger k-1}}\left( \frac{h_{k}^{\ast }}{h_{k}}-\frac{h_{k}^{\ast }}{%
h_{k}^{\dagger }}\right) \Pi _{DR,k,j}\left[ \eta _{j,DR}-\eta _{j}^{g}%
\right] \right\}
\end{eqnarray*}

\item 
\begin{eqnarray}
&&\dsum\limits_{k=1}^{K}E_{p}\left\{ \left. \frac{\pi ^{\ast k-1}}{\pi ^{k-1}%
}\left( \frac{h_{k}^{\ast }}{h_{k}}-\frac{h_{k}^{\ast }}{h_{k}^{\dagger }}%
\right) \left( \widetilde{\eta }_{k,MR}-\eta _{k}^{g}\right) \right\vert
L_{1}\right\}  \label{mr-theo1} \\
&=&\dsum\limits_{k=1}^{K}E_{p}\left\{ \left. \nabla _{0,k}\left( \eta
_{k,MR}-\eta _{k}^{g}\right) \right\vert L_{1}\right\}  \notag \\
&&+\dsum\limits_{\underset{r_{1}<r_{2}<...<r_{u}}{\emptyset \not=\left\{
r_{1},...,r_{u}\right\} \subseteq \left[ K-1\right] }%
}\sum_{k=r_{u}+1}^{K}E_{p}\left( \left. \nabla _{0,r_{1}}\Pi
_{MR,r_{1},r_{2},...,r_{u},k}^{\dag }\left[ \eta _{k,MR}-\eta _{k}^{g}\right]
\right\vert L_{1}\right)  \notag
\end{eqnarray}

\item 
\begin{eqnarray*}
a^{p}\left( h^{\dagger },\widetilde{\eta }_{MR}\right) &\equiv
&\dsum\limits_{k=1}^{K}E_{p}\left\{ \frac{\pi ^{\ast k-1}}{\widehat{\pi }%
^{k-1}}\left( \frac{h_{k}^{\ast }}{h_{k}}-\frac{h_{k}^{\ast }}{%
h_{k}^{\dagger }}\right) \left( \widetilde{\eta }_{k,MR}-\eta
_{k}^{g}\right) \right\} \\
&=&\dsum\limits_{k=1}^{K}E_{p}\left\{ \frac{\pi ^{\ast k-1}}{\pi ^{\dagger
k-1}}\left( \frac{h_{k}^{\ast }}{h_{k}}-\frac{h_{k}^{\ast }}{h_{k}^{\dagger }%
}\right) \left( \eta _{k,MR}-\eta _{k}^{g}\right) \right\} \\
&&+\dsum\limits_{\underset{r_{1}<r_{2}<...<r_{u}}{\emptyset \not=\left\{
r_{1},...,r_{u}\right\} \subseteq \left[ K-1\right] }%
}\sum_{k=r_{u}+1}^{K}E_{p}\left( \nabla _{0,r_{1}}\Pi
_{MR,r_{1},r_{2},...,r_{u},k}^{\dag }\left[ \eta _{k,MR}-\eta _{k}^{g}\right]
\right)
\end{eqnarray*}
\end{enumerate}

\bigskip

\textbf{Notational remark: }in parts (3) and (4) of the\ Theorem, the
summation $\dsum\limits_{\underset{r_{1}<r_{2}<...<r_{u}}{\emptyset
\not=\left\{ r_{1},...,r_{u}\right\} \subseteq \left[ K-1\right] }}$ is over
all non-empty subsets of $\left[ K-1\right] \equiv \left\{ 1,...,K-1\right\}
,$ where we denote the ordered elements of a subset with cardinality $u$
with $r_{1}<r_{2}<...<r_{u}.$

\medskip

In the special case in which $K=2,$ assertions (2) and (4) of the Theorem
reduce to

\begin{eqnarray}
a^{p}\left( h^{\dag },\widehat{\eta }_{DR}\right) &\equiv &E_{p}\left\{ 
\frac{\pi ^{\ast 1}}{\pi ^{\dag 1}}\left( \frac{h_{2}^{\ast }}{h_{2}}-\frac{%
h_{2}^{\ast }}{h_{2}^{\dag }}\right) \left( \eta _{2,DR}-\eta
_{2}^{g}\right) \right\}  \label{drift1} \\
&&+E_{p}\left\{ \left( \frac{h_{1}^{\ast }}{h_{1}}-\frac{h_{1}^{\ast }}{%
h_{1}^{\dag }}\right) \left( \eta _{1,DR}-\eta _{1}^{g}\right) \right\} 
\notag \\
&&+E_{p}\left\{ \left( \frac{h_{1}^{\ast }}{h_{1}}-\frac{h_{1}^{\ast }}{%
h_{1}^{\dag }}\right) \Pi ^{1}\left[ E_{p}\left\{ \left. \frac{h_{2}^{\ast }%
}{h_{2}}\left( \eta _{2,DR}-\eta _{2}^{g}\right) \right\vert A_{1},\overline{%
L}_{2}\right\} \right] \right\}  \notag
\end{eqnarray}%
and

\begin{eqnarray}
a^{p}\left( h^{\dag },\widetilde{\eta }_{MR}\right) &\equiv &E_{p}\left\{ 
\frac{\pi ^{\ast 1}}{\pi ^{\dag 1}}\left( \frac{h_{2}^{\ast }}{h_{2}}-\frac{%
h_{2}^{\ast }}{h_{2}^{\dag }}\right) \left( \eta _{2,MR}-\eta
_{2}^{g}\right) \right\}  \label{drift2} \\
&&+E_{p}\left\{ \left( \frac{h_{1}^{\ast }}{h_{1}}-\frac{h_{1}^{\ast }}{%
h_{1}^{\dag }}\right) \left( \eta _{1,MR}-\eta _{1}^{g}\right) \right\} 
\notag \\
&&+E_{p}\left[ \left( \frac{h_{1}^{\ast }}{h_{1}}-\frac{h_{1}^{\ast }}{%
h_{1}^{\dag }}\right) \Pi ^{1}\left[ E_{p}\left\{ \left. \left( \frac{%
h_{2}^{\ast }}{h_{2}}-\frac{h_{2}^{\ast }}{h_{2}^{\dag }}\right) \left( \eta
_{2,MR}-\eta _{2}^{g}\right) \right\vert A_{1},\overline{L}_{2}\right\} %
\right] \right]  \notag
\end{eqnarray}%
When $K=3,$ these formulae are%
\begin{eqnarray}
a^{p}\left( h^{\dag },\widehat{\eta }_{DR}\right) &\equiv
&\dsum\limits_{k=1}^{3}E_{p}\left\{ \frac{\pi ^{\ast k-1}}{\pi ^{\dag k-1}}%
\left( \frac{h_{k}^{\ast }}{h_{k}}-\frac{h_{k}^{\ast }}{h_{k}^{\dag }}%
\right) \left( \eta _{k,DR}-\eta _{k}^{g}\right) \right\}  \label{drift3} \\
&&+E_{p}\left\{ \frac{\pi ^{\ast 1}}{\pi ^{\dag 1}}\left( \frac{h_{2}^{\ast }%
}{h_{2}}-\frac{h_{2}^{\ast }}{h_{2}^{\dag }}\right) \Pi ^{2}\left[ \left.
E_{p}\left\{ \left. \frac{h_{3}^{\ast }}{h_{3}}\left( \eta _{3,DR}-\eta
_{3}^{g}\right) \right\vert \overline{A}_{2},\overline{L}_{3}\right\}
\right\vert \right] \right\}  \notag \\
&&+E_{p}\left\{ \left( \frac{h_{1}^{\ast }}{h_{1}}-\frac{h_{1}^{\ast }}{%
h_{1}^{\dag }}\right) \Pi ^{1}\left[ E_{p}\left\{ \left. \frac{h_{2}^{\ast }%
}{h_{2}}\left( \eta _{2,DR}-\eta _{2}^{g}\right) \right\vert A_{1},\overline{%
L}_{2}\right\} \right] \right\}  \notag \\
&&+E_{p}\left\{ \left( \frac{h_{1}^{\ast }}{h_{1}}-\frac{h_{1}^{\ast }}{%
h_{1}^{\dag }}\right) \Pi ^{1}\left[ E_{p}\left\{ \left. \frac{h_{2}^{\ast }%
}{h_{2}}\Pi ^{2}\left[ E_{p}\left\{ \left. \frac{h_{3}^{\ast }}{h_{3}}\left(
\eta _{3,DR}-\eta _{3}^{g}\right) \right\vert A_{2},\overline{L}_{3}\right\} %
\right] \right\vert A_{1},\overline{L}_{2}\right\} \right] \right\}  \notag
\end{eqnarray}%
and%
\begin{eqnarray}
&&\left. a^{p}\left( h^{\dag },\widetilde{\eta }_{MR}\right) \equiv \right.
\label{drift4} \\
&\equiv &\dsum\limits_{k=1}^{3}E_{p}\left[ \frac{\pi ^{\ast k-1}}{\pi ^{\dag
k-1}}\left( \frac{h_{k}^{\ast }}{h_{k}}-\frac{h_{k}^{\ast }}{h_{k}^{\dag }}%
\right) \left( \eta _{k,MR}-\eta _{k}^{g}\right) \right]  \notag \\
&&+E_{p}\left\{ \frac{\pi ^{\ast 1}}{\pi ^{\dag 1}}\left( \frac{h_{2}^{\ast }%
}{h_{2}}-\frac{h_{2}^{\ast }}{h_{2}^{\dag }}\right) \Pi ^{2}\left[
E_{p}\left\{ \left. \left( \frac{h_{3}^{\ast }}{h_{3}}-\frac{h_{3}^{\ast }}{%
h_{3}^{\dag }}\right) \left( \eta _{3,MR}-\eta _{3}^{g}\right) \right\vert
A_{2},\overline{L}_{3}\right\} \right] \right\}  \notag \\
&&+E_{p}\left\{ \left( \frac{h_{1}^{\ast }}{h_{1}}-\frac{h_{1}^{\ast }}{%
h_{1}^{\dag }}\right) \Pi ^{1}\left[ E_{p}\left\{ \left. \left( \frac{%
h_{2}^{\ast }}{h_{2}}-\frac{h_{2}^{\ast }}{h_{2}^{\dag }}\right) \left( \eta
_{2,MR}-\eta _{2}^{g}\right) \right\vert A_{1},\overline{L}_{2}\right\} %
\right] \right\}  \notag \\
&&+E_{p}\left\{ \left( \frac{h_{1}^{\ast }}{h_{1}}-\frac{h_{1}^{\ast }}{%
h_{1}^{\dag }}\right) \Pi ^{1}\left[ E_{p}\left\{ \left. \left( \frac{%
h_{2}^{\ast }}{h_{2}}-\frac{h_{2}^{\ast }}{h_{2}^{\dag }}\right) \Pi ^{2}%
\left[ E_{p}\left\{ \left. \left( \frac{h_{3}^{\ast }}{h_{3}}-\frac{%
h_{3}^{\ast }}{h_{3}^{\dag }}\right) \left( \eta _{3,MR}-\eta
_{3}^{g}\right) \right\vert A_{2},\overline{L}_{3}\right\} \right]
\right\vert A_{1},\overline{L}_{2}\right\} \right] \right\}  \notag \\
&&+E_{p}\left\{ \left( \frac{h_{1}^{\ast }}{h_{1}}-\frac{h_{1}^{\ast }}{%
h_{1}^{\dag }}\right) \Pi ^{1}\left[ E_{p}\left\{ \left. \frac{h_{2}^{\ast }%
}{h_{2}}\left( \frac{h_{3}^{\ast }}{h_{3}}-\frac{h_{3}^{\ast }}{h_{3}^{\dag }%
}\right) \left( \eta _{3,MR}-\eta _{3}^{g}\right) \right\vert A_{1},%
\overline{L}_{2}\right\} \right] \right\}  \notag
\end{eqnarray}

Theorem 1 can be applied to quantify the rates of convergence of the drifts
of $\widehat{\theta }_{DR,mach}$ and $\widehat{\theta }_{MR,mach}$ when the
machine learning algorithm used is a linear operator. Here we will apply it
to the special case in which the machine learning algorithms used are series
estimators. For any $k\in \left[ K\right] ,$ let $S_{k}=s_{k}\left( 
\overline{A}_{k},\overline{L}_{k}\right) $ be a vector valued function of $%
\left( \overline{A}_{k},\overline{L}_{k}\right) $ of dimension $%
m_{k}=m_{k}\left( n\right) $ which depends on the sample size $n.$ For a
given $s_{k}\left( \overline{A}_{k},\overline{L}_{k}\right) \in L_{2}\left(
P_{k}\right) ,$ define the operator $\Pi _{n,k}^{t}:L_{2}\left( Q_{k}\right)
\rightarrow L_{2}\left( P_{k}\right) $ 
\begin{equation*}
\Pi _{n,k}^{t}\left[ f_{k}\right] \equiv \widehat{\beta }s_{k}\left( \cdot
\right)
\end{equation*}%
where $\widehat{\beta }=\mathbb{P}_{n}^{t}\left[ f_{k}\left( \overline{A}%
_{k},\overline{L}_{k+1}\right) s_{k}\left( \overline{A}_{k},\overline{L}%
_{k}\right) ^{T}\right] \mathbb{P}_{n}^{t}\left[ s_{k}\left( \overline{A}%
_{k},\overline{L}_{k}\right) s_{k}\left( \overline{A}_{k},\overline{L}%
_{k}\right) ^{T}\right] ^{-1}$ is the least squares coefficient in the
regression of $f_{k}\left( \overline{A}_{k},\overline{L}_{k+1}\right) $ on $%
s_{k}\left( \overline{A}_{k},\overline{L}_{k}\right) \,\ $in the training
sample $\mathcal{S}_{t}.$

Notice that $\widehat{\eta }_{k,DR}$ defined in (b) above coincides with the
estimator $\widehat{\eta }_{k,mach}^{t,u}$ from step (c) of Algorithm 6 when
the machine learning algorithm is linear regression on $S_{k}$ and the
linear operator $\Pi ^{k}$ is $\Pi _{n,k}^{t}.$ Likewise, $\widetilde{\eta }%
_{k,MR}\,$defined in (c) above coincides with the estimator $\widetilde{\eta 
}_{k,mach}^{t,u}\,$\ of step (d) of Algorithm 6. We can then apply the
formula in part (2) of Theorem 1 evaluated at $\widehat{\eta }_{k,DR}=%
\widehat{\eta }_{k,mach}^{t,u}$ to compute the drift of $\widehat{\theta }%
_{DR,mach}$ and the formula in part (4) of Theorem 1 evaluated at $%
\widetilde{\eta }_{k,MR}=\widetilde{\eta }_{k,mach}^{t,u}$ to compute the
drift of $\widehat{\theta }_{MR,mach}$. We will now do so in the special
cases $K=2$ and $K=3.$ This will illustrate and clarify the relationship
between the drifts of $\widehat{\theta }_{DR,mach}$ and $\widehat{\theta }%
_{MR,mach}$ without unduly complicating the notation.

Using arguments analogous to those in the sections dealing with parametric
nuisance models, it can be shown that the drifts of the estimators $\widehat{%
\theta }_{DR,CF,mach,bang},\widehat{\theta }_{DR,CF,mach,reg}$ have the same
rate of convergence to 0 as the drift of $\widehat{\theta }_{DR,CF,mach},$
and the drifts of $\widehat{\theta }_{MR,CF,mach,bang}$ and $\widehat{\theta 
}_{MR,CF,mach,reg}$ have the same rate of convergence as $\widehat{\theta }%
_{MR,CF,mach}$. Hence, we will restrict our discussion to the analysis of $%
\widehat{\theta }_{DR,CF,mach}$ and $\widehat{\theta }_{MR,CF,mach}.$

In what follows we let $E^{v,\left( \overline{A}_{k},\overline{L}%
_{k+1}\right) }\left( \cdot \right) $ denote the conditional expectation
given $\left( \overline{A}_{k},\overline{L}_{k+1}\right) $ operator,
regarding the data in the training sample as fixed, i.e. non-random, e.g. 
\begin{eqnarray*}
&&E^{v,\left( \overline{A}_{2},\overline{L}_{3}\right) }\left\{ \left( \frac{%
h_{3}^{\ast }}{h_{3}}-\frac{h_{3}^{\ast }}{\widehat{h}_{3}}\right) \left( 
\widetilde{\eta }_{3,MR}-\eta _{3}^{g}\right) \right\} \\
&=&\dint \left( \frac{h_{3}^{\ast }\left( a_{3}|\overline{A}_{2},\overline{L}%
_{3}\right) }{h_{3}\left( a_{3}|\overline{A}_{2},\overline{L}_{3}\right) }-%
\frac{h_{3}^{\ast }\left( a_{3}|\overline{A}_{2},\overline{L}_{3}\right) }{%
\widehat{h}_{3}\left( a_{3}|\overline{A}_{2},\overline{L}_{3}\right) }%
\right) \left( \widetilde{\eta }_{3,MR}\left( a_{3},\overline{A}_{2},%
\overline{L}_{3}\right) -\eta _{3}^{g}\left( a_{3},\overline{A}_{2},%
\overline{L}_{3}\right) \right) h\left( a_{3}|\overline{A}_{2},\overline{L}%
_{3}\right) da_{3}
\end{eqnarray*}

For $K=2,$ applying the formula $\left( \ref{drift1}\right) $ with $\widehat{%
\eta }_{k,DR}=\widehat{\eta }_{k,mach}^{t,u}$ and $\Pi ^{k}=\Pi _{n,k}^{t},$
we conclude that the drift $E^{v}\left[ Q_{1}\left( \widehat{\overline{h}}%
^{t},\widehat{\overline{\eta }}_{mach}^{t}\right) \right] -\theta \left(
g\right) $ of $\widehat{\theta }_{DR,mach}$ is 
\begin{eqnarray*}
a^{p}\left( \widehat{\overline{h}}^{t},\widehat{\overline{\eta }}%
_{mach}^{t}\right) &=&E^{v}\left\{ \frac{\pi ^{\ast 1}}{\widehat{\pi }^{1}}%
\left( \frac{h_{2}^{\ast }}{h_{2}}-\frac{h_{2}^{\ast }}{\widehat{h}_{2}^{t}}%
\right) \left( \eta _{2,DR}-\eta _{2}^{g}\right) \right\} \\
&&+E^{v}\left\{ \left( \frac{h_{1}^{\ast }}{h_{1}}-\frac{h_{1}^{\ast }}{%
\widehat{h}_{1}^{t}}\right) \left( \eta _{1,DR}-\eta _{1}^{g}\right) \right\}
\\
&&+E^{v}\left\{ \left( \frac{h_{1}^{\ast }}{h_{1}}-\frac{h_{1}^{\ast }}{%
\widehat{h}_{1}^{t}}\right) \Pi _{n,1}^{t}\left\{ E^{v,\left( A_{1},%
\overline{L}_{2}\right) }\left\{ \frac{h_{2}^{\ast }}{h_{2}}\left( \eta
_{2,DR}-\eta _{2}^{g}\right) \right\} \right\} \right\}
\end{eqnarray*}%
and applying the formula $\left( \ref{drift2}\right) $ with $\widetilde{\eta 
}_{k,MR}=\widetilde{\eta }_{k,mach}^{t,u}$ and $\Pi ^{k}=\Pi _{n,k}^{t},$
the drift $E^{v}\left[ Q_{1}\left( \widehat{\overline{h}}^{t},\widetilde{%
\overline{\eta }}_{mach}^{t}\right) \right] -\theta \left( g\right) $ of $%
\widehat{\theta }_{MR,mach}$ is

\begin{eqnarray*}
a^{p}\left( \widehat{\overline{h}}^{t},\widetilde{\overline{\eta }}%
_{mach}^{t}\right) &\equiv &E^{v}\left\{ \frac{\pi ^{\ast 1}}{\widehat{\pi }%
^{1}}\left( \frac{h_{2}^{\ast }}{h_{2}}-\frac{h_{2}^{\ast }}{\widehat{h}%
_{2}^{t}}\right) \left( \eta _{2,MR}-\eta _{2}^{g}\right) \right\} \\
&&+E^{v}\left\{ \left( \frac{h_{1}^{\ast }}{h_{1}}-\frac{h_{1}^{\ast }}{%
\widehat{h}_{1}^{t}}\right) \left( \eta _{1,MR}-\eta _{1}^{g}\right) \right\}
\\
&&+E^{v}\left\{ \left( \frac{h_{1}^{\ast }}{h_{1}}-\frac{h_{1}^{\ast }}{%
\widehat{h}_{1}^{t}}\right) \Pi _{n,1}^{t}\left\{ E^{v,\left( A_{1},%
\overline{L}_{2}\right) }\left\{ \left( \frac{h_{2}^{\ast }}{h_{2}}-\frac{%
h_{2}^{\ast }}{\widehat{h}_{2}^{t}}\right) \left( \eta _{2,MR}-\eta
_{2}^{g}\right) \right\} \right\} \right\}
\end{eqnarray*}

For $K=3,$ formula $\left( \ref{drift3}\right) $ applied to $\widehat{\eta }%
_{k,DR}=\widehat{\eta }_{k,mach}^{t,u}$ and $\Pi ^{k}=\Pi _{n,k}^{t}$
implies that the drift $E^{v}\left[ Q_{1}\left( \widehat{\overline{h}}^{t},%
\widehat{\overline{\eta }}_{mach}^{t}\right) \right] -\theta \left( g\right)
\,\ $of $\widehat{\theta }_{DR,mach}$ is 
\begin{equation}
a^{p}\left( \widehat{\overline{h}}^{t},\widehat{\overline{\eta }}%
_{mach}^{t}\right) \equiv \dsum\limits_{k=1}^{3}E^{v}\left( \delta
_{k}^{DR}\right) +\dsum\limits_{1\leq k<j\leq 3}^{3}E^{v}\left( \xi
_{k,j}^{DR}\right)  \label{drift-dr}
\end{equation}%
and formula $\left( \ref{drift4}\right) $ applied to $\widetilde{\eta }%
_{k,MR}=\widetilde{\eta }_{k,mach}^{t,u}$ and $\Pi ^{k}=\Pi _{n,k}^{t}$
implies that the drift $E^{v}\left[ Q_{1}\left( \widehat{\overline{h}}^{t},%
\widetilde{\overline{\eta }}_{mach}^{t}\right) \right] -\theta \left(
g\right) \,\ $of $\widehat{\theta }_{MR,mach}$ is 
\begin{equation}
a^{p}\left( \widehat{\overline{h}}^{t},\widetilde{\overline{\eta }}%
_{mach}^{t}\right) =\dsum\limits_{k=1}^{3}E^{v}\left( \delta
_{k}^{MR}\right) +\dsum\limits_{1\leq k<j\leq 3}^{3}E^{v}\left( \xi
_{k,j}^{MR}\right) +E^{v}\left( \xi _{1,2,3}^{MR}\right)  \label{drift-mr}
\end{equation}

where%
\begin{eqnarray*}
\delta _{k}^{DR} &\equiv &\frac{\pi ^{\ast k-1}}{\widehat{\pi }^{k-1}}\left( 
\frac{h_{k}^{\ast }}{h_{k}}-\frac{h_{k}^{\ast }}{\widehat{h}_{k}^{t}}\right)
\left( \eta _{k,DR}-\eta _{k}^{g}\right) \text{,} \\
\delta _{k}^{MR} &\equiv &\frac{\pi ^{\ast k-1}}{\widehat{\pi }^{k-1}}\left( 
\frac{h_{k}^{\ast }}{h_{k}}-\frac{h_{k}^{\ast }}{\widehat{h}_{k}^{t}}\right)
\left( \eta _{k,MR}-\eta _{k}^{g}\right) ,
\end{eqnarray*}%
\begin{eqnarray*}
\xi _{1,2}^{DR} &\equiv &E^{v}\left\{ \left( \frac{h_{1}^{\ast }}{h_{1}}-%
\frac{h_{1}^{\ast }}{\widehat{h}_{1}^{t}}\right) \Pi _{n,1}^{t}\left\{
E^{v,\left( A_{1},\overline{L}_{2}\right) }\left\{ \frac{h_{2}^{\ast }}{h_{2}%
}\left( \eta _{2,DR}-\eta _{2}^{g}\right) \right\} \right\} \right\} , \\
\xi _{1,2}^{MR} &\equiv &E^{v}\left\{ \left( \frac{h_{1}^{\ast }}{h_{1}}-%
\frac{h_{1}^{\ast }}{\widehat{h}_{1}^{t}}\right) \Pi _{n,1}^{t}\left\{
E^{v,\left( A_{1},\overline{L}_{2}\right) }\left\{ \left( \frac{h_{2}^{\ast }%
}{h_{2}}-\frac{h_{2}^{\ast }}{\widehat{h}_{2}^{t}}\right) \left( \eta
_{2,MR}-\eta _{2}^{g}\right) \right\} \right\} \right\}
\end{eqnarray*}%
\begin{eqnarray*}
\xi _{2,3}^{DR} &\equiv &\frac{\pi ^{\ast 1}}{\widehat{\pi }^{1}}\left( 
\frac{h_{2}^{\ast }}{h_{2}}-\frac{h_{2}^{\ast }}{\widehat{h}_{2}^{t}}\right)
\Pi _{n,2}^{t}\left\{ E^{v,\left( \overline{A}_{2},\overline{L}_{3}\right)
}\left\{ \frac{h_{3}^{\ast }}{h_{3}}\left( \eta _{3,DR}-\eta _{3}^{g}\right)
\right\} \right\} \\
\xi _{2,3}^{MR} &\equiv &E^{v}\left\{ \frac{\pi ^{\ast 1}}{\widehat{\pi }^{1}%
}\left( \frac{h_{2}^{\ast }}{h_{2}}-\frac{h_{2}^{\ast }}{\widehat{h}_{2}^{t}}%
\right) \Pi _{n,2}^{t}\left\{ E^{v,\left( \overline{A}_{2},\overline{L}%
_{3}\right) }\left\{ \left( \frac{h_{3}^{\ast }}{h_{3}}-\frac{h_{3}^{\ast }}{%
\widehat{h}_{3}^{t}}\right) \left( \eta _{3,MR}-\eta _{3}^{g}\right)
\right\} \right\} \right\}
\end{eqnarray*}
\begin{eqnarray*}
\xi _{1,3}^{DR} &\equiv &E^{v}\left\{ \left( \frac{h_{1}^{\ast }}{h_{1}}-%
\frac{h_{1}^{\ast }}{\widehat{h}_{1}^{t}}\right) \Pi _{n,1}^{t}\left\{
E^{v,\left( A_{1},\overline{L}_{2}\right) }\left\{ \frac{h_{2}^{\ast }}{h_{2}%
}\Pi _{n,2}^{t}\left\{ E^{v,\left( \overline{A}_{2},\overline{L}_{3}\right)
}\left\{ \frac{h_{3}^{\ast }}{h_{3}}\left( \eta _{3,DR}-\eta _{3}^{g}\right)
\right\} \right\} \right\} \right\} \right\} \\
\xi _{1,3}^{MR} &\equiv &E^{v}\left\{ \left( \frac{h_{1}^{\ast }}{h_{1}}-%
\frac{h_{1}^{\ast }}{\widehat{h}_{1}^{t}}\right) \Pi _{n,1}^{t}\left\{
E^{v,\left( A_{1},\overline{L}_{2}\right) }\left\{ \frac{h_{2}^{\ast }}{h_{2}%
}\left( \frac{h_{3}^{\ast }}{h_{3}}-\frac{h_{3}^{\ast }}{\widehat{h}_{3}^{t}}%
\right) \left( \eta _{3,MR}-\eta _{3}^{g}\right) \right\} \right\} \right\}
\end{eqnarray*}%
and 
\begin{equation*}
\xi _{1,2,3}^{MR}\equiv E^{v}\left\{ \left( \frac{h_{1}^{\ast }}{h_{1}}-%
\frac{h_{1}^{\ast }}{\widehat{h}_{1}^{t}}\right) \Pi _{n,1}^{t}\left\{
E^{v,\left( A_{1},\overline{L}_{2}\right) }\left\{ \left( \frac{h_{2}^{\ast }%
}{h_{2}}-\frac{h_{2}^{\ast }}{\widehat{h}_{2}^{t}}\right) \Pi
_{n,2}^{t}\left\{ E^{v,\left( \overline{A}_{2},\overline{L}_{3}\right)
}\left\{ \left( \frac{h_{3}^{\ast }}{h_{3}}-\frac{h_{3}^{\ast }}{\widehat{h}%
_{3}^{t}}\right) \left( \eta _{3,MR}-\eta _{3}^{g}\right) \right\} \right\}
\right\} \right\} \right\} .
\end{equation*}

\bigskip

We will now compare, under assumptions (i) - (vi) listed below, the rate of
convergence of the drifts of the estimators $\widehat{\theta }_{DR,mach}$
and $\widehat{\theta }_{MR,mach}$ when the machine learning algorithms used
are series estimators:

For each $k\in \left\{ 1,2,3\right\} ,$

\begin{enumerate}
\item[i.] $A_{k}$ is discrete with finite sample space,

\item[ii.] $\overline{L}_{k}$ is absolutely continuous with respect to
Lebesgue measure with support on a compact set in $\mathbb{R}^{d_{k}}$

\item[iii.] $\eta _{k}^{g}\left( \overline{a}_{k},\overline{l}_{k}\right) $
and $h_{k}\left( \overline{a}_{k},\overline{l}_{k}\right) ,$ as functions of 
$\overline{l}_{k}$ for each fixed $\overline{a}_{k},$ lie in Holder balls
with exponents $\nu _{\eta ,k}$ and $\nu _{h,}{}_{k}$

\item[iv.] the machine learning algorithm procedure in steps (b.1) and (c.1)
of Algorithm 6 is least squares on\ the covariate vector 
\begin{equation*}
S_{k}^{\eta }=vec\left[ \overline{A}_{k}\otimes s_{k}^{\eta }\left( 
\overline{L}_{k}\right) ^{T}\right]
\end{equation*}%
where $s_{k}^{\eta }\left( \overline{L}_{k}\right) $ is the vector of the
first $m_{k}^{\eta }\left( n\right) $ elements of a complete basis having
optimal rates of approximation for Holder classes in $L_{r}\left( \mu
\right) ,1\leq r\leq \infty .$

\item[v.] the machine learning algorithm procedure in step (a) of Algorithm
6 is linear logistic regression with covariate vector 
\begin{equation*}
S_{k}^{h}=vec\left[ \overline{A}_{k}\otimes s_{k}^{h}\left( \overline{L}%
_{k}\right) ^{T}\right]
\end{equation*}%
where $s_{k}^{h}\left( \overline{L}_{k}\right) $ is the vector of the first $%
m_{k}^{h}\left( n\right) $ elements of a complete basis having optimal rates
of approximation for Holder classes in $L_{r}\left( \mu \right) ,1\leq r\leq
\infty .$

\item[vi.] the Holder exponents $\nu _{\eta ,k}$ and $\nu _{h,}{}_{k}$ are
known, $m_{k}^{\eta }\left( n\right) =n^{1/\left( 1+2\gamma _{\eta
,k}\right) }$ and $m_{k}^{h}\left( n\right) =n^{1/\left( 1+2\gamma
_{h,k}\right) },$ $k=1,2,3,$ where $\gamma _{\eta ,k}\equiv \nu _{\eta
,k}/d_{k}\,\ $and $\gamma _{h,k}\equiv \nu _{h,k}/d_{k}.$
\end{enumerate}

Note that in assumption (ii), $d_{k}>d_{k^{\prime }}$ if $k>k^{\prime }.$
Also, a function $f(\cdot )$ with compact domain in $\mathbb{R}^{d}$ is said
to belong to a H\"{o}lder ball $H(\nu ,C),$ with H\"{o}lder exponent $\nu >0$
and radius $C>0,$ if and only if $f\left( \cdot \right) $ is uniformly
bounded by $C$, all partial derivatives of $f(\cdot )$ up to order $%
\left\lfloor \nu \right\rfloor $ exist and are bounded, and all partial
derivatives $\nabla ^{\left\lfloor \nu \right\rfloor }$ of order $%
\left\lfloor \nu \right\rfloor $ satisfy 
\begin{equation*}
\sup_{x,x+\delta x\in \left[ 0,1\right] ^{d}}\left\vert \nabla
^{\left\lfloor \nu \right\rfloor }f(x+\delta x)-\nabla ^{\left\lfloor \nu
\right\rfloor }f(x)\right\vert \leq C||\delta x||^{\nu -\left\lfloor \nu
\right\rfloor }.
\end{equation*}

It is well known that under assumptions (i)-(iii) the optimal rates of
convergence for $\eta _{k}^{g}$ and $h_{k}$ are $n^{-\gamma _{\eta
,k}/\left( 1+2\gamma _{\eta ,k}\right) }$ and $n^{-\gamma _{h,k}/\left(
1+2\gamma _{h,k}\right) }$ in $L_{r}\left( \mu \right) $ norms, $1\leq
r<\infty .$ For estimation of $\eta _{k}^{g},$ the optimal rate of
convergence is obtained by least squares regression of $y_{\eta
_{k+1}^{g}}\left( \overline{A}_{k},\overline{L}_{k+1}\right) $ on $%
S_{k}^{\eta }$ where $s_{k}^{\eta }\left( \overline{L}_{k}\right) $ is the
vector of the first $n^{1/\left( 1+2\gamma _{\eta ,k}\right) }$ elements of
a complete basis having optimal rates of approximation for Holder classes in 
$L_{r}\left( \mu \right) ,1\leq r<\infty .$ Two examples of such basis are
B-splines with sufficient number of derivatives or Daubechies compact
wavelets of sufficient order.

Based on these facts and Theorem 1, we conjecture that the following result
holds:

\textbf{Result 1:} When $K=3$, if assumptions (i)-(vi) hold, the drift $E^{v}%
\left[ Q_{1}\left( \widehat{\overline{h}}^{t},\widehat{\overline{\eta }}%
_{mach}^{t}\right) \right] -\theta \left( g\right) $ of the $\widehat{\theta 
}_{DR,mach}$ is

\begin{eqnarray}
a^{p}\left( \widehat{\overline{h}}^{t},\widehat{\overline{\eta }}%
_{mach}^{t}\right) &=&O_{p}\left[ \max \left\{ \left( n^{-\left\{ \frac{%
\gamma _{h,1}}{1+2\gamma _{h,1}}+\frac{\gamma _{\eta ,2}}{1+2\gamma _{\eta
,2}}\right\} }\right) ,\max_{k\in \left\{ 1,2,3\right\} }\left( n^{-\left\{ 
\frac{\gamma _{h,k}}{1+2\gamma _{h,k}}+\frac{\gamma _{\eta ,k}}{1+2\gamma
_{\eta ,k}}\right\} }\right) ,\right. \right.  \label{drift-dr-result1} \\
&&\text{ \ \ \ \ }\left. \left. \max_{k\in \left\{ 1,2\right\} }\left(
n^{-\left\{ \frac{\gamma _{h,k}}{1+2\gamma _{h,k}}+\frac{\gamma _{\eta ,3}}{%
1+2\gamma _{\eta ,3}}\right\} }\right) \right\} \right]  \notag
\end{eqnarray}

and the drift $E^{v}\left[ Q_{1}\left( \widehat{\overline{h}}^{t},\widehat{%
\overline{\eta }}_{mach}^{t}\right) \right] -\theta \left( g\right) \,\ $of $%
\widehat{\theta }_{DR,mach}$ is%
\begin{eqnarray}
a^{p}\left( \widehat{\overline{h}}^{t},\widetilde{\overline{\eta }}%
_{mach}^{t}\right) &=&O_{p}\left[ \max \left\{ n^{-\left\{ \frac{\gamma
_{h,1}}{1+2\gamma _{h,1}}+\frac{\gamma _{h,2}}{1+2\gamma _{h,2}}+\frac{%
\gamma _{\eta ,2}}{1+2\gamma _{\eta ,2}}\right\} },n^{-\left\{ \left(
\sum_{k=1}^{3}\frac{\gamma _{h,k}}{1+2\gamma _{h,k}}\right) +\frac{\gamma
_{\eta ,3}}{1+2\gamma _{\eta ,3}}\right\} }\right. ,\right.
\label{drift-mr-result1} \\
&&\ \left. \left. \max_{k\in \left\{ 1,2,3\right\} }\left( n^{-\left\{ \frac{%
\gamma _{h,k}}{1+2\gamma _{h,k}}+\frac{\gamma _{\eta ,k}}{1+2\gamma _{\eta
,k}}\right\} }\right) ,\max_{k\in \left\{ 1,2\right\} }\left( n^{-\left\{ 
\frac{\gamma _{h,k}}{1+2\gamma _{h,k}}+\frac{\gamma _{h,3}}{1+2\gamma _{h,3}}%
+\frac{\gamma _{\eta ,3}}{1+2\gamma _{\eta ,3}}\right\} }\right) \right\} %
\right]  \notag
\end{eqnarray}

We provide here a sketch of the argument why we believe Result 1 is true and
towards the end of this argument we explain why we view it as a conjecture
and not as a theorem. Under assumption (iv) $\eta _{k,DR}=\Pi _{n,k}^{t,\eta
}\left[ y_{k+1,\eta _{k+1}^{g}}\left( \overline{A}_{k},\overline{L}%
_{k+1}\right) \right] $ where $\Pi _{n,k}^{t,\eta }$ is the projection
operator $\Pi _{n,k}^{t}$ but with $S_{k}^{\eta }=vec\left[ \overline{A}%
_{k}\otimes s_{k}^{\eta }\left( \overline{L}_{k}\right) ^{T}\right] $
instead of $s_{k}\left( \overline{L}_{k}\right) $. On the other hand, $\eta
_{k}^{g}=E_{g}\left[ y_{k+1,\eta _{k+1}^{g}}\left( \overline{A}_{k},%
\overline{L}_{k+1}\right) |\overline{A}_{k},\overline{L}_{k}\right] .$ Thus,
under assumptions (i)-(iv) and (vi), we have $\eta _{k,DR}-\eta
_{k}^{g}=O_{p}\left( n^{-\gamma _{\eta ,k}/\left( 1+2\gamma _{\eta
,k}\right) }\right) .$ Now, invoking Cauchy-Schwartz repeatedly in the right
hand side of formula $\left( \ref{drift-dr}\right) $, we obtain the rate of
convergence stated for the drift $a^{p}\left( \widehat{\overline{h}}^{t},%
\widehat{\overline{\eta }}_{mach}^{t}\right) $ of $\widehat{\theta }%
_{DR,mach}$ in Result 1. Next, for $k=1,2$ and 3, when $\widetilde{\eta }%
_{k,MR}=\widetilde{\eta }_{k,mach}^{t,u}$ and $\Pi ^{k}=\Pi _{n,k}^{t,\eta
}, $ the formula for $\eta _{k,MR}$ becomes

\begin{eqnarray*}
\eta _{k,MR} &\equiv &\eta _{k,MR}\left( \overline{A}_{k},\overline{L}%
_{k}\right) \\
&\equiv &\eta _{k,DR}+\Pi _{n,k}^{t,\eta }\left[ Q_{k+1}\left( \widehat{%
\overline{h}}_{k+1}^{3},\overline{\widetilde{\eta }}_{k+1,mach}^{3}\right)
-E^{v,\left( \overline{A}_{k},\overline{L}_{k+1}\right) }\left\{
Q_{k+1}\left( \widehat{\overline{h}}_{k+1}^{3},\overline{\widetilde{\eta }}%
_{k+1,mach}^{3}\right) \right\} \right]
\end{eqnarray*}%
where $Q_{k+1}\left( \widehat{\overline{h}}_{k+1}^{3},\overline{\widetilde{%
\eta }}_{k+1,mach}^{3}\right) =\psi \left( L_{4}\right) $ when $k=3.$ If $%
\widehat{h}_{j+1},\widetilde{\eta }_{j+1,mach},$ had been fixed functions,
i.e. they had not depended on the training sample data, then the difference $%
Q_{k+1}\left( \widehat{\overline{h}}_{k+1}^{3},\overline{\widetilde{\eta }}%
_{k+1,mach}^{3}\right) -E^{v,\left( \overline{A}_{k},\overline{L}%
_{k+1}\right) }\left\{ Q_{k+1}\left( \widehat{\overline{h}}_{k+1}^{3},%
\overline{\widetilde{\eta }}_{k+1,mach}^{3}\right) \right\} $ would have
been a mean zero random variable and $\Pi _{n,k}^{t,\eta }$ would have been
applied to i.i.d. mean zero random variables. Thus, $\Pi _{n,k}^{t,\eta }%
\left[ Q_{k+1}\left( \widehat{\overline{h}}_{k+1}^{3},\overline{\widetilde{%
\eta }}_{k+1,mach}^{3}\right) -E^{v,\left( \overline{A}_{k},\overline{L}%
_{k+1}\right) }\left\{ Q_{k+1}\left( \widehat{\overline{h}}_{k+1}^{3},%
\overline{\widetilde{\eta }}_{k+1,mach}^{3}\right) \right\} \right] $ would
have been of order $O_{p}\left( m_{k}^{\eta }\left( n\right) /n\right) =$ $%
O_{p}\left( n^{-\gamma _{\eta ,k}/\left( 1+2\gamma _{\eta ,k}\right)
}\right) ,$ which would then have proved Result 1. However, in the training
sample to which $\Pi _{n,k}^{t,\eta }$ is applied to, the random variables $%
Q_{k+1}\left( \widehat{\overline{h}}_{k+1}^{3},\overline{\widetilde{\eta }}%
_{k+1,mach}^{3}\right) -E^{v,\left( \overline{A}_{k},\overline{L}%
_{k+1}\right) }\left\{ Q_{k+1}\left( \widehat{\overline{h}}_{k+1}^{3},%
\overline{\widetilde{\eta }}_{k+1,mach}^{3}\right) \right\} $ are neither
mean zero nor i.i.d. because the functions $\widehat{h}_{j+1},\widetilde{%
\eta }_{j+1,mach}$ are not fixed, but rather they depend on data from that
training sample. As a consequence we view Result 1 as a conjecture, although
we expect it to be true. As an alternative to Algorithm 6, in the Appendix
(section 7.5) we provide two multi-layer cross-fit algorithms that avoid the
within training sample dependence described above.

Even if true, Result 1 is of no direct practical application because, in
reality, one does not know the Holder exponents $\nu _{\eta ,k}$ and $\nu
_{h,}{}_{k}$. However, it is known that the rates of convergence $n^{-\gamma
_{\eta ,k}/\left( 1+2\gamma _{\eta ,k}\right) }$ and $n^{-\gamma
_{h,k}/\left( 1+2\gamma _{h,k}\right) }$ for estimation of $\eta _{k}$ and $%
h_{k}$ can be achieved, up to log factors, \bigskip even if the smoothness
of the functions is unknown. For example, such adaption to unknown
smoothness can be achieved by choosing the number of basis functions by
cross-validation (Dudoit and van der Laan, 2003). This leads to the
following conjecture. \medskip

\textbf{Result 2}: Result 1 holds if we replace assumption (vi) with the
following assumption:

(vi') $m_{k}^{\eta }\left( n\right) $ and $m_{k}^{h}\left( n\right) $ are
chosen by V-fold cross-validation using empirical $L_{2}-$ loss, $k=1,2,3.$

To proceed with the discussion, we will assume henceforth that Results 1 and
2 hold. The formula $\left( \ref{drift-dr-result1}\right) $ for the order of
the drift of $\widehat{\theta }_{DR,mach}$ involves the maximum over six
second order terms corresponding to the six terms in the right hand side of $%
\left( \ref{drift-dr}\right) .$ On the other hand, formula $\left( \ref%
{drift-mr-result1}\right) $ for the order of the drift of $\widehat{\theta }%
_{MR,mach}$ involves the maximum over three second order terms
(corresponding to the terms $E^{v}\left( \delta _{k}^{MR}\right) ,k=1,2,3$
in $\left( \ref{drift-mr}\right) ),$ three third order terms and one fourth
order term.

Under assumptions (i)-(vi), or assumptions (i)-(v) and (vi'), for each $%
k=1,2,3,$ $E^{v}\left( \delta _{k}^{DR}\right) $ and $E^{v}\left( \delta
_{k}^{MR}\right) $ are of the same order, namely $O_{p}$ $\left( n^{-\frac{%
\gamma _{h,k}}{1+2\gamma _{h,k}}-\frac{\gamma _{\eta ,k}}{1+2\gamma _{\eta
,k}}}\right) $. Also, for each $\left( i,j\right) \in \left\{ \left(
1,2\right) ,\left( 1,3\right) ,\left( 2,3\right) \right\} ,E^{v}\left( \xi
_{i.j}^{DR}\right) $ converges to 0 slower than $E^{v}\left( \xi
_{i,j}^{MR}\right) $ because $E^{v}\left( \xi _{i,j}^{MR}\right) $ is a
third order term that involves the expectation of the product of three
differences, two of which agree with the differences in $E^{v}\left( \xi
_{i,j}^{DR}\right) .$ By the same reasoning, $E^{v}\left( \xi
_{1,2,3}^{MR}\right) $ converges to 0 faster than $E^{v}\left( \xi
_{1,3}^{MR}\right) $ and $E^{v}\left( \xi _{2,3}^{MR}\right) .$

In general, one might expect that the terms $E^{v}\left( \delta
_{3}^{DR}\right) $ and $E^{v}\left( \delta _{3}^{MR}\right) $ would be the
dominating terms in the drifts of $\widehat{\theta }_{DR,mach}$ and $%
\widehat{\theta }_{MR,mach}$, i.e. the terms with the slowest rates of
convergence, because (1) these terms involve two regressions on the
covariates $\left( \overline{A}_{3},\overline{L}_{3}\right) $ and (2) these
covariates are a superset of the covariates conditioned upon in the
regressions involved in all other terms that appear in the right hand sides
of $\left( \ref{drift-dr}\right) $ and $\left( \ref{drift-mr}\right) $. By
the same reasoning, one might expect that the second largest term of the
drift $\widehat{\theta }_{DR,mach}$ should be $E^{v}\left( \xi
_{2,3}^{DR}\right) $ with rate of convergence $O_{p}\left( n^{-\frac{\gamma
_{h,2}}{1+2\gamma _{h,2}}+\frac{\gamma _{\eta ,3}}{1+2\gamma _{\eta ,3}}%
}\right) $. However, it could happen that at the particular law that
generated the data, $\gamma _{h,2}=\nu _{h,2}/d_{2}$ could be less than $%
\gamma _{h,3}=\nu _{h,3}/d_{3}$ even though $d_{3}$ is greater than $d_{2}.$
If $E^{v}\left( \xi _{2,3}^{DR}\right) $ were, in fact, the dominating term
of the drift of $\widehat{\theta }_{DR,mach}$ then, in view of the
comparisons of the orders of the terms of the two drifts made above, the
drift of $\widehat{\theta }_{MR,mach}$ would have a faster rate of
convergence to 0 than that of $\widehat{\theta }_{DR,mach}$. Thus, it could
be the case that the drift of $\widehat{\theta }_{MR,mach}$ is $o_{p}\left(
n^{-1/2}\right) $ but the drift of $\widehat{\theta }_{DR,mach}$ is not, in
which case, by the analysis of section \ref{prelim-backg}, $\widehat{\theta }%
_{MR,mach}$ would be an asymptotically linear estimator of $\theta \left(
g\right) $ but $\widehat{\theta }_{DR,mach}$ would not.

In the next subsection we will need to refer to the following additional
result which follows from arguments analogous to those used to establish
Results 1 and 2.

\textbf{Result 3}. Under assumptions (i)-(vi) or, assumptions (i)-(v) and
(vi'), \emph{\ }$\widehat{\eta }_{k,mach}^{t}-E_{g_{k}}\left\{ \left. y_{k+1,%
\widehat{\eta }_{k+1,mach}^{t}}\left( \overline{A}_{k},\overline{L}%
_{k+1}\right) \right\vert \overline{A}_{k},\overline{L}_{k}\right\} $\emph{,}%
$\widetilde{\eta }_{k,mach}^{t}-E_{g_{k}}\left\{ \left. \widetilde{Q}%
_{k+1,mach}\right\vert \overline{A}_{k},\overline{L}_{k}\right\} ,\eta
_{k,DR}-\eta _{k}^{g}$\emph{\ }and\emph{\ }$\eta _{k,MR}-\eta _{k}^{g}\ $\
all converge to 0 at the same rate.

\subsubsection{Analysis when the ML algorithms are arbitrary. \label{any-ML}}

In this section we consider estimators $\widehat{\theta }_{MR,mach}$ and $%
\widehat{\theta }_{DR,mach}$ that use arbitrary machine learning algorithms
to estimate the nuisance functions. In order to analyze the rates of
convergence to 0 of the drifts of $\widehat{\theta }_{DR,mach}$ and $%
\widehat{\theta }_{MR,mach}$ we return to the formulae $\left( \ref{driftDR}%
\right) $ and $\left( \ref{driftMR}\right) .$ To facilitate the discussion
we write formula $\left( \ref{driftDR}\right) $ for the drift of $\widehat{%
\theta }_{DR,mach}$ as \emph{\ }%
\begin{eqnarray*}
E^{v}\left[ Q_{1}\left( \widehat{\overline{h}}^{t},\widehat{\overline{\eta }}%
_{mach}^{t}\right) \right] -\theta \left( g\right) &=&\sum_{k=1}^{K}E^{v}%
\left[ \frac{\pi ^{\ast K}}{\pi ^{k-1}\widehat{\pi }_{k+1}^{K}}\left\{ \frac{%
h_{k}^{\ast }}{h_{k}}-\frac{h_{k}^{\ast }}{\widehat{h}_{k}}\right\} \mathbf{R%
}_{DR,k}\right] \\
&&+\sum_{1\leq j<k\leq K}E^{v}\left[ \frac{\pi ^{\ast K}}{\pi ^{j-1}\widehat{%
\pi }_{j+1}^{K}}\left\{ \frac{h_{j}^{\ast }}{h_{j}}-\frac{h_{j}^{\ast }}{%
\widehat{h}_{j}}\right\} \mathbf{R}_{DR,k}\right] \\
&\equiv &\sum_{k=1}^{K}\rho _{k}^{DR}+\sum_{1\leq j<k\leq K}\chi _{j,k}^{DR}
\\
&\equiv &\rho ^{DR}+\chi ^{DR}
\end{eqnarray*}%
and formula $\left( \ref{driftMR}\right) $ for the drift of\emph{\ }$%
\widehat{\theta }_{MR,mach}$ as\emph{\ }%
\begin{eqnarray*}
E^{v}\left[ Q_{1}\left( \widehat{\overline{h}}^{t},\widetilde{\overline{\eta 
}}_{mach}^{t}\right) \right] -\theta \left( g\right) &=&\sum_{k=1}^{K}E^{v}%
\left[ \frac{\pi ^{\ast \left( k-1\right) }}{\pi ^{\left( k-1\right) }}%
\left( \frac{h_{k}^{\ast }}{h_{k}}-\frac{h_{k}^{\ast }}{\widehat{h}_{k}}%
\right) \mathbf{R}_{MR,k}\right] \\
&\equiv &\sum_{k=1}^{K}\rho _{k}^{MR} \\
&\equiv &\rho ^{MR}
\end{eqnarray*}%
where%
\begin{equation*}
\mathbf{R}_{DR,k}\equiv \widehat{\eta }_{k,mach}^{t}-E_{g_{k}}\left\{ \left.
y_{k+1,\widehat{\eta }_{k+1,mach}^{t}}\left( \overline{A}_{k},\overline{L}%
_{k+1}\right) \right\vert \overline{A}_{k},\overline{L}_{k}\right\} \text{ }
\end{equation*}%
and 
\begin{equation*}
\mathbf{R}_{MR,k}\equiv \widetilde{\eta }_{k,mach}^{t}-E_{g_{k}}\left\{
\left. \widetilde{Q}_{k+1,mach}\right\vert \overline{A}_{k},\overline{L}%
_{k}\right\}
\end{equation*}

Suppose it were the case that, as with the linear machine learning
algorithm, the rate of convergence to 0 of \emph{\ }$\mathbf{R}_{DR,k}$\emph{%
\ }and\emph{\ }$\mathbf{R}_{MR,k}$\emph{\ }were the same. Then $\rho ^{MR}$
and $\rho ^{DR}$ would converge to 0 at the same rate. Now, one would
generally expect that the terms $\rho _{1,K}^{DR}$\ and $\rho _{1,K}^{MR}$
would be the dominating terms in the drifts of $\widehat{\theta }_{DR,mach}$
and $\widehat{\theta }_{MR,mach}$, i.e. the terms with the slowest rate of
convergence, because (1) $\rho _{1,K}^{DR}$ and $\rho _{1,K}^{MR}$\ involve
two regressions on the covariates $\left( \overline{A}_{K},\overline{L}%
_{K}\right) $\ and (2) these covariates are a superset of the covariates
conditioned upon in the regressions involved in all other terms. However,
again it could happen that at the particular law that generated the data,
one of the $K\left( K-1\right) /2$\ terms $\chi _{j,k}^{DR}$ in $\chi ^{DR}$%
\ dominates $\rho ^{DR},$ i.e. it converges to 0 slower than any of the
terms in $\rho ^{DR}.$\ In such case, the drift $\rho ^{MR}$ of $\widehat{%
\theta }_{MR,mach}$ would have a faster rate of convergence to 0 than the
drift of $\widehat{\theta }_{DR,mach}$. \ In particular, it could happen
that $\widehat{\theta }_{MR,mach}$\ is an asymptotically linear estimator of 
$\theta \left( g\right) \ $even though $\widehat{\theta }_{DR,mach}$\ is
not. The frequency with which the law generating the data has the drift of $%
\widehat{\theta }_{MR,mach}$ converging to 0 faster than the drift of $%
\widehat{\theta }_{DR,mach}$ may be greater for $K$\ large\ because the
ratio of the number of terms in $\chi ^{DR}$\ to that in $\rho ^{DR}$\
increases linearly with $K,$ providing an increasing number of opportunities
for $\chi ^{DR}$ to dominate the drift of $\widehat{\theta }_{DR,mach}.$ Of
course this discussion must be tempered by the fact that, for non-linear
machine learning algorithms, we have no guarantee that the residuals $%
\mathbf{R}_{MR,k}$ converge to 0 as fast or faster than the residuals $%
\mathbf{R}_{DR,k}$.

\section{References}

Ayyagari, R. (2010), Applications of Influence Functions to Semiparametric
Regression Models. Harvard School of Public Health, Department of
Biostatistics Doctoral Thesis

Babino, L., Rotnitzky, A. and Robins, J. (2017). Multiple Robust Estimation
of Marginal Structural Mean Models. Submitted to Biometrics.

Bang, H., and Robins, J. M. (2005). Doubly robust estimation in missing data
and causal inference models. \emph{Biometrics} \textbf{61}, 962--972.

Belloni, Alexandre, Chen, Daniel, Chernozhukov, Victor and Hansen,
Christian. (2010, 2012). Sparse Models and Methods for Optimal Instruments
with an Application to Eminent Domain', arXiv (2010) preprint, Econometrica
(2012) 80, 2369-2429

Chernozhukov, V. et al. Double machine learning for treatment and causal
parameters. arXiv:1608.00060, 1\{37 (2016).

Dudoit, S, and van der Laan, M. (2003) "Asymptotics of cross-validated risk
estimation in model selection and performance assessment." \textit{UC
Berkeley Division of Biostatistics Working Paper Series} 126.

Molina, J., Rotnitzky, A., Sued, M. and Robins, J.M. (2017). Multiple
robustness in factorized likelihood models. \emph{Biometrika.} To appear.

Petersen M1, Schwab J1, Gruber S2, Blaser N3, Schomaker M4, van der Laan M1.
(2014) Targeted Maximum Likelihood Estimation for Dynamic and Static
Longitudinal Marginal Structural Working Models. \textit{J Causal Inference}%
. Jun 18;2(2):147-185.

Robins JM. (1986). A new approach to causal inference in mortality studies
with sustained exposure periods - Application to control of the healthy
worker survivor effect. \emph{Mathematical Modelling}, \textbf{7(9-12)},
1393-1512.

Robins, J. M. (1987). Addendum to \textquotedblleft a new approach to causal
inference in mortality studies with a sustained exposure
period---application to control of the healthy worker survivor
effect\textquotedblright . \emph{Computers \& Mathematics with Applications}%
, \textbf{14(9)}, 923-945.

Robins JM. (1993). Analytic methods for estimating HIV treatment and
cofactor effects. \emph{Methodological Issues of AIDS Mental Health Research}%
. Eds: Ostrow D.G., Kessler R. New York: Plenum Publishing. pp. 213-290

Robins, J. M. (1997). Causal inference from complex longitudinal data. \emph{%
In Latent variable modeling and applications to causality}. Springer New
York.

Robins J. M. (2000). Robust estimation in sequentially ignorable missing
data and causal inference models. \emph{Proceedings of the American
Statistical Association Section on Bayesian Statistical Science}, 6-10.

Robins JM. (2002). Commentary on "Using inverse weighting and predictive
inference to estimate the effects of time-varying treatments on the
discrete-time hazard." \textit{Statistics in Medicine}. 21:1663-1680

Robins JM, Li L, Tchetgen E, van der Vaart A. (2008). Higher order influence
functions and minimax estimation of nonlinear functionals. \emph{Probability
and Statistics: Essays in Honor of David A. Freedman} 2:335-421.

Robins JM, Rotnitzky A, Zhao LP. (1994) Estimation of regression
coefficients when some regressors are not always observed. \emph{Journal of
the American Statistical Association.} 89:846-866

Robins JM, Zhang P, Ayyagari R, Logan R, Tchetgen ET, Li L, Lumley T, van
der Vaart A, HEI Health Review Committee. (2013) \emph{New statistical
approaches to semiparametric regression with application to air pollution
research.} \textit{Res Rep Health Eff Inst.} 175:3-129.

Tchegen Tchegen, E. J. (2009). A Commentary on G. Molenberghs's Review of
Missing Data Methods. \emph{Drug Information Journal} \textbf{43(4)},
433--435.

van der Laan MJ and Rose (2011). \textit{Targeted Learning Causal Inference
for Observational and Experimental Data. }Springer-Verlag. New York.

van der Laan MJ, Gruber S. (2012) Targeted minimum loss based estimation of
causal effects of multiple time point interventions. \textit{Int J Biostat.}%
; 8(1) Article 8

\section{Appendix}

\subsection{Proof of Lemma 1}

By definition, and the absolute continuity of $gh^{\ast }$ with respect to $%
gh,$ we have that for $k\in \left[ K-1\right] ,$ 
\begin{eqnarray*}
y_{k+1,\eta _{k+1}}\left( \overline{A}_{k},\overline{L}_{k+1}\right)
&=&E_{h_{k+1}^{\ast }}\left( \left. \eta _{k+1}\right\vert \overline{A}%
_{k+1},\overline{L}_{k+1}\right) \\
&=&E_{h_{k+1}}\left( \left. \frac{h_{k+1}^{\ast }}{h_{k+1}}\eta
_{k+1}\right\vert \overline{A}_{k+1},\overline{L}_{k+1}\right) ,
\end{eqnarray*}%
where $h_{k}^{\ast }\equiv h_{k}^{\ast }\left( A_{k}|\overline{A}_{k-1},%
\overline{L}_{k}\right) ,h_{k}\equiv h_{k}\left( A_{k}|\overline{A}_{k-1},%
\overline{L}_{k}\right) $ and $\eta _{k}\equiv \eta _{k}\left( \overline{A}%
_{k},\overline{L}_{k}\right) $. Thus, for $k\in \left[ K-1\right] ,$%
\begin{eqnarray}
\frac{\Delta _{k}\left( \eta _{k},\eta _{k+1};g_{k}\right) }{\pi ^{k}} &=&%
\frac{\pi ^{\ast k}}{\pi ^{k}}\left\{ \eta _{k}-E_{g_{k}}\left[ \left.
E_{h_{k+1}}\left( \left. \frac{h_{k+1}^{\ast }}{h_{k+1}}\eta
_{k+1}\right\vert \overline{A}_{k+1},\overline{L}_{k+1}\right) \right\vert 
\overline{A}_{k},\overline{L}_{k}\right] \right\}  \label{deltak} \\
&=&\frac{\pi ^{\ast k}}{\pi ^{k}}\left\{ \eta _{k}-E_{g_{k},h_{k+1}}\left(
\left. \frac{h_{k+1}^{\ast }}{h_{k+1}}\eta _{k+1}\right\vert \overline{A}%
_{k},\overline{L}_{k}\right) \right\} .  \notag
\end{eqnarray}

Consequently, 
\begin{equation*}
E_{\overline{g}_{k-1},\overline{h}_{k}}\left\{ \frac{\Delta _{k}\left( \eta
_{k},\eta _{k+1};g_{k}\right) }{\pi ^{k}}\right\} =E_{\overline{g}_{k-1},%
\overline{h}_{k}}\left( \frac{\pi ^{\ast k}}{\pi ^{k}}\eta _{k}\right) -E_{%
\overline{g}_{k},\overline{h}_{k+1}}\left( \frac{\pi ^{\ast k+1}}{\pi ^{k+1}}%
\eta _{k+1}\right) .
\end{equation*}

In addition,%
\begin{eqnarray*}
&&\left. E_{\overline{g}_{K-1},\overline{h}_{K}}\left\{ \frac{\Delta
_{K}\left( \eta _{K},\eta _{K+1};g_{K}\right) }{\pi ^{K}}\right\} =\right. \\
&=&E_{\overline{g}_{K-1},\overline{h}_{K}}\left( \frac{\pi ^{\ast K}}{\pi
^{K}}\eta _{K}\right) -E_{\overline{g}_{K-1},\overline{h}_{K}}\left[ \frac{%
\pi ^{\ast K}}{\pi ^{K}}E_{g_{k}}\left\{ \left. \psi \left( \overline{L}%
_{K+1}\right) \right\vert \overline{A}_{K},\overline{L}_{K}\right\} \right]
\\
&=&E_{\overline{g}_{K-1},\overline{h}_{K}}\left( \frac{\pi ^{\ast K}}{\pi
^{K}}\eta _{K}\right) -E_{\overline{g}_{K},\overline{h}_{K}}\left\{ \frac{%
\pi ^{\ast K}}{\pi ^{K}}\psi \left( \overline{L}_{K+1}\right) \right\} \\
&=&E_{\overline{g}_{K-1},\overline{h}_{K}}\left( \frac{\pi ^{\ast K}}{\pi
^{K}}\eta _{K}\right) -\theta \left( g\right) .
\end{eqnarray*}

Consequently,%
\begin{eqnarray*}
&&\left. \sum_{k=1}^{K}E_{\overline{g}_{k-1},\overline{h}_{k}}\left\{ \frac{%
\Delta _{k}\left( \eta _{k},\eta _{k+1};g_{k}\right) }{\pi ^{k}}\right\}
=\right. \\
&=&\sum_{k=1}^{K-1}\left\{ E_{\overline{g}_{k-1},\overline{h}_{k}}\left( 
\frac{\pi ^{\ast k}}{\pi ^{k}}\eta _{k}\right) -E_{\overline{g}_{k},%
\overline{h}_{k+1}}\left( \frac{\pi ^{\ast k+1}}{\pi ^{k+1}}\eta
_{k+1}\right) \right\} +E_{\overline{g}_{K-1},\overline{h}_{K}}\left( \frac{%
\pi ^{\ast K}}{\pi ^{K}}\eta _{K}\right) -\theta \left( g\right) \\
&=&E_{g_{0},h_{1}}\left( \frac{h_{1}^{\ast }}{h_{1}}\eta _{1}\right) -\theta
\left( g\right) \\
&=&E_{g_{1}}\left\{ y_{1,\eta _{1}}\left( L_{1}\right) \right\} -\theta
\left( g\right) ,
\end{eqnarray*}%
as we wished to show.

\bigskip

\subsection{Proof of Lemma 2}

The identity $\left( \ref{bias1}\right) $ coincides with $\left( \ref{bias00}%
\right) $ for $j=0$ if $\overline{A}_{j-1}$ and $\overline{L}_{j-1}$ are
defined as nill when $j=1.$ It thus suffices to show $\left( \ref{bias00}%
\right) $ for an arbitrary $j\in \left\{ 0,...,K\right\} .$ We prove it by
reverse induction. For $j=K$ the result holds by definition of $\eta
_{K}^{g}\left( \overline{A}_{K},\overline{L}_{K}\right) $ since $%
Q_{K+1}\left( \overline{h}_{K+1}^{\dag K},\overline{\eta }_{K+1}^{\dag
K}\right) =\psi \left( \overline{L}_{K+1}\right) .$ Suppose now that $\left( %
\ref{bias00}\right) $ holds for a given $j\in \left[ K\right] ,$ we want to
show that it also holds for $j-1.$ Now, 
\begin{eqnarray*}
&&\left. E_{\underline{g}_{j-1},\underline{h}_{j}}\left\{ \left. Q_{j}\left( 
\overline{h}_{j}^{\dag K},\overline{\eta }_{j}^{\dag K}\right) \right\vert 
\overline{A}_{j-1},\overline{L}_{j-1}\right\} -\eta _{j-1}^{g}\left( 
\overline{A}_{j-1},\overline{L}_{j-1}\right) =\right. \\
&=&E_{\underline{g}_{j-1},\underline{h}_{j}}\left[ \left. \frac{h_{j}^{\ast }%
}{h_{j}^{\dag }}\left\{ Q_{j+1}\left( \overline{h}_{j+1}^{\dag K},\overline{%
\eta }_{j+1}^{\dag K}\right) -\eta _{j}^{\dag }\right\} +y_{j,\eta
_{j}^{\dag }}\left( \overline{A}_{j-1},\overline{L}_{j}\right) \right\vert 
\overline{A}_{j-1},\overline{L}_{j-1}\right] -\eta _{j-1}^{g}\left( 
\overline{A}_{j-1},\overline{L}_{j-1}\right) \\
&=&E_{g_{j-1},h_{j}}\left[ \left. \frac{h_{j}^{\ast }}{h_{j}^{\dag }}E_{%
\underline{g}_{j},\underline{h}_{j+1}}\left\{ \left. Q_{j+1}\left( \overline{%
h}_{j+1}^{\dag K},\overline{\eta }_{j+1}^{\dag K}\right) \right\vert 
\overline{A}_{j},\overline{L}_{j}\right\} \right\vert \overline{A}_{j-1},%
\overline{L}_{j-1}\right] \\
&&-E_{g_{j-1},h_{j}}\left\{ \left. \frac{h_{j}^{\ast }}{h_{j}^{\dag }}\eta
_{j}^{\dag }\left( \overline{A}_{j},\overline{L}_{j}\right) \right\vert 
\overline{A}_{j-1},\overline{L}_{j-1}\right\} \\
&&+E_{g_{j-1}}\left[ \left. y_{j,\eta _{j}^{\dag }}\left( \overline{A}_{j-1},%
\overline{L}_{j}\right) \right\vert \overline{A}_{j-1},\overline{L}_{j-1}%
\right] -\eta _{j-1}^{g}\left( \overline{A}_{j-1},\overline{L}_{j-1}\right)
\\
&=&E_{g_{j-1},h_{j}}\left[ \left. \frac{h_{j}^{\ast }}{h_{j}^{\dag }}\left[
\sum_{k=j+1}^{K}E_{\underline{g}_{j},\underline{h}_{j+1}}\left\{ \left. 
\frac{\pi _{j+1}^{\ast \left( k-1\right) }}{\pi _{j+1}^{\dag \left(
k-1\right) }}\left( \eta _{k}^{\dag }-\eta _{k}^{g}\right) \left( \frac{1}{%
h_{k}}-\frac{1}{h_{k}^{\dag }}\right) \right\vert \overline{A}_{j},\overline{%
L}_{j}\right\} +\eta _{j}^{g}\right] \right\vert \overline{A}_{j-1},%
\overline{L}_{j-1}\right] \\
&&-E_{g_{j-1},h_{j}}\left( \left. \frac{h_{j}^{\ast }}{h_{j}^{\dag }}\eta
_{j}^{\dag }\right\vert \overline{A}_{j-1},\overline{L}_{j-1}\right)
+E_{g_{j-1}}\left\{ \left. E_{h_{j}}\left( \left. \frac{h_{j}^{\ast }}{h_{j}}%
\eta _{j}^{\dag }\right\vert \overline{A}_{j-1},\overline{L}_{j}\right)
\right\vert \overline{A}_{j-1},\overline{L}_{j-1}\right\} -\eta
_{j-1}^{g}\left( \overline{A}_{j-1},\overline{L}_{j-1}\right) \\
&=&\sum_{k=j+1}^{K}E_{\underline{g}_{j-1},\underline{h}_{j}}\left\{ \left. 
\frac{\pi _{j}^{\ast \left( k-1\right) }}{\pi _{j}^{\dag \left( k-1\right) }}%
\left( \eta _{k}^{\dag }-\eta _{k}^{g}\right) \left( \frac{1}{h_{k}}-\frac{1%
}{h_{k}^{\dag }}\right) \right\vert \overline{A}_{j-1},\overline{L}%
_{j-1}\right\} +E_{g_{j-1},h_{j}}\left( \left. \frac{h_{j}^{\ast }}{%
h_{j}^{\dag }}\eta _{j}^{g}\right\vert \overline{A}_{j-1},\overline{L}%
_{j-1}\right) \\
&&+E_{g_{j-1},h_{j}}\left( \left. h_{j}^{\ast }\left( \frac{1}{h_{j}}-\frac{1%
}{h_{j}^{\dag }}\right) \eta _{j}^{\dag }\right\vert \overline{A}_{j-1},%
\overline{L}_{j-1}\right) -E_{g_{j-1},h_{j}}\left( \left. \frac{h_{j}^{\ast }%
}{h_{j}}\eta _{j}^{g}\right\vert \overline{A}_{j-1},\overline{L}_{j-1}\right)
\\
&=&\sum_{k=j+1}^{K}E_{\underline{g}_{j-1},\underline{h}_{j}}\left\{ \left. 
\frac{\pi _{j}^{\ast \left( k-1\right) }}{\pi _{j}^{\dag \left( k-1\right) }}%
\left( \eta _{k}^{\dag }-\eta _{k}^{g}\right) \left( \frac{1}{h_{k}}-\frac{1%
}{h_{k}^{\dag }}\right) \right\vert \overline{A}_{j-1},\overline{L}%
_{j-1}\right\} \\
&&+E_{g_{j-1},h_{j}}\left\{ \left. h_{j}^{\ast }\left( \frac{1}{h_{j}}-\frac{%
1}{h_{j}^{\dag }}\right) \left( \eta _{j}^{\dag }-\eta _{j}^{g}\right)
\right\vert \overline{A}_{j-1},\overline{L}_{j-1}\right\} ,
\end{eqnarray*}%
where the third equality is by the inductive hypothesis. This concludes the
proof.

\bigskip

\subsection{Proof of Lemma 3}

We prove (1) by reverse induction that for $k\in \left\{ 0,1...,K\right\} ,$

\begin{equation}
\pi ^{\ast k}\left\{ \eta _{k}^{\dag }-\eta _{k}^{g}\right\} =\Gamma
_{k}+\sum_{s=k+1}^{K}E_{gh}\left\{ \left. \frac{1}{\pi _{k+1}^{s-1}}\left( 
\frac{1}{h_{s}}-\frac{1}{h_{s}^{\dag }}\right) \Gamma _{s}\right\vert 
\overline{A}_{k},\overline{L}_{k}\right\}  \label{c-iden}
\end{equation}%
where to simplify notation we use the shortcut $\Gamma _{k}\equiv \Gamma
_{k}\left( \overline{h}_{k+1}^{\dag K},\overline{\eta }_{k}^{\dag
K};g_{k}\right) $ and $\sum_{s=K+1}^{K}\left( \cdot \right) \equiv 0.$

Applying this equality to $k=0$ with 
\begin{equation}
\eta _{0}^{\dag }=E_{gh}\left\{ Q_{1}\left( \overline{h}_{1}^{\dag K},%
\overline{\eta }_{1}^{\dag K}\right) \right\}  \label{eta0}
\end{equation}%
we obtain that 
\begin{equation*}
\pi ^{\ast 0}\left[ E_{gh}\left\{ Q_{1}\left( \overline{h}_{1}^{\dag K},%
\overline{\eta }_{1}^{\dag K}\right) \right\} -\eta _{0}^{g}\right] =\Gamma
_{0}+\sum_{s=1}^{K}E_{gh}\left\{ \left. \frac{1}{\pi _{1}^{s-1}}\left( \frac{%
1}{h_{s}}-\frac{1}{h_{s}^{\dag }}\right) \Gamma _{s}\right\vert \overline{A}%
_{0},\overline{L}_{0}\right\}
\end{equation*}%
Recalling that $\pi ^{\ast 0}\equiv 1,\eta _{0}^{g}=\theta \left( g\right) ,$
$\left( \overline{A}_{0},\overline{L}_{0}\right) \equiv $ nill, and that
with $\eta _{0}^{\dag }$ defined as in $\left( \ref{eta0}\right) ,$ $\Gamma
_{0}\equiv \eta _{0}^{\dag }-E_{gh}\left\{ Q_{1}\left( \overline{h}%
_{1}^{\dag K},\overline{\eta }_{1}^{\dag K}\right) \right\} =0,$ we conclude
that 
\begin{equation*}
E_{gh}\left\{ Q_{1}\left( \overline{h}_{1}^{\dag K},\overline{\eta }%
_{1}^{\dag K}\right) \right\} -\theta \left( g\right)
=\sum_{s=1}^{K}E_{gh}\left\{ \frac{1}{\pi ^{s-1}}\left( \frac{1}{h_{s}}-%
\frac{1}{h_{s}^{\dag }}\right) \Gamma _{s}\right\} \equiv b^{p}\left(
h^{\dag },\eta ^{\dag }\right)
\end{equation*}%
which, invoking Lemma 2, proves that $b^{p}\left( h^{\dag },\eta ^{\dag
}\right) =a^{p}\left( h^{\dag },\eta ^{\dag }\right) .$

We now prove identity $\left( \ref{c-iden}\right) $ by induction.

For $k=K,$ $\left( \ref{c-iden}\right) $ holds because by definition 
\begin{eqnarray*}
\pi ^{\ast K}\left( \eta _{K}^{\dag }-\eta _{K}^{g}\right) &\equiv &\pi
^{\ast K}\left[ \eta _{K}^{\dag }-E_{g_{K}}\left\{ \left. \psi \left( 
\overline{L}_{K+1}\right) \right\vert \overline{A}_{K},\overline{L}%
_{K}\right\} \right] \\
&\equiv &\pi ^{\ast K}\left[ \eta _{K}^{\dag }-E_{g_{K}}\left\{ \left.
Q_{K+1}\left( \overline{h}_{K+1}^{\dag K},\overline{\eta }_{K+1}^{\dag
K}\right) \right\vert \overline{A}_{K},\overline{L}_{K}\right\} \right] \\
&\equiv &\Gamma _{K}.
\end{eqnarray*}

Suppose $\left( \ref{c-iden}\right) $ holds for $k=K,...,j+1.\,\ $We will
show that it holds for $k=j.$

By Lemma 2 we have 
\begin{eqnarray*}
\pi ^{\ast j}\left( \eta _{j}^{\dag }-\eta _{j}^{g}\right) &=&\pi ^{\ast j} 
\left[ \eta _{j}^{\dag }-E_{\underline{g}_{j},\underline{h}_{j+1}}\left\{
Q_{j+1}\left( \overline{h}_{j+1}^{\dag K},\overline{\eta }_{j+1}^{\dag
K}\right) |\overline{A}_{j},\overline{L}_{j}\right\} \right] \\
&&+\pi ^{\ast j}\left[ E_{\underline{g}_{j},\underline{h}_{j+1}}\left\{
Q_{j+1}\left( \overline{h}_{j+1}^{\dag K},\overline{\eta }_{j+1}^{\dag
K}\right) |\overline{A}_{j},\overline{L}_{j}\right\} -\eta _{j}^{g}\right] \\
&=&\Gamma _{j}+\sum_{k=j+1}^{K}E_{\underline{g}_{j},\underline{h}%
_{j+1}}\left\{ \left. \frac{\pi _{j+1}^{\ast \left( k-1\right) }}{\pi
_{j+1}^{\dag \left( k-1\right) }}\left( \frac{h_{k}^{\ast }}{h_{k}}-\frac{%
h_{k}^{\ast }}{h_{k}^{\dag }}\right) \left( \eta _{k}^{\dag }-\eta
_{k}^{g}\right) \right\vert \overline{A}_{j},\overline{L}_{j}\right\} \\
&=&\Gamma _{j}+\sum_{k=j+1}^{K}E_{\underline{g}_{j},\underline{h}%
_{j+1}}\left\{ \left. \frac{1}{\pi _{j+1}^{\dag \left( k-1\right) }}\left( 
\frac{1}{h_{k}}-\frac{1}{h_{k}^{\dag }}\right) \pi ^{\ast k}\left( \eta
_{k}^{\dag }-\eta _{k}^{g}\right) \right\vert \overline{A}_{j},\overline{L}%
_{j}\right\}
\end{eqnarray*}%
Then, invoking the inductive assumption we obtain 
\begin{eqnarray*}
\pi ^{\ast j}\left( \eta _{j}^{\dag }-\eta _{j}^{g}\right) &=&\Gamma
_{j}+\sum_{k=j+1}^{K}E_{\underline{g}_{j},\underline{h}_{j+1}}\left\{ \left. 
\frac{1}{\pi _{j+1}^{\dag \left( k-1\right) }}\left( \frac{1}{h_{k}}-\frac{1%
}{h_{k}^{\dag }}\right) \Gamma _{k}\right\vert \overline{A}_{j},\overline{L}%
_{j}\right\} \\
&&+\sum_{k=j+1}^{K}\sum_{s=k+1}^{K}E_{\underline{g}_{j},\underline{h}_{j+1}}%
\left[ \left. \frac{1}{\pi _{j+1}^{\dag \left( k-1\right) }}\left( \frac{1}{%
h_{k}}-\frac{1}{h_{k}^{\dag }}\right) \left\{ \frac{1}{\pi _{k+1}^{s-1}}%
\left( \frac{1}{h_{s}}-\frac{1}{h_{s}^{\dag }}\right) \Gamma _{s}\right\}
\right\vert \overline{A}_{j},\overline{L}_{j}\right]
\end{eqnarray*}%
Now, rearranging the terms in the double-sum we obtain 
\begin{eqnarray*}
&&\sum_{k=j+1}^{K}\sum_{s=k+1}^{K}E_{\underline{g}_{j},\underline{h}_{j+1}}%
\left[ \left. \frac{1}{\pi _{j+1}^{\dag \left( k-1\right) }}\left( \frac{1}{%
h_{k}}-\frac{1}{h_{k}^{\dag }}\right) \left\{ \frac{1}{\pi _{k+1}^{s-1}}%
\left( \frac{1}{h_{s}}-\frac{1}{h_{s}^{\dag }}\right) \Gamma _{s}\right\}
\right\vert \overline{A}_{j},\overline{L}_{j}\right] \\
&=&\sum_{s=j+2}^{K}E_{\underline{g}_{j},\underline{h}_{j+1}}\left[ \left( 
\frac{1}{h_{s}}-\frac{1}{h_{s}^{\dag }}\right) \Gamma _{s}\left.
\sum_{k=j+1}^{s-1}\left\{ \frac{1}{\pi _{j+1}^{\dag \left( k-1\right) }}%
\left( \frac{1}{h_{k}}-\frac{1}{h_{k}^{\dag }}\right) \frac{1}{\pi
_{k+1}^{s-1}}\right\} \right\vert \overline{A}_{j},\overline{L}_{j}\right]
\end{eqnarray*}%
and we prove below that%
\begin{equation}
\sum_{k=j+1}^{s-1}\left\{ \frac{1}{\pi _{j+1}^{\dag \left( k-1\right) }}%
\left( \frac{1}{h_{k}}-\frac{1}{h_{k}^{\dag }}\right) \frac{1}{\pi
_{k+1}^{s-1}}\right\} =\frac{1}{\pi _{j+1}^{s-1}}-\frac{1}{\pi _{j+1}^{\dag
\left( s-1\right) }}  \label{algebra}
\end{equation}%
Thus,

\begin{eqnarray*}
\pi ^{\ast j}\left( \eta _{j}^{\dag }-\eta _{j}^{g}\right) &=&\Gamma
_{j}+\sum_{k=j+1}^{K}E_{\underline{g}_{j},\underline{h}_{j+1}}\left\{ \left. 
\frac{1}{\pi _{j+1}^{\dag \left( k-1\right) }}\left( \frac{1}{h_{k}}-\frac{1%
}{h_{k}^{\dag }}\right) \Gamma _{k}\right\vert \overline{A}_{j},\overline{L}%
_{j}\right\} \\
&&+\sum_{s=j+2}^{K}E_{\underline{g}_{j},\underline{h}_{j+1}}\left\{ \left( 
\frac{1}{h_{s}}-\frac{1}{h_{s}^{\dag }}\right) \Gamma _{s}\left. \left( 
\frac{1}{\pi _{j+1}^{s-1}}-\frac{1}{\pi _{j+1}^{\dag \left( s-1\right) }}%
\right) \right\vert \overline{A}_{j},\overline{L}_{j}\right\} \\
&=&\Gamma _{j}+E_{\underline{g}_{j},\underline{h}_{j+1}}\left\{ \left.
\Gamma _{j+1}\left( \frac{1}{h_{j+1}}-\frac{1}{h_{j+1}^{\dag }}\right)
\right\vert \overline{A}_{j},\overline{L}_{j}\right\} \\
&&+\sum_{k=j+2}^{K}E_{\underline{g}_{j},\underline{h}_{j+1}}\left\{ \left. 
\frac{1}{\pi _{j+1}^{\dag \left( k-1\right) }}\left( \frac{1}{h_{k}}-\frac{1%
}{h_{k}^{\dag }}\right) \Gamma _{k}\right\vert \overline{A}_{j},\overline{L}%
_{j}\right\} \\
&&+\sum_{s=j+2}^{K}E_{\underline{g}_{j},\underline{h}_{j+1}}\left\{ \left( 
\frac{1}{h_{s}}-\frac{1}{h_{s}^{\dag }}\right) \Gamma _{s}\left. \left( 
\frac{1}{\pi _{j+1}^{s-1}}-\frac{1}{\pi _{j+1}^{\dag \left( s-1\right) }}%
\right) \right\vert \overline{A}_{j},\overline{L}_{j}\right\} \\
&=&\Gamma _{j}+\sum_{s=j+1}^{K}E_{\underline{g}_{j},\underline{h}%
_{j+1}}\left\{ \left. \frac{1}{\pi _{j+1}^{s-1}}\left( \frac{1}{h_{s}}-\frac{%
1}{h_{s}^{\dag }}\right) \Gamma _{s}\right\vert \overline{A}_{j},\overline{L}%
_{j}\right\}
\end{eqnarray*}%
as we wish to show.

We now show $\left( \ref{algebra}\right) .$

\begin{eqnarray*}
&&\sum_{k=j+1}^{s-1}\frac{1}{\pi _{j+1}^{\dag \left( k-1\right) }}\left( 
\frac{1}{h_{k}}-\frac{1}{h_{k}^{\dag }}\right) \frac{1}{\pi _{k+1}^{s-1}} \\
&=&\sum_{k=j+1}^{s-1}\frac{1}{\pi _{j+1}^{\dag \left( k-1\right) }}\frac{1}{%
\pi _{k}^{s-1}}-\sum_{k=j+1}^{s-1}\frac{1}{\pi _{j+1}^{\dag k}}\frac{1}{\pi
_{k+1}^{s-1}} \\
&=&\frac{1}{\pi _{j+1}^{s-1}}+\sum_{k=j+2}^{s-1}\frac{1}{\pi _{j+1}^{\dag
\left( k-1\right) }}\frac{1}{\pi _{k}^{s-1}}-\sum_{k=j+1}^{s-2}\frac{1}{\pi
_{j+1}^{\dag k}}\frac{1}{\pi _{k+1}^{s-1}}-\frac{1}{\pi _{j+1}^{\dag s-1}} \\
&=&\frac{1}{\pi _{j+1}^{s-1}}-\frac{1}{\pi _{j+1}^{\dag \left( s-1\right) }}
\end{eqnarray*}

This concludes the proof that $b^{p}\left( h^{\dag },\eta ^{\dag }\right)
=a^{p}\left( h^{\dag },\eta ^{\dag }\right) $.

We now prove that $c^{p}\left( h^{\dag },\eta ^{\dag }\right) =a^{p}\left(
h^{\dag },\eta ^{\dag }\right) $.

\bigskip For any $\left( \overline{A}_{k},\overline{L}_{k}\right) $ such
that $\pi ^{\ast k}>0$ define 
\begin{equation*}
\delta _{k}\left( \eta _{k},\eta _{k+1};g_{k}\right) \equiv \eta _{k}\left( 
\overline{A}_{k},\overline{L}_{k}\right) -E_{g_{k}}\left\{ \left.
y_{k+1,\eta _{k+1}}\left( \overline{A}_{k},\overline{L}_{k+1}\right)
\right\vert \overline{A}_{k},\overline{L}_{k}\right\}
\end{equation*}

Then, conditioning in $\left( \overline{A}_{k},\overline{L}_{k}\right) ,$
the proof of Lemma 1 can be immediately adapted to show that 
\begin{equation*}
E_{g_{k}}\left\{ \left. y_{k+1,\eta _{k+1}^{\dag }}\left( \overline{A}_{k},%
\overline{L}_{k+1}\right) \right\vert \overline{A}_{k},\overline{L}%
_{k}\right\} -\eta _{k}^{g}\left( \overline{A}_{k},\overline{L}_{k}\right)
=\sum_{j=k+1}^{K}E_{\overline{g}_{k}^{K},\overline{h}_{k+1}^{K}}\left\{
\left. \frac{\pi _{k+1}^{\ast j}}{\pi _{k+1}^{j}}\delta _{j}\left( \eta
_{j}^{\dag },\eta _{j+1}^{\dag };g_{j}\right) \right\vert \overline{A}_{k},%
\overline{L}_{k}\right\} ,
\end{equation*}%
from where we deduce that

\begin{eqnarray}
\eta _{k}^{\dag }\left( \overline{A}_{k},\overline{L}_{k}\right) -\eta
_{k}^{g}\left( \overline{A}_{k},\overline{L}_{k}\right) &=&\eta _{k}^{\dag
}\left( \overline{A}_{k},\overline{L}_{k}\right) -E_{g_{k}}\left\{ \left.
y_{k+1,\eta _{k+1}^{\dag }}\left( \overline{A}_{k},\overline{L}_{k+1}\right)
\right\vert \overline{A}_{k},\overline{L}_{k}\right\}  \label{etag} \\
&&+E_{g_{k}}\left\{ \left. y_{k+1,\eta _{k+1}^{\dag }}\left( \overline{A}%
_{k},\overline{L}_{k+1}\right) \right\vert \overline{A}_{k},\overline{L}%
_{k}\right\} -\eta _{k}^{g}\left( \overline{A}_{k},\overline{L}_{k}\right) 
\notag \\
&=&\sum_{j=k}^{K}E_{\overline{g}_{k}^{K},\overline{h}_{k+1}^{K}}\left\{
\left. \frac{\pi _{k+1}^{\ast j}}{\pi _{k+1}^{j}}\delta _{j}\left( \eta
_{j}^{\dag },\eta _{j+1}^{\dag };g_{j}\right) \right\vert \overline{A}_{k},%
\overline{L}_{k}\right\} .  \notag
\end{eqnarray}

Then,%
\begin{eqnarray*}
a^{p}\left( h^{\dag },\eta ^{\dag }\right) &\equiv
&\sum_{k=1}^{K}E_{gh}\left\{ \frac{\pi ^{\ast \left( k-1\right) }}{\pi
^{\dag \left( k-1\right) }}\left( \frac{h_{k}^{\ast }}{h_{k}}-\frac{%
h_{k}^{\ast }}{h_{k}^{\dag }}\right) \left( \eta _{k}^{\dag }-\eta
_{k}^{g}\right) \right\} \\
&=&\sum_{k=1}^{K}E_{gh}\left[ \frac{\pi ^{\ast \left( k-1\right) }}{\pi
^{\dag \left( k-1\right) }}\left( \frac{h_{k}^{\ast }}{h_{k}}-\frac{%
h_{k}^{\ast }}{h_{k}^{\dag }}\right) \sum_{j=k}^{K}\left\{ \frac{\pi
_{k+1}^{\ast j}}{\pi _{k+1}^{j}}\delta _{j}\left( \eta _{j},\eta
_{j+1};g_{j}\right) \right\} \right] \\
&=&\sum_{j=1}^{K}E_{gh}\left[ \delta _{j}\left( \eta _{j},\eta
_{j+1};g_{j}\right) \sum_{k=1}^{j}\left\{ \frac{\pi ^{\ast \left( k-1\right)
}}{\pi ^{\dag \left( k-1\right) }}\left( \frac{h_{k}^{\ast }}{h_{k}}-\frac{%
h_{k}^{\ast }}{h_{k}^{\dag }}\right) \frac{\pi _{k+1}^{\ast j}}{\pi
_{k+1}^{j}}\right\} \right] .
\end{eqnarray*}

The result $c^{p}\left( h^{\dag },\eta ^{\dag }\right) =a^{p}\left( h^{\dag
},\eta ^{\dag }\right) $ is then proved if we show that 
\begin{equation}
\sum_{k=1}^{j}\left\{ \frac{\pi ^{\ast \left( k-1\right) }}{\pi ^{\dag
\left( k-1\right) }}\left( \frac{h_{k}^{\ast }}{h_{k}}-\frac{h_{k}^{\ast }}{%
h_{k}^{\dag }}\right) \frac{\pi _{k+1}^{\ast j}}{\pi _{k+1}^{j}}\right\} =%
\frac{\pi ^{\ast j}}{\pi ^{j}}-\frac{\pi ^{\ast j}}{\pi ^{\dag j}}.
\label{algebra2}
\end{equation}

Now, 
\begin{equation*}
\sum_{k=1}^{j}\left\{ \frac{\pi ^{\ast \left( k-1\right) }}{\pi ^{\dag
\left( k-1\right) }}\left( \frac{h_{k}^{\ast }}{h_{k}}-\frac{h_{k}^{\ast }}{%
h_{k}^{\dag }}\right) \frac{\pi _{k+1}^{\ast j}}{\pi _{k+1}^{j}}\right\}
=\pi ^{\ast j}\sum_{k=1}^{j}\left\{ \frac{1}{\pi ^{\dag \left( k-1\right) }}%
\left( \frac{1}{h_{k}}-\frac{1}{h_{k}^{\dag }}\right) \frac{1}{\pi _{k+1}^{j}%
}\right\} ,
\end{equation*}%
so $\left( \ref{algebra2}\right) $ follows from $\left( \ref{algebra}\right) 
$ by evaluating in $\left( \ref{algebra}\right) $ $j$ at 0 and $s$ at $j+1.$
This concludes the proof.\bigskip

\subsection{Proof of Theorem 1}

The proof of Theorem 1 invokes the following Lemma.

\textbf{Lemma A.1.}

For any $j\in \left[ K\right] ,$%
\begin{eqnarray}
&&\left. E_{\underline{g}_{j},\underline{h}_{j}}\left\{ \left. Q_{j}\left( 
\overline{h}_{j}^{\dag K},\overline{\eta }_{j}^{\dag K}\right) \right\vert 
\overline{A}_{j-1},\overline{L}_{j}\right\} -E_{h_{j}}\left( \left. \frac{%
h_{j}^{\ast }}{h_{j}}\eta _{j}^{g}\right\vert A_{j-1},\overline{L}%
_{j}\right) =\right.  \label{bias-new} \\
&=&\sum_{k=j}^{K}E_{\underline{g}_{j},\underline{h}_{j}}\left\{ \left. \frac{%
\pi _{j}^{\ast \left( k-1\right) }}{\pi _{j}^{\dag \left( k-1\right) }}%
\left( \eta _{k}^{\dag }-\eta _{k}^{g}\right) \left( \frac{h_{k}^{\ast }}{%
h_{k}}-\frac{h_{k}^{\ast }}{h_{k}^{\dag }}\right) \right\vert \overline{A}%
_{j-1},\overline{L}_{j}\right\} .  \notag
\end{eqnarray}

\textbf{Proof of Lemma A.1.}

We prove it by reverse induction.

For $j=K\,\ $we have%
\begin{eqnarray*}
&&\left. E_{\underline{g}_{K},\underline{h}_{K}}\left\{ \left. Q_{K}\left( 
\overline{h}_{K}^{\dag K},\overline{\eta }_{K}^{\dag K}\right) \right\vert 
\overline{A}_{K-1},\overline{L}_{K}\right\} -E_{h_{K}}\left( \left. \frac{%
h_{K}^{\ast }}{h_{K}}\eta _{K}^{g}\right\vert A_{K-1},\overline{L}%
_{K}\right) =\right. \\
&=&E_{\underline{g}_{K},\underline{h}_{K}}\left\{ \left. \frac{h_{K}^{\ast }%
}{h_{K}^{\dag }}\left( \psi \left( \overline{L}_{K+1}\right) -\eta
_{K}^{\dag }\right) +y_{K,\eta _{K}^{\dag }}\left( \overline{A}_{K-1},%
\overline{L}_{K}\right) \right\vert \overline{A}_{K-1},\overline{L}%
_{K}\right\} -E_{h_{K}}\left( \left. \frac{h_{K}^{\ast }}{h_{K}}\eta
_{K}^{g}\right\vert A_{K-1},\overline{L}_{K}\right) \\
&=&E_{\underline{g}_{K},\underline{h}_{K}}\left\{ \left. \frac{h_{K}^{\ast }%
}{h_{K}^{\dag }}\left( \psi \left( \overline{L}_{K+1}\right) -\eta
_{K}^{\dag }\right) \right\vert \overline{A}_{K-1},\overline{L}_{K}\right\}
+E_{h_{K}}\left( \left. \frac{h_{K}^{\ast }}{h_{K}}\eta _{K}^{\dag
}\right\vert A_{K-1},\overline{L}_{K}\right) \\
&&-E_{h_{K}}\left( \left. \frac{h_{K}^{\ast }}{h_{K}}\eta
_{K}^{g}\right\vert A_{K-1},\overline{L}_{K}\right) \\
&=&E_{\underline{g}_{K},\underline{h}_{K}}\left[ \left. \frac{h_{K}^{\ast }}{%
h_{K}^{\dag }}\left\{ E_{g_{K}}\left\{ \psi \left( \overline{L}_{K+1}\right)
|\overline{A}_{K},\overline{L}_{K}\right\} -\eta _{K}^{\dag }\right\}
\right\vert \overline{A}_{K-1},\overline{L}_{K}\right] \\
&&+E_{h_{K}}\left( \left. \frac{h_{K}^{\ast }}{h_{K}}\eta _{K}^{\dag
}\right\vert A_{K-1},\overline{L}_{K}\right) -E_{h_{K}}\left( \left. \frac{%
h_{K}^{\ast }}{h_{K}}\eta _{K}^{g}\right\vert A_{K-1},\overline{L}_{K}\right)
\\
&=&E_{\underline{g}_{K},\underline{h}_{K}}\left\{ \left. \frac{h_{K}^{\ast }%
}{h_{K}^{\dag }}\left( \eta _{K}^{g}-\eta _{K}^{\dag }\right) \right\vert 
\overline{A}_{K-1},\overline{L}_{K}\right\} +E_{h_{K}}\left\{ \left. \frac{%
h_{K}^{\ast }}{h_{K}}\left( \eta _{K}^{\dag }-\eta _{K}^{g}\right)
\right\vert A_{K-1},\overline{L}_{K}\right\} \\
&=&E_{\underline{g}_{K},\underline{h}_{K}}\left\{ \left. \left( \frac{%
h_{K}^{\ast }}{h_{K}}-\frac{h_{K}^{\ast }}{h_{K}^{\dag }}\right) \left( \eta
_{K}^{\dag }-\eta _{K}^{g}\right) \right\vert A_{K-1},\overline{L}%
_{K}\right\}
\end{eqnarray*}

Suppose now that $\left( \ref{bias-new}\right) $ holds for a given $j\in %
\left[ K\right] ,$ we want to show that it also holds for $j-1.$ Now, 
\begin{eqnarray*}
&&\left. E_{\underline{g}_{j},\underline{h}_{j}}\left\{ \left. Q_{j}\left( 
\overline{h}_{j}^{\dag K},\overline{\eta }_{j}^{\dag K}\right) \right\vert 
\overline{A}_{j-1},\overline{L}_{j}\right\} -E_{h_{j}}\left( \left. \frac{%
h_{j}^{\ast }}{h_{j}}\eta _{j}^{g}\right\vert A_{j-1},\overline{L}%
_{j}\right) =\right. \\
&=&E_{\underline{g}_{j},\underline{h}_{j}}\left[ \left. \frac{h_{j}^{\ast }}{%
h_{j}^{\dag }}\left\{ Q_{j+1}\left( \overline{h}_{j+1}^{\dag K},\overline{%
\eta }_{j+1}^{\dag K}\right) -\eta _{j}^{\dag }\right\} +y_{j,\eta
_{j}^{\dag }}\left( \overline{A}_{j-1},\overline{L}_{j}\right) \right\vert 
\overline{A}_{j-1},\overline{L}_{j}\right] -E_{h_{j}}\left( \left. \frac{%
h_{j}^{\ast }}{h_{j}}\eta _{j}^{g}\right\vert A_{j-1},\overline{L}_{j}\right)
\\
&=&E_{\underline{g}_{j},\underline{h}_{j}}\left[ \left. \frac{h_{j}^{\ast }}{%
h_{j}^{\dag }}E_{\underline{g}_{j+1},\underline{h}_{j+1}}\left\{ \left.
Q_{j+1}\left( \overline{h}_{j+1}^{\dag K},\overline{\eta }_{j+1}^{\dag
K}\right) \right\vert \overline{A}_{j},\overline{L}_{j+1}\right\}
\right\vert \overline{A}_{j-1},\overline{L}_{j}\right] \\
&&-E_{h_{j}}\left( \left. \frac{h_{j}^{\ast }}{h_{j}^{\dag }}\eta _{j}^{\dag
}\right\vert \overline{A}_{j-1},\overline{L}_{j}\right) +y_{j,\eta
_{j}^{\dag }}\left( \overline{A}_{j-1},\overline{L}_{j}\right)
-E_{h_{j}}\left( \left. \frac{h_{j}^{\ast }}{h_{j}}\eta _{j}^{g}\right\vert
A_{j-1},\overline{L}_{j}\right) \\
&=&E_{\underline{g}_{j},\underline{h}_{j}}\left[ \left. \frac{h_{j}^{\ast }}{%
h_{j}^{\dag }}\left[ E_{\underline{g}_{j+1},\underline{h}_{j+1}}\left\{
\left. Q_{j+1}\left( \overline{h}_{j+1}^{\dag K},\overline{\eta }%
_{j+1}^{\dag K}\right) \right\vert \overline{A}_{j},\overline{L}%
_{j+1}\right\} -E_{h_{j+1}}\left( \left. \frac{h_{j+1}^{\ast }}{h_{j+1}}\eta
_{j+1}^{g}\right\vert A_{j},\overline{L}_{j+1}\right) \right] \right\vert 
\overline{A}_{j-1},\overline{L}_{j}\right] \\
&&+E_{\underline{g}_{j},\underline{h}_{j}}\left\{ \left. \frac{h_{j}^{\ast }%
}{h_{j}^{\dag }}E_{h_{j+1}}\left( \left. \frac{h_{j+1}^{\ast }}{h_{j+1}}\eta
_{j+1}^{g}\right\vert A_{j},\overline{L}_{j+1}\right) \right\vert \overline{A%
}_{j-1},\overline{L}_{j}\right\} \\
&&-E_{h_{j}}\left( \left. \frac{h_{j}^{\ast }}{h_{j}^{\dag }}\eta _{j}^{\dag
}\right\vert \overline{A}_{j-1},\overline{L}_{j}\right) +E_{h_{j}}\left(
\left. \frac{h_{j}^{\ast }}{h_{j}}\eta _{j}^{\dag }\right\vert \overline{A}%
_{j-1},\overline{L}_{j}\right) -E_{h_{j}}\left( \left. \frac{h_{j}^{\ast }}{%
h_{j}}\eta _{j}^{g}\right\vert A_{j-1},\overline{L}_{j}\right) \\
&=&E_{\underline{g}_{j},\underline{h}_{j}}\left[ \left. \frac{h_{j}^{\ast }}{%
h_{j}^{\dag }}\left[ \sum_{k=j+1}^{K}E_{\underline{g}_{j+1},\underline{h}%
_{j+1}}\left\{ \left. \frac{\pi _{j+1}^{\ast \left( k-1\right) }}{\pi
_{j+1}^{\dag \left( k-1\right) }}\left( \eta _{k}^{\dag }-\eta
_{k}^{g}\right) \left( \frac{h_{k}^{\ast }}{h_{k}}-\frac{h_{k}^{\ast }}{%
h_{k}^{\dag }}\right) \right\vert \overline{A}_{j},\overline{L}%
_{j+1}\right\} \right] \right\vert \overline{A}_{j-1},\overline{L}_{j}\right]
\\
&&+E_{\underline{g}_{j},\underline{h}_{j}}\left\{ \left. \frac{h_{j}^{\ast }%
}{h_{j}^{\dag }}E_{h_{j+1}}\left( \left. \frac{h_{j+1}^{\ast }}{h_{j+1}}\eta
_{j+1}^{g}\right\vert A_{j},\overline{L}_{j+1}\right) \right\vert \overline{A%
}_{j-1},\overline{L}_{j}\right\} \\
&&-E_{h_{j}}\left( \left. \frac{h_{j}^{\ast }}{h_{j}^{\dag }}\eta _{j}^{\dag
}\right\vert \overline{A}_{j-1},\overline{L}_{j}\right) +E_{h_{j}}\left(
\left. \frac{h_{j}^{\ast }}{h_{j}}\eta _{j}^{\dag }\right\vert \overline{A}%
_{j-1},\overline{L}_{j}\right) -E_{h_{j}}\left( \left. \frac{h_{j}^{\ast }}{%
h_{j}}\eta _{j}^{g}\right\vert A_{j-1},\overline{L}_{j}\right) \\
&=&\sum_{k=j+1}^{K}E_{\underline{g}_{j},\underline{h}_{j}}\left\{ \left. 
\frac{\pi _{j}^{\ast \left( k-1\right) }}{\pi _{j}^{\dag \left( k-1\right) }}%
\left( \eta _{k}^{\dag }-\eta _{k}^{g}\right) \left( \frac{h_{k}^{\ast }}{%
h_{k}}-\frac{h_{k}^{\ast }}{h_{k}^{\dag }}\right) \right\vert \overline{A}%
_{j-1},\overline{L}_{j}\right\} +E_{h_{j}}\left( \left. \frac{h_{j}^{\ast }}{%
h_{j}^{\dag }}\eta _{j}^{g}\right\vert \overline{A}_{j-1},\overline{L}%
_{j}\right) \\
&&+E_{h_{j}}\left\{ \left. \left( \frac{h_{j}^{\ast }}{h_{j}}-\frac{%
h_{j}^{\ast }}{h_{j}^{\dag }}\right) \eta _{j}^{\dag }\right\vert \overline{A%
}_{j-1},\overline{L}_{j}\right\} -E_{h_{j}}\left( \left. \frac{h_{j}^{\ast }%
}{h_{j}}\eta _{j}^{g}\right\vert \overline{A}_{j-1},\overline{L}_{j}\right)
\\
&=&\sum_{k=j+1}^{K}E_{\underline{g}_{j},\underline{h}_{j}}\left\{ \left. 
\frac{\pi _{j}^{\ast \left( k-1\right) }}{\pi _{j}^{\dag \left( k-1\right) }}%
\left( \eta _{k}^{\dag }-\eta _{k}^{g}\right) \left( \frac{h_{k}^{\ast }}{%
h_{k}}-\frac{h_{k}^{\ast }}{h_{k}^{\dag }}\right) \right\vert \overline{A}%
_{j-1},\overline{L}_{j}\right\} \\
&&+E_{h_{j}}\left\{ \left. \left( \frac{h_{j}^{\ast }}{h_{j}}-\frac{%
h_{j}^{\ast }}{h_{j}^{\dag }}\right) \left( \eta _{j}^{\dag }-\eta
_{j}^{g}\right) \right\vert \overline{A}_{j-1},\overline{L}_{j}\right\} \\
&=&\sum_{k=j}^{K}E_{\underline{g}_{j},\underline{h}_{j}}\left\{ \left. \frac{%
\pi _{j}^{\ast \left( k-1\right) }}{\pi _{j}^{\dag \left( k-1\right) }}%
\left( \eta _{k}^{\dag }-\eta _{k}^{g}\right) \left( \frac{h_{k}^{\ast }}{%
h_{k}}-\frac{h_{k}^{\ast }}{h_{k}^{\dag }}\right) \right\vert \overline{A}%
_{j-1},\overline{L}_{j}\right\}
\end{eqnarray*}

This concludes the proof of the Lemma A.1.

\bigskip

\textbf{Proof of Theorem 1:}

We prove part (1) by induction. Part (2) follows immediately.

For $k=K,$ $\left( \ref{dr-theo1}\right) $ is true because $\widehat{\eta }%
_{K,DR}=\eta _{K,DR}$ since $y_{K+1,\eta _{K+1}^{g}}\left( \overline{A}_{K},%
\overline{L}_{K+1}\right) \equiv y_{K+1,\widehat{\eta }_{K+1,DR}}\left( 
\overline{A}_{j},\overline{L}_{j+1}\right) \equiv \psi \left( \overline{L}%
_{K+1}\right) .$

Suppose $\left( \ref{dr-theo1}\right) $ is true for $k=K,...,j+1.$ We will
show it is true for $k=j.$

\begin{eqnarray*}
\widehat{\eta }_{j,DR}-\eta _{j}^{g} &=&\Pi ^{j}\left[ y_{j+1,\widehat{\eta }%
_{j+1,DR}}\left( \overline{A}_{j},\overline{L}_{j+1}\right) \right] -\eta
_{j}^{g} \\
&=&\left( \eta _{j,DR}-\eta _{j}^{g}\right) +\Pi ^{j}\left[ y_{j+1,\widehat{%
\eta }_{j+1,DR}}\left( \overline{A}_{j},\overline{L}_{j+1}\right) \right]
-\eta _{j,DR} \\
&=&\left( \eta _{j,DR}-\eta _{j}^{g}\right) +\Pi ^{j}\left[ y_{j+1,\widehat{%
\eta }_{j+1,DR}}\left( \overline{A}_{j},\overline{L}_{j+1}\right)
-y_{j+1,\eta _{j+1,DR}^{g}}\left( \overline{A}_{j},\overline{L}_{j+1}\right) %
\right] \\
&=&\left( \eta _{j,DR}-\eta _{j}^{g}\right) +\Pi ^{j}\left[
E_{h_{j+1}}\left\{ \left. \frac{h_{j+1}^{\ast }}{h_{j+1}}\left( \widehat{%
\eta }_{j+1,DR}-\eta _{j+1}^{g}\right) \right\vert \overline{A}_{j},%
\overline{L}_{j+1}\right\} \right] \\
&=&\left( \eta _{j,DR}-\eta _{j}^{g}\right) +\Pi _{DR}^{j}\left[ \widehat{%
\eta }_{j+1,DR}-\eta _{j+1}^{g}\right] \\
&=&\left( \eta _{j,DR}-\eta _{j}^{g}\right) +\Pi _{DR}^{j}\left[ \eta
_{j+1,DR}-\eta _{j+1}^{g}+\dsum\limits_{k=j+2}^{K}\Pi _{DR,j+1,k}\left[ \eta
_{k,DR}-\eta _{k}^{g}\right] \right] \\
&=&\left( \eta _{j,DR}-\eta _{j}^{g}\right) +\Pi _{DR}^{j}\left[ \eta
_{j+1,DR}-\eta _{j+1}^{g}\right] +\dsum\limits_{k=j+2}^{K}\Pi _{DR}^{j}\left[
\Pi _{DR,j+1,k}\left[ \eta _{k,DR}-\eta _{k}^{g}\right] \right] \\
&=&\left( \eta _{j,DR}-\eta _{j}^{g}\right) +\dsum\limits_{k=j+1}^{K}\Pi
_{DR,j,k}\left[ \eta _{k,DR}-\eta _{k}^{g}\right]
\end{eqnarray*}%
The third to last equality is by the inductive hypothesis and the second to
last is by the assumed linearity of the operator $\Pi ^{j}$ which induces
linearity of the operator $\Pi _{DR}^{j}.$

This concludes the proof of part (1).

We now prove part (3) by induction in $K.$ Part (4) follows immediately.

First we show $\left( \ref{mr-theo1}\right) $ is true when $K=1.$

For $K=1,$ we have 
\begin{eqnarray*}
&&\left. E_{p}\left\{ \left. \left( \frac{h_{1}^{\ast }}{h_{1}}-\frac{%
h_{1}^{\ast }}{h_{1}^{\dagger }}\right) \left( \widetilde{\eta }_{1,MR}-\eta
_{1}^{g}\right) \right\vert L_{1}\right\} =\right. \\
&=&E_{p}\left\{ \left. \left( \frac{h_{1}^{\ast }}{h_{1}}-\frac{h_{1}^{\ast }%
}{h_{1}^{\dagger }}\right) \left( \eta _{1,MR}-\eta _{1}^{g}\right)
\right\vert L_{1}\right\} \\
&=&E_{p}\left\{ \left. \nabla _{0,1}\left( \eta _{1,MR}-\eta _{1}^{g}\right)
\right\vert L_{1}\right\} \\
&=&\dsum\limits_{k=1}^{K}E_{p}\left\{ \left. \nabla _{0,k}\left( \eta
_{k,MR}-\eta _{k}^{g}\right) \right\vert L_{1}\right\} \\
&&+\dsum\limits_{1\leq r_{1}<r_{2}<...<r_{u}\leq
K-1}\sum_{k=r_{u}+1}^{K}E_{p}\left( \left. \nabla _{0,r_{1}}\Pi
_{MR,r_{1},r_{2},...,r_{u},k}\left[ \eta _{k,MR}-\eta _{k}^{g}\right]
\right\vert L_{1}\right)
\end{eqnarray*}%
where the first equality follows because when $K=1,Q_{2}\left( \overline{h}%
_{2}^{\dag 1},\overline{\widetilde{\eta }}_{2,MR}^{1}\right) \equiv \psi
\left( \overline{L}_{2}\right) $ so $\,\ \widetilde{\eta }_{1,MR}\equiv \Pi
^{1}\left[ Q_{2}\left( \overline{h}_{2}^{\dag 1},\overline{\widetilde{\eta }}%
_{2,MR}^{1}\right) |S_{1}\right] \equiv \Pi ^{1}\left[ \psi \left( \overline{%
L}_{2}\right) |S_{1}\right] $ $\equiv \eta _{1,MR}$ and the third equality
is true because $\dsum\limits_{1\leq r_{1}<r_{2}<...<r_{u}\leq 0}\left(
\cdot \right) \equiv 0$ $.$ This proves $\left( \ref{mr-theo1}\right) $ for $%
K=1.$

Next, assume $\left( \ref{mr-theo1}\right) $ is true for $K-1,$ we will show
it is true for $K.$

If $\left( \ref{mr-theo1}\right) $ is true for $K-1,$ then it holds that 
\begin{eqnarray*}
&&\dsum\limits_{k=2}^{K}E_{p}\left\{ \left. \frac{\pi _{2}^{\ast k-1}}{%
\widehat{\pi }_{2}^{k-1}}\left( \frac{h_{k}^{\ast }}{h_{k}}-\frac{%
h_{k}^{\ast }}{h_{k}^{\dagger }}\right) \left( \widetilde{\eta }_{k,MR}-\eta
_{k}^{g}\right) \right\vert \overline{A}_{1},\overline{L}_{2}\right\} \\
&=&\dsum\limits_{k=2}^{K}E_{p}\left\{ \left. \nabla _{1,k}\left( \widetilde{%
\eta }_{k,MR}-\eta _{k}^{g}\right) \right\vert \overline{A}_{1},\overline{L}%
_{2}\right\} \\
&=&\dsum\limits_{k=2}^{K}E_{p}\left\{ \left. \nabla _{1,k}\left( \eta
_{k,MR}-\eta _{k}^{g}\right) \right\vert \overline{A}_{1},\overline{L}%
_{2}\right\} \\
&&+\dsum\limits_{2\leq r_{1}<r_{2}<...<r_{u}\leq
K-1}\sum_{k=r_{u}+1}^{K}E_{p}\left\{ \left. \nabla _{1,r_{1}}\Pi
_{MR,r_{1},r_{2},...,r_{u},k}\left[ \eta _{k,MR}-\eta _{k}^{g}\right]
\right\vert \overline{A}_{1},\overline{L}_{2}\right\}
\end{eqnarray*}

Note that in the preceding expression we used the inductive hypothesis
pretending that our study started at cycle 2 instead of cycle 1, i.e. with $%
\left( \overline{A}_{1},\overline{L}_{2}\right) $ playing the role of $L_{1}$%
, with each $\left( L_{j},A_{j}\right) $ playing the role of $\left(
L_{j-1},A_{j-1}\right) ,$ and with $\nabla _{1,k}$ playing the role of $%
\nabla _{0,k}.$

We also have that 
\begin{eqnarray*}
\widetilde{\eta }_{1,MR}-\eta _{1}^{g} &=&\Pi ^{1}\left[ Q_{2}\left( 
\overline{h}_{2}^{\dag K},\overline{\widetilde{\eta }}_{2,MR}^{K}\right)
-E_{p}\left\{ \left. Q_{2}\left( \overline{h}_{2}^{\dag K},\overline{%
\widetilde{\eta }}_{2,MR}^{K}\right) \right\vert A_{1},\overline{L}%
_{2}\right\} \right] \\
&&+\Pi ^{1}\left[ y_{2,\eta _{2}^{g}}\left( \overline{A}_{1},\overline{L}%
_{2}\right) \right] -\eta _{1}^{g} \\
&&+\Pi ^{1}\left[ E_{p}\left[ \left. Q_{2}\left( \overline{h}_{2}^{\dag K},%
\overline{\widetilde{\eta }}_{2,MR}^{K}\right) \right\vert A_{1},\overline{L}%
_{2}\right] -y_{2,\eta _{2}^{g}}\left( \overline{A}_{1},\overline{L}%
_{2}\right) \right] \\
&=&\left( \eta _{1,MR}-\eta _{1}^{g}\right) \\
&&+\Pi ^{1}\left[ \sum_{r=2}^{K}E_{p}\left\{ \left. \frac{\pi _{2}^{\ast
\left( r-1\right) }}{\pi _{2}^{\dag \left( r-1\right) }}\left( \widetilde{%
\eta }_{r,MR}-\eta _{r}^{g}\right) \left( \frac{h_{r}^{\ast }}{h_{r}}-\frac{%
h_{r}^{\ast }}{h_{r}^{\dag }}\right) \right\vert \overline{A}_{1},\overline{L%
}_{2}\right\} \right] \\
&=&\left( \eta _{1,MR}-\eta _{1}^{g}\right) +\Pi ^{1}\left[
\sum_{r=2}^{K}E_{p}\left\{ \left. \nabla _{1,r}\left( \widetilde{\eta }%
_{r,MR}-\eta _{r}^{g}\right) \right\vert \overline{A}_{1},\overline{L}%
_{2}\right\} \right] \\
&=&\left( \eta _{1,MR}-\eta _{1}^{g}\right) +\Pi ^{1}\left[
\sum_{k=2}^{K}E_{p}\left[ \left. \nabla _{1,k}\left( \eta _{k,MR}-\eta
_{k}^{g}\right) \right\vert \overline{A}_{1},\overline{L}_{2}\right] \right]
\\
&&+\dsum\limits_{2\leq r_{1}<r_{2}<...<r_{u}\leq K-1}\sum_{k=r_{u}+1}^{K}\Pi
^{1}\left[ E_{p}\left\{ \left. \nabla _{1,r_{1}}\Pi
_{MR,r_{1},r_{2},...,r_{u},k}\left[ \eta _{k,MR}-\eta _{k}^{g}\right]
\right\vert \overline{A}_{1},\overline{L}_{2}\right\} \right] \\
&=&\left( \eta _{1,MR}-\eta _{1}^{g}\right) +\sum_{k=2}^{K}\Pi _{MR,1,k} 
\left[ \eta _{k,MR}-\eta _{k}^{g}\right] \\
&&+\dsum\limits_{2\leq r_{1}<r_{2}<...<r_{u}\leq K-1}\sum_{k=r_{u}+1}^{K}\Pi
_{MR,1,r_{1},r_{2},...,r_{u},k}\left[ \eta _{k,MR}-\eta _{k}^{g}\right]
\end{eqnarray*}

where the second equality follows after invoking Lemma A.1. So,

\begin{eqnarray*}
&&\dsum\limits_{k=1}^{K}E_{p}\left\{ \left. \frac{\pi ^{\ast k-1}}{\pi
^{\dag k-1}}\left( \frac{h_{k}^{\ast }}{h_{k}}-\frac{h_{k}^{\ast }}{%
h_{k}^{\dag }}\right) \left( \widetilde{\eta }_{k,MR}-\eta _{k}^{g}\right)
\right\vert L_{1}\right\} \\
&=&\dsum\limits_{k=2}^{K}E_{p}\left[ \left. \frac{h_{1}^{\ast }}{h_{1}^{\dag
}}E_{p}\left\{ \left. \frac{\pi _{2}^{\ast k-1}}{\pi _{2}^{\dag k-1}}\left( 
\frac{h_{k}^{\ast }}{h_{k}}-\frac{h_{k}^{\ast }}{h_{k}^{\dag }}\right)
\left( \widetilde{\eta }_{k,MR}-\eta _{k}^{g}\right) \right\vert \overline{A}%
_{1},\overline{L}_{2}\right\} \right\vert L_{1}\right] \\
&&+E_{p}\left\{ \left. \left( \frac{h_{1}^{\ast }}{h_{1}}-\frac{h_{1}^{\ast }%
}{h_{1}^{\dag }}\right) \left( \widetilde{\eta }_{1,MR}-\eta _{1}^{g}\right)
\right\vert L_{1}\right\} \\
&=&E_{p}\left[ \left. \frac{h_{1}^{\ast }}{h_{1}^{\dag }}\left[
\dsum\limits_{k=2}^{K}E_{p}\left\{ \left. \frac{\pi _{2}^{\ast k-1}}{\pi
_{2}^{\dag k-1}}\left( \frac{h_{k}^{\ast }}{h_{k}}-\frac{h_{k}^{\ast }}{%
h_{k}^{\dag }}\right) \left( \widetilde{\eta }_{k,MR}-\eta _{k}^{g}\right)
\right\vert \overline{A}_{1},\overline{L}_{2}\right\} \right] \right\vert
L_{1}\right] \\
&&+E_{p}\left\{ \left. \left( \frac{h_{1}^{\ast }}{h_{1}}-\frac{h_{1}^{\ast }%
}{h_{1}^{\dag }}\right) \left( \widetilde{\eta }_{1,MR}-\eta _{1}^{g}\right)
\right\vert L_{1}\right\} \\
&=&E_{p}\left[ \left. \frac{h_{1}^{\ast }}{h_{1}^{\dag }}\dsum%
\limits_{k=2}^{K}E_{p}\left\{ \left. \nabla _{1,k}\left( \eta _{k,MR}-\eta
_{k}^{g}\right) \right\vert \overline{A}_{1},\overline{L}_{2}\right\}
\right\vert L_{1}\right] \\
&&+E_{p}\left[ \left. \frac{h_{1}^{\ast }}{h_{1}^{\dag }}\left[
\dsum\limits_{2\leq r_{1}<r_{2}<...<r_{u}\leq
K-1}\sum_{k=r_{u}+1}^{K}E_{p}\left\{ \left. \nabla _{1,r_{1}}\Pi
_{MR,r_{1},r_{2},...,r_{u},k}\left[ \eta _{k,MR}-\eta _{k}^{g}\right]
\right\vert \overline{A}_{1},\overline{L}_{2}\right\} \right] \right\vert
L_{1}\right] \\
&&+E_{p}\left\{ \left. \left( \frac{h_{1}^{\ast }}{h_{1}}-\frac{h_{1}^{\ast }%
}{h_{1}^{\dag }}\right) \left( \eta _{1,MR}-\eta _{1}^{g}\right) \right\vert
L_{1}\right\} \\
&&+E_{p}\left\{ \left. \left( \frac{h_{1}^{\ast }}{h_{1}}-\frac{h_{1}^{\ast }%
}{h_{1}^{\dag }}\right) \sum_{k=2}^{K}\Pi _{MR,1,k}\left[ \eta _{k,MR}-\eta
_{k}^{g}\right] \right\vert L_{1}\right\} \\
&&+E_{p}\left\{ \left. \left( \frac{h_{1}^{\ast }}{h_{1}}-\frac{h_{1}^{\ast }%
}{h_{1}^{\dag }}\right) \dsum\limits_{2\leq r_{1}<r_{2}<...<r_{u}\leq
K-1}\sum_{k=r_{u}+1}^{K}\Pi _{MR,1,r_{1},r_{2},...,r_{u},k}\left[ \eta
_{k,MR}-\eta _{k}^{g}\right] \right\vert L_{1}\right\} \\
&=&\dsum\limits_{k=1}^{K}E_{p}\left\{ \left. \nabla _{0,k}\left( \eta
_{k,MR}-\eta _{k}^{g}\right) \right\vert L_{1}\right\} \\
&&+\dsum\limits_{2\leq r_{1}<r_{2}<...<r_{u}\leq
K-1}\sum_{k=r_{u}+1}^{K}E_{p}\left( \left. \nabla _{0,r_{1}}\Pi
_{MR,r_{1},r_{2},...,r_{u},k}\left[ \eta _{k,MR}-\eta _{k}^{g}\right]
\right\vert L_{1}\right) \\
&&+\sum_{k=2}^{K}E_{p}\left\{ \left. \nabla _{0,1}\Pi _{MR,1,k}\left[ \eta
_{k,MR}-\eta _{k}^{g}\right] \right\vert L_{1}\right\} \\
&&+\dsum\limits_{2\leq r_{1}<r_{2}<...<r_{u}\leq
K-1}\sum_{k=r_{u}+1}^{K}E_{p}\left( \left. \nabla _{0,1}\Pi
_{MR,1,r_{1},r_{2},...,r_{u},k}\left[ \eta _{k,MR}-\eta _{k}^{g}\right]
\right\vert L_{1}\right) \\
&=&\dsum\limits_{k=1}^{K}E_{p}\left\{ \left. \nabla _{0,k}\left( \eta
_{k,MR}-\eta _{k}^{g}\right) \right\vert L_{1}\right\} \\
&&+\dsum\limits_{1\leq r_{1}<r_{2}<...<r_{u}\leq
K-1}\sum_{k=r_{u}+1}^{K}E_{p}\left( \left. \nabla _{0,r_{1}}\Pi
_{MR,r_{1},r_{2},...,r_{u},k}\left[ \eta _{k,MR}-\eta _{k}^{g}\right]
\right\vert L_{1}\right)
\end{eqnarray*}%
This concludes the proof of Theorem 1.

\bigskip

\bigskip

\bigskip \newpage

\bigskip

\subsection{Multi-layer cross-fitting MR machine learning algorithms}

\bigskip

In this section we describe two algorithms for multiple robust estimation of 
$\theta $ in which, not only $\theta $ but also the nuisance functions $%
h_{k}\,\ $and $\eta _{k}$ are estimated by cross-fit sample splitting, thus
avoiding the within sample dependence problem discussed after Result 1 of
section 5.2.1. The first is a two-layer cross-fit algorithm and the second
is a multi-layer cross-fit algorithm. The first is simpler to implement, as
it involves (exponentially in $K)$ fewer estimation steps, but it ignores
part of the data for estimation of each $h_{k}\,\ $and $\eta _{k}.$These
algorithms are new, having been developed in April 2017; in contrast, all
other results in the paper are from the period 2012-2014.

In order to describe the algorithms it is convenient to establish the
following notation and definitions.

Given a finite set $\mathcal{S\subseteq }\mathbb{N},$ a \textit{random
partition of size} $\mathbf{U}$ of $\mathcal{S}$ is a collection 
\begin{equation*}
\left\{ \mathcal{S}_{u}\mathcal{\subseteq S}:1\leq u\leq \mathbf{U,}\cup
_{u=1}^{\mathbf{U}}\mathcal{S}_{u}=\mathcal{S},\mathcal{S}_{u}\cap \mathcal{S%
}_{u^{\prime }}=\emptyset \text{ if }u\not=u^{\prime }\right\}
\end{equation*}%
where for each $u,$ $\mathcal{S}_{u}$ is a random subset of $\mathcal{S}$.
The random partition of size $\mathbf{U}$ is generated from the uniform
distribution of size $\mathbf{U}$ if all possible partitions of size $%
\mathbf{U}$ are equally likely. Each subset $\mathcal{S}_{u}$ of a random
partition is called \textit{a random split}, or simply a \textit{split-sample%
}, of the partition. We call the complement set 
\begin{equation*}
\mathcal{S}_{u}^{c}\equiv \mathcal{S}\backslash \mathcal{S}_{u}
\end{equation*}
the $u^{th}$-random \textit{c-split}, or simply, a $u^{th}$-\textit{%
c-split-sample. }In the sequel, the word \textit{partition} stands for 
\textit{a random partition generated from the uniform distribution of a
given size}.

For $k=0,...,K^{\ast },$ where $\,K^{\ast }$ is any non-negative integer and
positive integers $\mathbf{U}_{1},...,\mathbf{U}_{K^{\ast }},$ define the
random c-splits $\mathcal{S}_{u_{\left[ 1\right] },...,u_{\left[ k\right]
}}^{c},k=1,...,K^{\ast }$ recursively as follows.

\begin{enumerate}
\item The set $\mathcal{S}_{u_{\left[ 1\right] }}^{c}$ is the $u_{\left[ 1%
\right] }^{th}$ c-split-sample of a partition of $\left\{ 1,...,n\right\} $
of size $\mathbf{U}_{1},$ for $u_{\left[ 1\right] }=1,...,\mathbf{U}_{1}.$

\item Given $\mathcal{S}_{u_{\left[ 1\right] },...,u_{\left[ k-1\right]
}}^{c},$the set $\mathcal{S}_{u_{\left[ 1\right] },...,u_{\left[ k\right]
}}^{c}$ is the $u_{\left[ k\right] }^{th}$ c-split-sample of a partition of $%
\mathcal{S}_{u_{\left[ 1\right] },...,u_{\left[ k-1\right] }}^{c}$ of size $%
\mathbf{U}_{k},$ for $u_{\left[ k\right] }=1,...,\mathbf{U}_{k}.$ That is, 
\begin{equation*}
\mathcal{S}_{u_{\left[ 1\right] },...,u_{\left[ k\right] }}^{c}=\mathcal{S}%
_{u_{\left[ 1\right] },...,u_{\left[ k-1\right] }}^{c}\backslash \mathcal{S}%
_{u_{\left[ 1\right] },...,u_{\left[ k\right] }}
\end{equation*}%
where $\mathcal{S}_{u_{\left[ 1\right] },...,u_{\left[ k\right] }}$ is the $%
u_{\left[ k\right] }^{th}$ split-sample of a partition of $\mathcal{S}_{u_{%
\left[ 1\right] },...,u_{\left[ k-1\right] }}^{c}$ of size $\mathbf{U}_{k}.$
\end{enumerate}

Given $n$ iid copies $\left( \overline{A}_{K,i},\overline{L}_{K+1,i}\right)
,i=1,...,n,$ of $\left( \overline{A}_{K},\overline{L}_{K+1}\right) $ we call
the subsample comprised by the units with indexes $i$ in $\mathcal{S}_{u_{%
\left[ 1\right] },...,u_{\left[ k\right] }}$ the $\left( u_{\left[ 1\right]
},...,u_{\left[ k\right] }\right) ^{th}\,\ $\textit{split-sample} and the
subsample comprised by the units in $\mathcal{S}_{u_{\left[ 1\right]
},...,u_{\left[ k\right] }}^{c}$ the $\left( u_{\left[ 1\right] },...,u_{%
\left[ k\right] }\right) ^{th}$ \textit{c-split-sample}. In an abuse of
notation, split-samples are denoted by the sets of indexes associated with
the units in the sample. Thus, for instance $\mathcal{S}_{u_{\left[ 0\right]
},u_{\left[ 1\right] },...,u_{\left[ k\right] }}$ denotes a specific subset
of $\left\{ i_{j}:j=1,...,J\right\} $ of $\left\{ 1,...,n\right\} $ as well
as the subsample $\left\{ \left( \overline{A}_{K,i_{j}},\overline{L}%
_{K+1,i_{j}}\right) :j=1,...,J\right\} $ of the sample $\left\{ \left( 
\overline{A}_{K,i},\overline{L}_{K+1,i}\right) ,i=1,...,n\right\} .$

\bigskip

\textbf{MR two-layer cross-fit ALGORITHM with first layer sample split of
size }$\mathbf{U}$

Compute the split samples $\mathcal{S}_{u_{\left[ 1\right] }}$ and $\mathcal{%
S}_{u_{\left[ 1\right] },u_{\left[ 2\right] }},1\leq u_{\left[ 1\right]
}\leq \mathbf{U}$ and $1\leq u_{\left[ 2\right] }\leq K,$ corresponding to
random splits of sizes $\mathbf{U}$ and $K$ respectively$\mathbf{.}$

\begin{description}
\item Let $\widetilde{Q}_{K+1,mach}\equiv \psi \left( \overline{L}%
_{K+1}\right) .$ For $u_{\left[ 1\right] }=1,2,...,\mathbf{U,}$

\item \{

\begin{description}
\item for $k=K,K-1,...,1\mathbf{,}$

\item \{

\begin{description}
\item[i)] using the units in the $\left( u_{\left[ 1\right] },u_{\left[ 2%
\right] }=k\right) ^{th}$ split-sample $\mathcal{S}_{u_{\left[ 1\right] },u_{%
\left[ 2\right] }=k}$ compute $\widehat{h}_{k},$ the output from a preferred
machine learning algorithm for estimating $h_{k}.$ For $r\in \left\{
k,k+1,...,K\right\} ,$ define $\widehat{\pi }_{k}^{r}\equiv
\prod\limits_{j=k}^{r}\widehat{h}_{j}.\,\ $Also, for units in the $\left( u_{%
\left[ 1\right] },u_{\left[ 2\right] }=k\right) ^{th}$ split-sample $%
\mathcal{S}_{u_{\left[ 1\right] },u_{\left[ 2\right] }=k}$ that have $\pi
^{\ast k}>0$ compute $\widetilde{\eta }_{k}\left( \cdot ,\cdot \right) ,$
the output of a preferred machine learning algorithm for estimating $E\left(
\left. \widetilde{Q}_{k+1,mach}\right\vert \overline{A}_{k},\overline{L}%
_{k}\right) .$

\item[ii)] for each unit in the $\left( u_{\left[ 1\right] },u_{\left[ 2%
\right] }=k-1\right) ^{th}$ split-sample $\mathcal{S}_{u_{\left[ 1\right]
},u_{\left[ 2\right] }=k-1}$ that has $\pi ^{\ast k-1}>0,$ compute 
\begin{eqnarray*}
\widetilde{Y}_{k} &\equiv &y_{k,\widetilde{\eta }_{k}}\left( \overline{A}%
_{k-1},\overline{L}_{k}\right) \\
&\equiv &\int h_{k}^{\ast }\left( a_{k}|\overline{A}_{k-1},\overline{L}%
_{k}\right) \widetilde{\eta }_{k}\left( \overline{A}_{k-1},a_{k},\overline{L}%
_{k}\right) d\mu _{k}\left( a_{k}\right) .
\end{eqnarray*}%
and 
\begin{eqnarray*}
\widetilde{Q}_{k} &\equiv &Q_{k}\left( \overline{\widehat{h}}_{k}^{K},%
\overline{\widetilde{\eta }}_{k}^{K}\right) \\
&\equiv &\frac{\pi _{k}^{\ast K}}{\widehat{\pi }_{k}^{K}}\psi \left( 
\overline{L}_{K+1}\right) - \\
&&\left\{ \sum_{j=k}^{K}\frac{\pi _{k}^{\ast j}}{\widehat{\pi }_{k}^{j}}%
\widetilde{\eta }_{j,mach}\left( \overline{A}_{j},\overline{L}_{j}\right) -%
\frac{\pi _{k}^{\ast j-1}}{\widehat{\pi }_{k}^{j-1}}y_{j,\widetilde{\eta }%
_{j}}\left( \overline{A}_{j-1},\overline{L}_{j}\right) \right\} \\
&\equiv &y_{k,\widetilde{\eta }_{k,mach}}\left( \overline{A}_{k-1},\overline{%
L}_{k}\right) + \\
&&\sum_{j=k}^{K}\frac{\pi _{k}^{\ast j}}{\widehat{\pi }_{k}^{j}}\left\{
y_{j+1,\widetilde{\eta }_{j+1}}\left( \overline{A}_{j},\overline{L}%
_{j+1}\right) -\widetilde{\eta }_{j}\right\}
\end{eqnarray*}%
where $\pi _{k}^{\ast k-1}\equiv 1$ and $\widehat{\pi }_{k}^{k-1}\equiv 1$
and the $\left( u_{\left[ 1\right] },u_{\left[ 2\right] }=0\right) ^{th}$
split-sample $\mathcal{S}_{u_{\left[ 1\right] },u_{\left[ 2\right] }=0}$ is
defined to be equal to the $u_{\left[ 1\right] }^{th}$ split-sample $%
\mathcal{S}_{u_{\left[ 1\right] }}.$
\end{description}

\item \}

\item Let $\widehat{\theta }_{MR,two-layer}^{u_{\left[ 1\right] }}$ be the
average of $\widetilde{Q}_{1}$ based on units in the $u_{\left[ 1\right]
}^{th}$ split sample $\mathcal{S}_{u_{\left[ 1\right] }}.$
\end{description}

\item \}
\end{description}

\bigskip

Finally, compute 
\begin{equation*}
\widehat{\theta }_{MR,two-layer}\equiv \frac{1}{\boldsymbol{U}}\sum_{u_{%
\left[ 1\right] }=1}^{\boldsymbol{U}}\widehat{\theta }_{MR,two-layer}^{u_{%
\left[ 1\right] }}
\end{equation*}

\bigskip

\textbf{MR multi-layer cross-fit ALGORITHM with sample splits at each layer
of size }$\mathbf{U}$\textbf{\ }

For $k=1,...,K,$ recursively calculate the random split samples $\mathcal{S}%
_{u_{\left[ 1\right] },u_{\left[ 2\right] },...,u_{\left[ k\right] }},$ $%
\left( u_{\left[ 1\right] },u_{\left[ 2\right] },...,u_{\left[ k\right]
}\right) \in \left\{ 1,...,\mathbf{U}\right\} ^{k}.$

\begin{description}
\item[a)] For each $\left( u_{\left[ 1\right] },...,u_{\left[ K\right]
}\right) \in \left\{ 1,2,...,\mathbf{U}\right\} ^{K}\mathbf{,}$

\begin{description}
\item 
\begin{description}
\item[i) ] using the units in the $\left( u_{\left[ 1\right] },...,u_{\left[
K\right] }\right) ^{th}$ c-split-sample $\mathcal{S}_{u_{\left[ 1\right]
},...,u_{\left[ K\right] }}^{c},$ compute $\widehat{h}_{K}^{\left( u_{\left[
1\right] },...,u_{\left[ K\right] }\right) },$ the output from a preferred
machine learning algorithm for estimating $h_{K}.$ Define $\widehat{\pi }%
_{K}^{\left( u_{\left[ 1\right] },...,u_{\left[ K\right] }\right) K}\equiv 
\widehat{h}_{K}^{\left( u_{\left[ 1\right] },...,u_{\left[ K\right] }\right)
}.$ Also for any $k\in \left\{ 1,....,K-1\right\} ,$ compute $\widehat{h}%
_{K}^{\left( u_{\left[ 1\right] },...,u_{\left[ k\right] }\right) }\equiv 
\frac{1}{\mathbf{U}^{K-k}}\sum_{i_{k+1},i_{k+2}...,i_{K}=1}^{\mathbf{U}}%
\widehat{h}_{K}^{\left( u_{\left[ 1\right] },...u_{\left[ k\right] },u_{%
\left[ k+1\right] =i_{k+1}},...,u_{\left[ K\right] =i_{K}}\right) }.$

\item[ii)] using the units in the $\left( u_{\left[ 1\right] },...,u_{\left[
K\right] }\right) ^{th}$ c-split-sample $\mathcal{S}_{u_{\left[ 1\right]
},...,u_{\left[ K\right] }}^{c}$ that have $\pi ^{\ast K}>0,$ compute $%
\widetilde{\eta }_{K}^{\left( u_{\left[ 1\right] },...,u_{\left[ K\right]
}\right) },$ the output from a preferred machine learning algorithm for
estimating $E\left( \left. \psi \left( \overline{L}_{K+1}\right) \right\vert 
\overline{A}_{K},\overline{L}_{K}\right) .$ Also for any $k\in \left\{
1,....,K-1\right\} ,$ compute $\widetilde{\eta }_{K}^{\left( u_{\left[ 1%
\right] },...,u_{\left[ k\right] }\right) }\equiv \frac{1}{\mathbf{U}^{K-k}}%
\sum_{i_{k+1},i_{k+2}...,i_{K}=1}^{\mathbf{U}}\widetilde{\eta }_{K}^{\left(
u_{\left[ 1\right] },...u_{\left[ k\right] },u_{\left[ k+1\right]
=i_{k+1}},...,u_{\left[ K\right] =i_{K}}\right) }.$
\end{description}
\end{description}

\item[b)] For $k=K,...,1,$

\item \{

\begin{description}
\item[ ] for each $\left( u_{\left[ 1\right] },...,u_{\left[ k\right]
}\right) \in \left\{ 1,2,...,\mathbf{U}\right\} ^{k}\mathbf{,}$

\item \{

\begin{description}
\item[i)] if $k\not=$ 1, using the units in the $\left( u_{\left[ 1\right]
},...,u_{\left[ k\right] }\right) ^{th}$ split-sample $\mathcal{S}_{u_{\left[
1\right] },...,u_{\left[ k\right] }},$ compute $\widehat{h}_{k-1}^{\left( u_{%
\left[ 1\right] },...,u_{\left[ k\right] }\right) },$ the output from a
preferred machine learning algorithm for estimating $h_{k-1}.$ Also, if $k<K$
then for $\,r\in \left\{ k,k+1,...,K-1\right\} ,$ compute $\widehat{h}%
_{r}^{\left( u_{\left[ 1\right] },...,u_{\left[ k\right] }\right) }=\frac{1}{%
\mathbf{U}^{r-k+1}}\sum_{i_{k+1},i_{k+2}...,i_{r},i_{r+1}=1}^{\mathbf{U}}%
\widehat{h}_{r}^{\left( u_{\left[ 1\right] },...u_{\left[ k\right] },u_{%
\left[ k+1\right] =i_{k+1}},...,u_{\left[ r+1\right] =i_{r+1}}\right) }$and
for $k-1\leq s\leq r\leq K,$ compute $\widehat{\pi }_{s}^{\left( u_{\left[ 1%
\right] },...,u_{\left[ k\right] }\right) ,r}=\prod\limits_{j=s}^{r}\widehat{%
h}_{j}^{\left( u_{\left[ 1\right] },...,u_{\left[ k\right] }\right) }$.

\item[ii)] For each unit in the $\left( u_{\left[ 1\right] },...,u_{\left[ k%
\right] }\right) ^{th}$ split sample $\mathcal{S}_{u_{\left[ 1\right]
},...,u_{\left[ k\right] }}$ that has $\pi ^{\ast k}>0,$ compute 
\begin{eqnarray*}
\widetilde{Y}_{k}^{\left( u_{\left[ 1\right] },...,u_{\left[ k\right]
}\right) } &\equiv &y_{k,\widetilde{\eta }_{k}^{\left( u_{\left[ 1\right]
},...,u_{\left[ k\right] }\right) }}\left( \overline{A}_{k-1},\overline{L}%
_{k}\right) \\
&\equiv &\int h_{k}^{\ast }\left( a_{k}|\overline{A}_{k-1},\overline{L}%
_{k}\right) \widetilde{\eta }_{k}^{\left( u_{\left[ 1\right] },...,u_{\left[
k\right] }\right) }\left( \overline{A}_{k-1},a_{k},\overline{L}_{k}\right)
d\mu _{k}\left( a_{k}\right) .
\end{eqnarray*}%
and 
\begin{eqnarray*}
\widetilde{Q}_{k}^{\left( u_{\left[ 1\right] },...,u_{\left[ k\right]
}\right) } &\equiv &Q_{k}\left( \overline{\widehat{h}}_{k}^{\left( u_{\left[
1\right] },...,u_{\left[ k\right] }\right) ,K},\overline{\widetilde{\eta }}%
_{k}^{\left( u_{\left[ 1\right] },...,u_{\left[ k\right] }\right) ,K}\right)
\\
&\equiv &\frac{\pi _{k}^{\ast K}}{\widehat{\pi }_{k}^{\left( u_{\left[ 1%
\right] },...,u_{\left[ k\right] }\right) ,K}}\psi \left( \overline{L}%
_{K+1}\right) \\
&&-\sum_{j=k}^{K}\left\{ \frac{\pi _{k}^{\ast j}}{\widehat{\pi }_{k}^{\left(
u_{\left[ 1\right] },...,u_{\left[ k\right] }\right) ,j}}\widetilde{\eta }%
_{j}^{\left( u_{\left[ 1\right] },...,u_{\left[ k\right] }\right) }\left( 
\overline{A}_{j},\overline{L}_{j}\right) -\right. \\
&&\text{ \ \ \ \ \ \ \ \ \ \ \ }\left. \frac{\pi _{k}^{\ast j-1}}{\widehat{%
\pi }_{k}^{\left( u_{\left[ 1\right] },...,u_{\left[ k\right] }\right) ,j-1}}%
y_{j,\widetilde{\eta }_{j}^{\left( u_{\left[ 1\right] },...,u_{\left[ k%
\right] }\right) }}\left( \overline{A}_{j-1},\overline{L}_{j}\right) \right\}
\\
&\equiv &y_{k,\widetilde{\eta }_{k}^{\left( u_{\left[ 1\right] },...,u_{%
\left[ k\right] }\right) }}\left( \overline{A}_{k-1},\overline{L}_{k}\right)
\\
&&+\sum_{j=k}^{K}\frac{\pi _{k}^{\ast j}}{\widehat{\pi }_{k}^{\left( u_{%
\left[ 1\right] },...,u_{\left[ k\right] }\right) ,j}}\left\{ y_{j+1,%
\widetilde{\eta }_{j+1}^{\left( u_{\left[ 1\right] },...,u_{\left[ k\right]
}\right) }}\left( \overline{A}_{j},\overline{L}_{j+1}\right) -\widetilde{%
\eta }_{j}^{\left( u_{\left[ 1\right] },...,u_{\left[ k\right] }\right)
}\right\}
\end{eqnarray*}%
where $\pi _{k}^{\ast k-1}\equiv 1$ and $\widehat{\pi }_{k}^{\left( u_{\left[
1\right] },...,u_{\left[ k\right] }\right) ,k-1}\equiv 1.$

\item[iii)] If $k=1,$ then using data in the $u_{\left[ 1\right] }^{th}$
split sample $\mathcal{S}_{u_{\left[ 1\right] }},$ compute $\widehat{\theta }%
_{MR,multi-layer}^{u_{\left[ 1\right] }}=\mathbb{P}_{n}^{u_{\left[ 1\right]
}}\left\{ \widetilde{Q}_{1}^{\left( u_{\left[ 1\right] }\right) }\right\} $,
the average of $\widetilde{Q}_{1}^{\left( u_{\left[ 1\right] }\right) }$ in
the $u_{\left[ 1\right] }^{th}$ split sample $\mathcal{S}_{u_{\left[ 1\right]
}}$; otherwise using data in the $\left( u_{\left[ 1\right] },...,u_{\left[ k%
\right] }\right) ^{th}$ split sample $\mathcal{S}_{u_{\left[ 1\right] },u_{%
\left[ 1\right] },...,u_{\left[ k\right] }},$ compute $\widetilde{\eta }%
_{k-1}^{\left( u_{\left[ 1\right] },...,u_{\left[ k\right] }\right) },$ the
output of a preferred machine learning algorithm for estimating $E\left(
\left. \widetilde{Q}_{k}^{\left( u_{\left[ 1\right] },...,u_{\left[ k\right]
}\right) }\right\vert \overline{A}_{k-1},\overline{L}_{k-1}\right) .$ Also,
if $1<k<K,$ then for $\,r\in \left\{ k,k+1,...,K-1\right\} ,$ compute $%
\widetilde{\eta }_{r}^{\left( u_{\left[ 1\right] },...,u_{\left[ k\right]
}\right) }=\frac{1}{\mathbf{U}^{r-k+1}}%
\sum_{i_{k+1},i_{k+2}...,i_{r},i_{r+1}=1}^{\mathbf{U}}\widetilde{\eta }%
_{r}^{\left( u_{\left[ 1\right] },...u_{\left[ k\right] },u_{\left[ k+1%
\right] =i_{k+1}},...,u_{\left[ r+1\right] =i_{r+1}}\right) }.$
\end{description}

\item \}
\end{description}

\item \}
\end{description}

Finally, compute 
\begin{equation*}
\widehat{\theta }_{MR,multi-layer}=\frac{1}{\mathbf{U}}\sum_{u_{\left[ 1%
\right] }=1}^{\mathbf{U}}\widehat{\theta }_{MR,multi-layer}^{u_{\left[ 1%
\right] }}
\end{equation*}%
\newpage

\bigskip

\begin{equation*}
\text{Figure 1: Illustration of the two-layer cross-fit algorithm for }K=3%
\text{ and }\mathbf{U}=5.
\end{equation*}

\begin{figure}[ph]
\begin{center}
\includegraphics[scale=0.80]{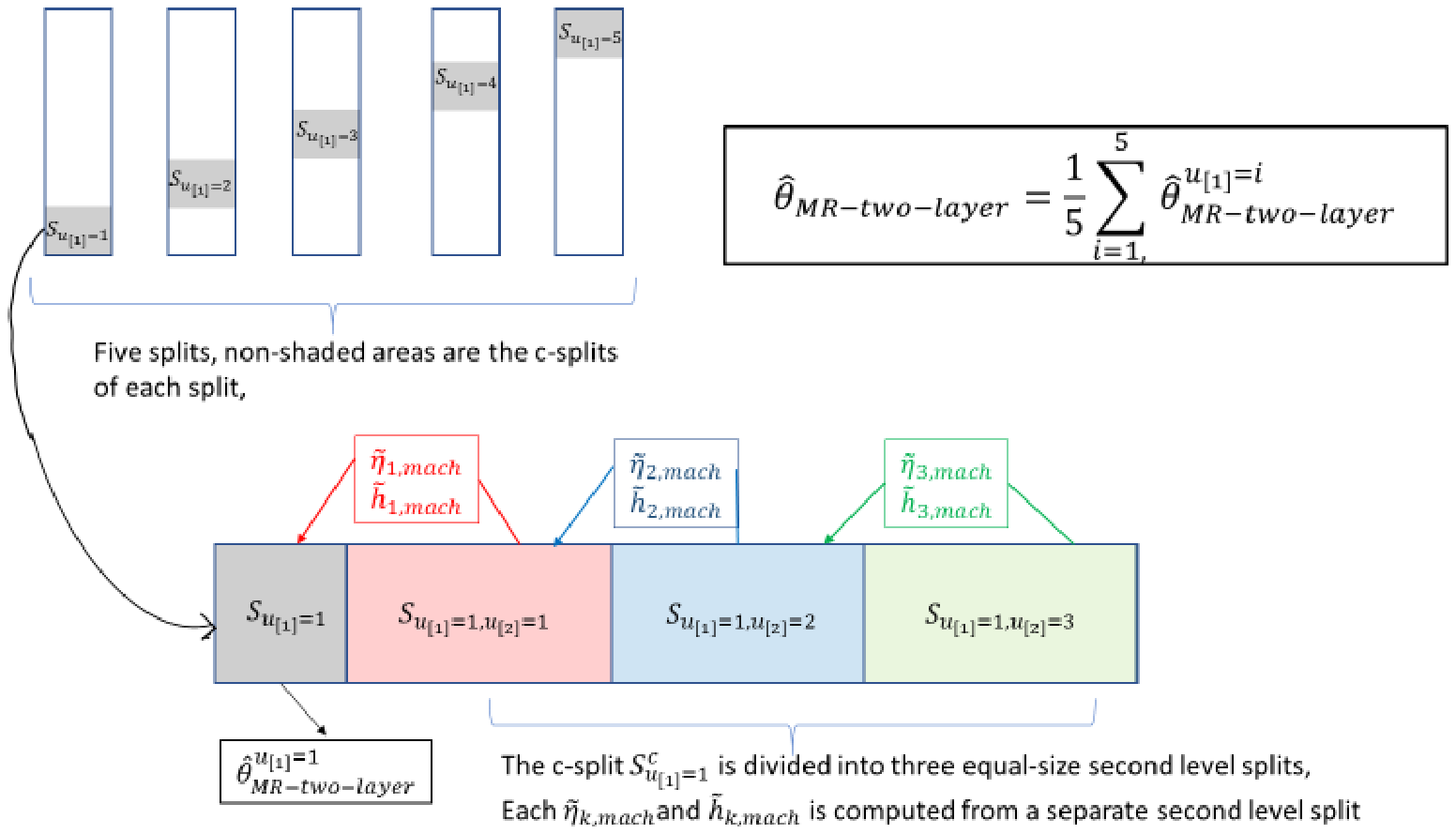} 
\end{center}
\end{figure}

\newpage

\begin{equation*}
\text{Figure 2: Illustration of multi-layer cross-fit algorithm for }K=3%
\text{ and }\mathbf{U}=5.
\end{equation*}

\begin{figure}[hp]
\begin{center}
\includegraphics[scale=0.8]{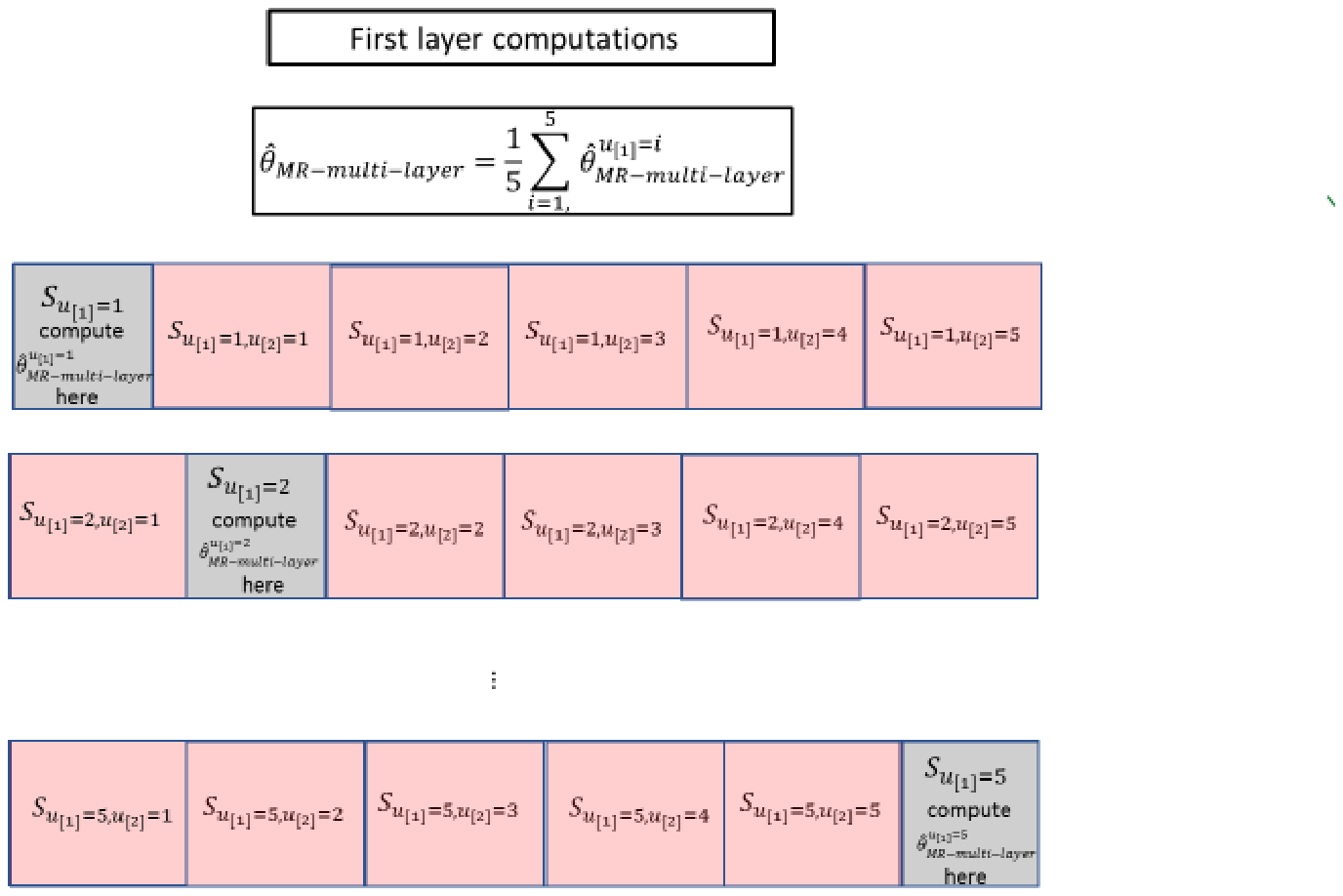} 
\end{center}
\end{figure}

\begin{figure}[hp]
\begin{center}
\includegraphics[scale=0.8]{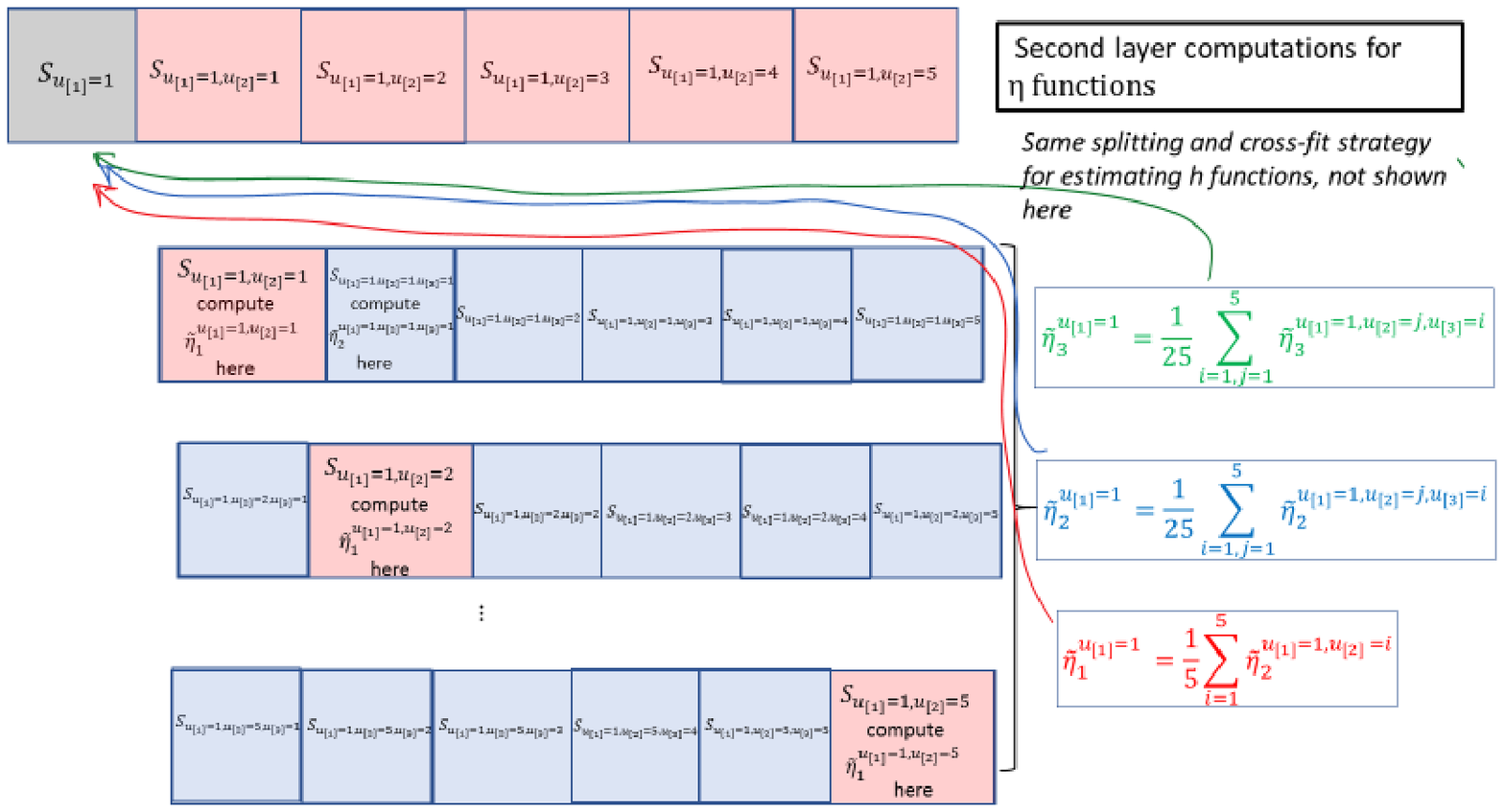} 
\end{center}
\end{figure}

\begin{figure}[tbp]
\begin{center}
\includegraphics[scale=0.8]{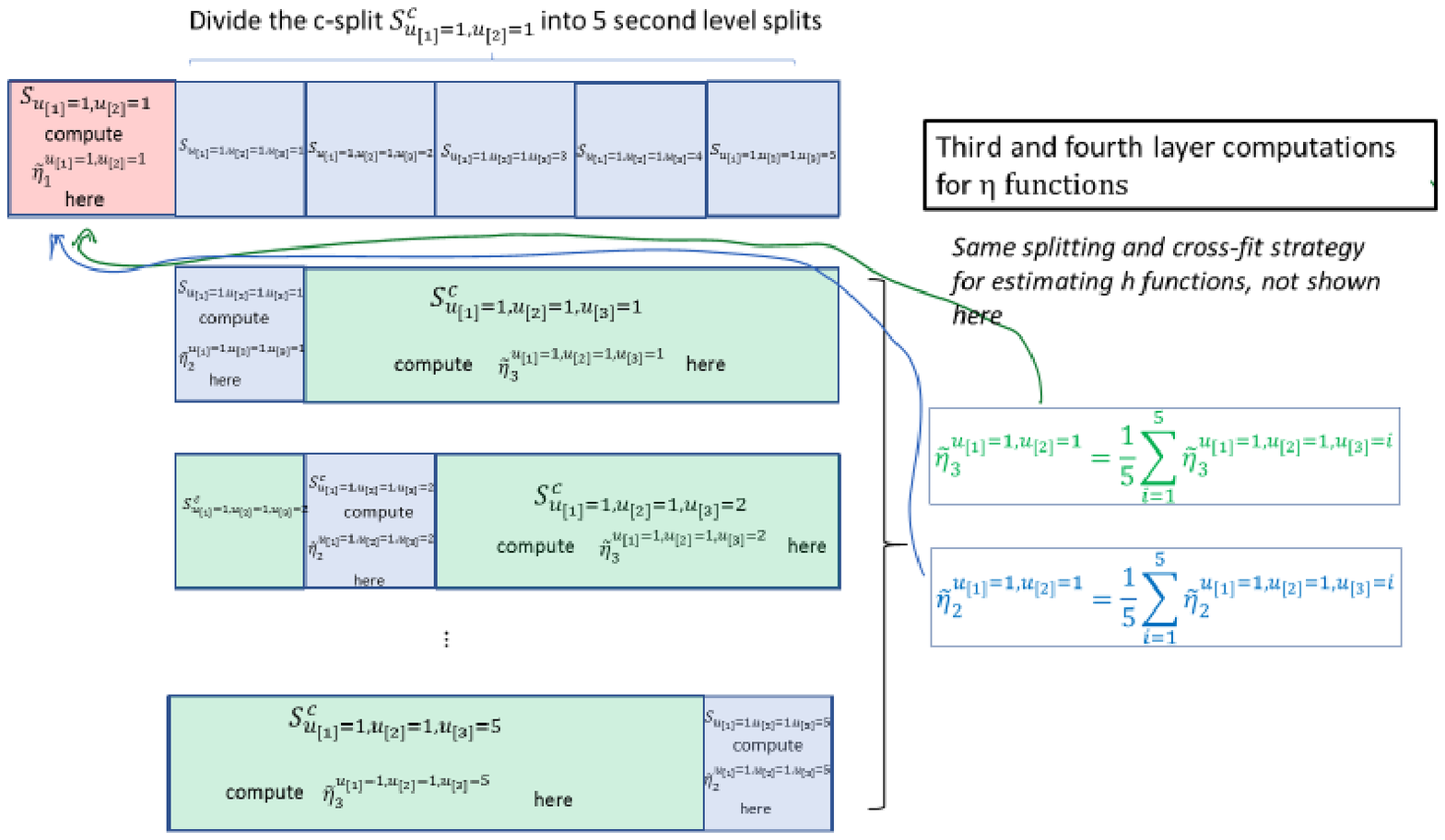}
\end{center}
\end{figure}

\end{document}